\documentclass[preprints,article,accept,pdftex,moreauthors]{Definitions/mdpi} 

\firstpage{1} 
\pubvolume{1}
\issuenum{1}
\articlenumber{0}
\doinum{}
\pubyear{2026}
\copyrightyear{2026}
\datereceived{11 February 2026} 
\daterevised{29 March 2026} 
\dateaccepted{31 March 2026} 
\datepublished{} 
\hreflink{https://doi.org/} 

\Title{A Few Years Later: Revisiting Period Variations of Eclipsing Binaries in the Northern Continuous Viewing Zone of TESS}

\Author{Tam\'as Borkovits $^{1,2,3,}$*
\orcidA{}, Tibor Mitnyan $^{2,4}$\orcidB{}, Don\'at R. Czavalinga
 $^{2,4}$\orcidC{}, Saul A. Rappaport $^{2,5}$\orcidD} 

\AuthorNames{Tam\'as Borkovits, Tibor Mitnyan, Don\'at R. Czavalinga, Saul A. Rappaport}

\address{%
$^{1}$ \quad Baja Astronomical Observatory of University of Szeged,  Szegedi \'ut, Kt. 766, H-6500 Baja, Hungary\\
$^{2}$ \quad HUN-REN---SZTE Stellar Astrophysics Research Group,  Szegedi \'ut, Kt. 766, H-6500 Baja, Hungary; mtibor@titan.physx.u-szeged.hu (T.M.); donat.czavalinga@gmail.com (D.R.C.); sar@mit.edu (S.A.R.)
\\
$^{3}$ \quad Konkoly Observatory,HUN-REN Research Centre for Astronomy and Earth Sciences,  Konkoly Thege Mikl\'os \'ut 15-17, H-1121 Budapest, Hungary \\
$^{4}$ \quad Department of Experimental Physics, University of Szeged, D\'om t\'er 9, H-6720 Szeged, Hungary \\
$^{5}$ \quad \textls[-25]{Department of Physics, Kavli Institute for Astrophysics and Space Research, M.I.T., Cambridge, MA 02139, USA}}

\corres{Correspondence: borko@bajaobs.hu}

\abstract{In our previous analysis of the eclipse timing variation patterns of eclipsing binaries located in and near the Northern Continuous Viewing Zone (NCVZ) of the TESS space telescope, 135 hierarchical triple star candidates were found. Now, two additional years of TESS observations are available and, hence, we have extended the former analysis with the use of the new observational data.  We now detect 168 triple star candidates in the updated and reanalyzed sample. The majority ($\sim74\%$) of them are identical to the former triples candidates. For many of them, our new solutions are more certain than the original ones. Therefore, we can now conclude that we have identified at least 66 short-period hierarchical triple stellar systems in the NCVZ with full confidence. In the case of the majority of the remaining systems in our sample, the presence of a close third stellar component appears to be very likely. We also identify additional, longer timescale period variations in 34 systems ($20\%$ of the total sample) and conclude that in at least three systems the presence of a fourth stellar component is quite plausible. Finally, we report the complete disappearance of the eclipses in two former EBs and detect eclipse depth variations in seven other EBs as well. We interpret this effect as the consequence of changing orbital inclination caused by a non-coplanar third body.}


\begin{document}

\section{Introduction}

Triple and multiple stars are quite common objects as, according to our current knowledge, they are natural end products of the various star formation processes (for recent reviews, see, e.g., \citet{tokovinin21} and \citet{offneretal23} and references therein). 

For dynamical reasons, mass-points (which are very good approximations to real stars, during almost their entire lifetimes), where the two most massive objects in the system have a mass ratio between $\sim$$10^{+2}$ and $10^{-2}$, are able to maintain long-term dynamical stability only by forming hierarchical configurations.  Here, we consider only hierarchical triple star systems, which are the smallest and most basic building block of any higher-order stellar hierarchies.  Here, `hierarchical' means that the separation of one of the three stars from the other two stars is continuously and always much larger than the separation of these latter two stars. The mathematical consequence is that the motion of the triple can be described approximately by two Keplerian orbits; that is, the triple can be considered as if it consisted of two binaries. One of these two, the so-called inner binary, is formed by the close pair of stars, while the wider or outer binary is formed by the center of mass of the inner pair, with the mass of the pair, and the third, more distant component. When even the pericenter distance of this third star is large enough, the two binaries can be considered to be independent of each other or, mathematically, the orbits remain nearly pure, unperturbed Keplerian motions. When the separation of the third star, at least during some segments of its motion, becomes smaller, the motion can be described as two perturbed Keplerian orbits or, in other words, the mutual gravitational effects of the inner and outer stars on their orbits can be handled with perturbation methods.

The characteristic size, that is, the length of the major axis of the outermost orbit in a hierarchical multiple star system, can span several orders of magnitudes from fractions of an au to 1--1.5 pc (the latter of which is the average half-distance between neighboring but unbound stars). While the smallest, or most compact, hierarchical systems carry information on the formation processes of the stars and also about their later-time evolutions including end stages, the widest hierarchies are mainly interesting for exploring properties of the potential field, as well as the mass distribution of the entire Galaxy.

In what follows, we concentrate on hierarchical systems having small characteristic dimensions. The observations of such systems, in contrast to the wider ones, have serious advantages and some disadvantages, too. Let us consider, for example, a moderately compact triple star system where three, let us say, solar-type stars are located within the average Sun--Jupiter distance. In other words, the semi-major axis of the outer orbit is about $a_\mathrm{out}\approx5.2$\,au. One can easily calculate, using Kepler's third law, that, in this case, the outer period would be $P_\mathrm{out}\approx\frac{1}{\sqrt{3}}P_\mathrm{Jup}\approx6.8$\,yr. Therefore, the motion in such an orbit can be well observed within a very short timescale, much shorter than a human scholar's active life and, moreover, even the quantitative properties of such an orbital motion can be determined with high accuracy. On the other hand, however, the angular separation of stars orbiting each other in such a small orbit becomes quickly very small with the system's increasing distance from the Earth and, therefore, the vast majority of such systems remain spatially unresolvable. Consequently, it is obvious that the most compact hierarchical triple and multiple systems, in general, cannot be studied in the classical way of astrometry, but different, practically indirect methods are needed.   

These methods are mainly based on either spectroscopic observations, more specifically radial velocity (RV) data, or photometric timing measurements, as they were reviewed, e.g., in Section 3 of \citet{borkovits22}. Between these two kinds of detections and investigations of compact hierarchical multiple and, especially, triple systems, the current study is based on the second method and, therefore, here we discuss only this latter case below.

When the inner binary subsystem of a compact triple (or multiple) star is serendipitously an eclipsing binary (EB), the time intervals between consecutive eclipses, for different reasons, no longer remain constant, but exhibit different, but characteristic, patterns from which, in theory, not only the presence but even several orbital (and/or dynamical) parameters of the third star can be inferred. Unfortunately, however, we must say, `in theory', because there are several other effects which may cause varying time intervals between the eclipses, sometimes either mimicking or even overlapping and hiding such effects which were caused by a third body.

In what follows, we briefly summarize a powerful tool which can be used to investigate the variations of times between consecutive eclipsing events (or, eclipse timing variations---ETVs) and which, in general, is called (the analysis of the) `observed minus calculated' (i.e., O-C) curve or diagram. Then, we also show what kinds of ETVs may be present in different types of EBs in addition to the ones induced by third stars. We also give the mathematical descriptions of these different types of ETVs.

Then, in the second part of the current work, we illustrate all of these kinds of phenomena with examples taken from the several year-long and quasi-continuous, highly accurate space-borne observations of the TESS satellite. We use the sample of the EBs observed by TESS in its northern continuous viewing zone (NCVZ). As we discuss, the period variations of these NCVZ EBs were investigated in detail formerly in \citet{mitnyanetal24}, but we now illustrate how the revision of the old study, with the use of the newer TESS observations, may lead to substantial new results.


\section{Orbital Period Variations in EBs}
\label{sec:analysis}

In this work, we primarily investigate the period variations of EBs in the NCVZ. Before doing this, we should clarify what is meant by the word `period' and what kinds of periods can and will be investigated. Most EBs produce two different kinds of fadings in flux during one orbital revolution. In the case of two stars revolving around each other either on a circular orbit or one viewed exactly edge-on, the event when the star with the lower surface brightness transits in front of the disk of its companion having the higher surface brightness (i.e., hotter in temperature) is called `primary eclipse' or primary minimum (for its deeper fading).  And, vice versa, when the hotter star is in front of the disk of its cooler companion, the event is called a `secondary eclipse' (or secondary minimum). In the case of an EB in an elliptical orbit that is viewed non-edge-on, this no longer will be necessarily true, as is illustrated clearly in the different panels of Figure 7 of \citet{borkovitsmitnyan23}. In what follows, however, for simplicity, we consider the frequent case when the star occulted during the primary eclipse has the larger surface brightness (i.e., the hotter one).

Naturally, when both stars are in or near to the zero-age main sequence (ZAMS), the hotter star is also the more massive one and, therefore, in this case, the photometric primary/secondary stars are the same as the spectroscopic primary/secondary ones.

What can be observed very easily and accurately from photometric observations is the time that has elapsed between the center-points of consecutive primary (or secondary) eclipses\endnote{Note also that in the case of elliptical orbits, the center-point of each eclipse is not necessarily the same as the point of the maximal fading. But, we also omit this usually small difference.}. The time interval which can be determined in such a way is called the `eclipsing period'. Naturally, to the extent that the two stars can be considered to be in perfect two-body Keplerian motion, which is realized for an orbit which is absolutely fixed in space and, moreover, the position of the observer is also fixed in space, and the velocity of light would also be infinitely large, one can say that this eclipsing period is identical with both the anomalistic and sidereal periods of the binary star.  The latter period is sometimes simply called `the orbital period'.

Naturally, this idealistic case never occurs.  The motion of the observer and its combination with the finite speed of light can be corrected out easily by converting the observations into a system fixed to either our Sun (heliocentric correction) or to the barycenter of the solar system (barycentric correction). Some other effects, such as internal light-travel time within a short-period EB\endnote{However, in some exceptional cases this latter effect should be (and was) taken into account (see, e.g., \citet{fabrycky09}).}, are so small that, in most cases, they are safely neglected. However, there are other effects which should also be taken into consideration, as they produce well-observable variations in the time intervals between consecutive mid-eclipse times over both shorter and longer timescales. 

The observed eclipsing period may vary for three different main reasons. (1) When the anomalistic period of a binary star no longer remains constant (whatever process can be in the background), it is traditionally called a `physical' or `real' period variation. (2) The eclipsing period, however, may also change for geometric reasons, when only the location or orientation of the orbit varies relative to the observer without any change in the anomalistic period. These are the so-called `apparent' period variations. (3) Finally, the never-before-seen accuracies of recent satellite photometry make it necessary to introduce a third kind of period variation. This occurred recently in \citet{borkovitsetal25}, where such variations, at least in the nomenclature of these latter authors, are named `spurious' period variations. In this latter case, there are neither physical nor apparent period variations but some periodic distortions in the accurate photometric light curves that may lead to (usually quasi-cyclic) virtual variations in the eclipsing periods.  The exact cause of these spurious period variations is not yet understood.

Without further discussion of these three kinds of period variations, we simply refer to the short review of them given in Section 2 of \citet{borkovitsetal25}.

\subsection{The Mathematical Tool Used for the Current Analysis.}

As is well known, the investigation of the eclipsing period variations in EBs can be traced back to the variation of the time that elapses between consecutive eclipses or, in other words, the analysis of the ETVs\endnote{In the case of transiting exoplanets, the equivalent of ETV is TTV, that is, `transit timing variations'.}. Such an analysis is called either the `O-C' method or simply the `ETV' method. Formerly, before the era of the planet-hunter space telescopes, such diagrams, in which the difference of an observed mid-eclipse time and a calculated one (predicted using a strictly constant period) was plotted against the cycle number (which can be easily converted to time), was uniquely called an $O-C$ diagram. It is easy to show that using such a diagram, or curve, the actual eclipsing or orbital period of the system under investigation is equal to the slope of a continuous function, which can be fitted to the ETV (or O-C) curve at a given local point. As a consequence, the varying (eclipsing) period is equivalent to the varying (local) slope of the ETV curve and, hence, any curvature in the shape of the curve directly indicates period variations, including any of the three kinds mentioned above\endnote{Naturally, the O-C curve can be, and is actually, used for the investigation of period variations in other types of strictly periodic variables, such as, e.g., pulsating stars. In these cases, the times of the maximum brightnesses of a classic pulsational variable (e.g., RR~Lyrae-type or classic Cepheid star) are generally used for forming the O-C diagrams because the times of pulsational maxima can be determined more easily and accurately than those of the pulsational minima. In such cases, naturally, the nomenclature `ETV' diagram (or curve) should not be used.}. Details of the `O--C' (or `ETV') method can be found in the conference volume of \citet{sterken05} and, moreover, this subject was recently reviewed in \citet{rovithis-livaniou20}.

In the current work, similar to our previous related studies \citep{borkovitsetal16,borkovitsetal25,mitnyanetal24}, from a mathematical point of view, we describe the analytical form of an ETV curve as follows:
\begin{equation}
T(E)-(T_0+E\times P) = \Delta,
\end{equation}
where $T(E)$ is the measured mid-eclipse time of the $E^{th}$ eclipse (where $E$ is the cycle number, which is an integer for primary eclipses and half-integer for secondary eclipses), $T_0$ is the mid-eclipse time at the $0^{th}$ cycle (that is, the base time with which the ETV curve is generated), and $P$ is a constant, which can be considered as an `average' eclipsing period. (We discuss briefly below why this constant is called `average' period.)

\begin{figure}[H]
\begin{adjustwidth}{-\extralength}{0cm}
\centering
{\includegraphics[width=0.35\textwidth]{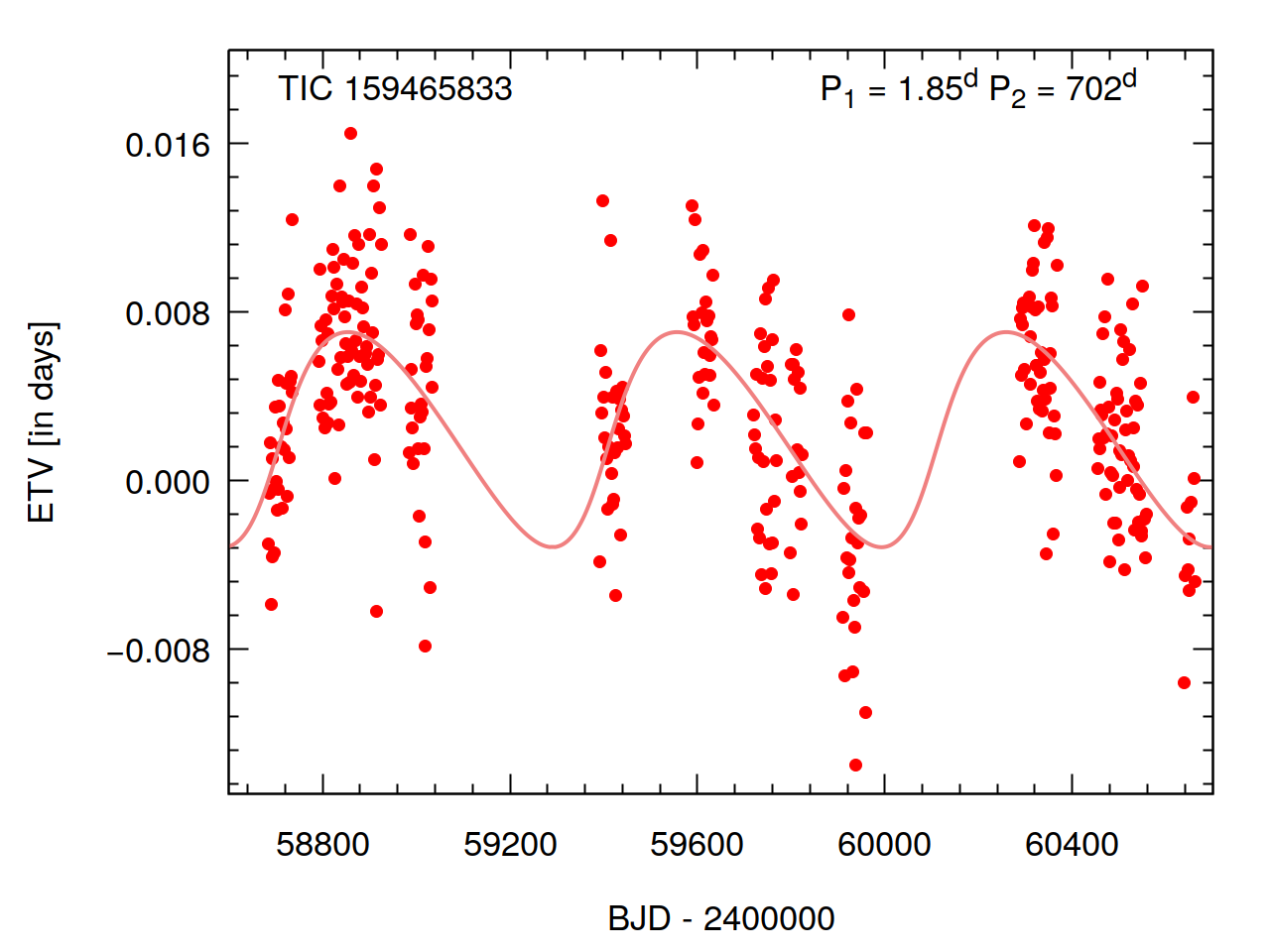}}\includegraphics[width=0.35\textwidth]{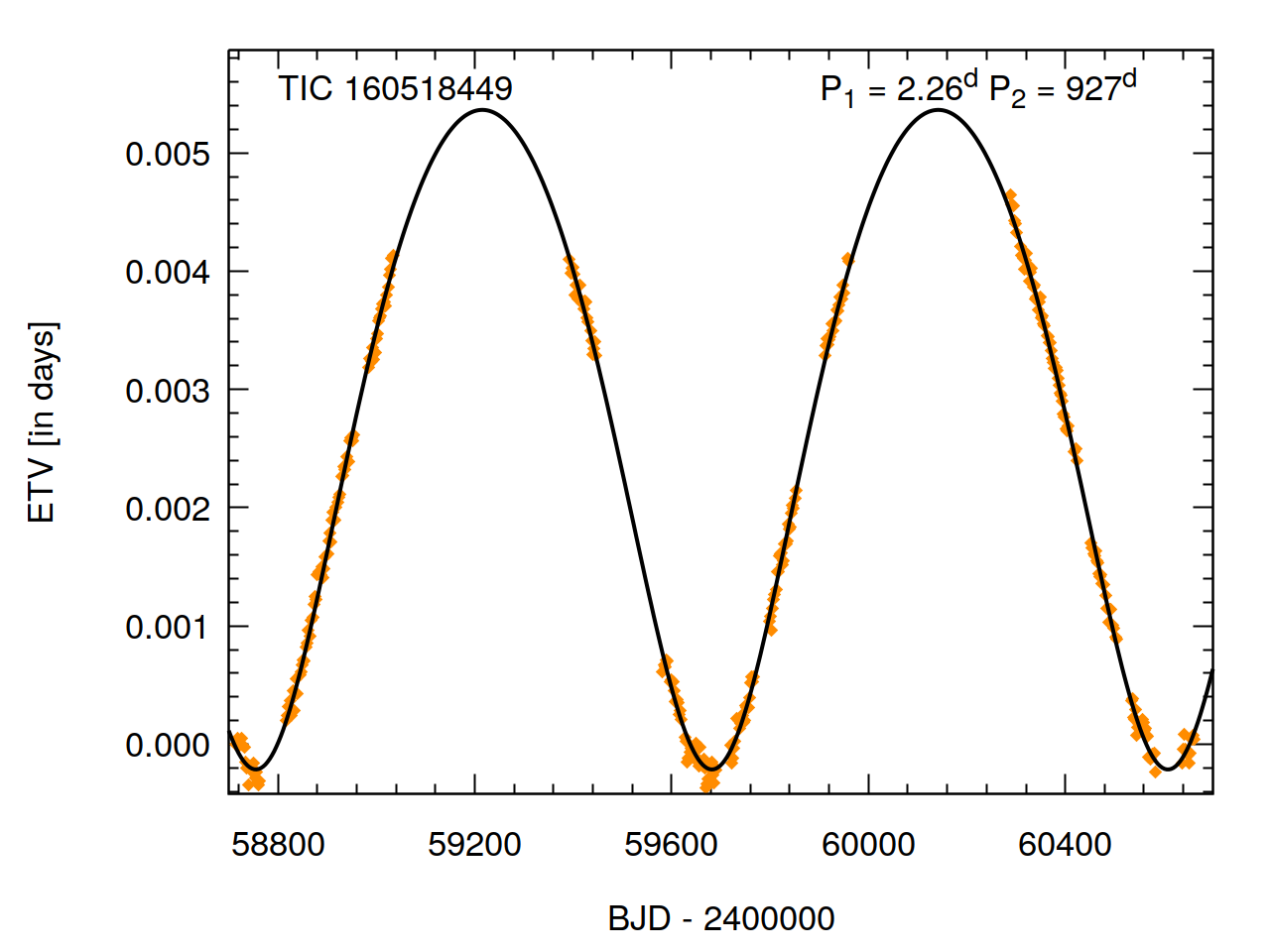}\includegraphics[width=0.35\textwidth]{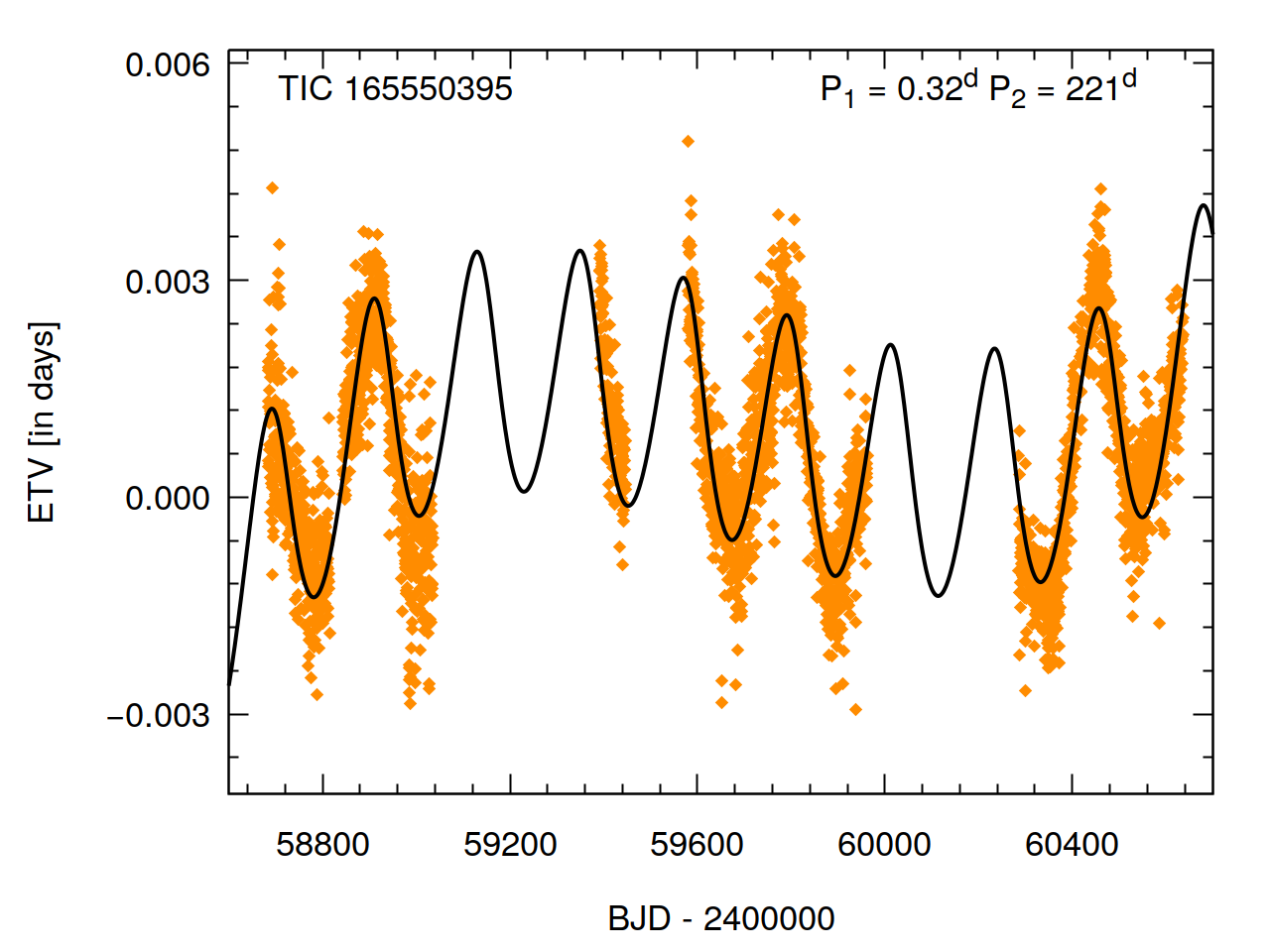}
\includegraphics[width=0.35\textwidth]{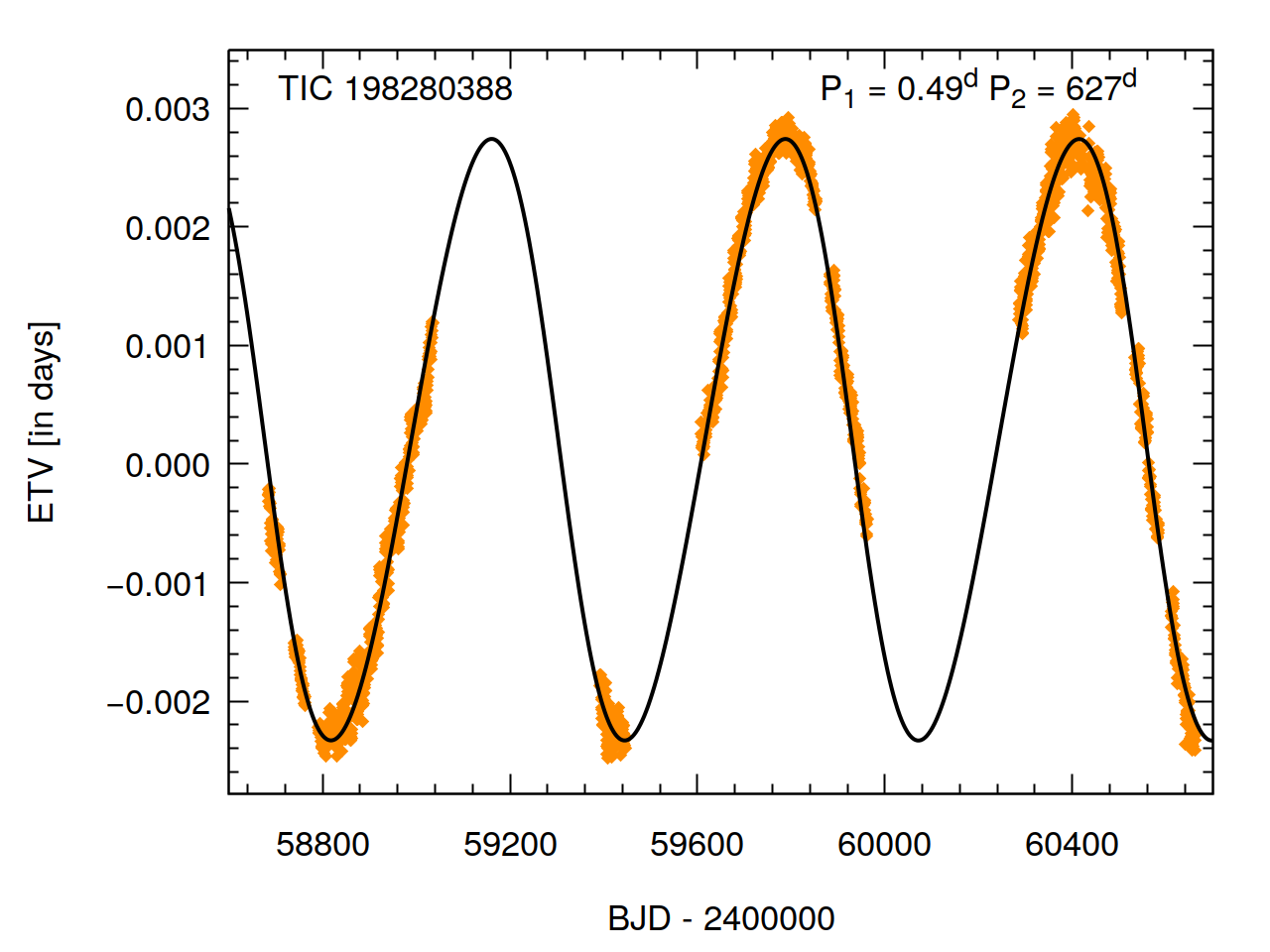}\includegraphics[width=0.35\textwidth]{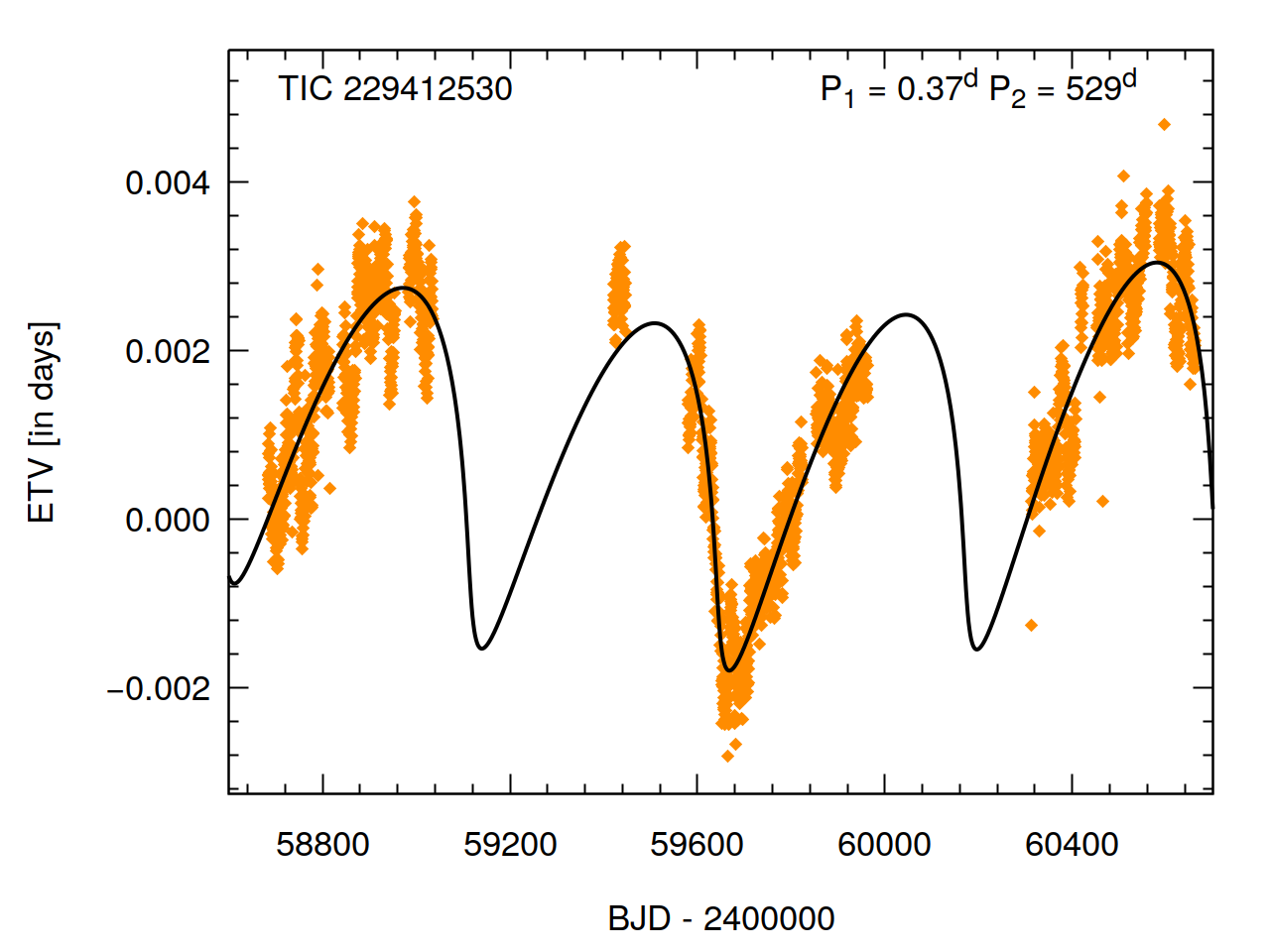}\includegraphics[width=0.35\textwidth]{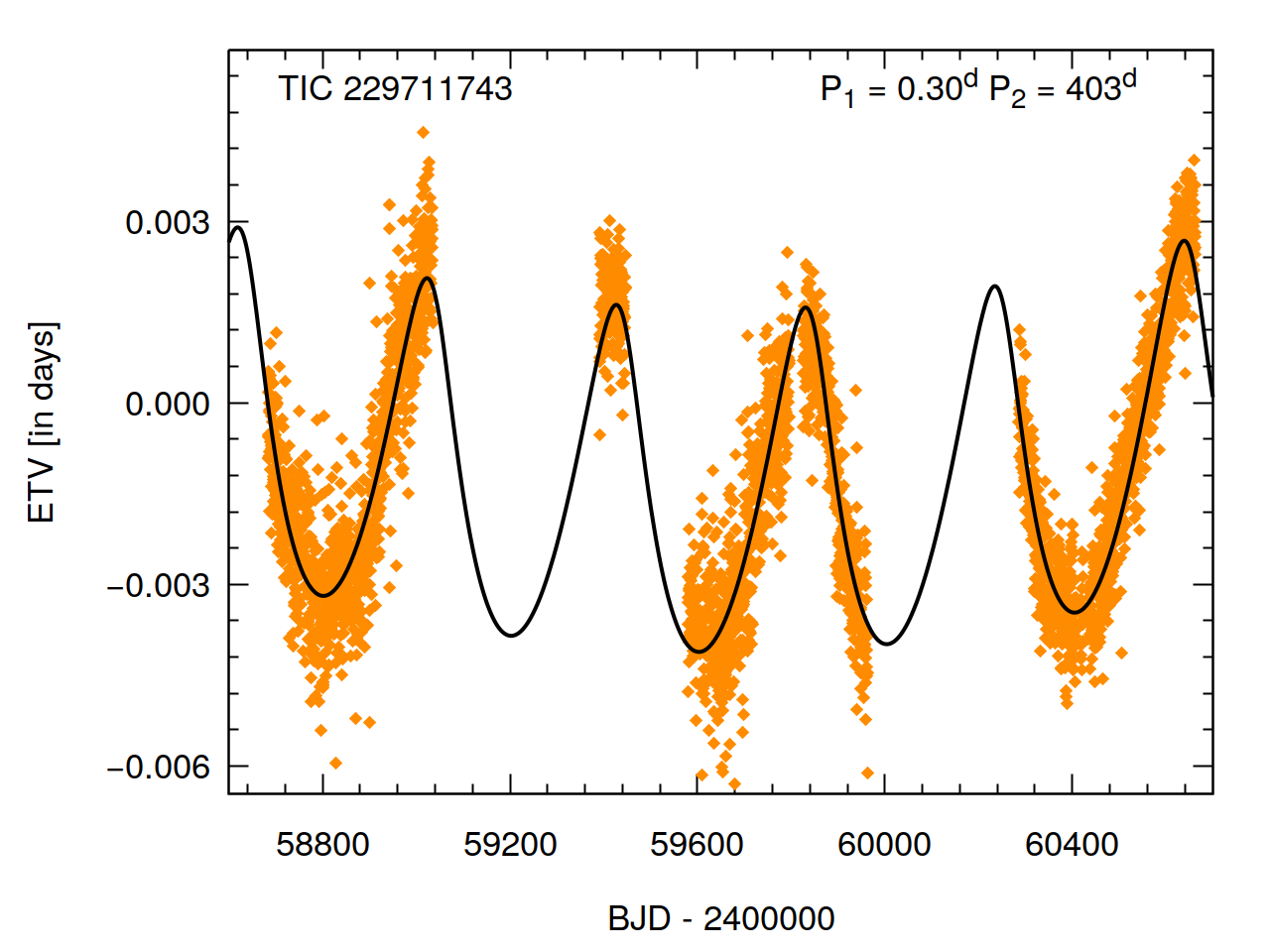}
\includegraphics[width=0.35\textwidth]{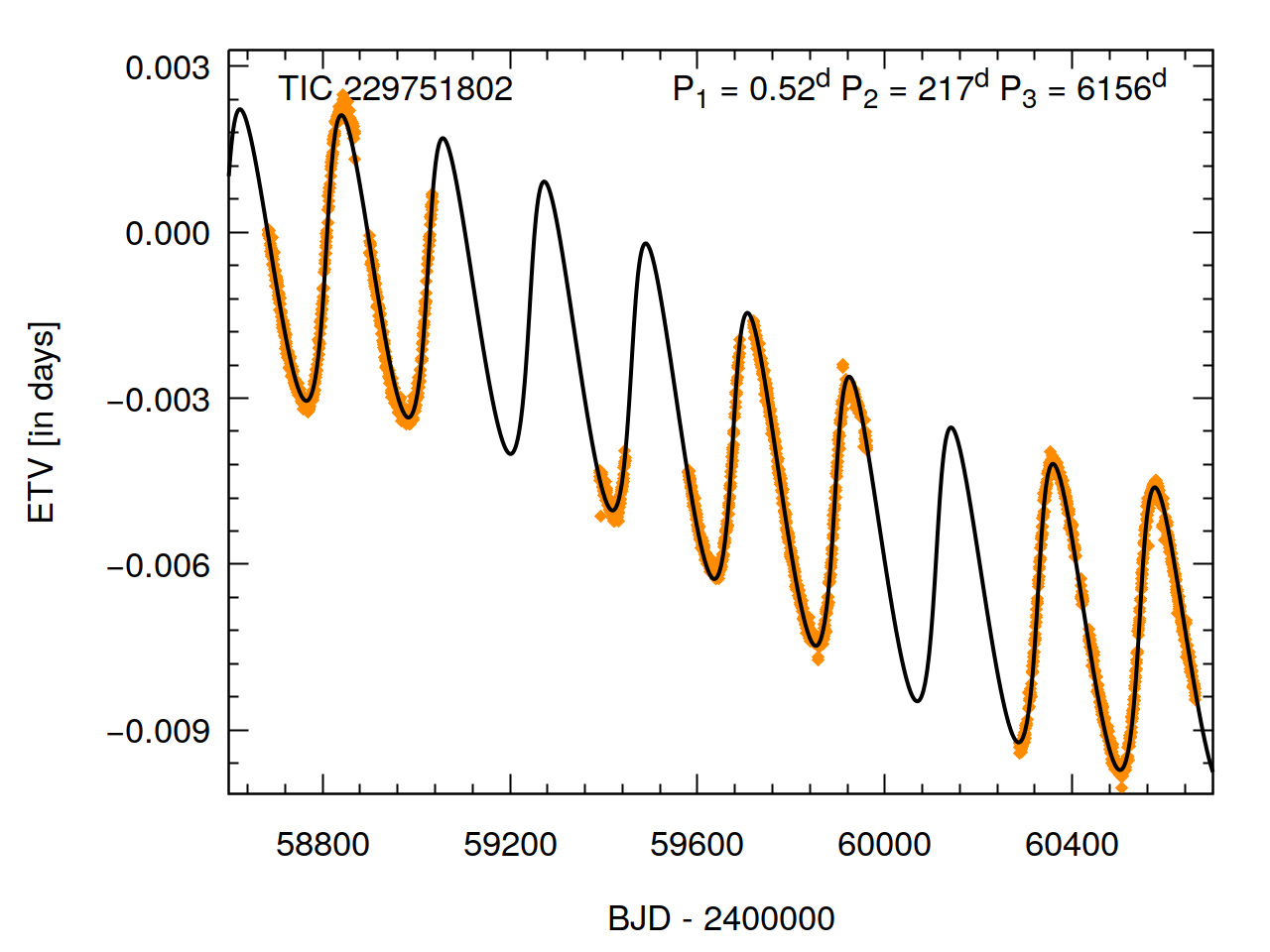}\includegraphics[width=0.35\textwidth]{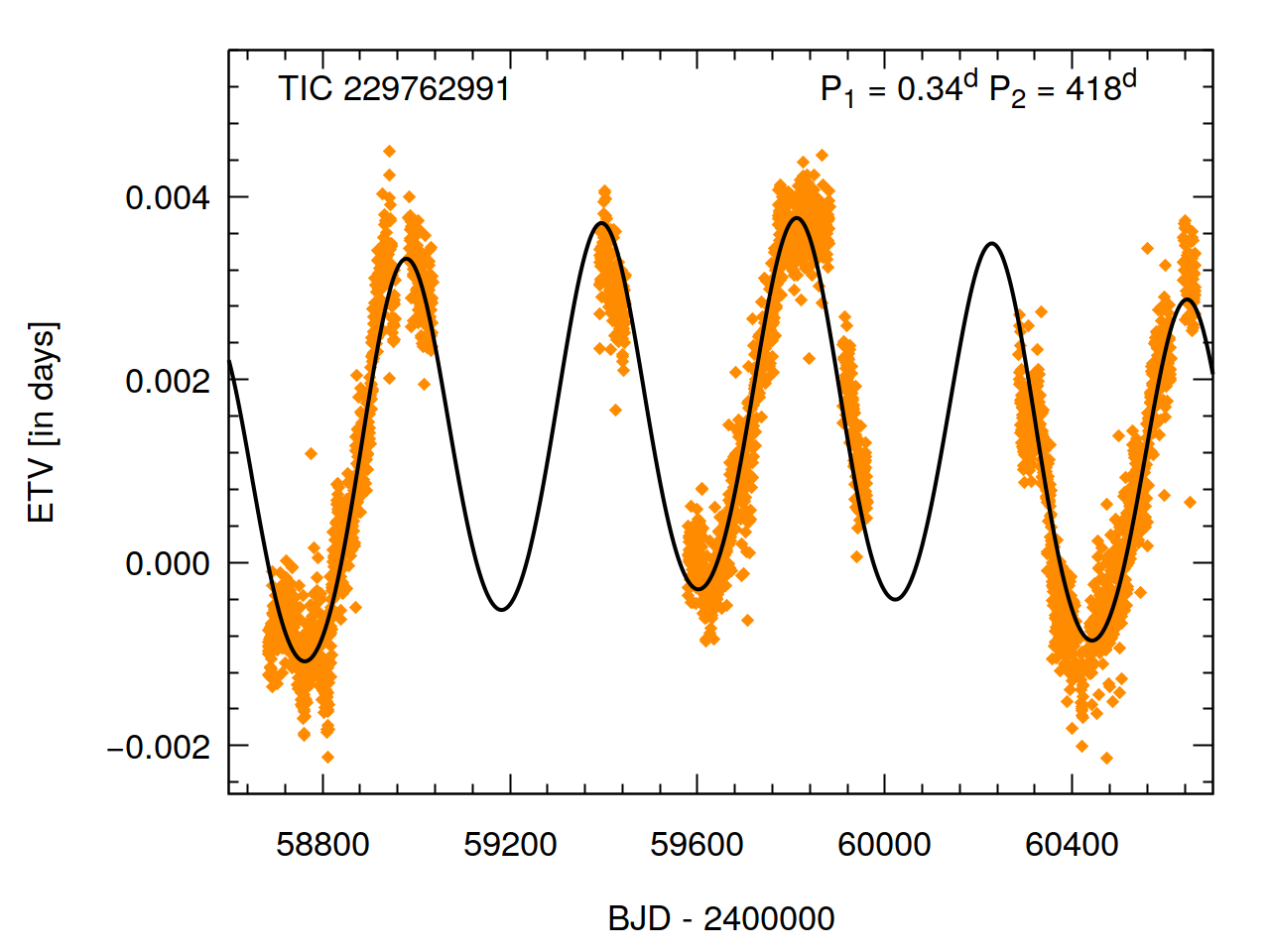}\includegraphics[width=0.35\textwidth]{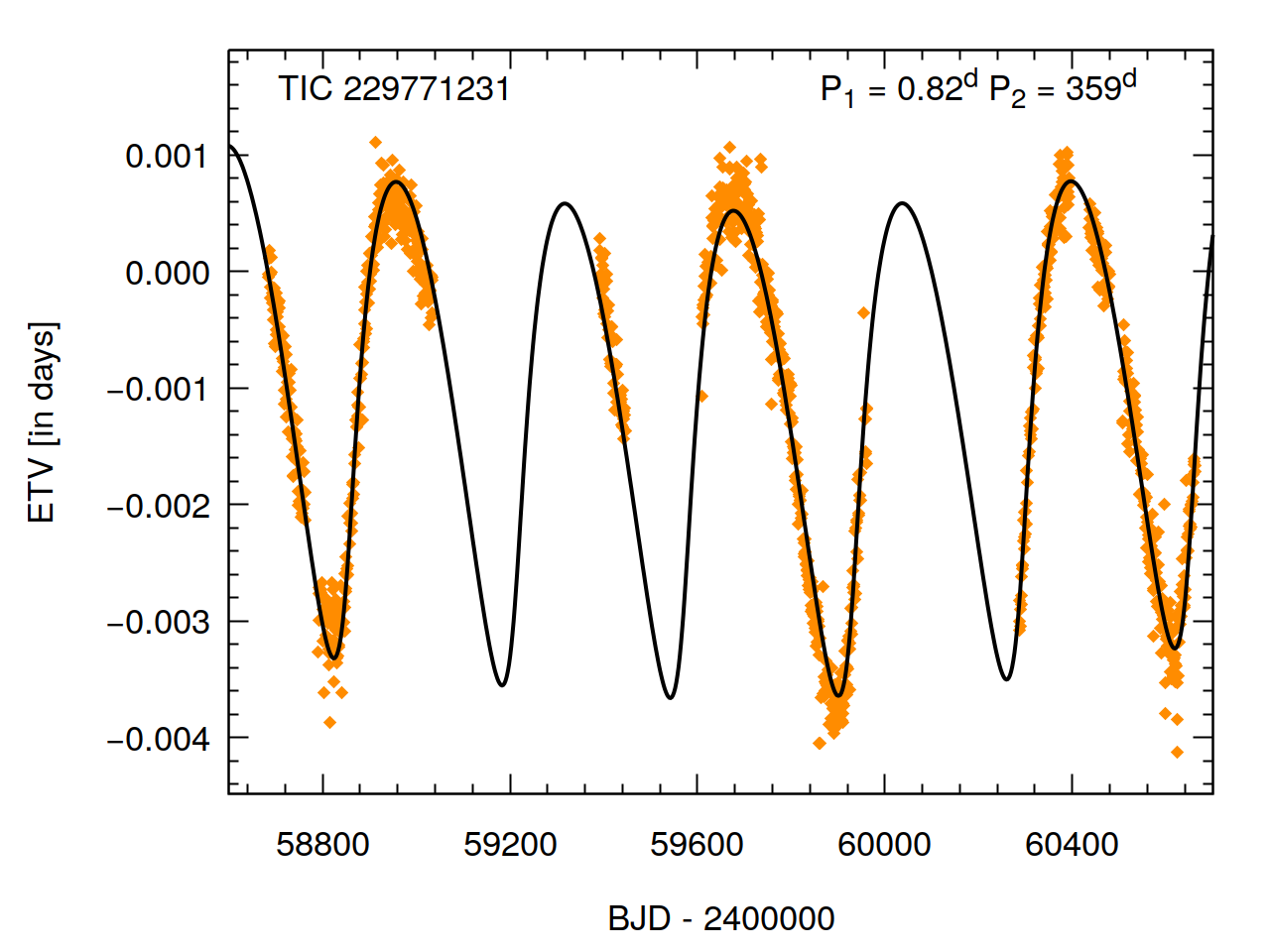}
\includegraphics[width=0.35\textwidth]{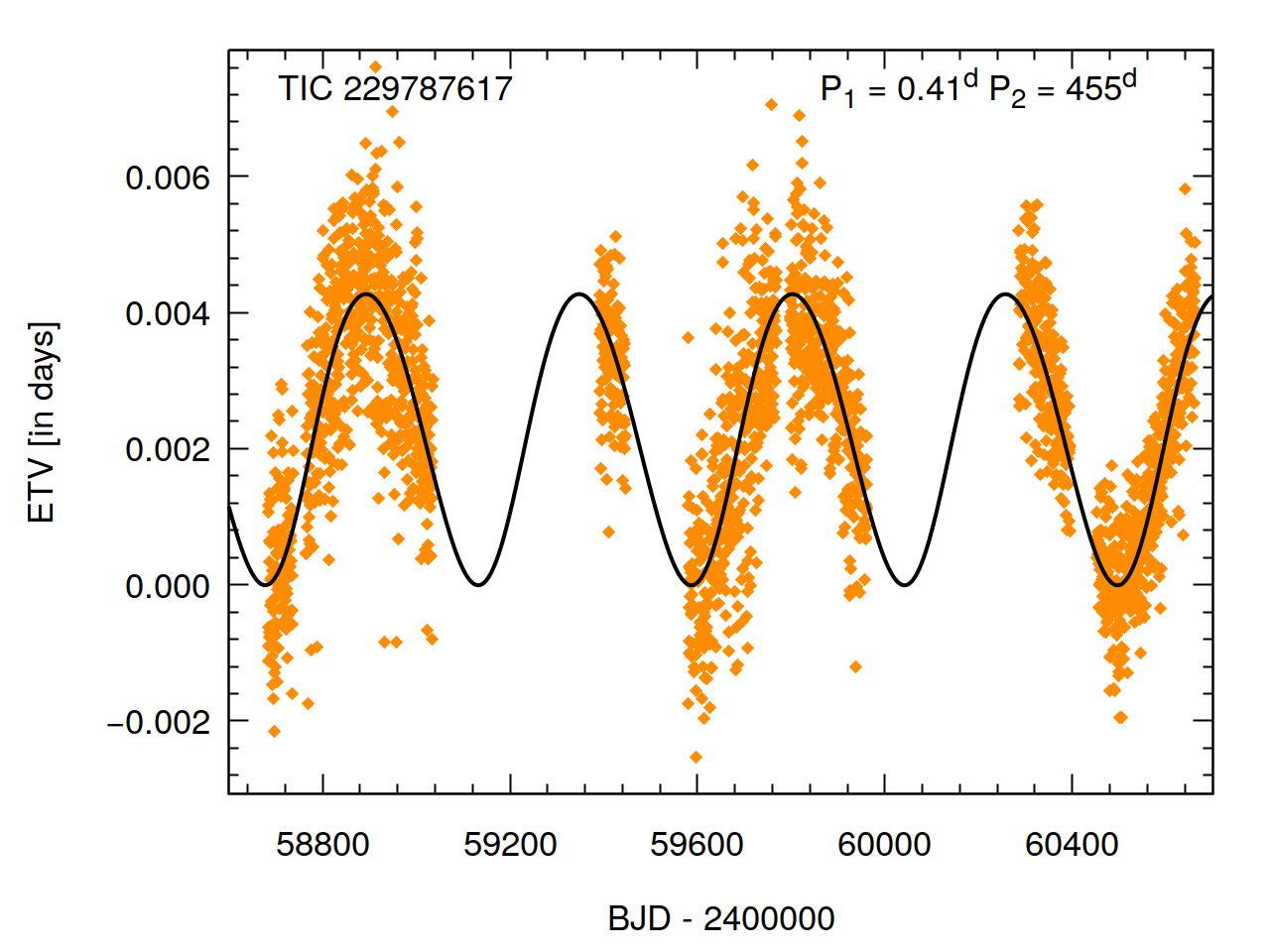}\includegraphics[width=0.35\textwidth]{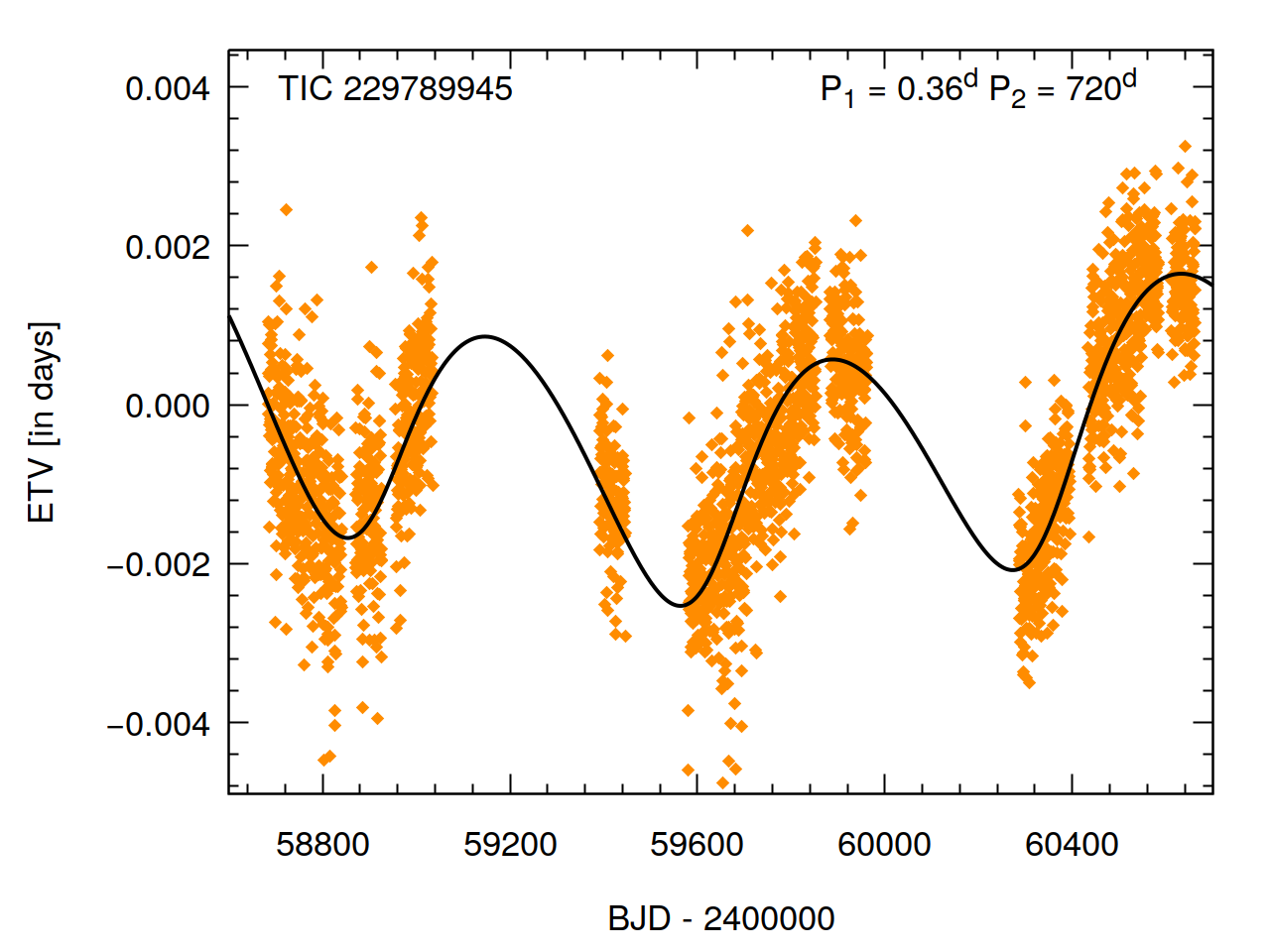}\includegraphics[width=0.35\textwidth]{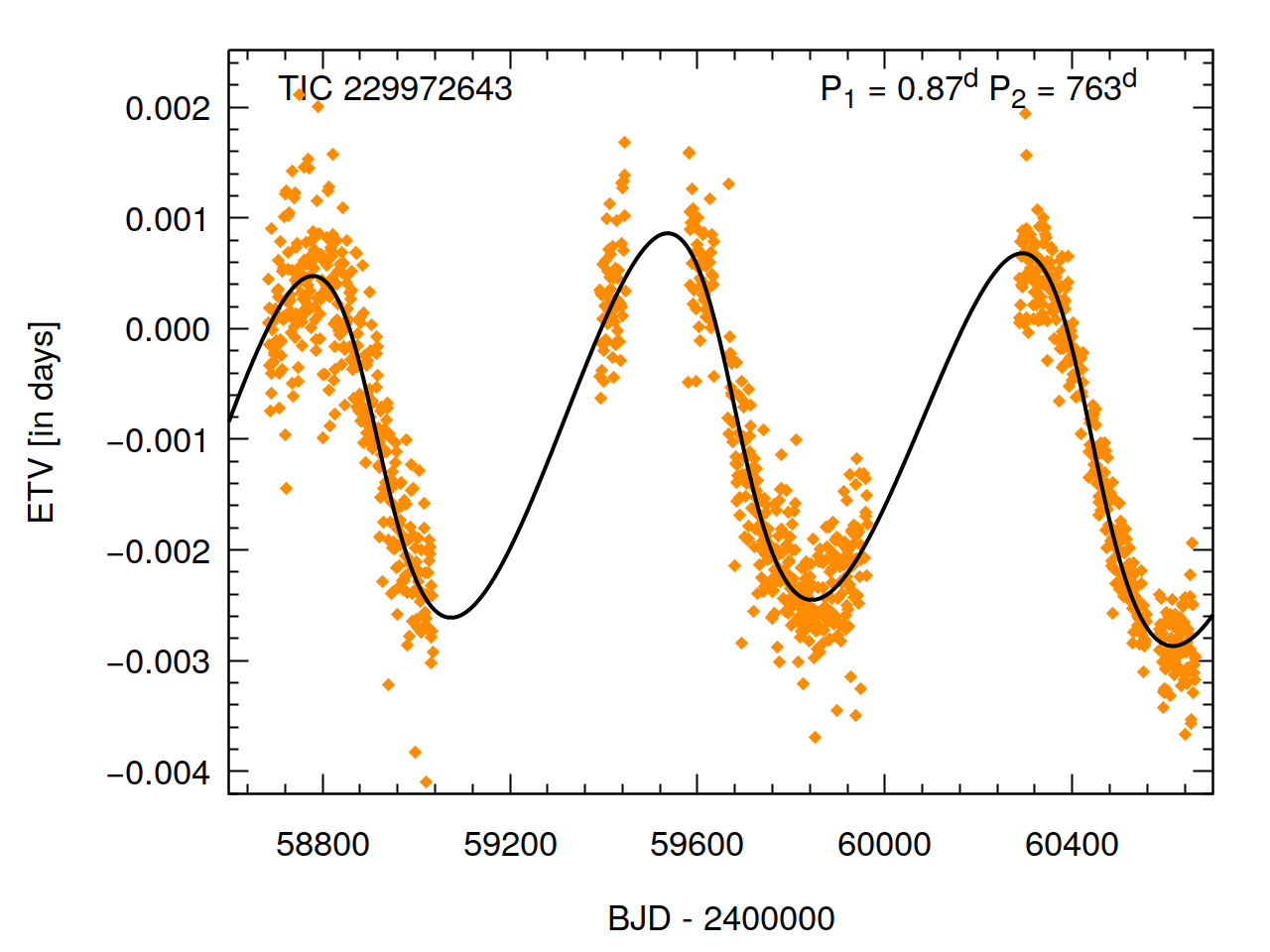}
\includegraphics[width=0.35\textwidth]{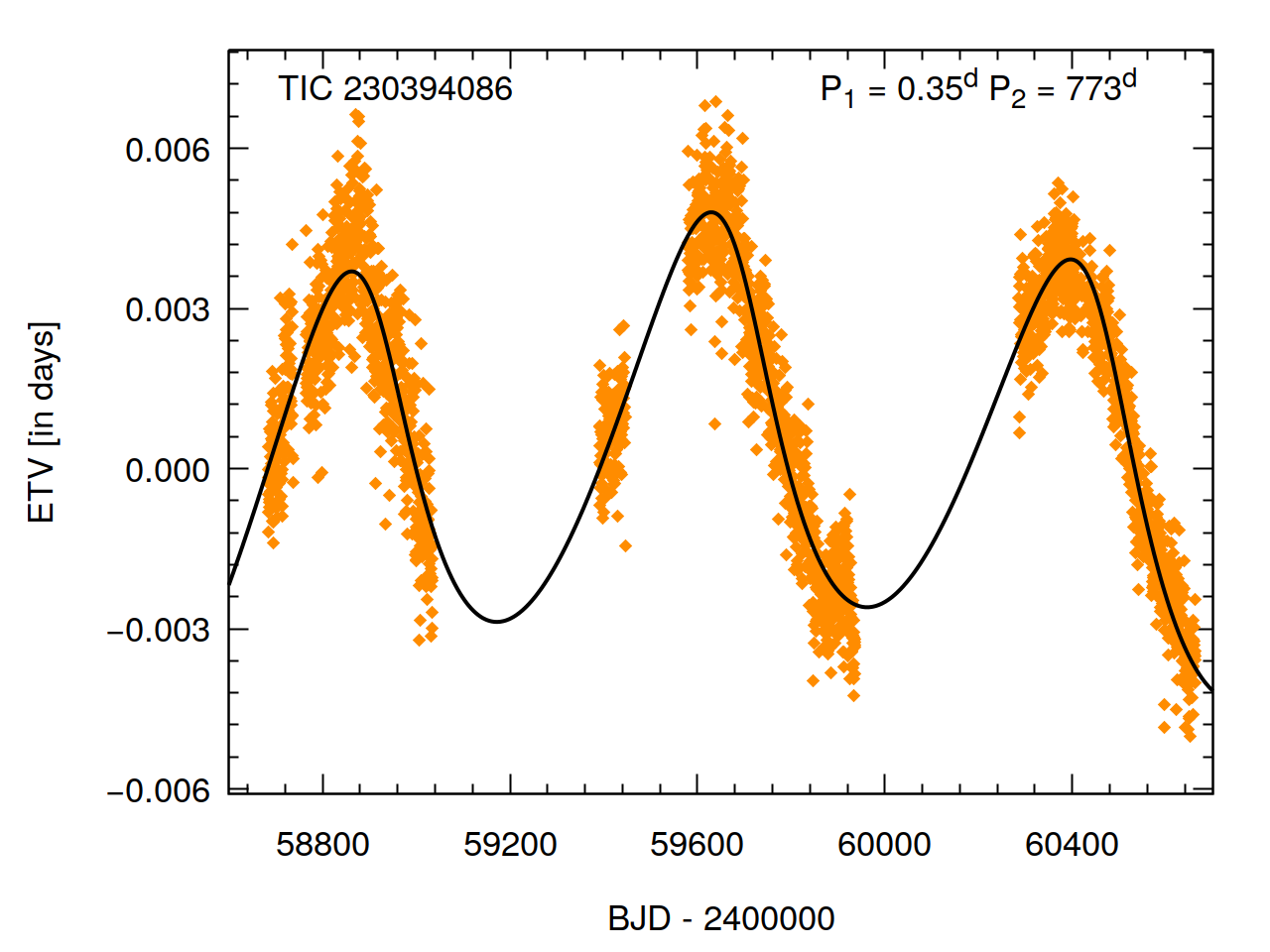}\includegraphics[width=0.35\textwidth]{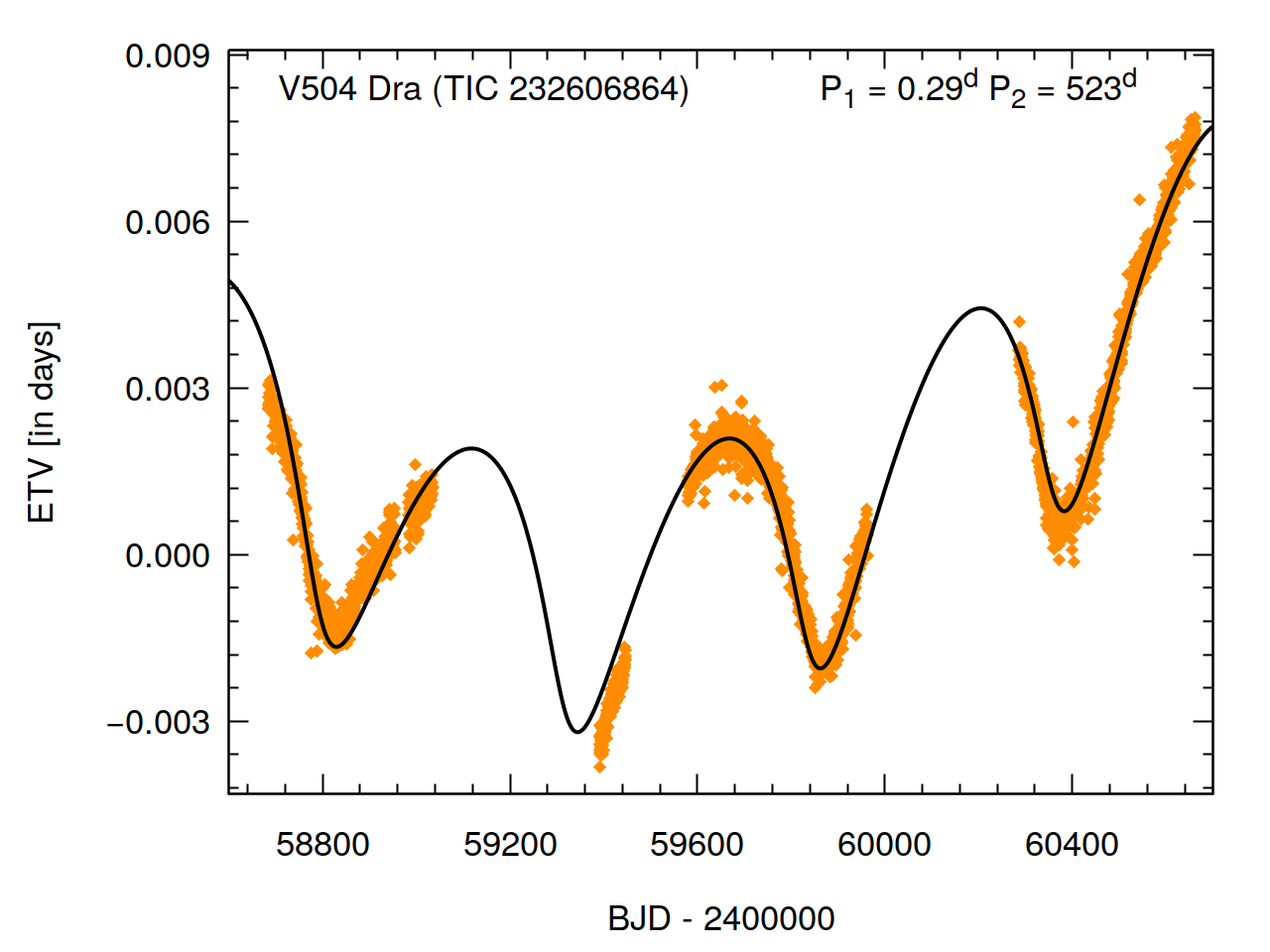}\includegraphics[width=0.35\textwidth]{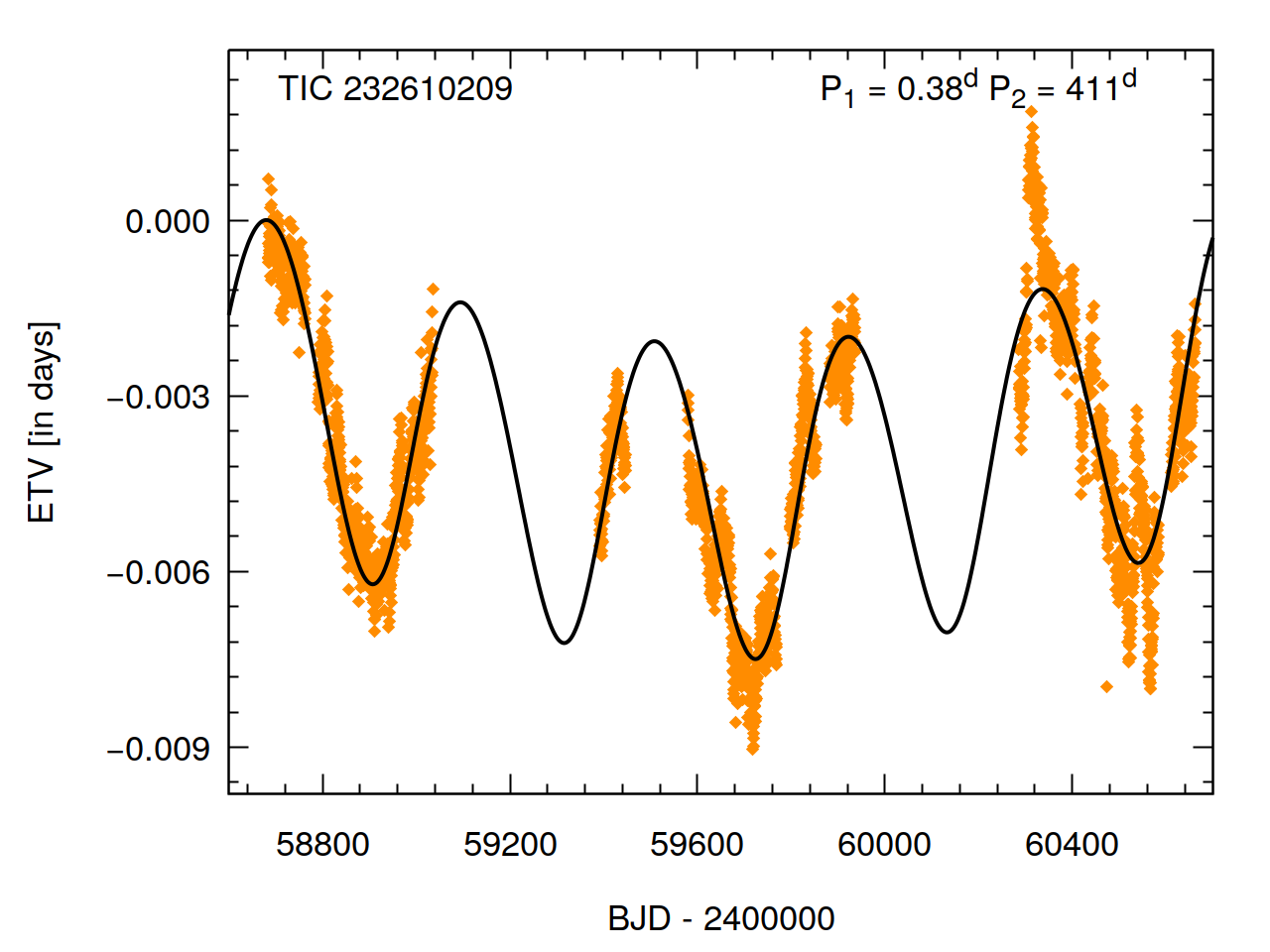}
\end{adjustwidth}
\caption{The first 15 ETVs of the most certain (group $L_1$) pure LTTE third body solutions. With the sole exception of TIC 159465833 (uppermost left panel), the primary and secondary ETV curves are of comparable quality and the binary eccentricity is nearly zero, and, therefore, we fit (and display) only the average of the two ETV curves (orange diamonds). In the case of TIC 159465833, the quality of the primary ETV curve is significantly better than that of the secondary curve and, thus, we present only the plot and the fit for the primary eclipses (with red circles). Note also that in the case of TIC 229751802 (third row, first column), a second LTTE solution, due to a potential fourth body, was also fitted simultaneously (see text for further details). The analytic, pure LTTE solutions are plotted as smooth curves with red for primary, blue for secondary, and black for the average of the two ETV curves, and we use the same convention in all of the following figures. (Finally, we note that the use of quadratic or cubic terms is not indicated for these systems; for these and other details, see Table~\ref{Tab:Orbelem_LTTE1}.)}
\label{Fig:ETVs_L1a}
\end{figure}

\begin{figure}[H]
\begin{adjustwidth}{-\extralength}{0cm}
\centering
\includegraphics[width=0.41\textwidth]{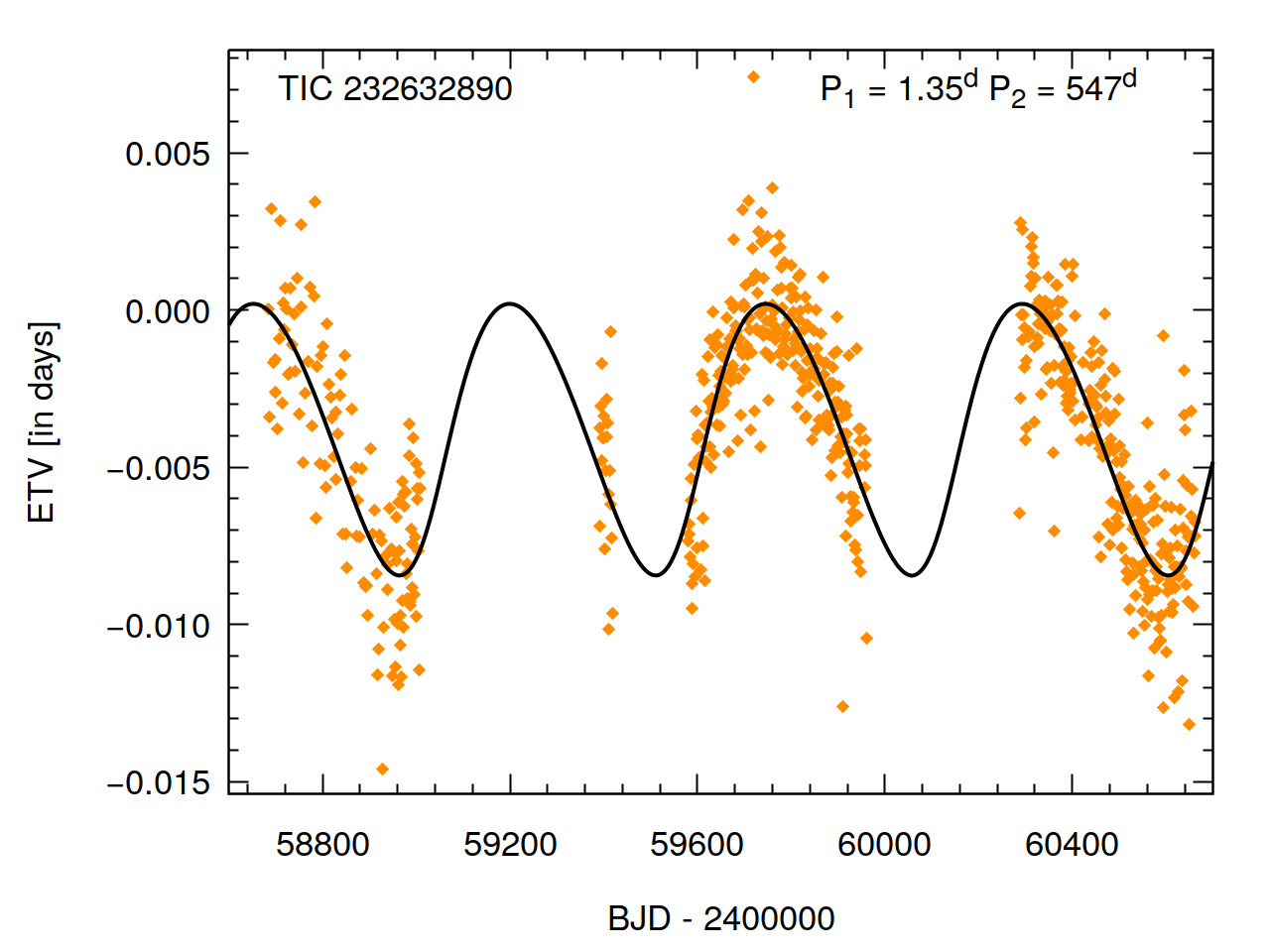}\includegraphics[width=0.41\textwidth]{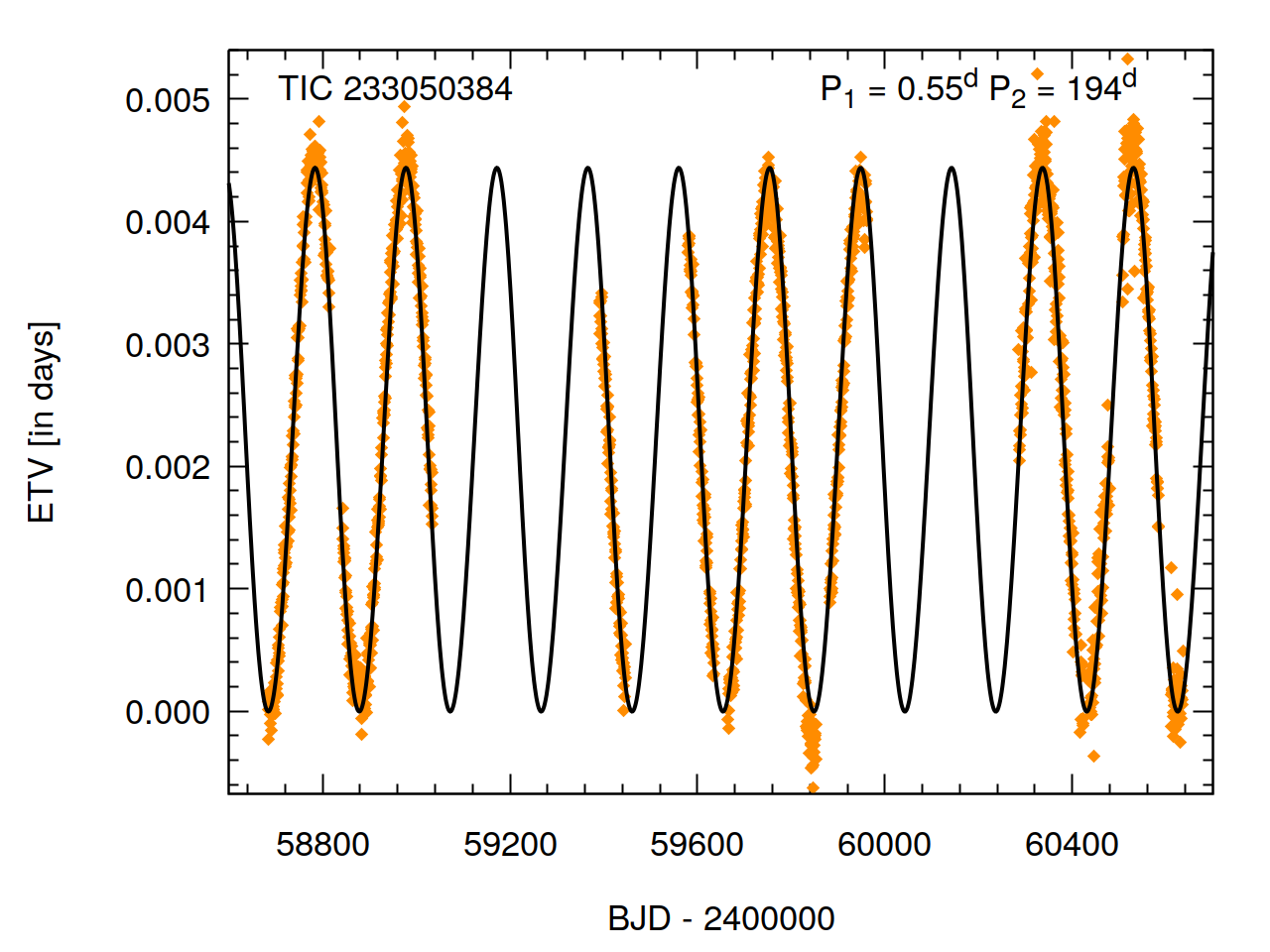}\includegraphics[width=0.41\textwidth]{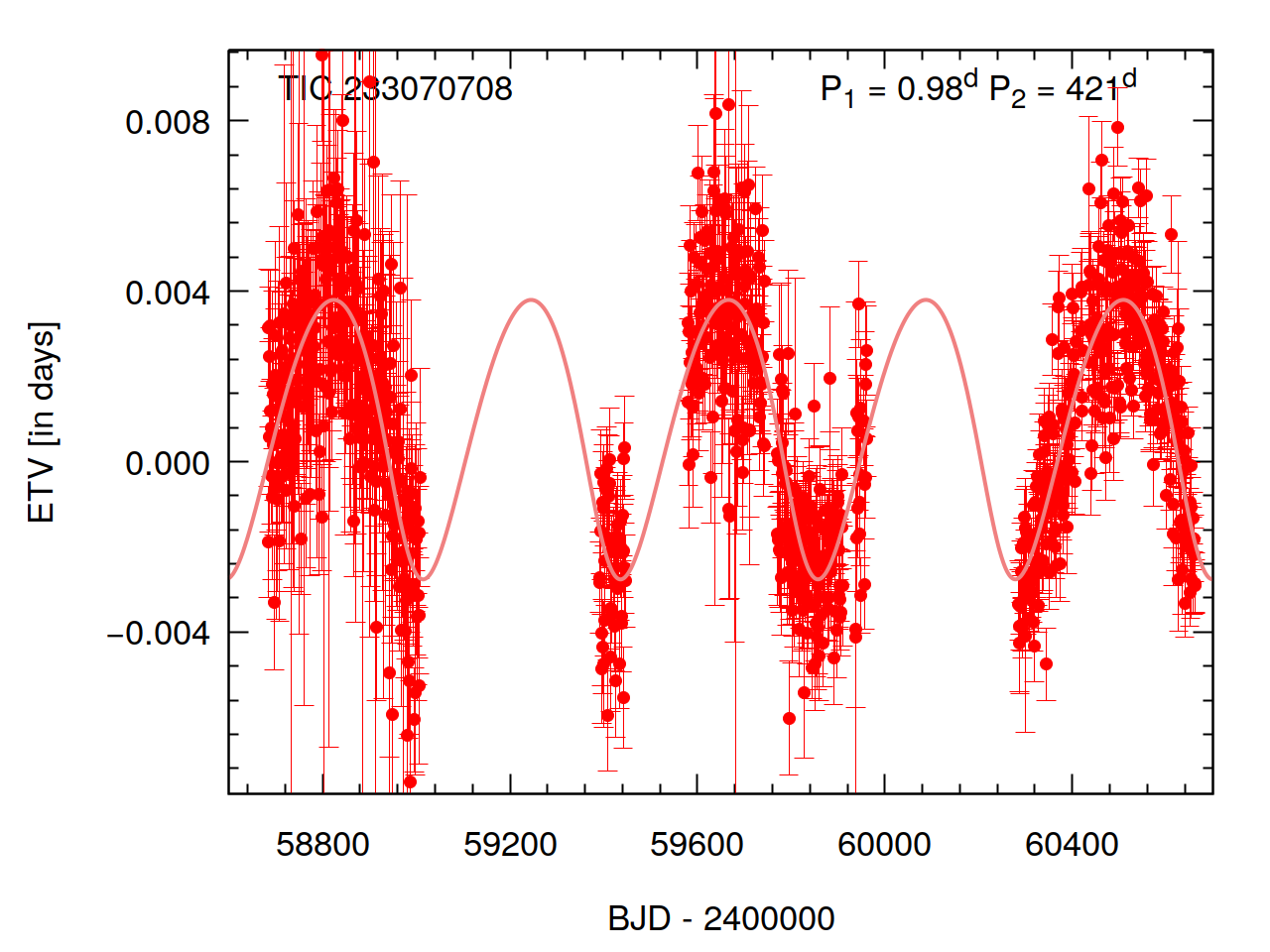}
\includegraphics[width=0.41\textwidth]{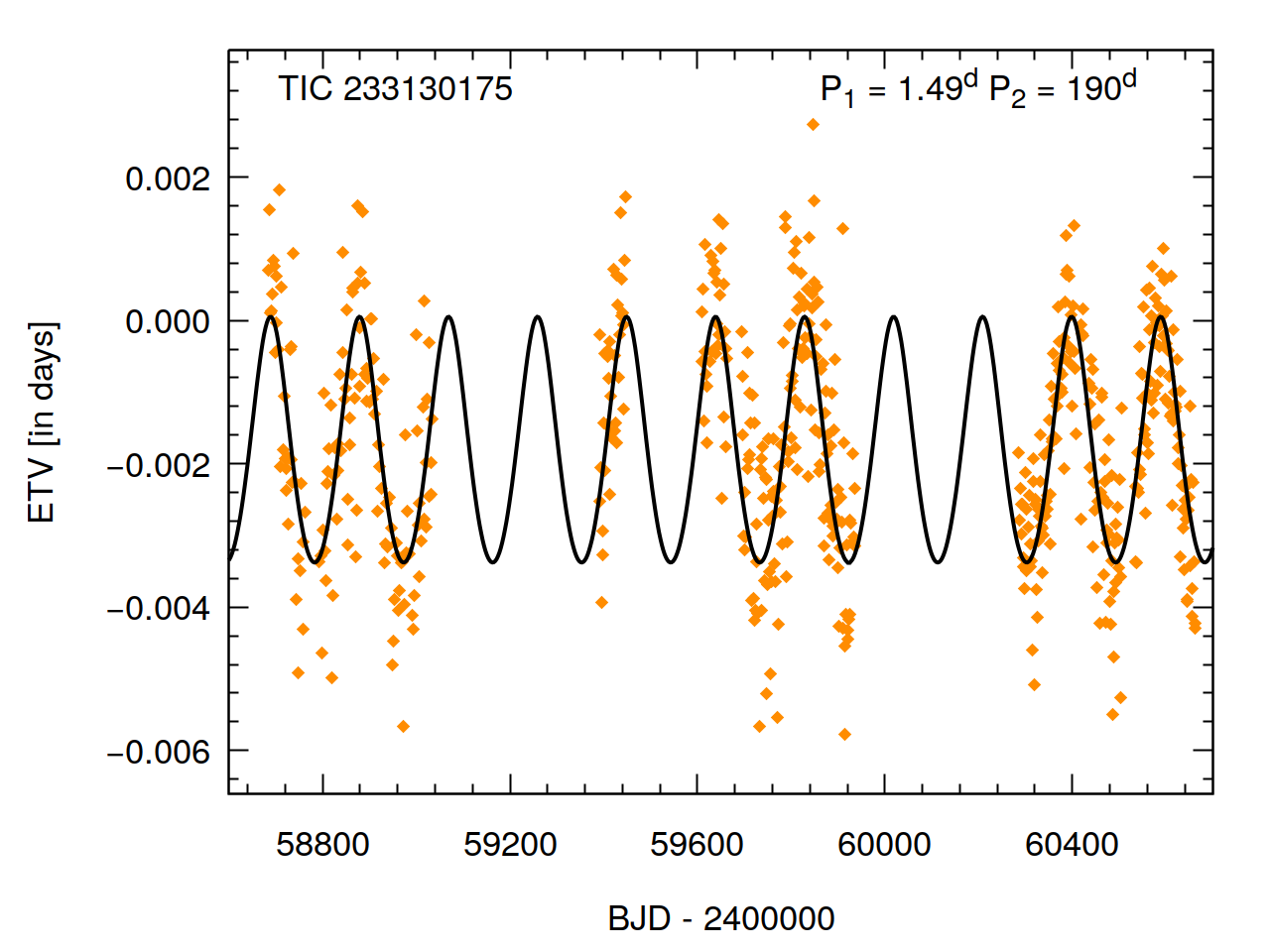}\includegraphics[width=0.41\textwidth]{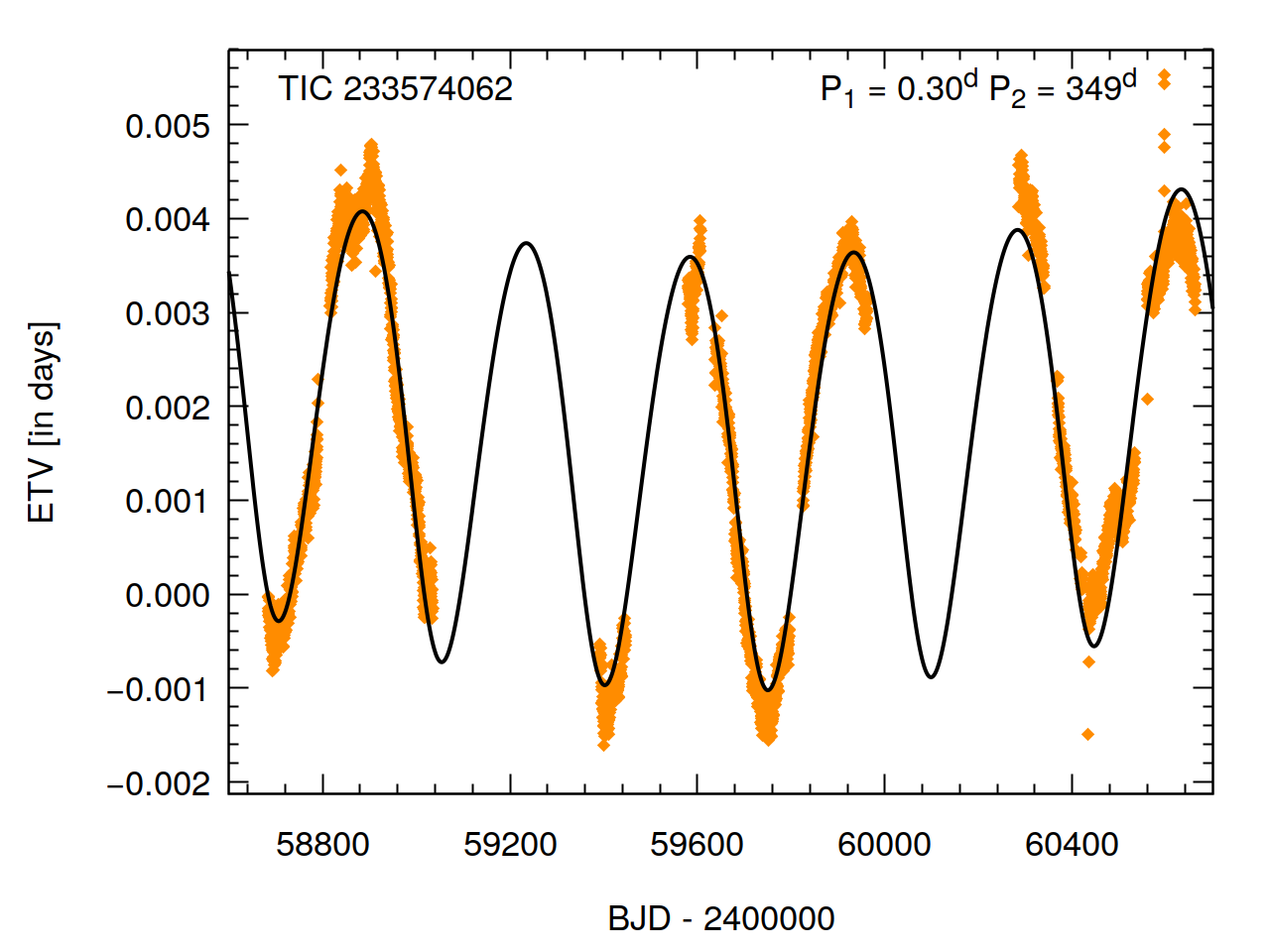}\includegraphics[width=0.41\textwidth]{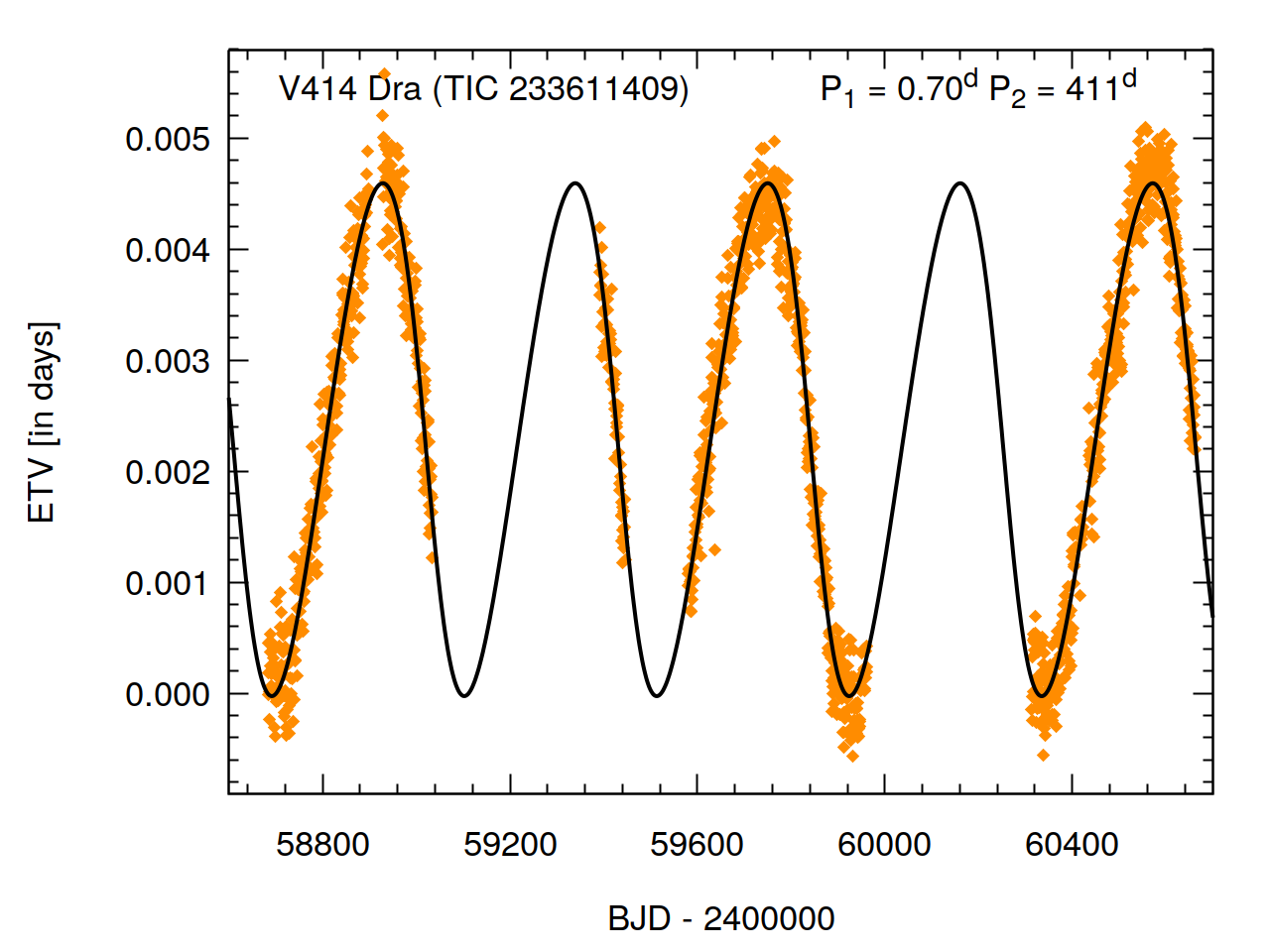}
\includegraphics[width=0.41\textwidth]{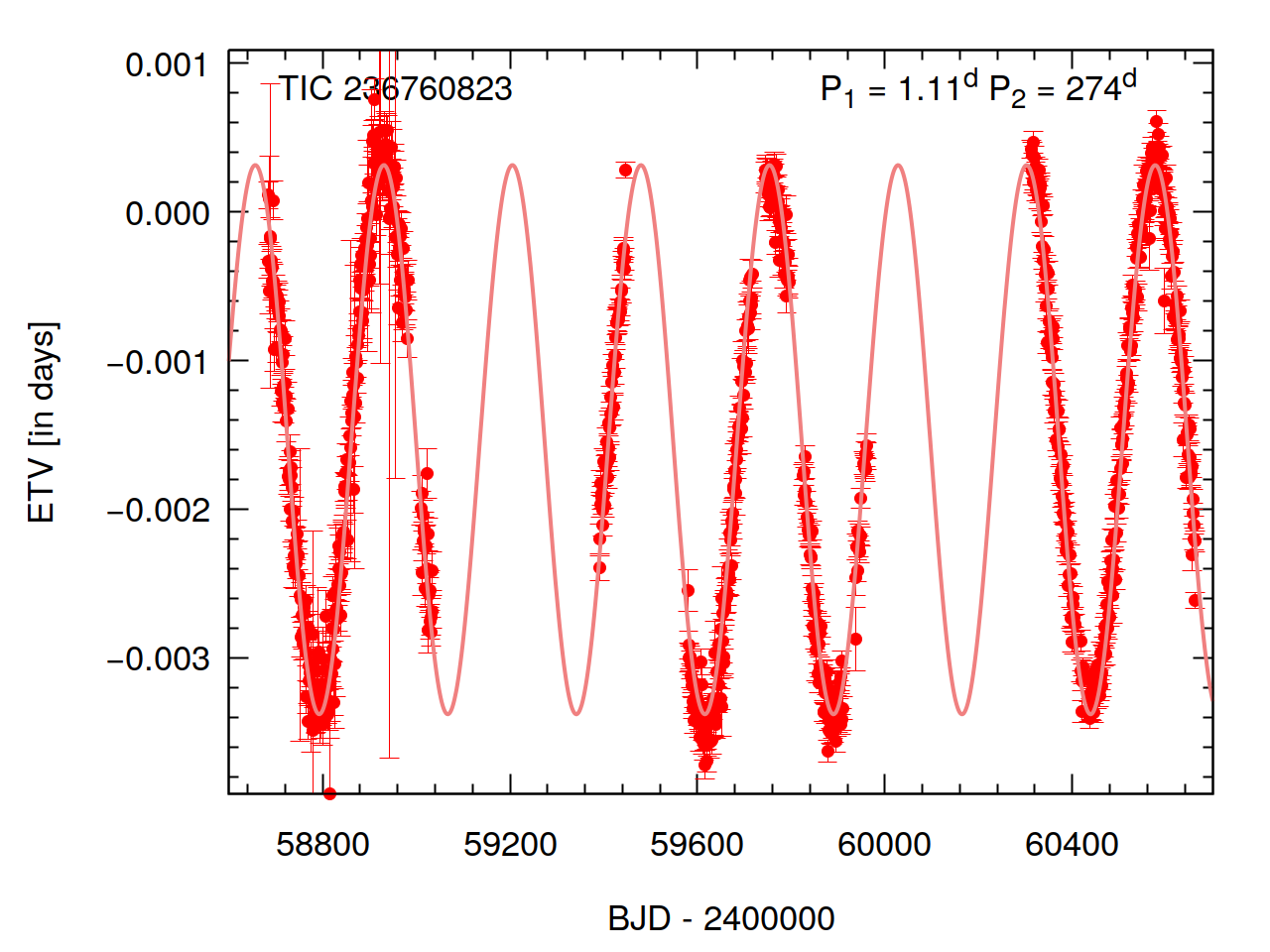}\includegraphics[width=0.41\textwidth]{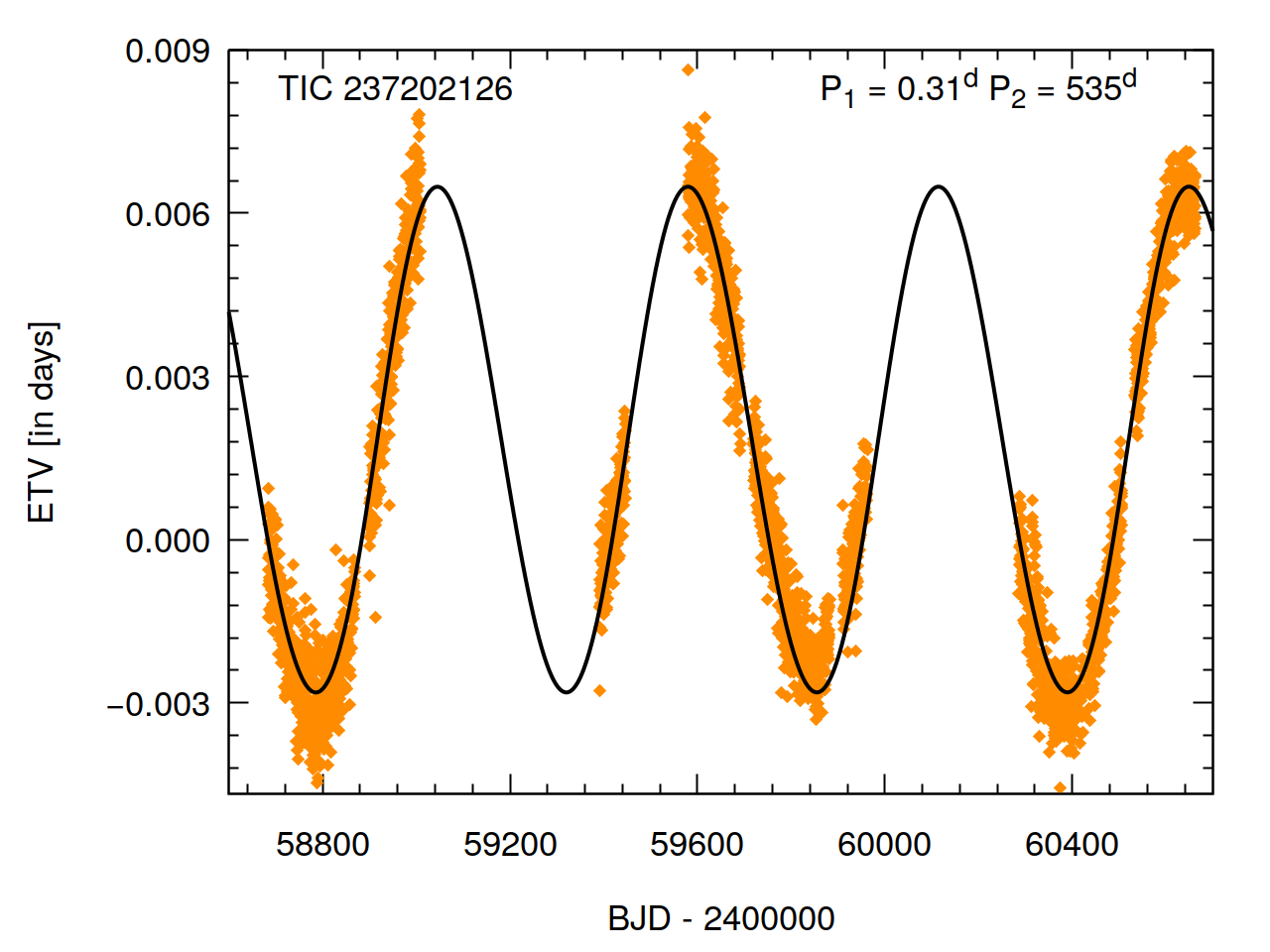}\includegraphics[width=0.41\textwidth]{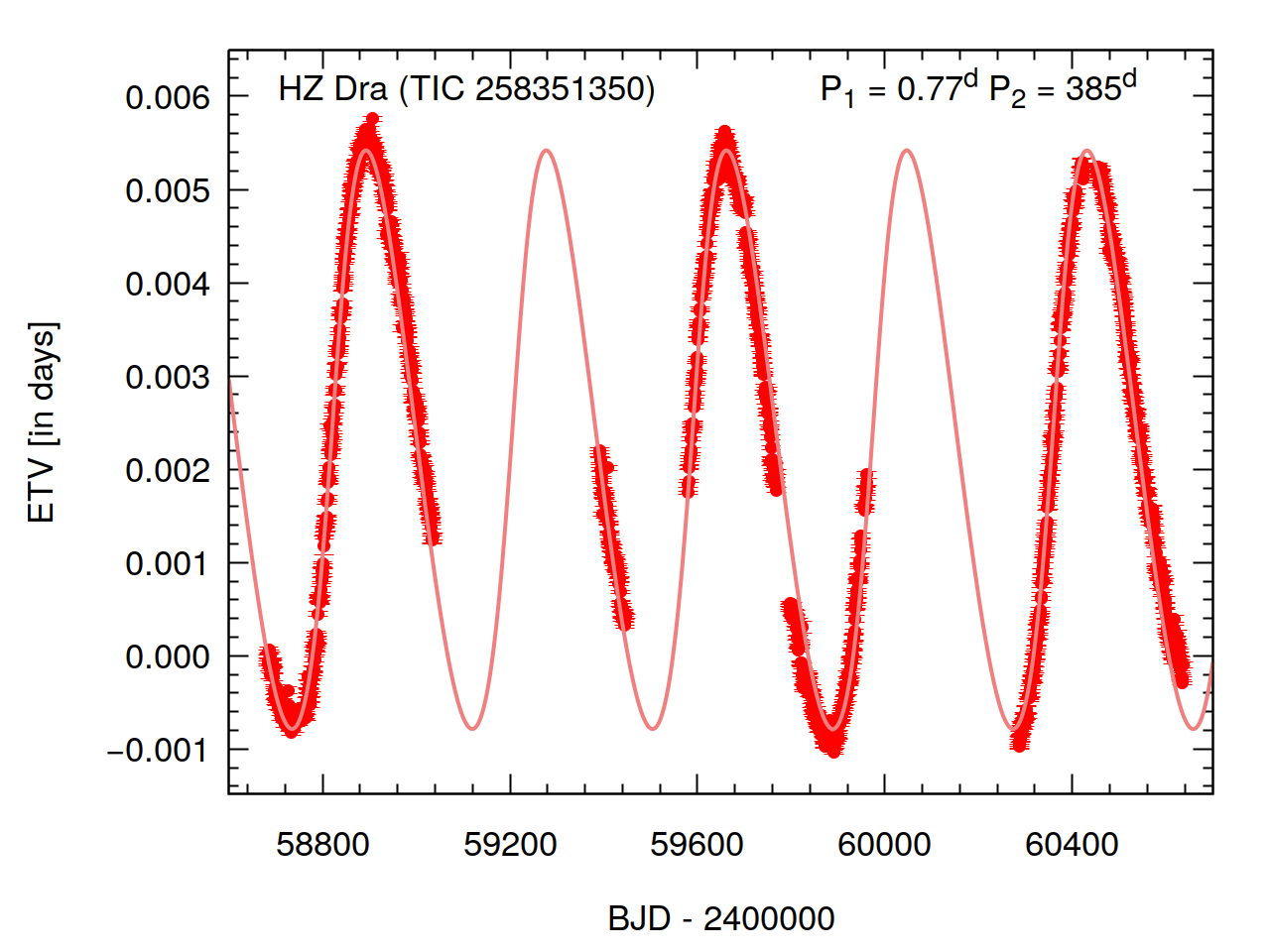}
\includegraphics[width=0.41\textwidth]{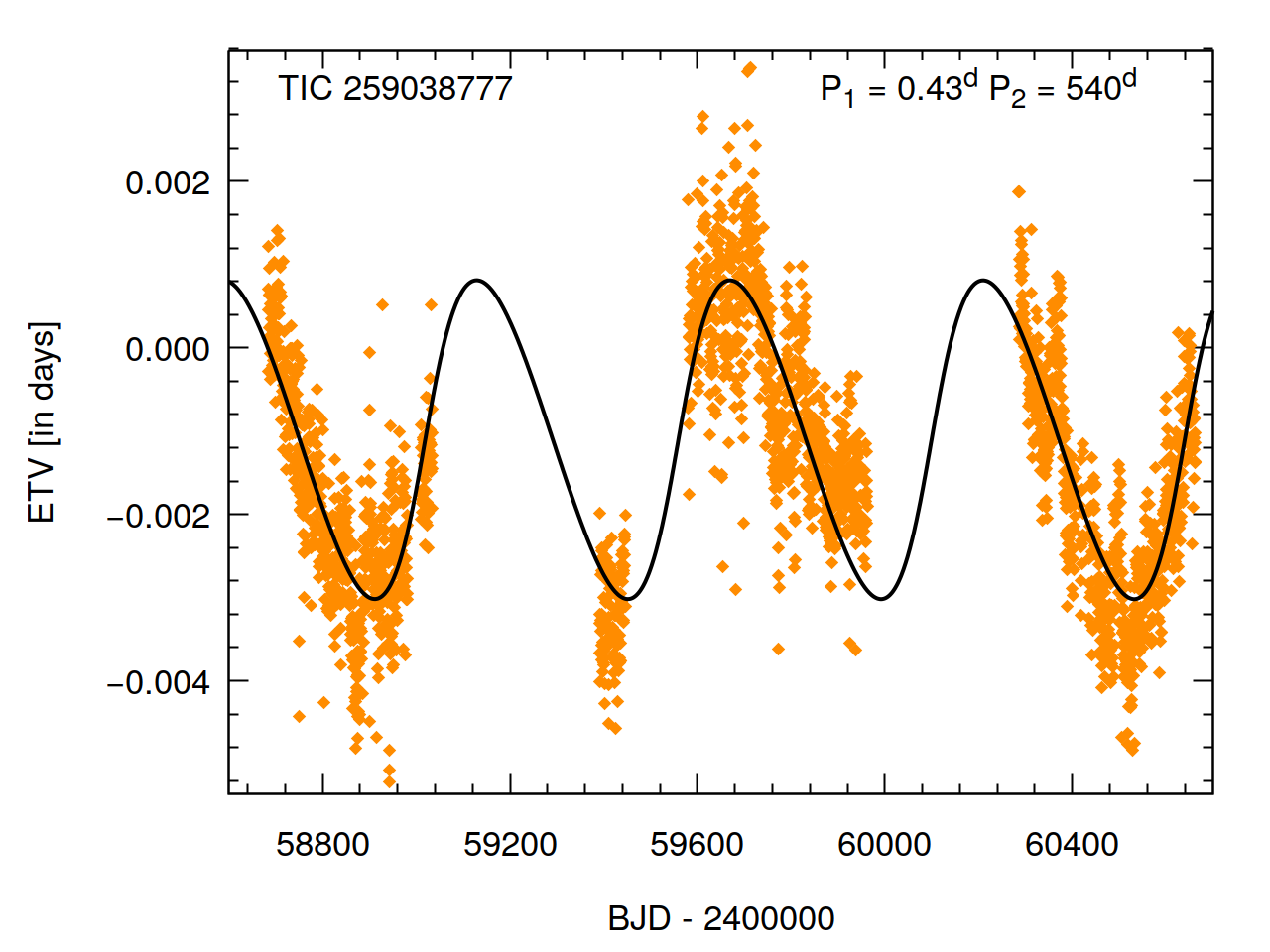}\includegraphics[width=0.41\textwidth]{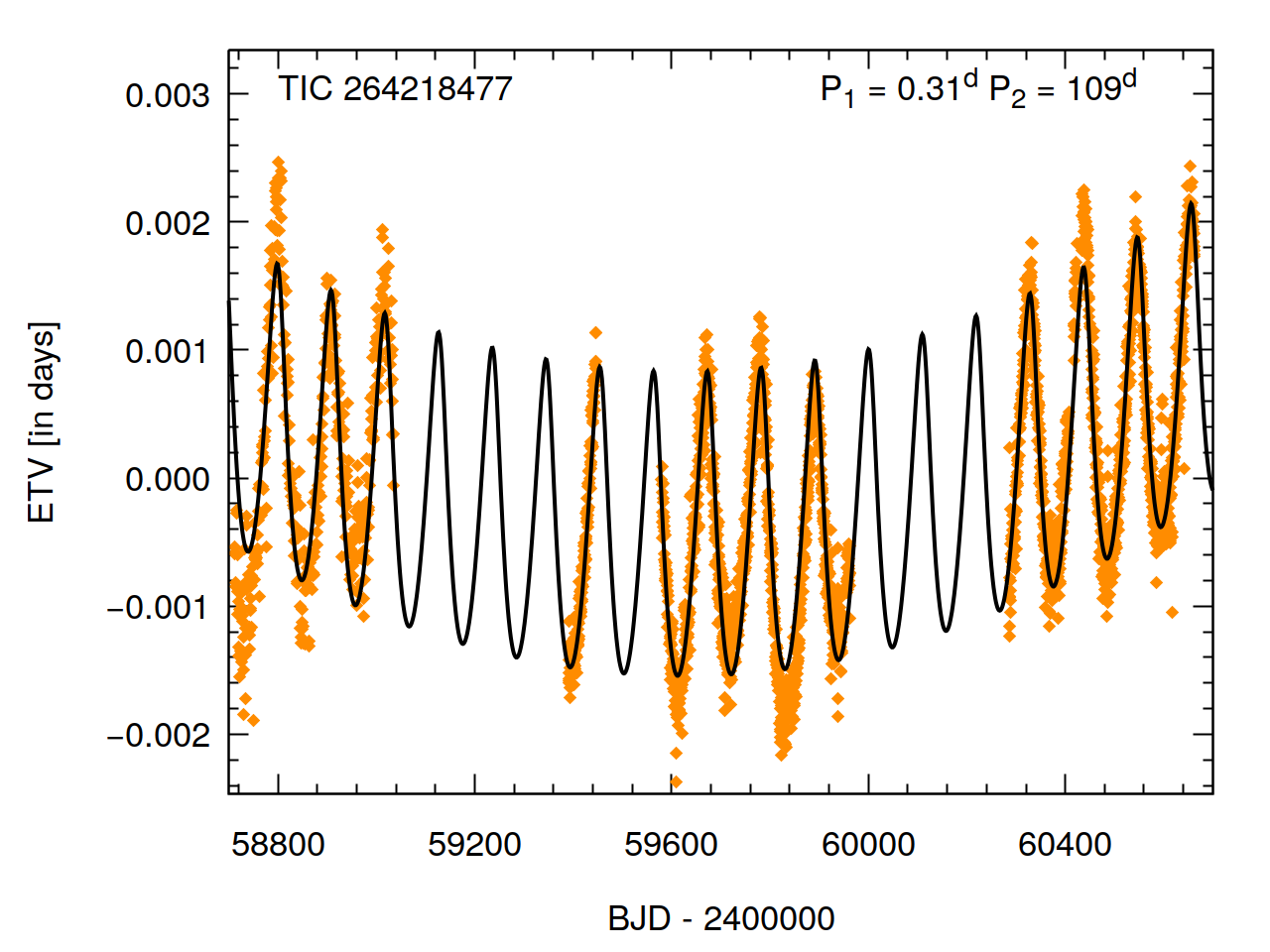}\includegraphics[width=0.41\textwidth]{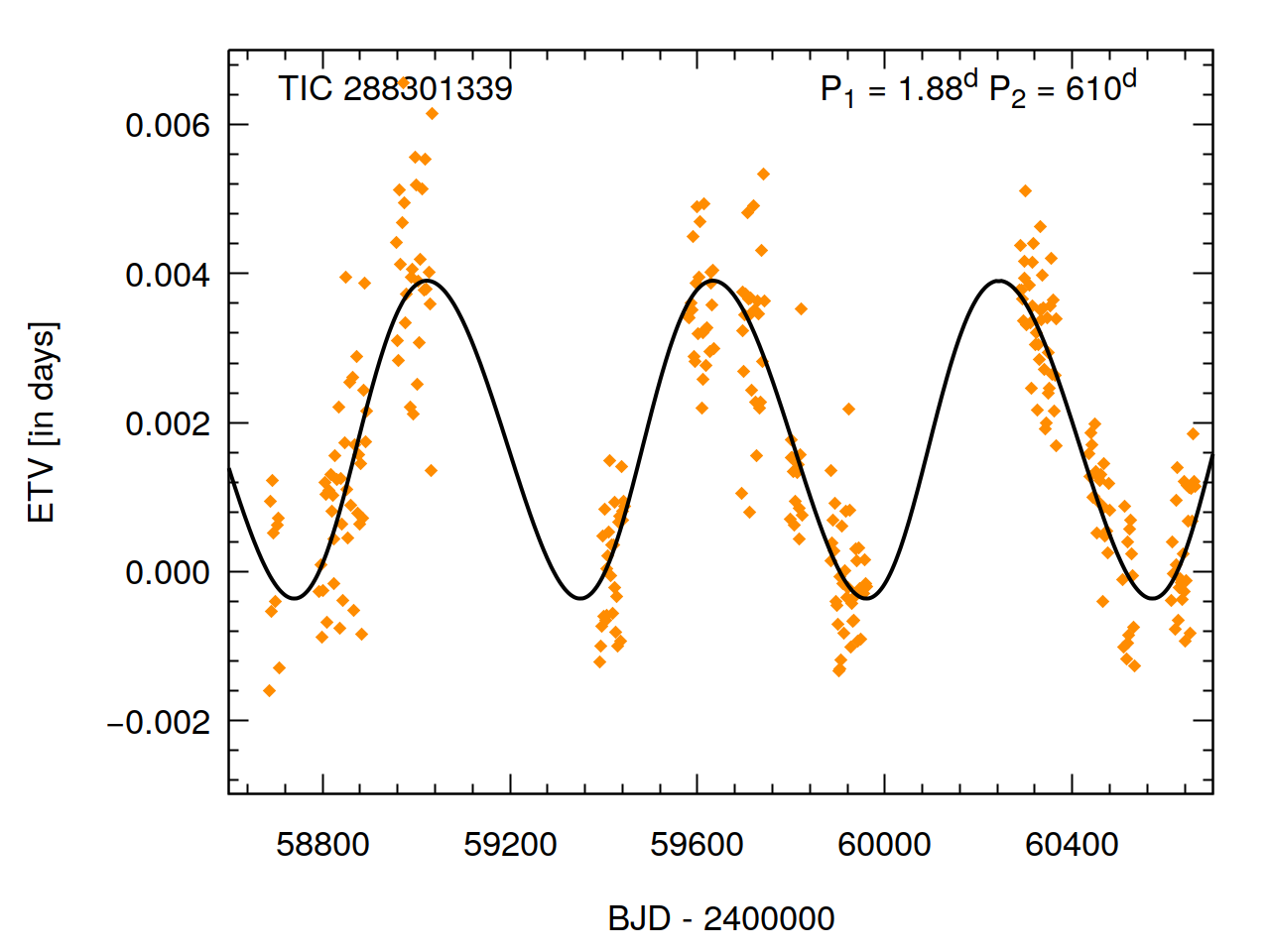}
\includegraphics[width=0.41\textwidth]{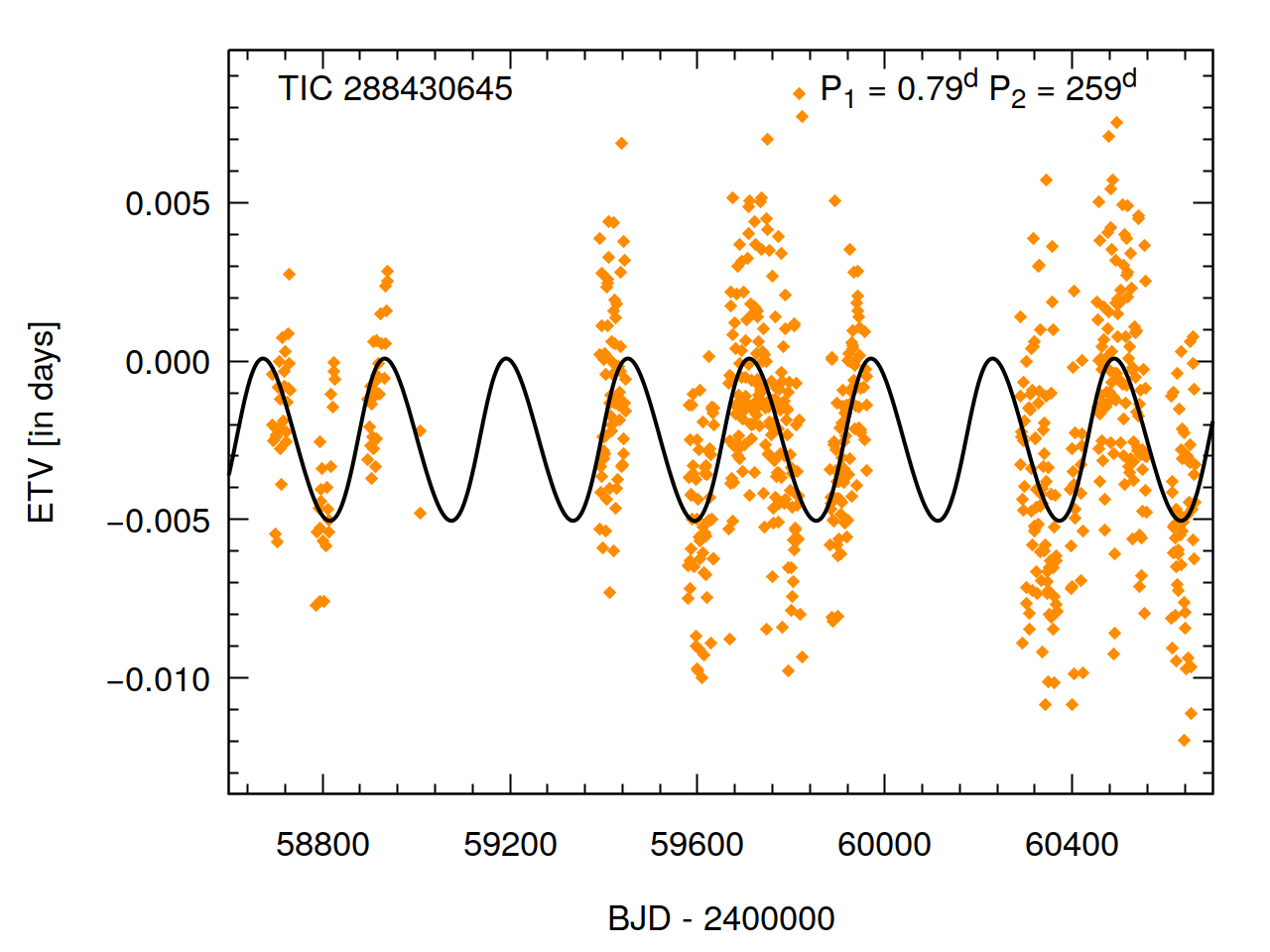}\includegraphics[width=0.41\textwidth]{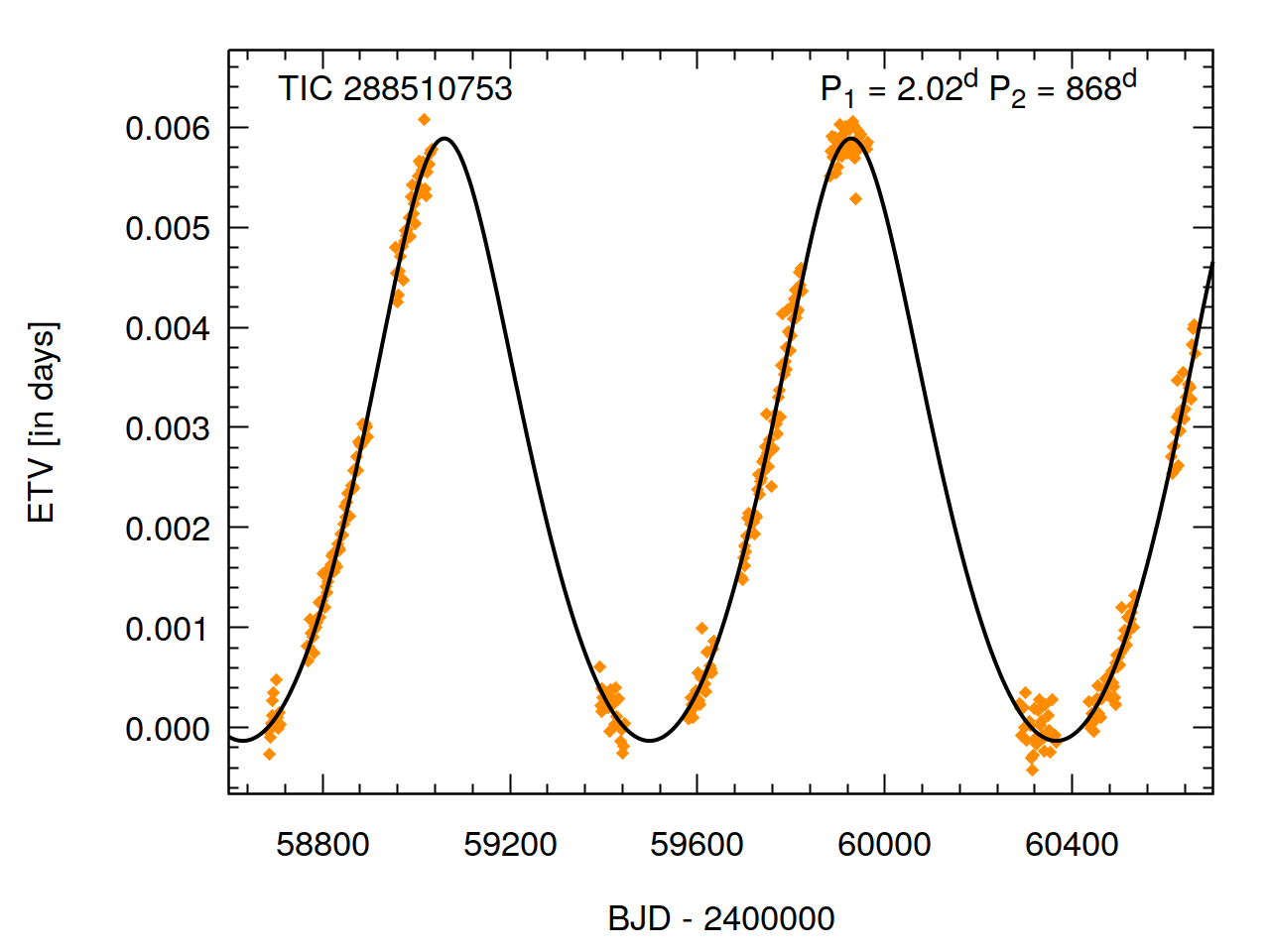}\includegraphics[width=0.41\textwidth]{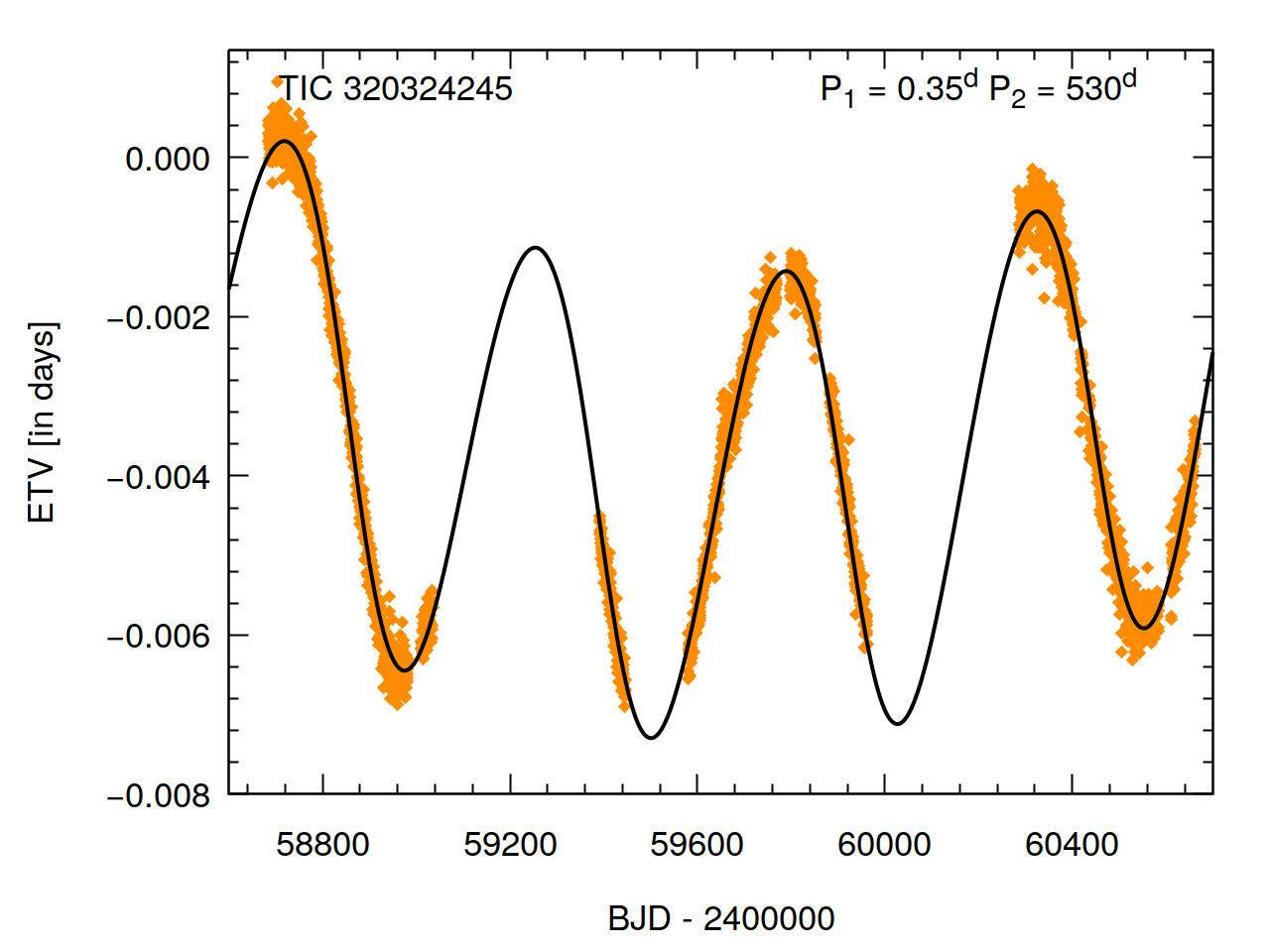}
\end{adjustwidth}
\caption{The next 15 ETVs of the most certain (group $L_1$) pure LTTE third body solutions. In most cases, again, the average of the primary and secondary ETVs are fitted. Exceptions are the plotted ETVs and pure LTTE solutions for TICs 233070708, 236760823, and 258351350, where we fit only the primary ETV curves. (In these cases, the uncertainty of each individual mid-eclipse time is also indicated with a vertical error bar.) For further details, see, again,  Table~\ref{Tab:Orbelem_LTTE1}.}
\label{Fig:ETVs_L1b}
\end{figure}

\begin{figure}[H]
\begin{adjustwidth}{-\extralength}{0cm}
\centering
\includegraphics[width=0.44\textwidth]{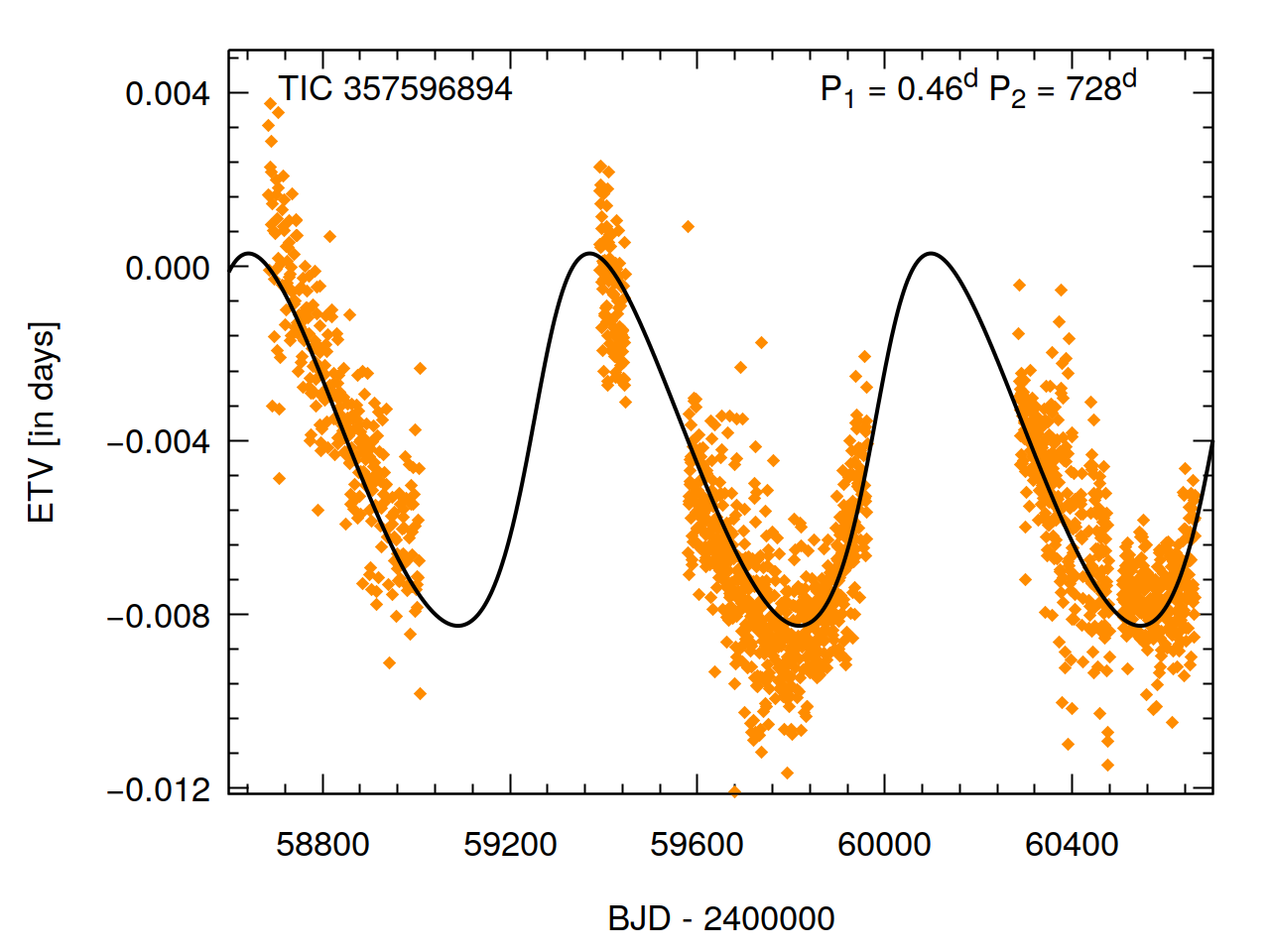}\includegraphics[width=0.44\textwidth]{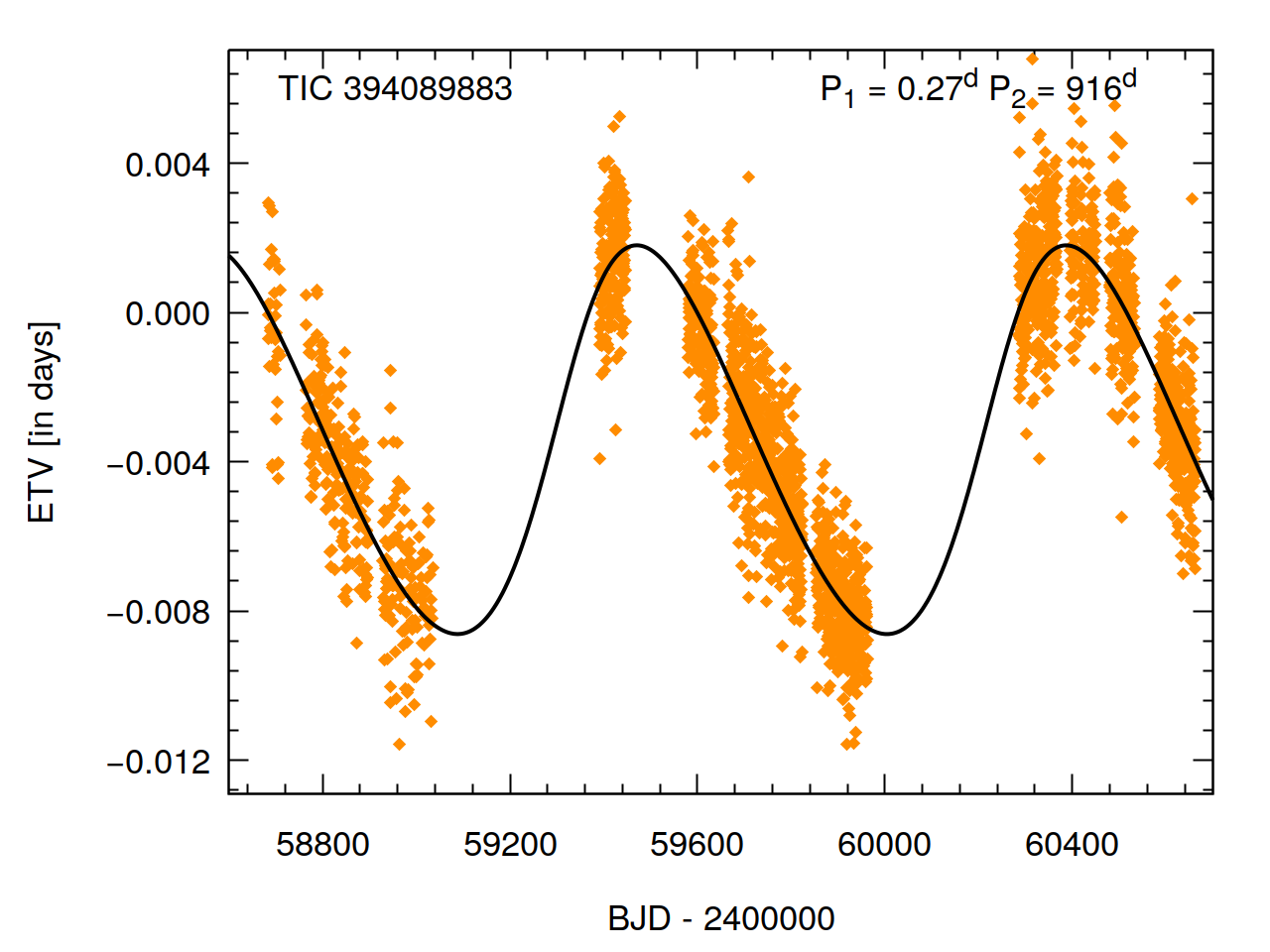}\includegraphics[width=0.44\textwidth]{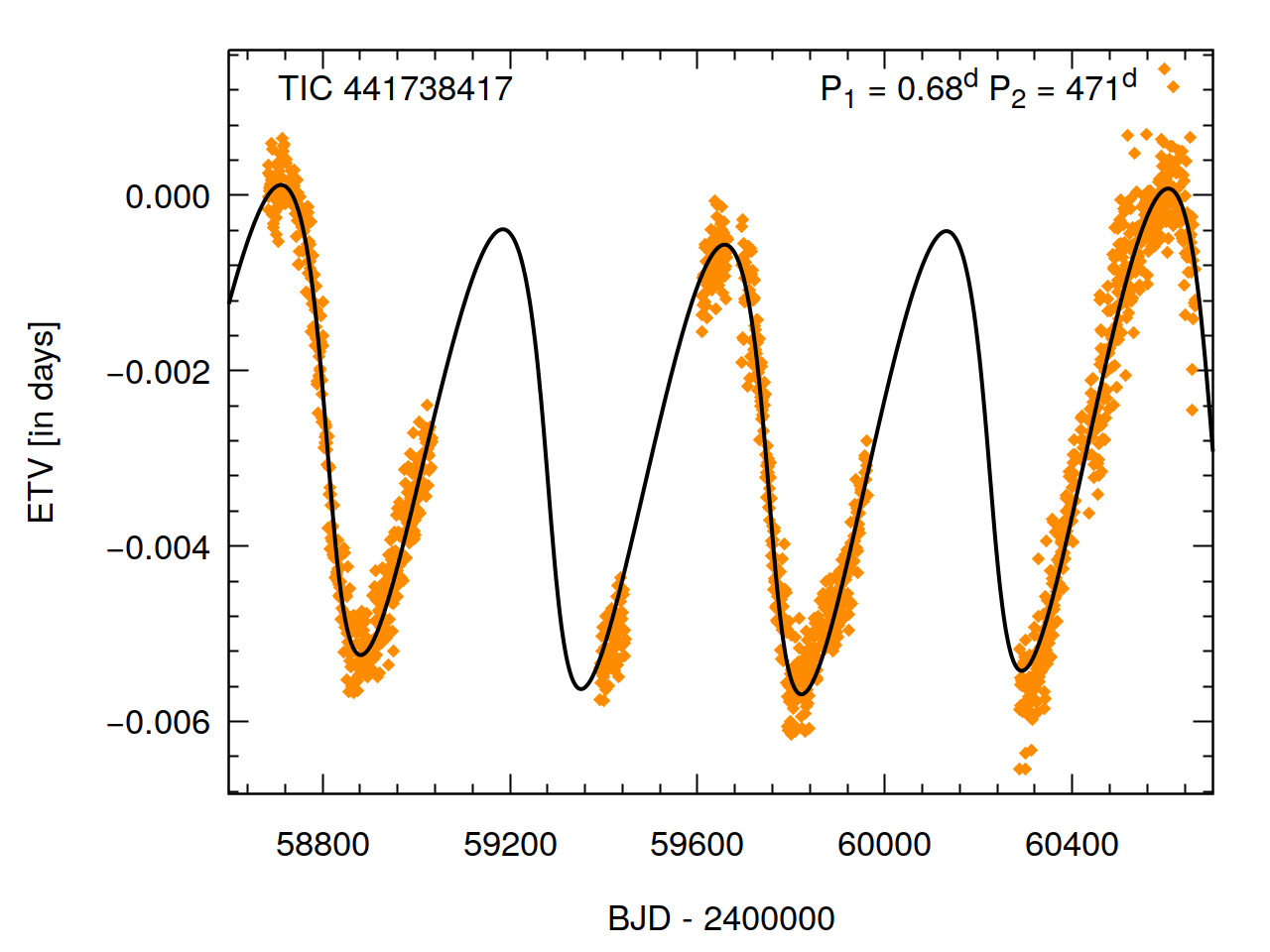}
\includegraphics[width=0.44\textwidth]{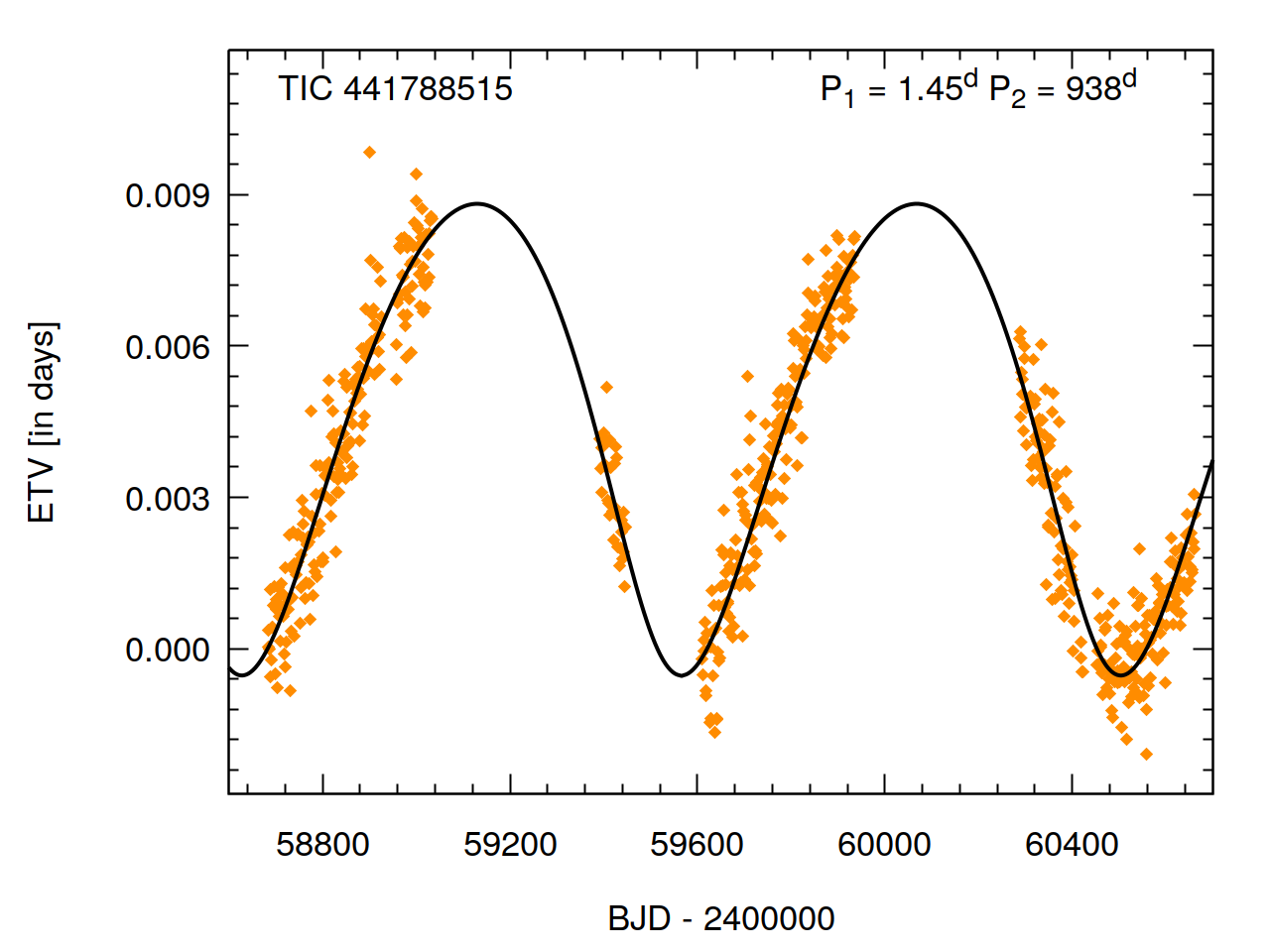}\includegraphics[width=0.44\textwidth]{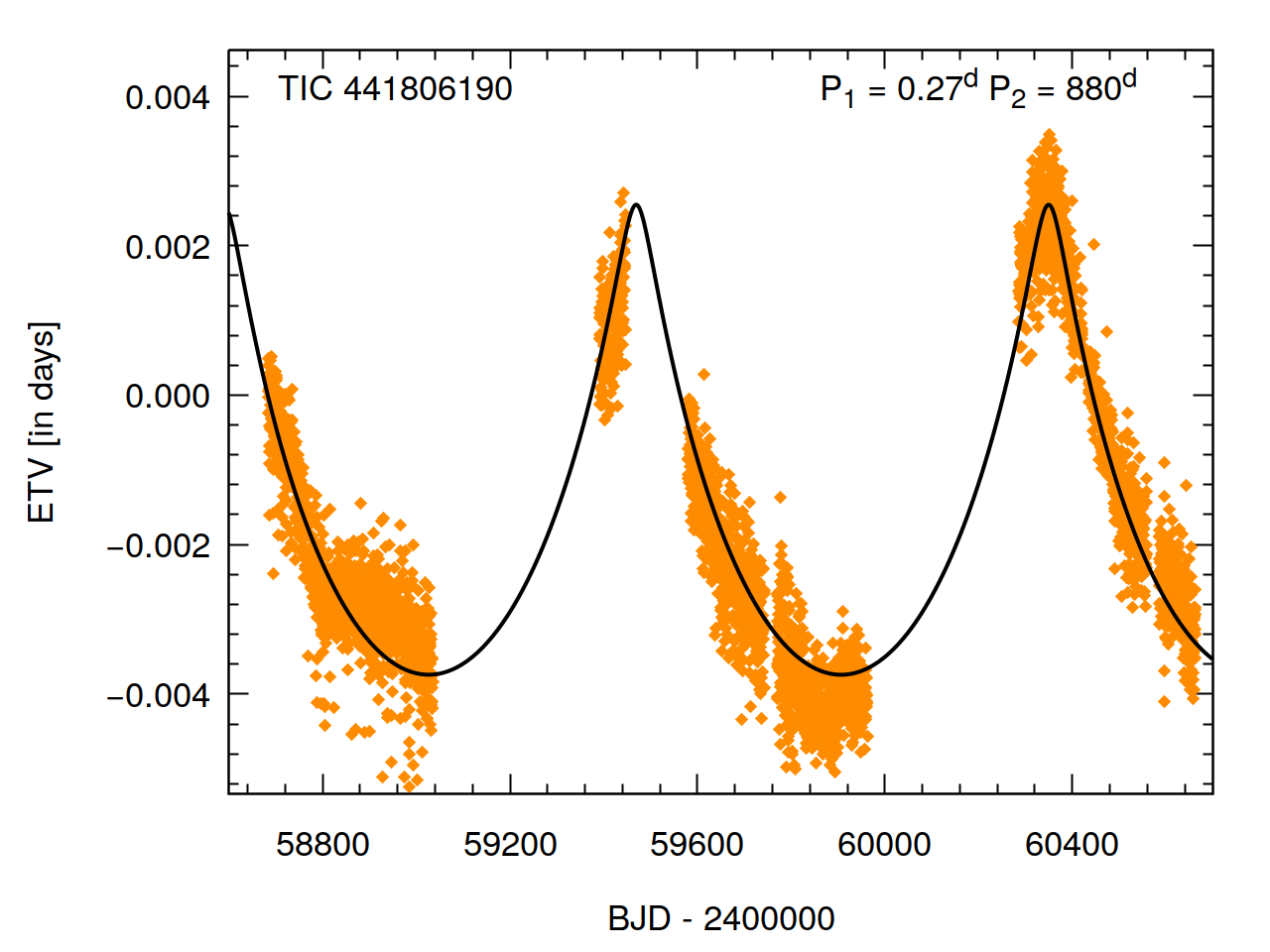}
\end{adjustwidth}
\caption{The last 5 of 35 triple star candidates for which the ETVs are ranked into the most certain (group $L_1$) pure LTTE systems. In all five cases, again, the averages of the primary and secondary ETVs are fitted. See Table~\ref{Tab:Orbelem_LTTE1} for further details.}
\label{Fig:ETVs_L1c}
\end{figure}

\begin{figure}[H]

\begin{adjustwidth}{-\extralength}{0cm}
\centering
\includegraphics[width=0.41\textwidth]{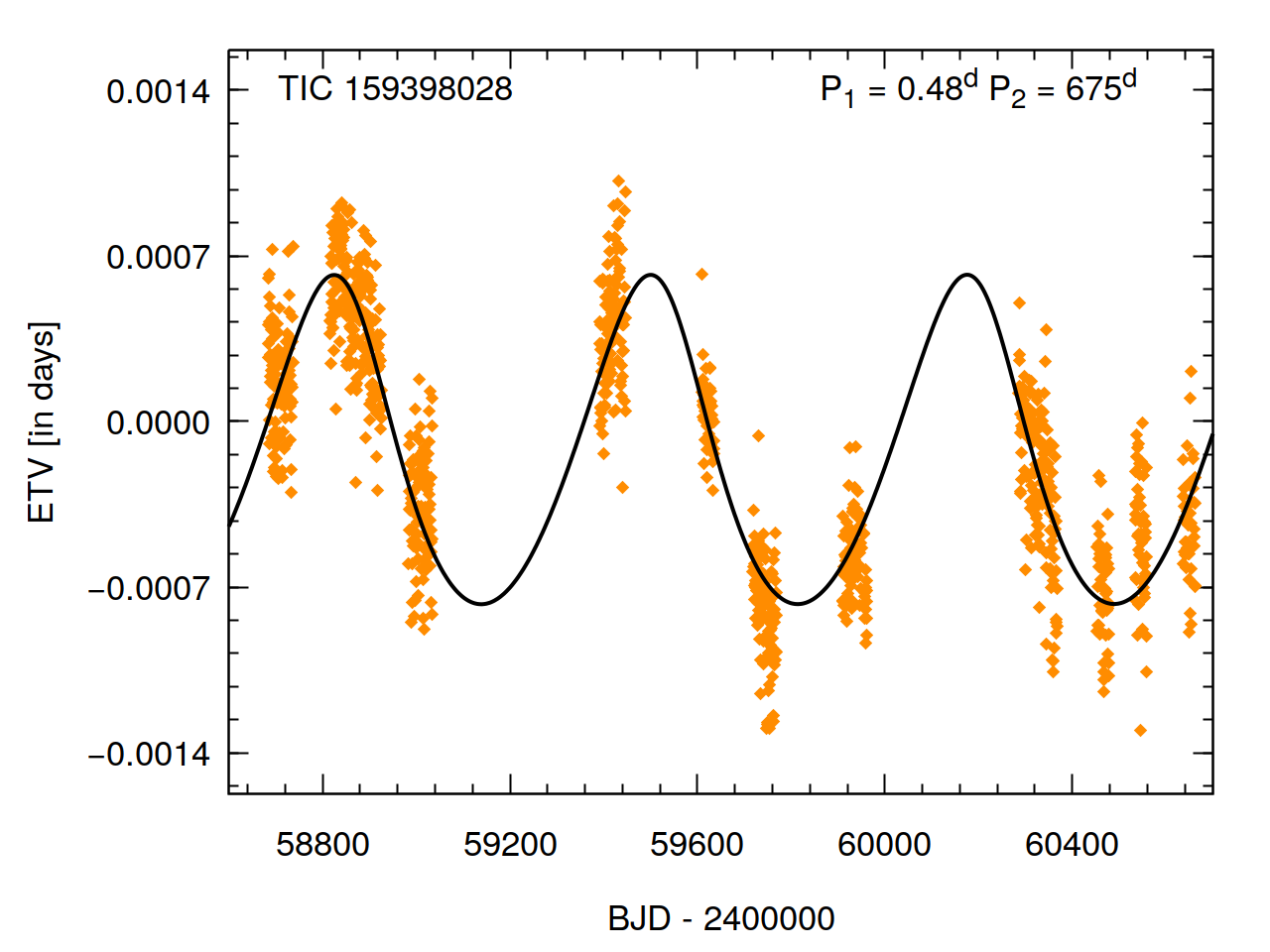}\includegraphics[width=0.41\textwidth]{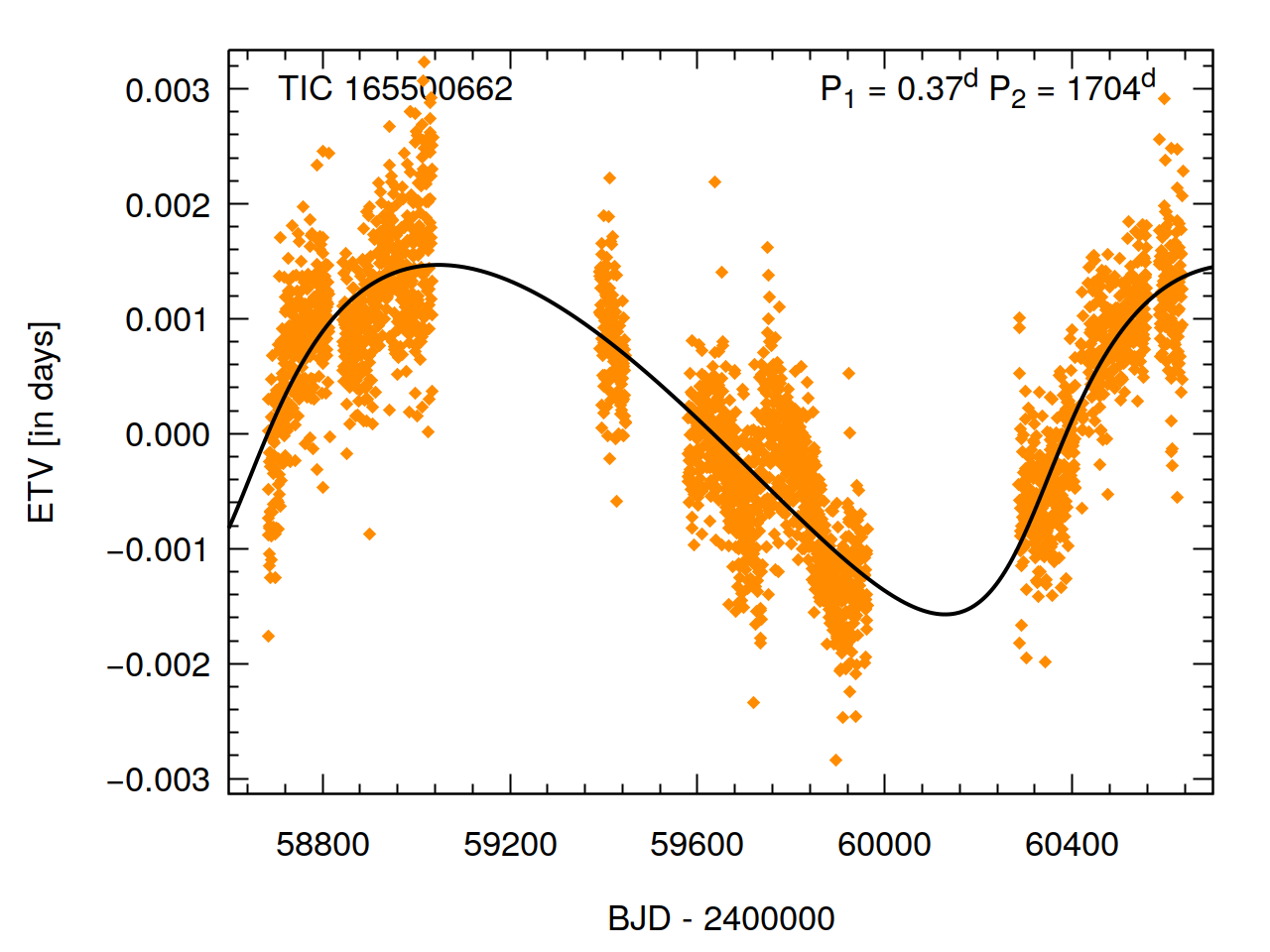}\includegraphics[width=0.41\textwidth]{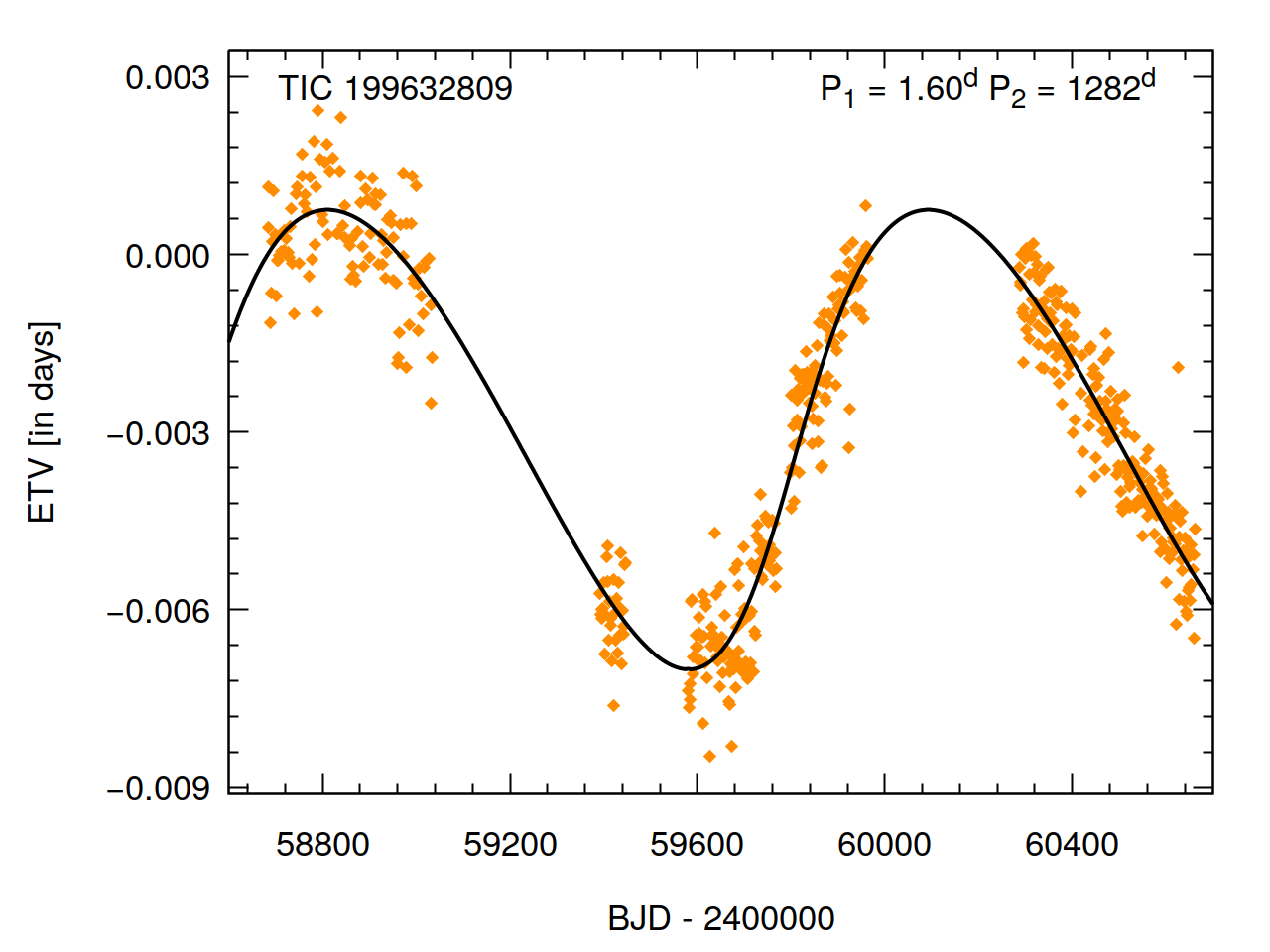}
\includegraphics[width=0.41\textwidth]{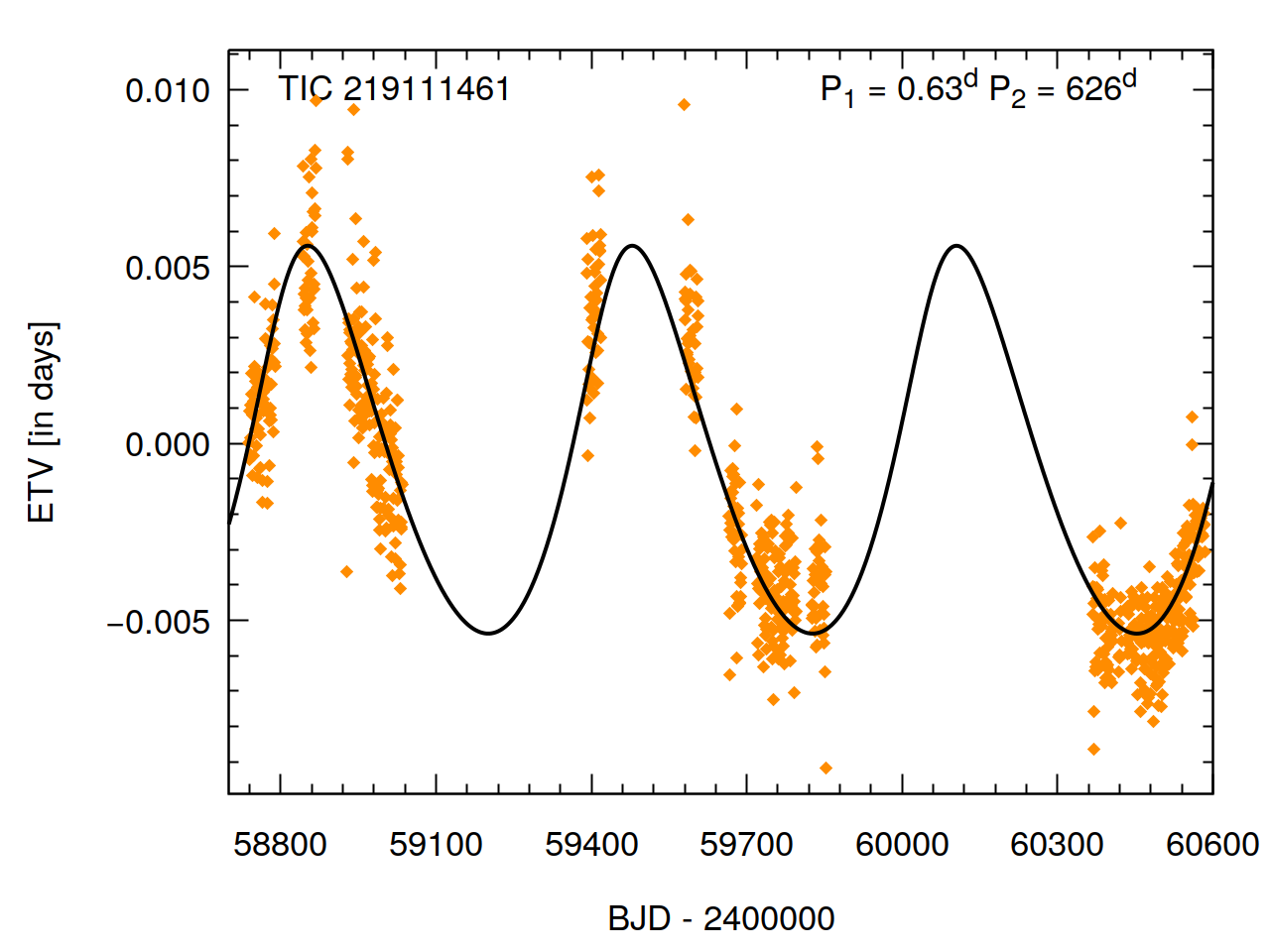}\includegraphics[width=0.41\textwidth]{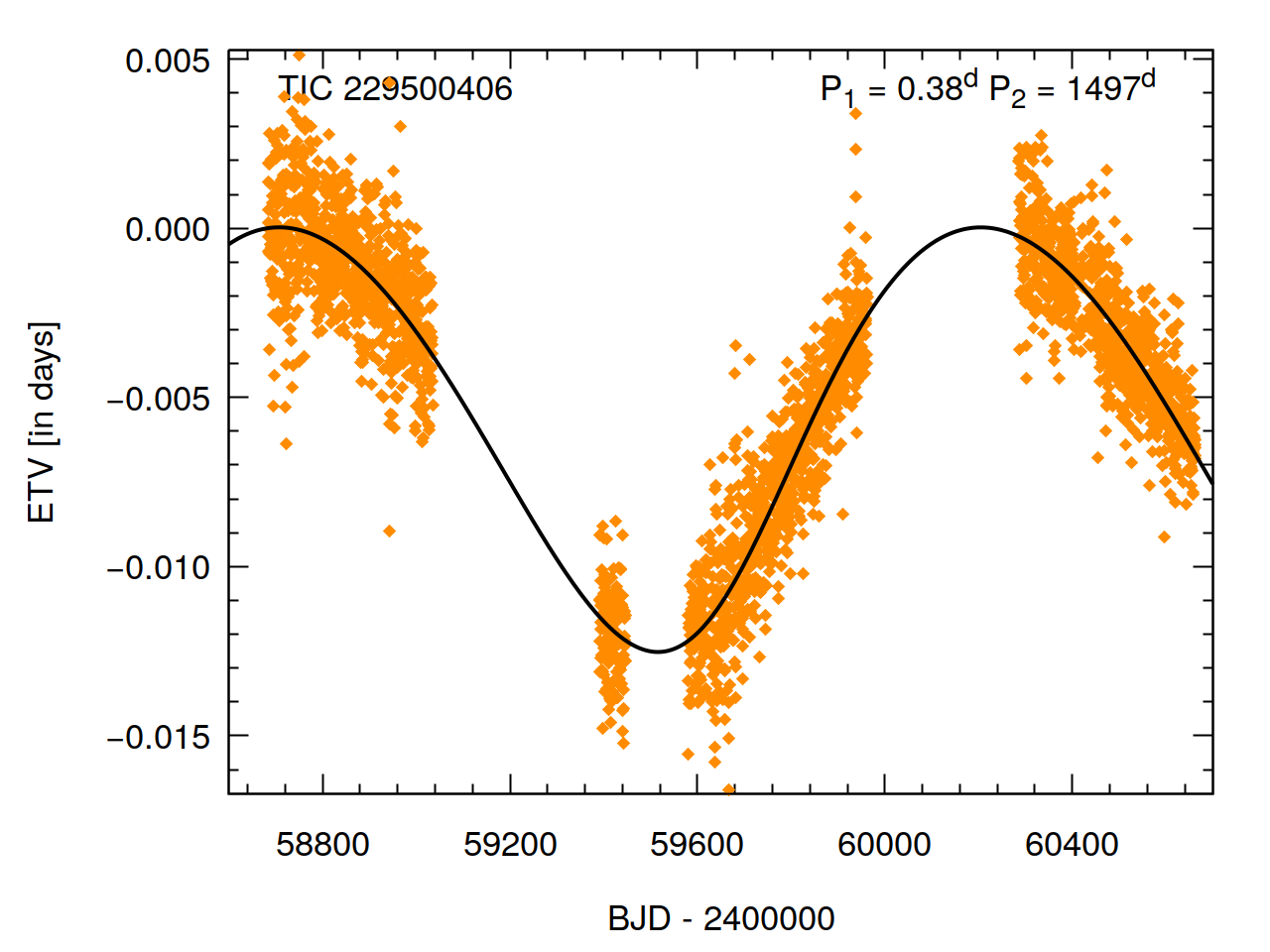}\includegraphics[width=0.41\textwidth]{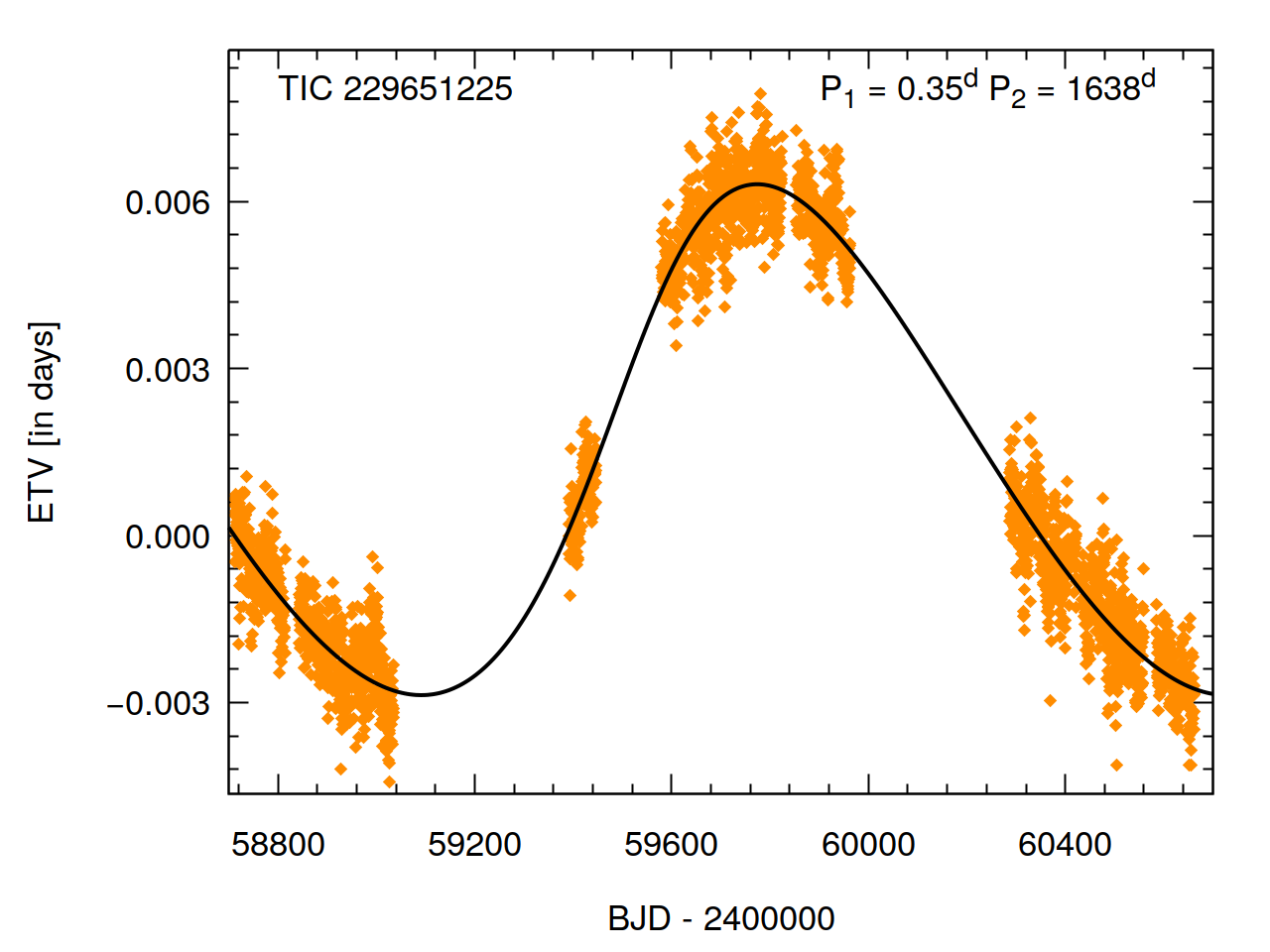}
\includegraphics[width=0.41\textwidth]{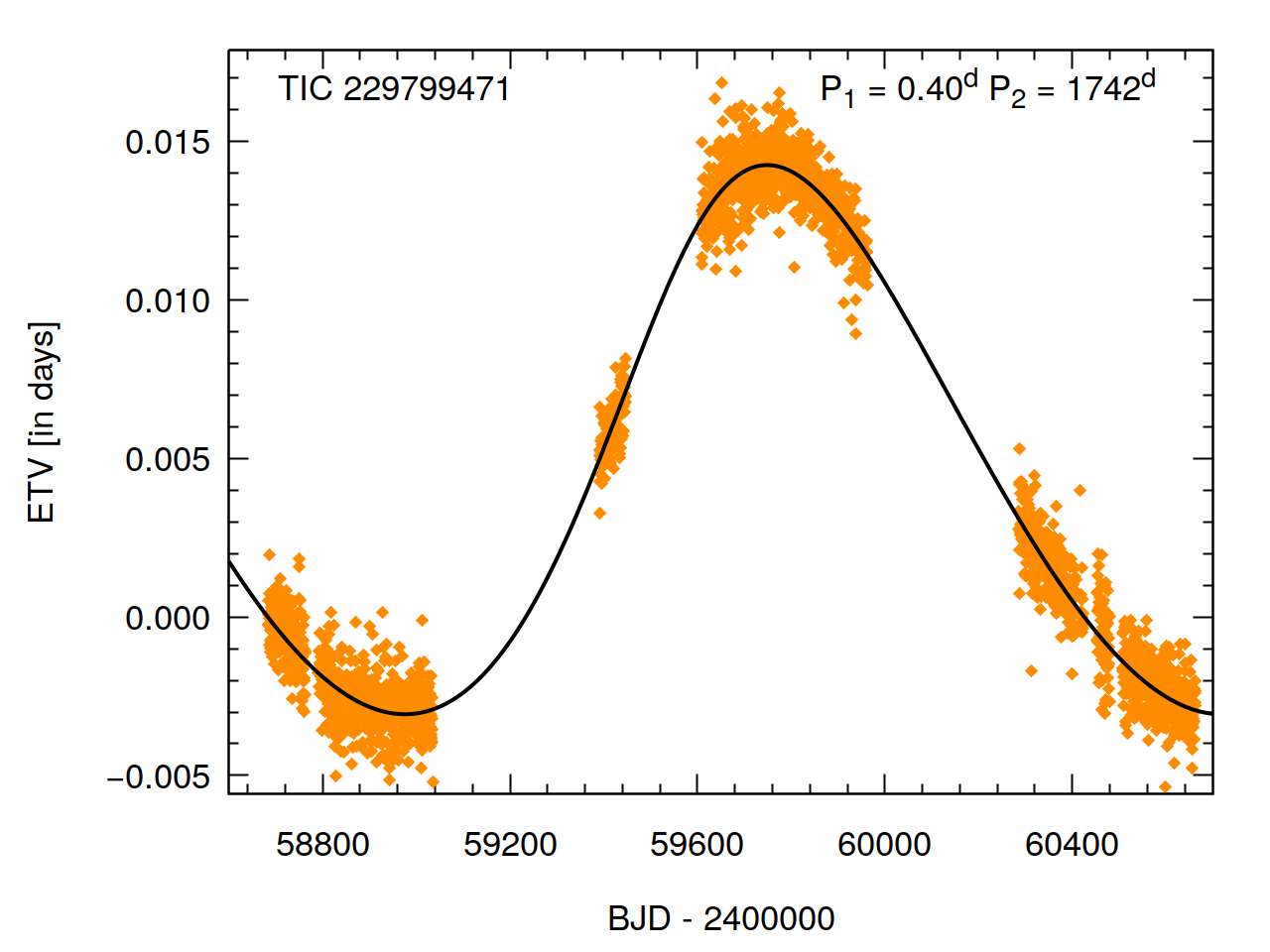}\includegraphics[width=0.41\textwidth]{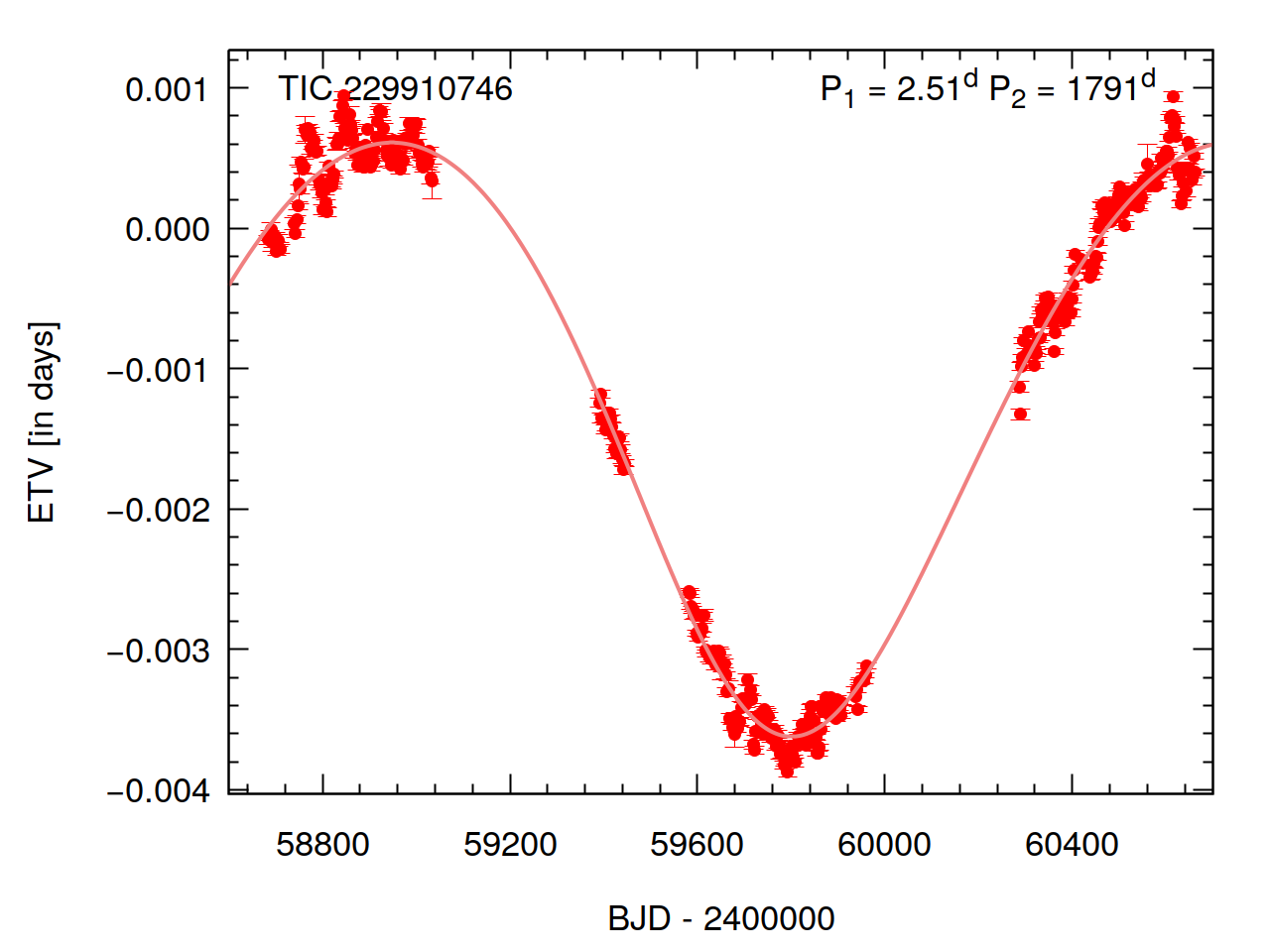}\includegraphics[width=0.41\textwidth]{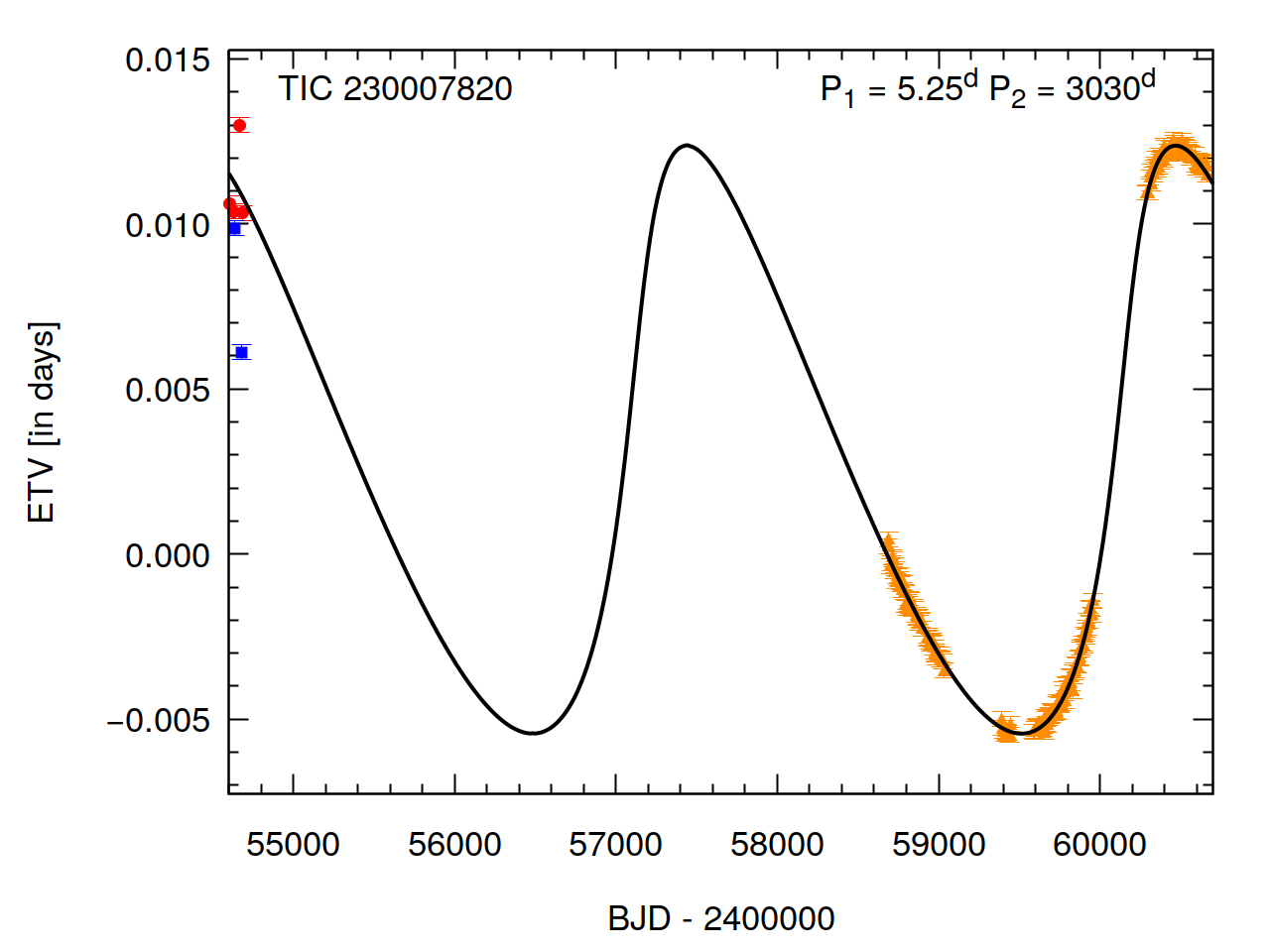}
\includegraphics[width=0.41\textwidth]{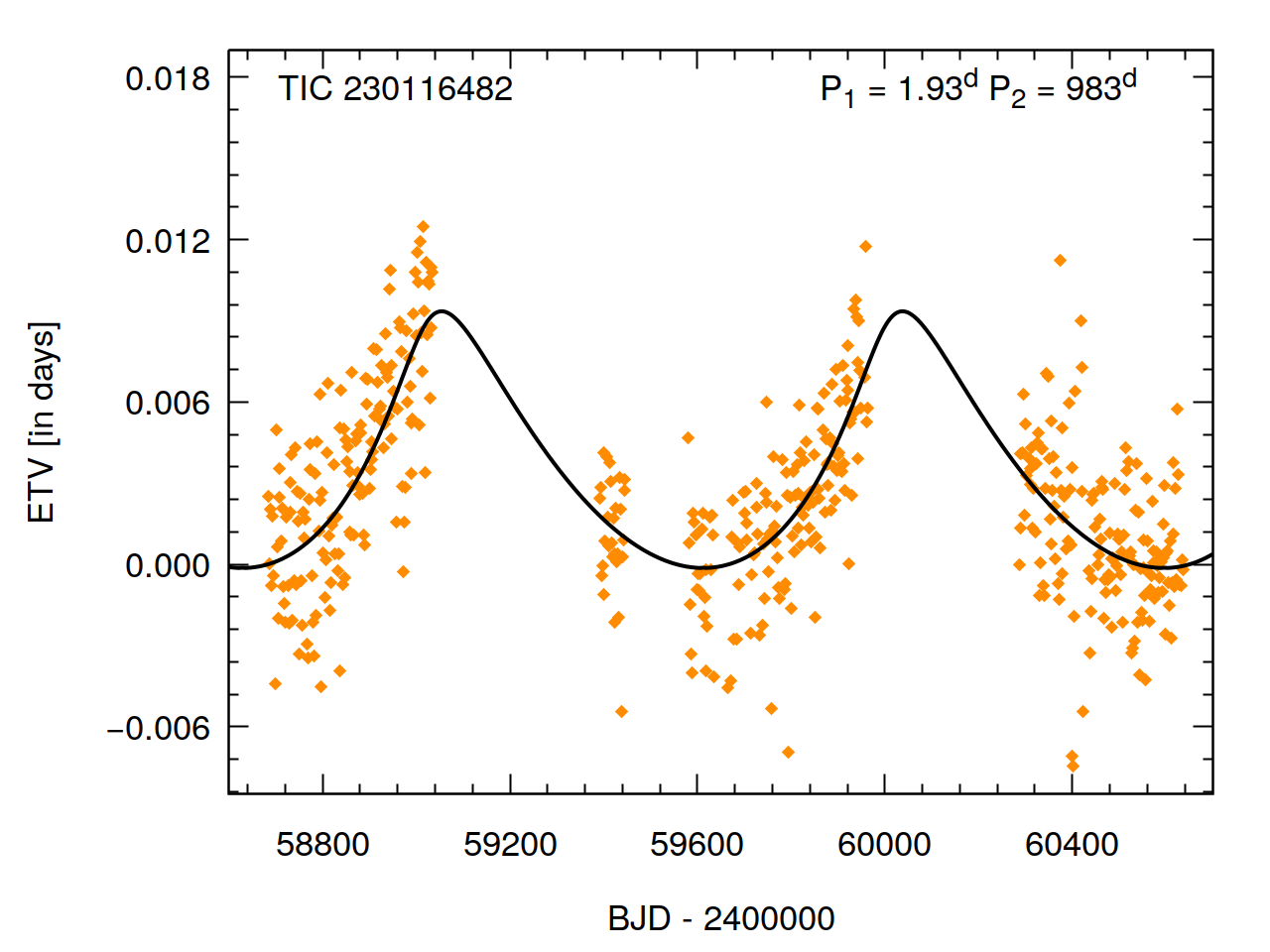}\includegraphics[width=0.41\textwidth]{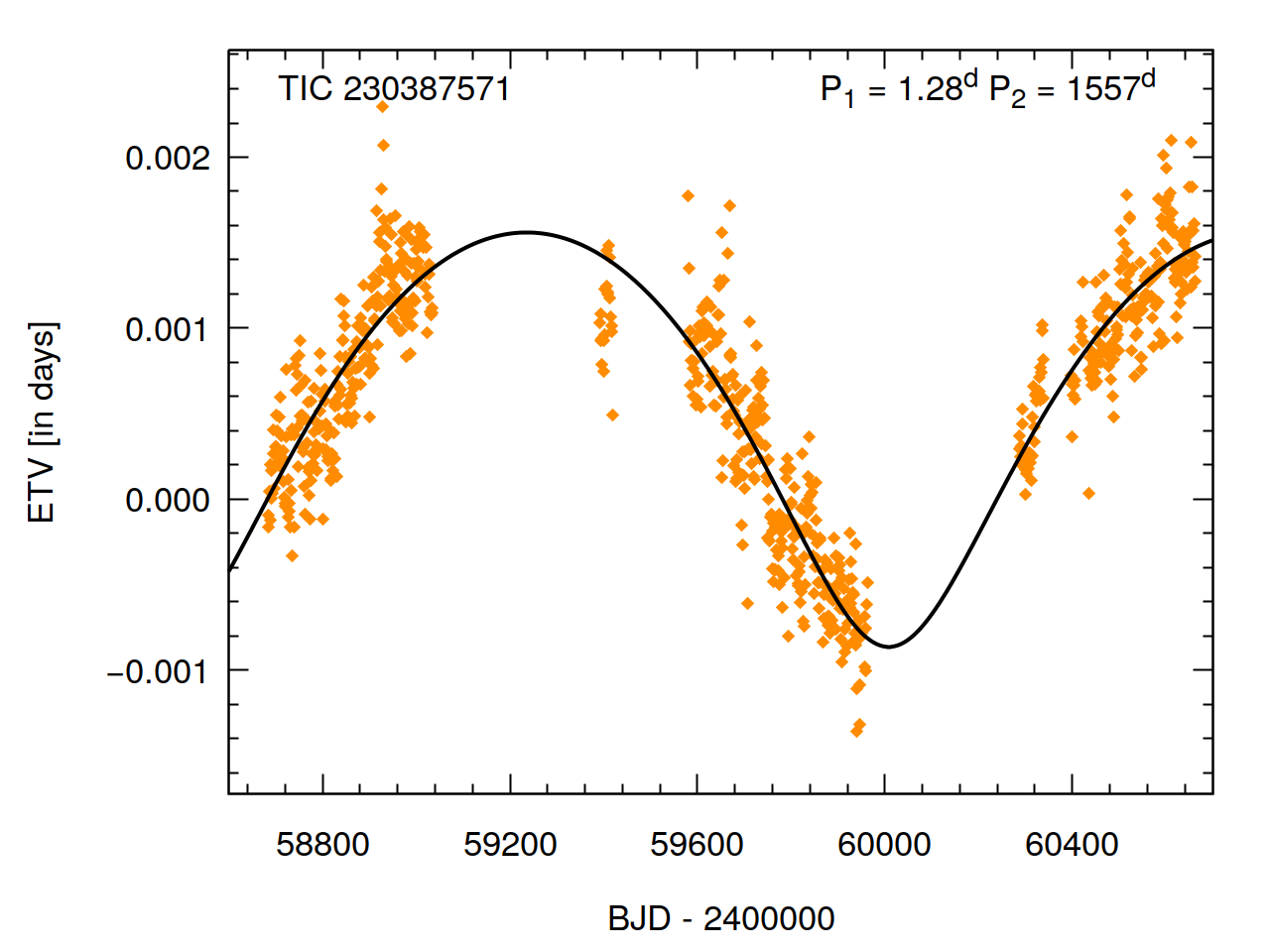}\includegraphics[width=0.41\textwidth]{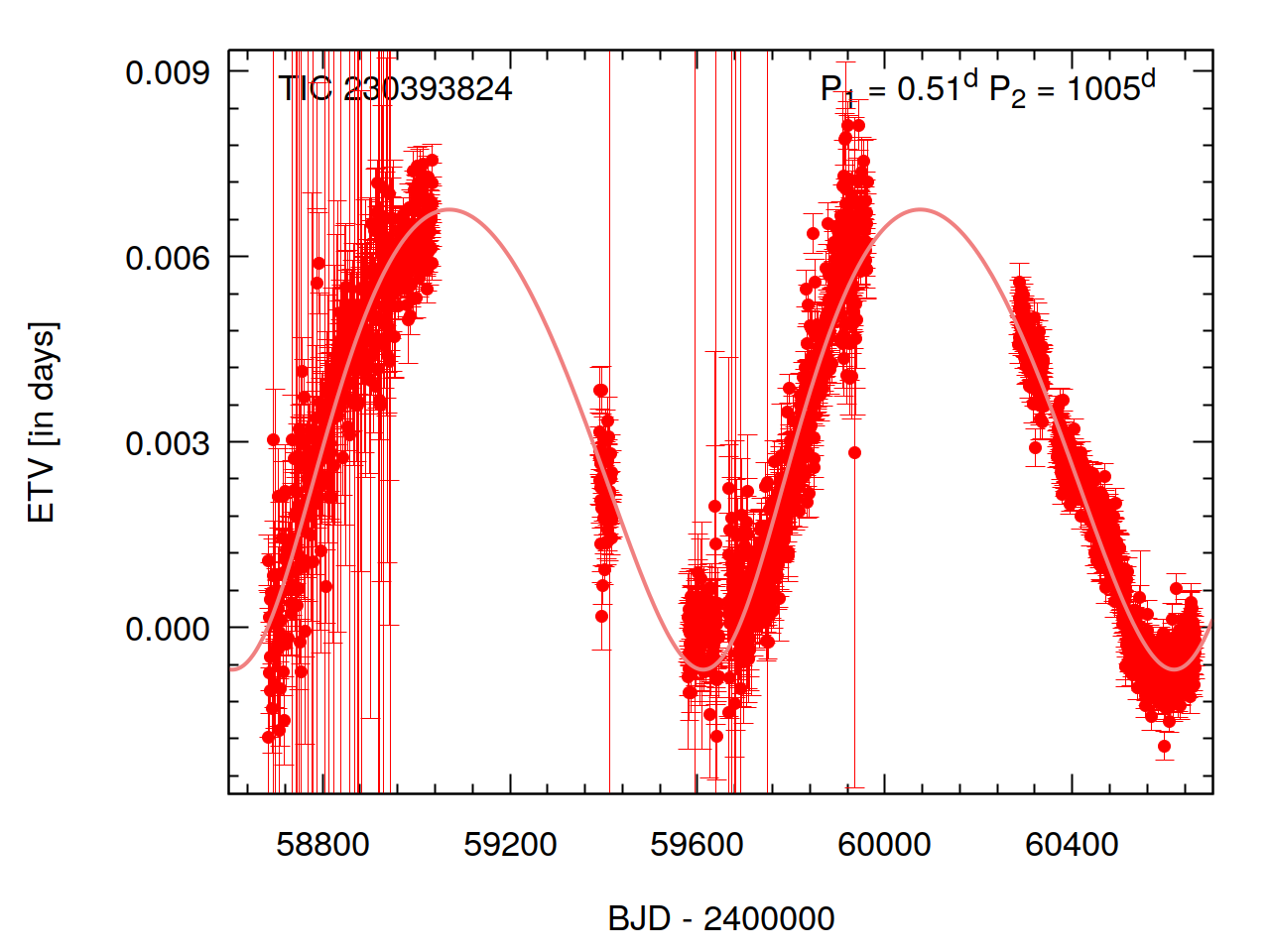}
\includegraphics[width=0.41\textwidth]{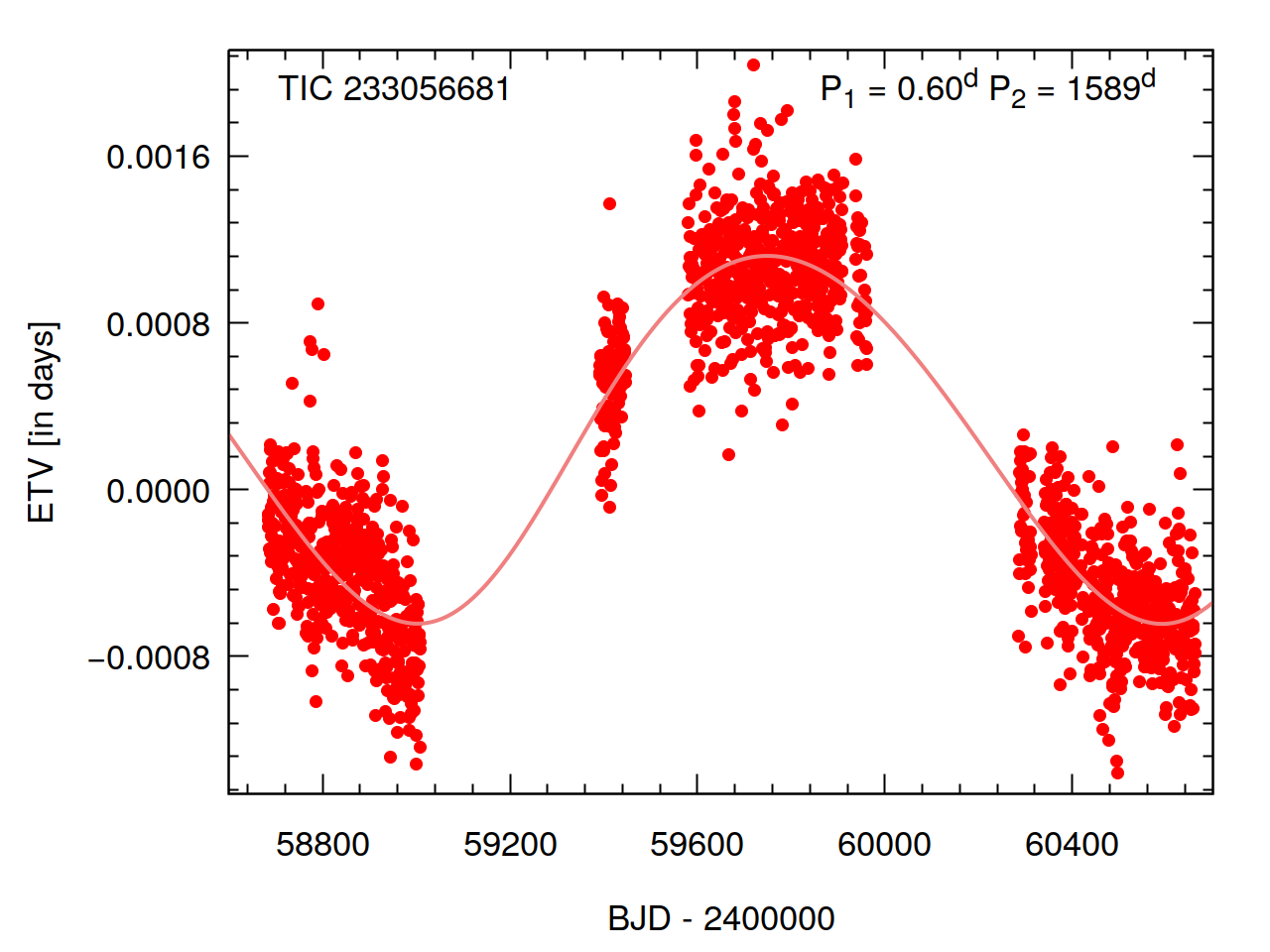}\includegraphics[width=0.41\textwidth]{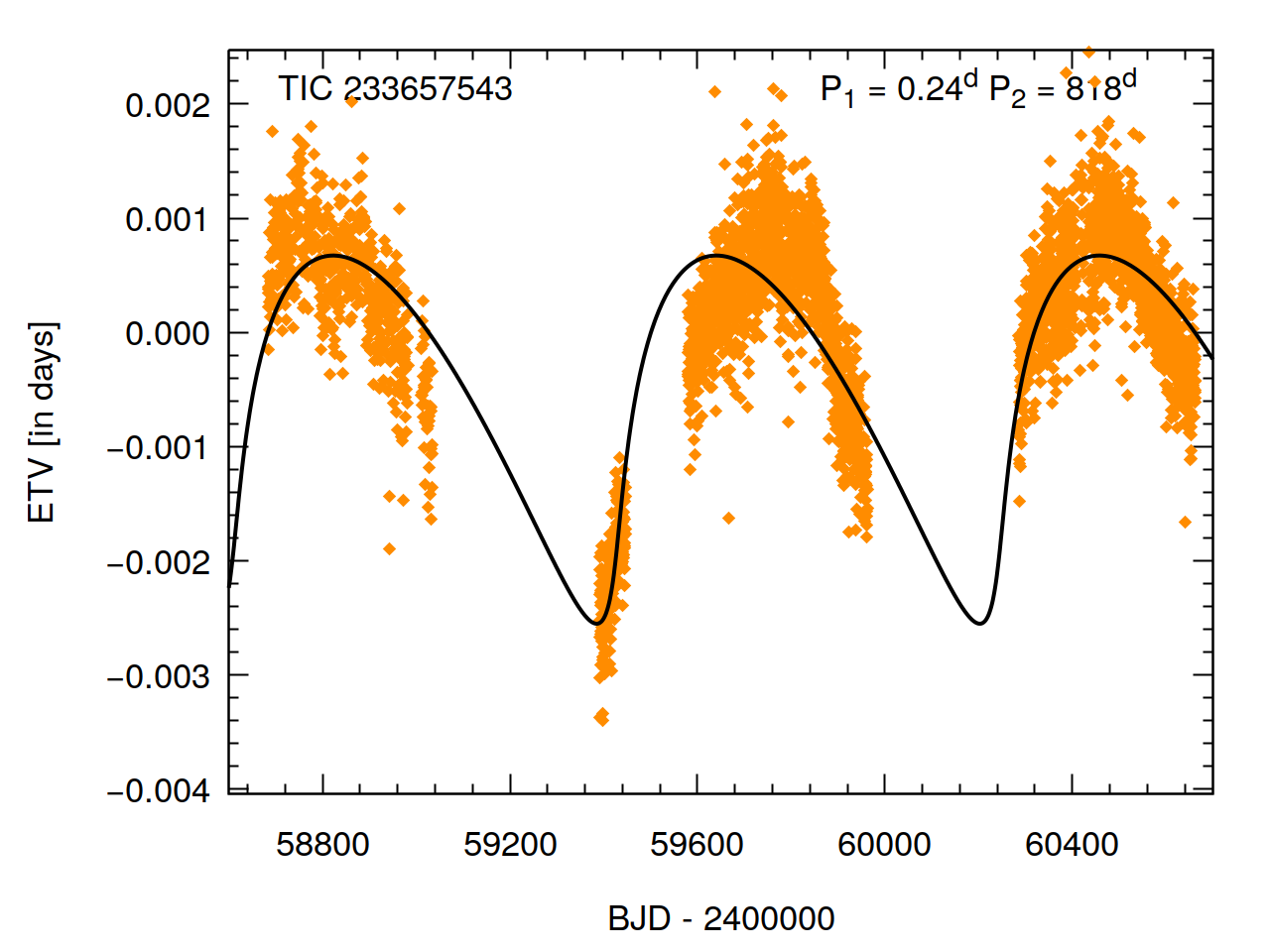}\includegraphics[width=0.41\textwidth]{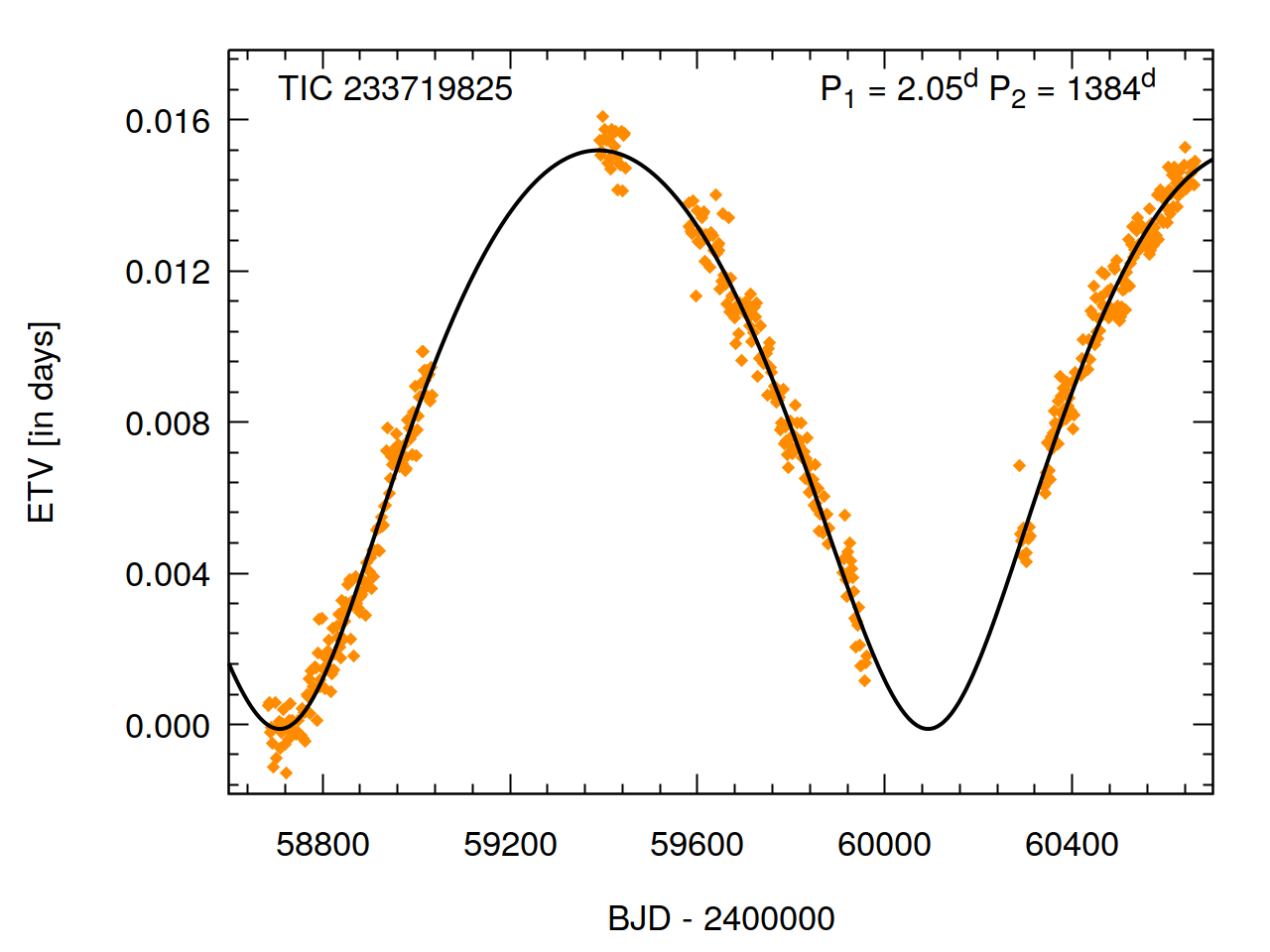}
\end{adjustwidth}
\caption{The first 15 ETVs of those third body candidates which are ranked into the not-so-certain, but likely (group $L_2$), pure LTTE third body solutions. Orange diamonds, red circles, and the smooth curves, as well as the error bars, have the same meanings as were explained in the former figure captions. In the case of the ETV plot for TIC 230007820 (right panel in the third row), some earlier ground-based ETV points are also plotted, together with error bars. Here, red upward triangles represent primary minima, while blue downward triangles stand for times of secondary minima. For further details, see Table~\ref{Tab:Orbelem_LTTE2}.}
\label{Fig:ETVs_L2a}
\end{figure}

\begin{figure}[H]
\begin{adjustwidth}{-\extralength}{0cm}
\centering
\includegraphics[width=0.44\textwidth]{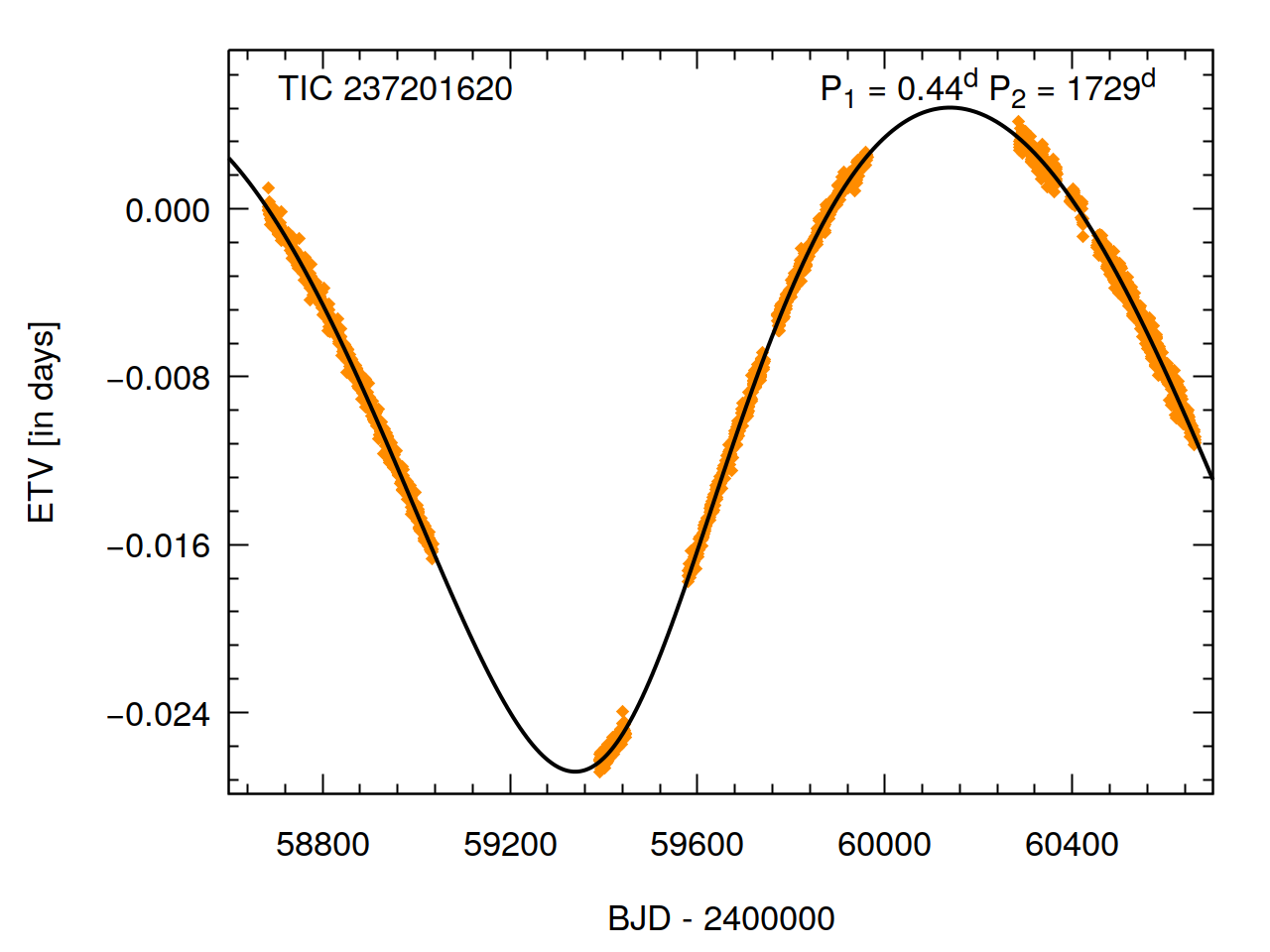}\includegraphics[width=0.44\textwidth]{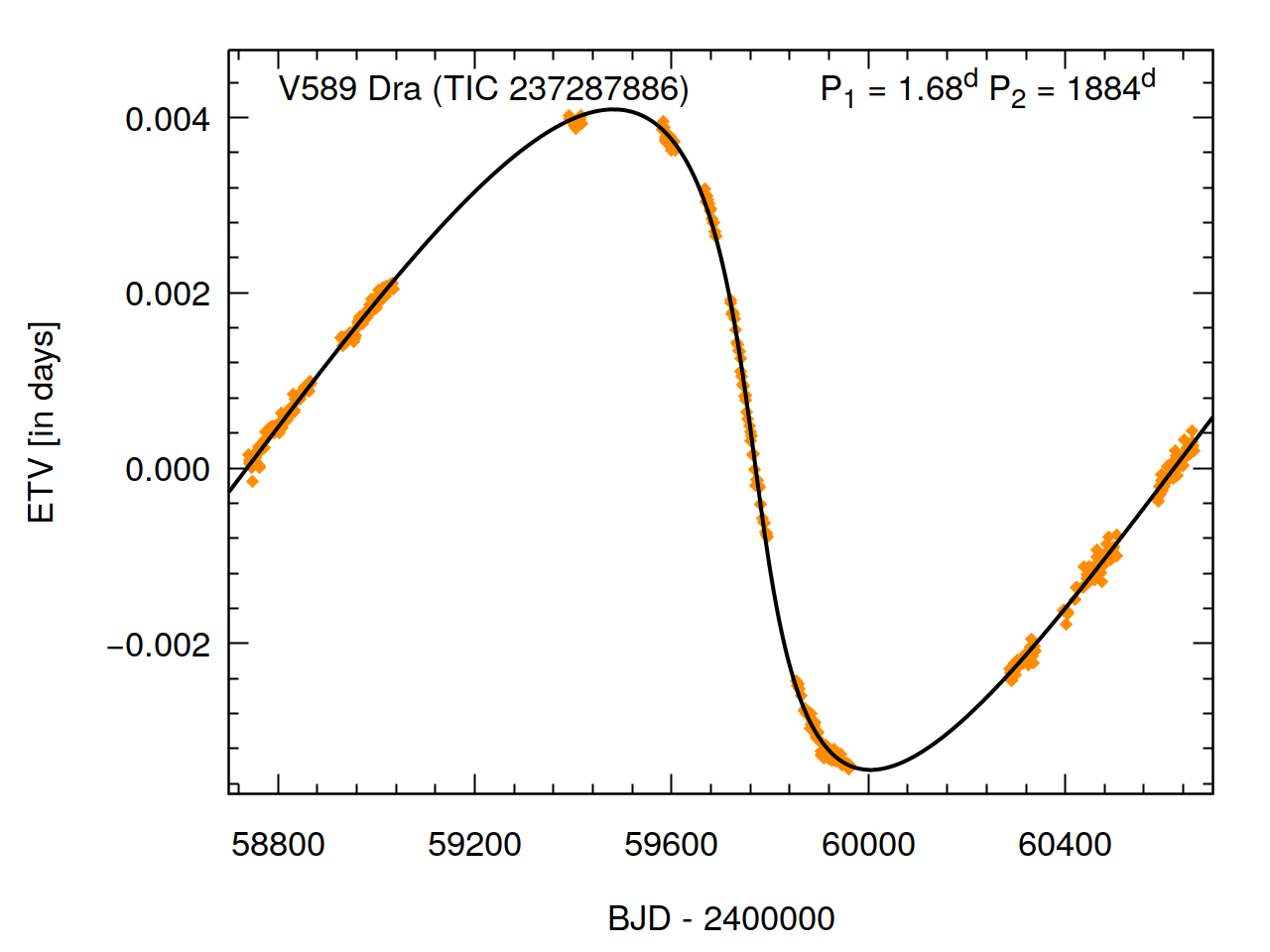}\includegraphics[width=0.44\textwidth]{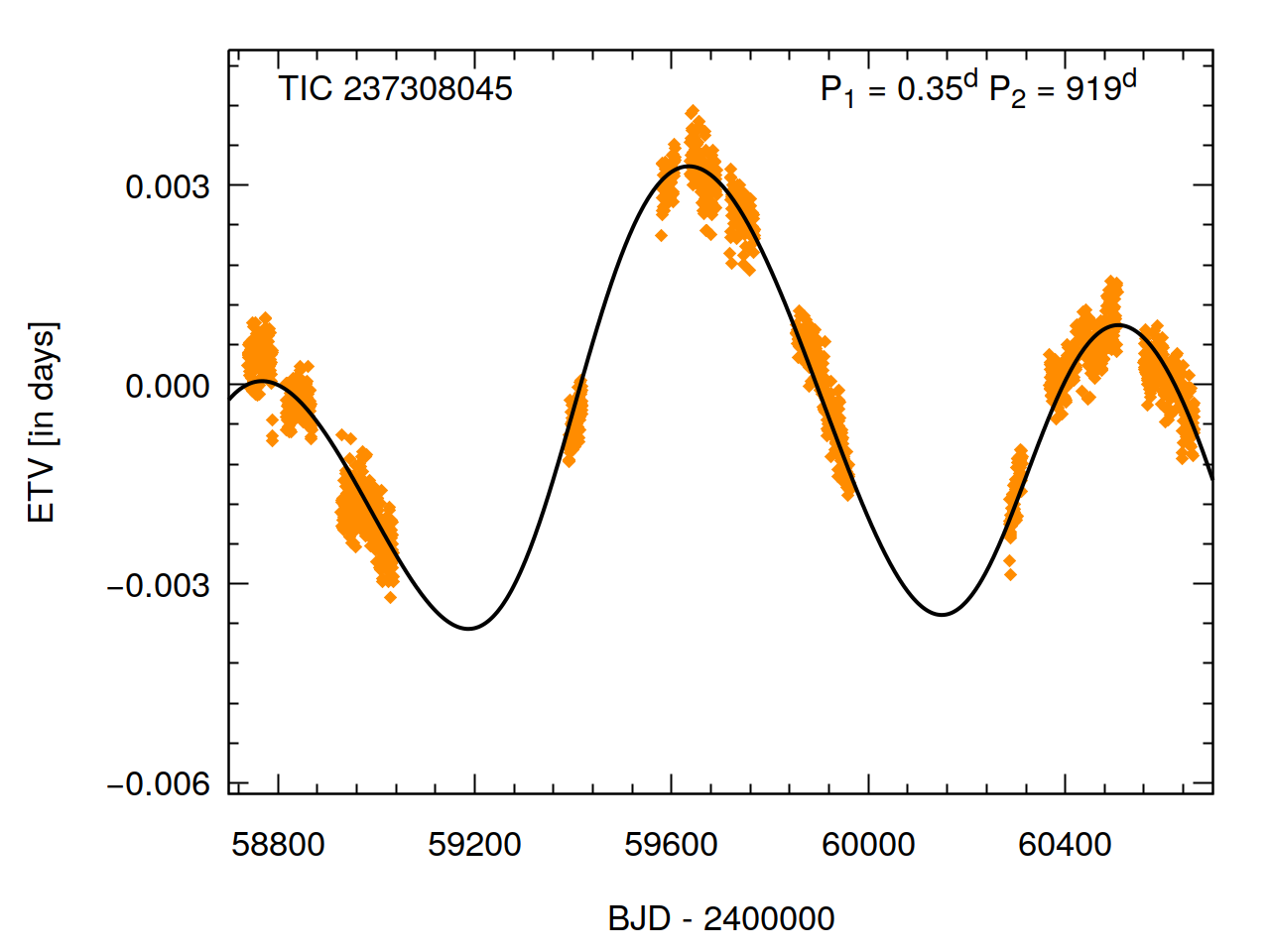}
\includegraphics[width=0.44\textwidth]{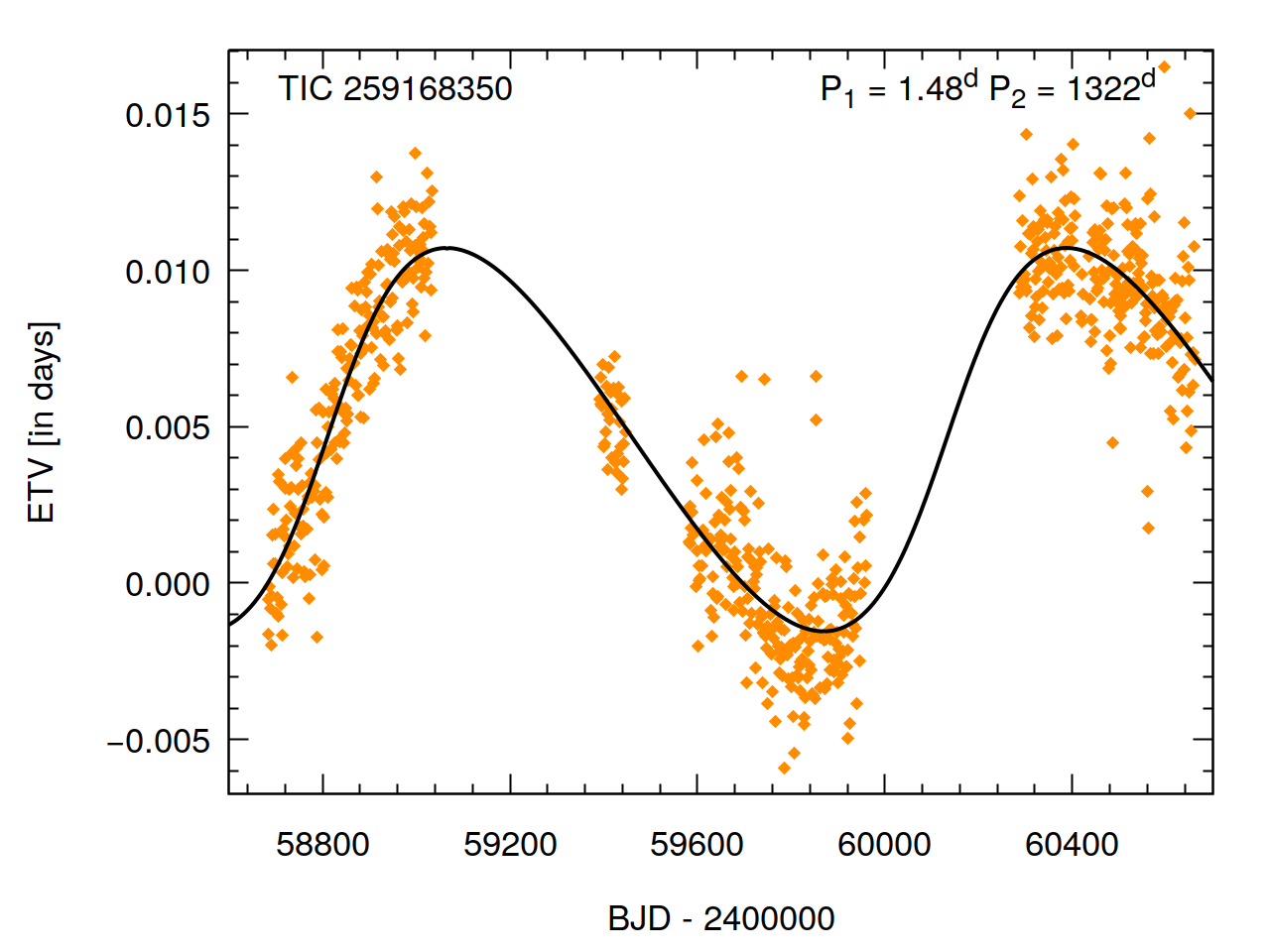}\includegraphics[width=0.44\textwidth]{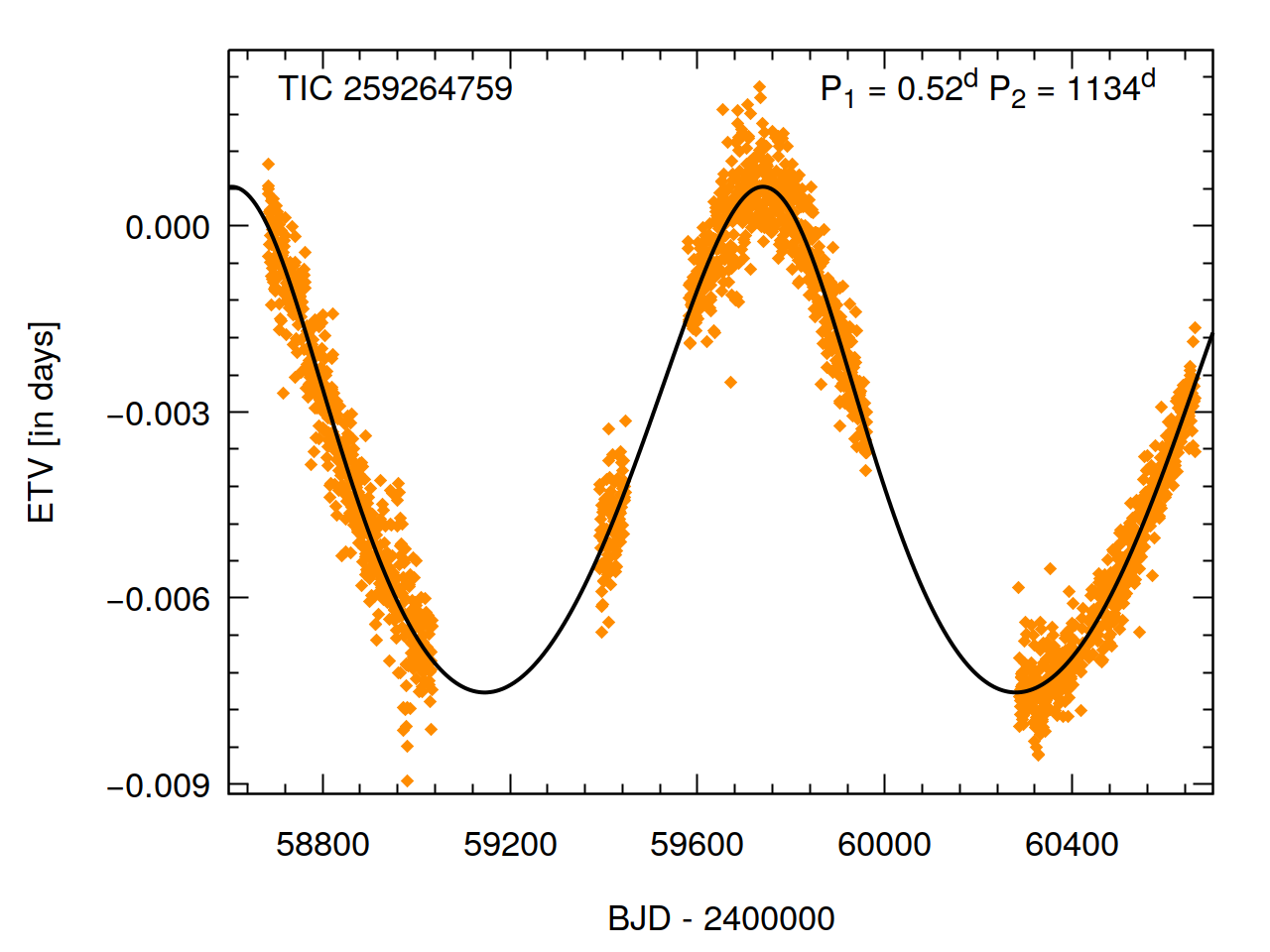}\includegraphics[width=0.44\textwidth]{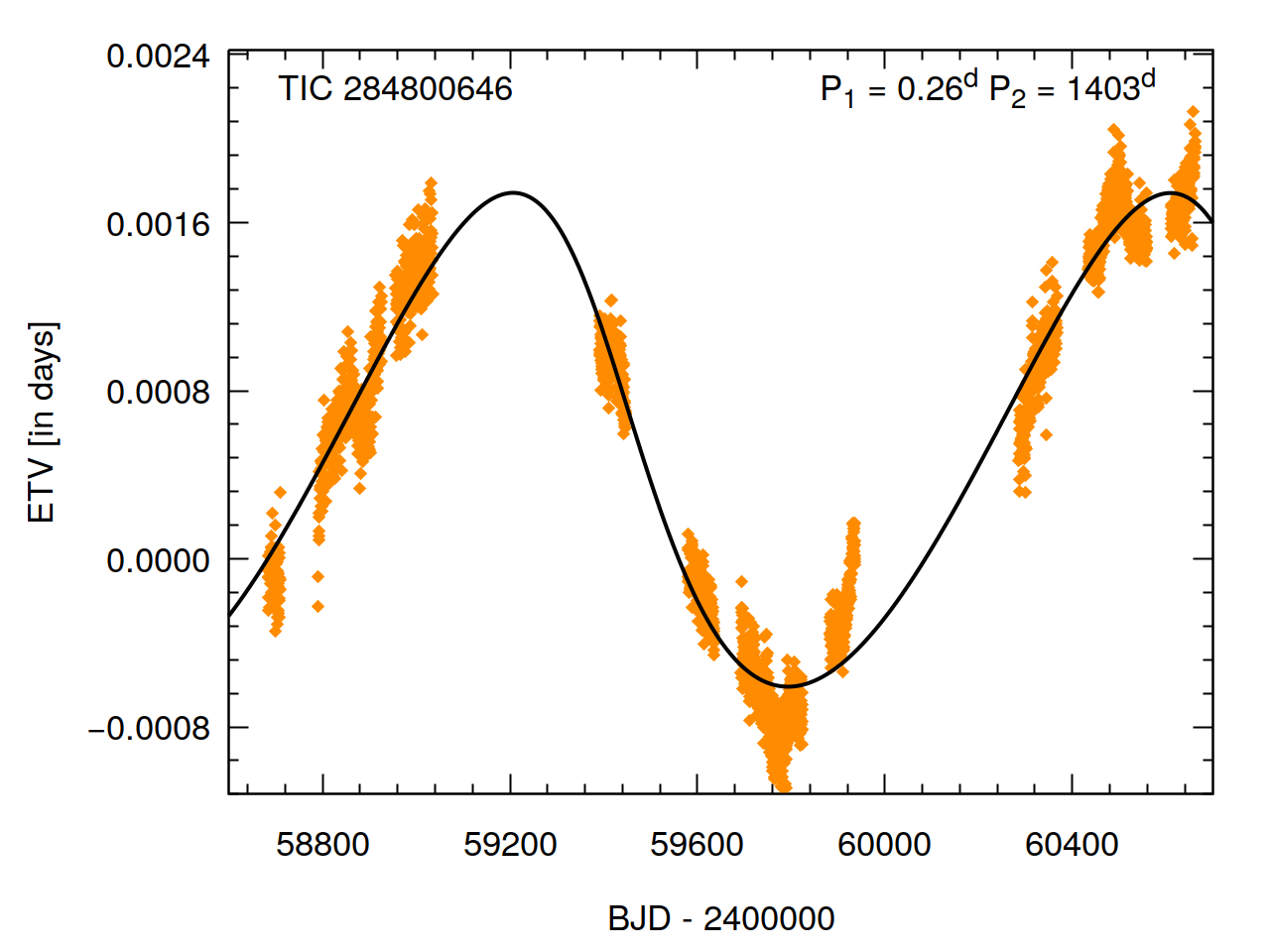}
\includegraphics[width=0.44\textwidth]{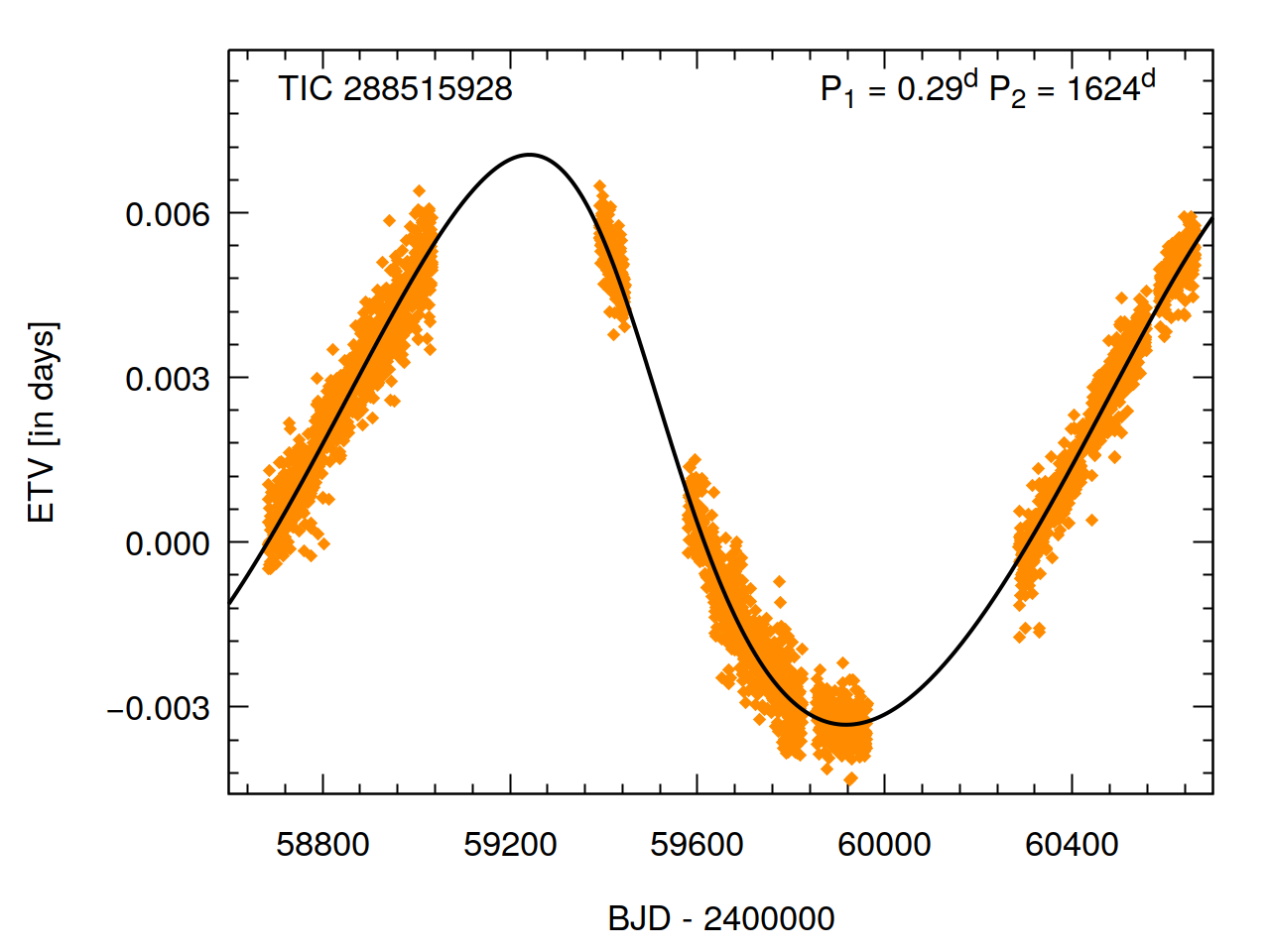}\includegraphics[width=0.44\textwidth]{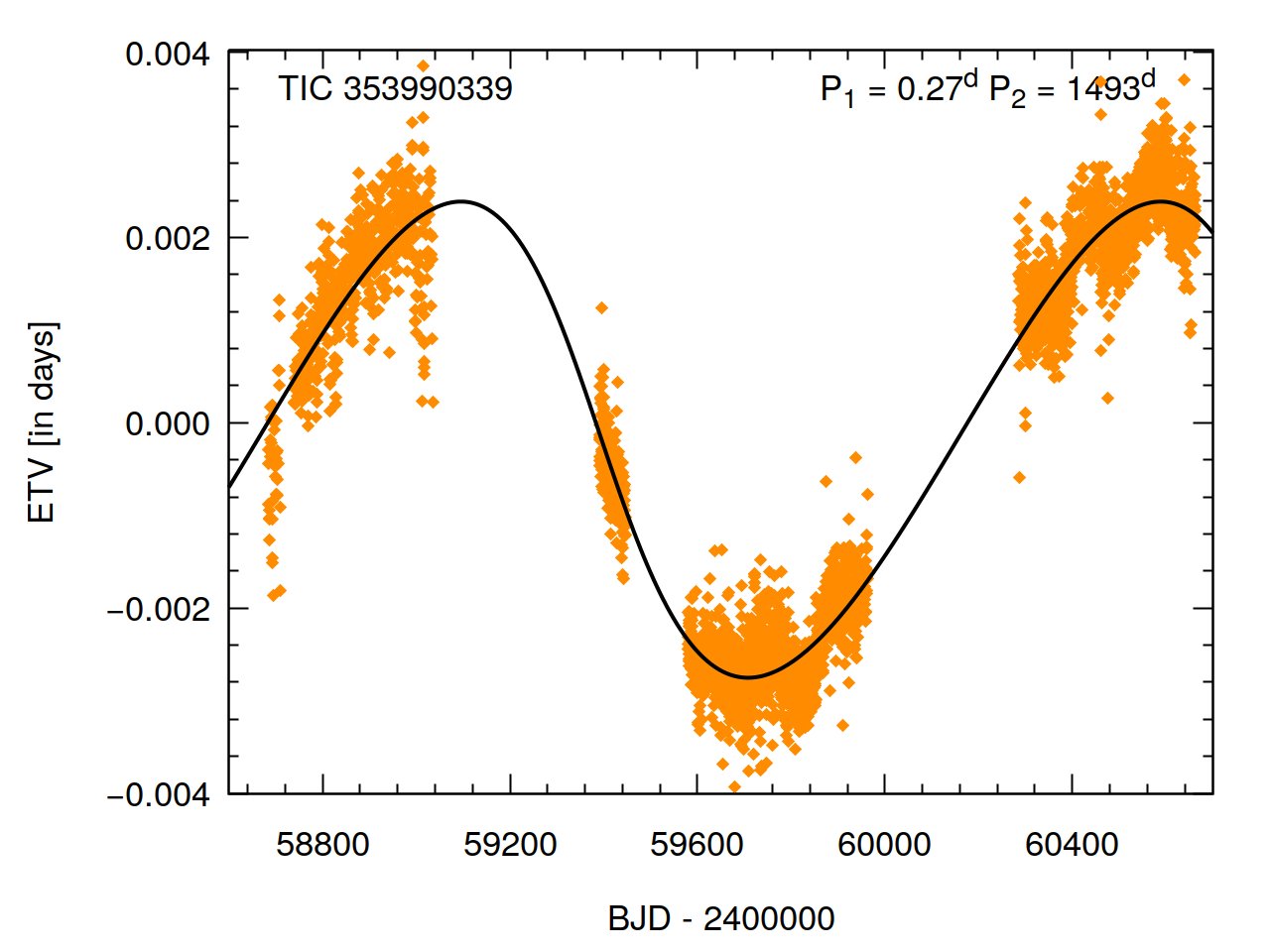}\includegraphics[width=0.44\textwidth]{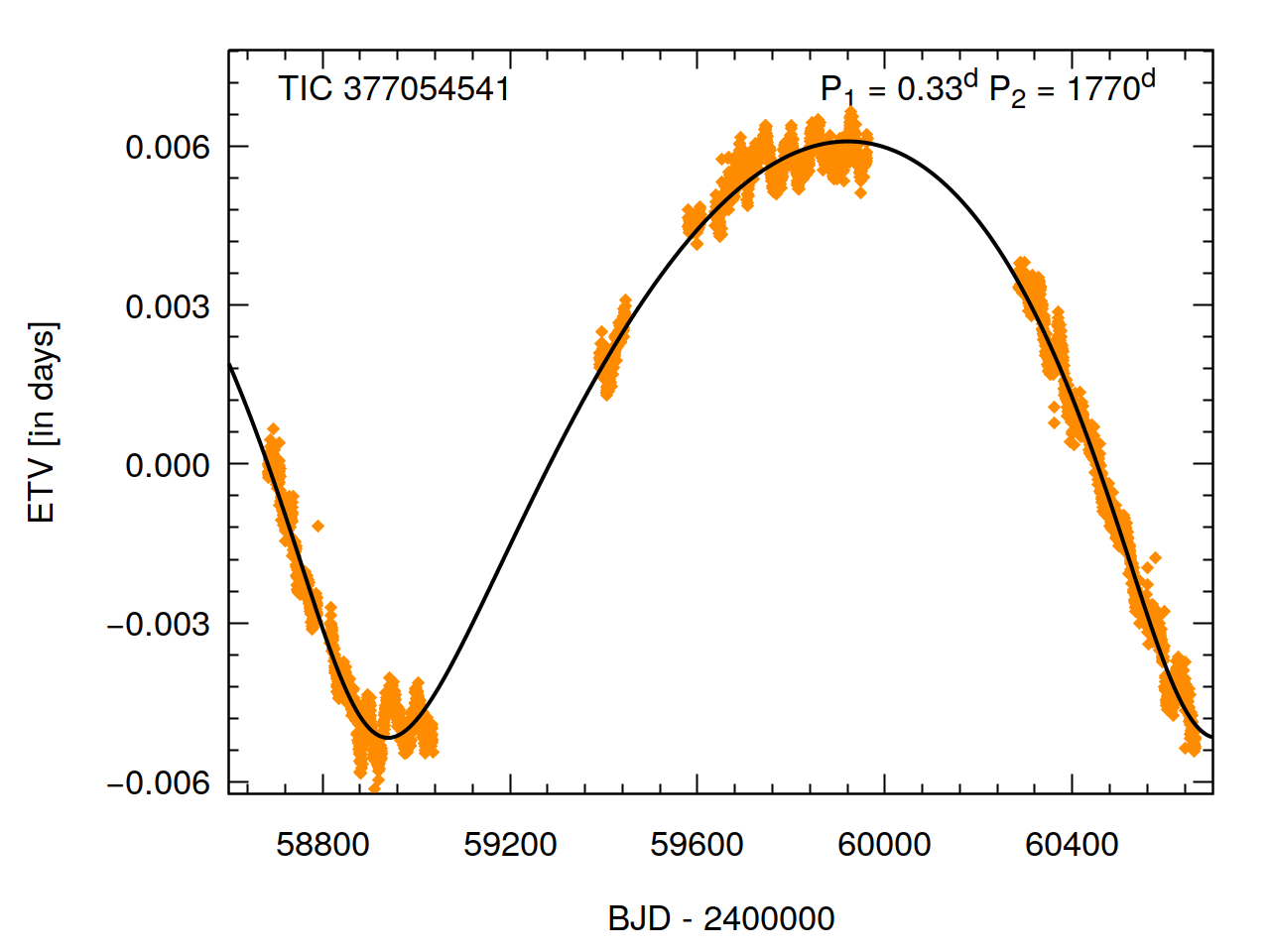}
\includegraphics[width=0.44\textwidth]{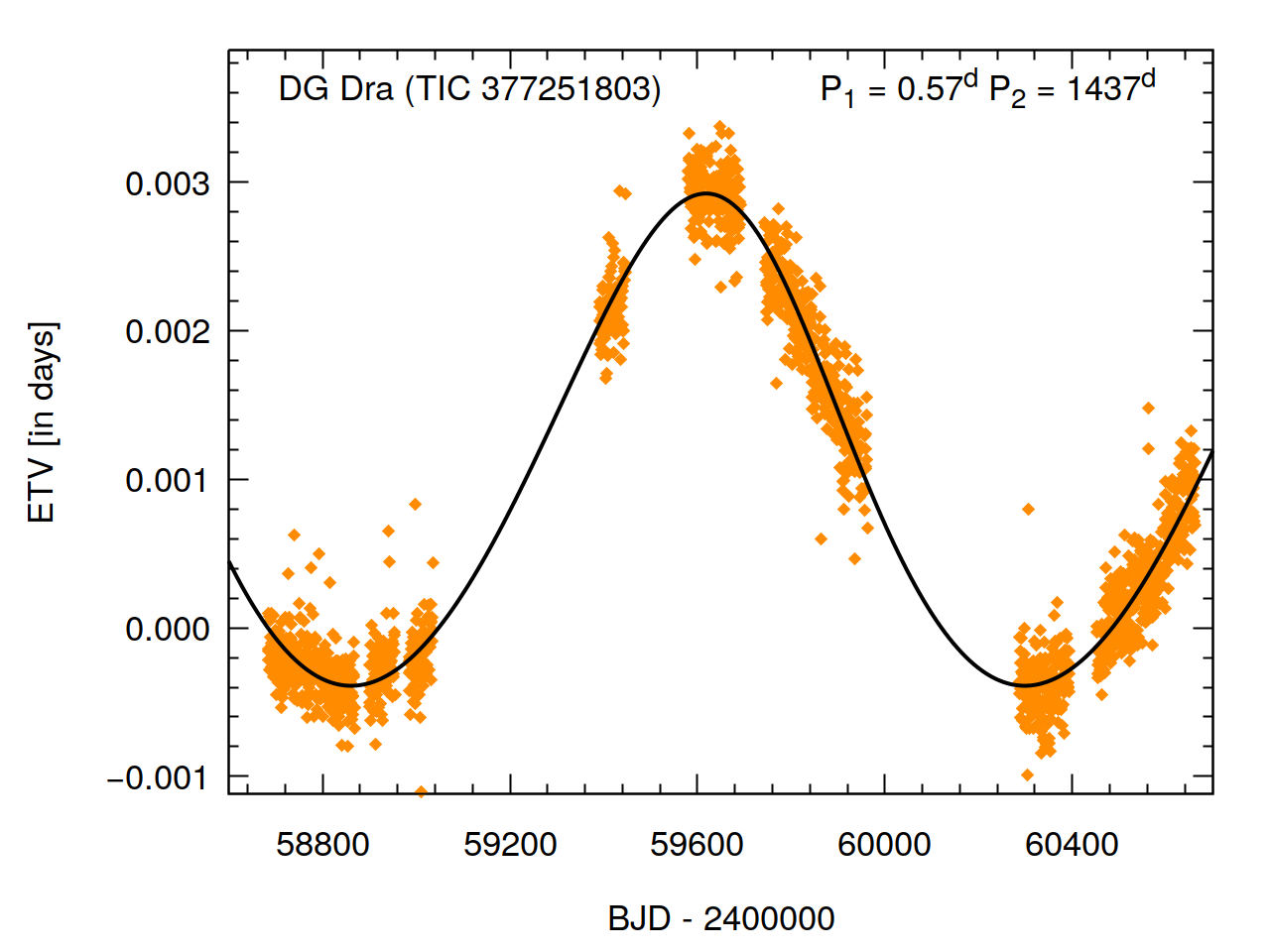}\includegraphics[width=0.44\textwidth]{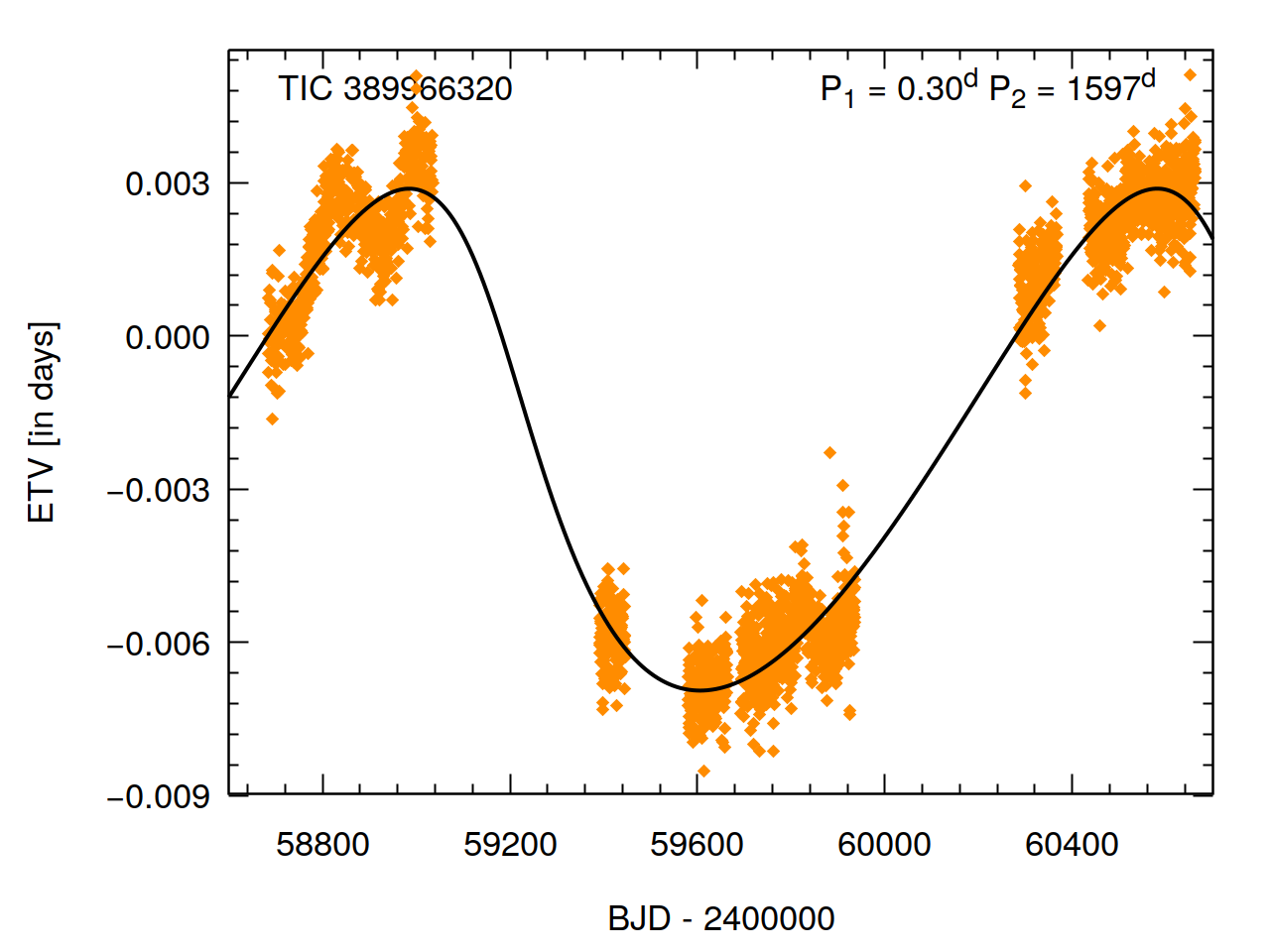}\includegraphics[width=0.44\textwidth]{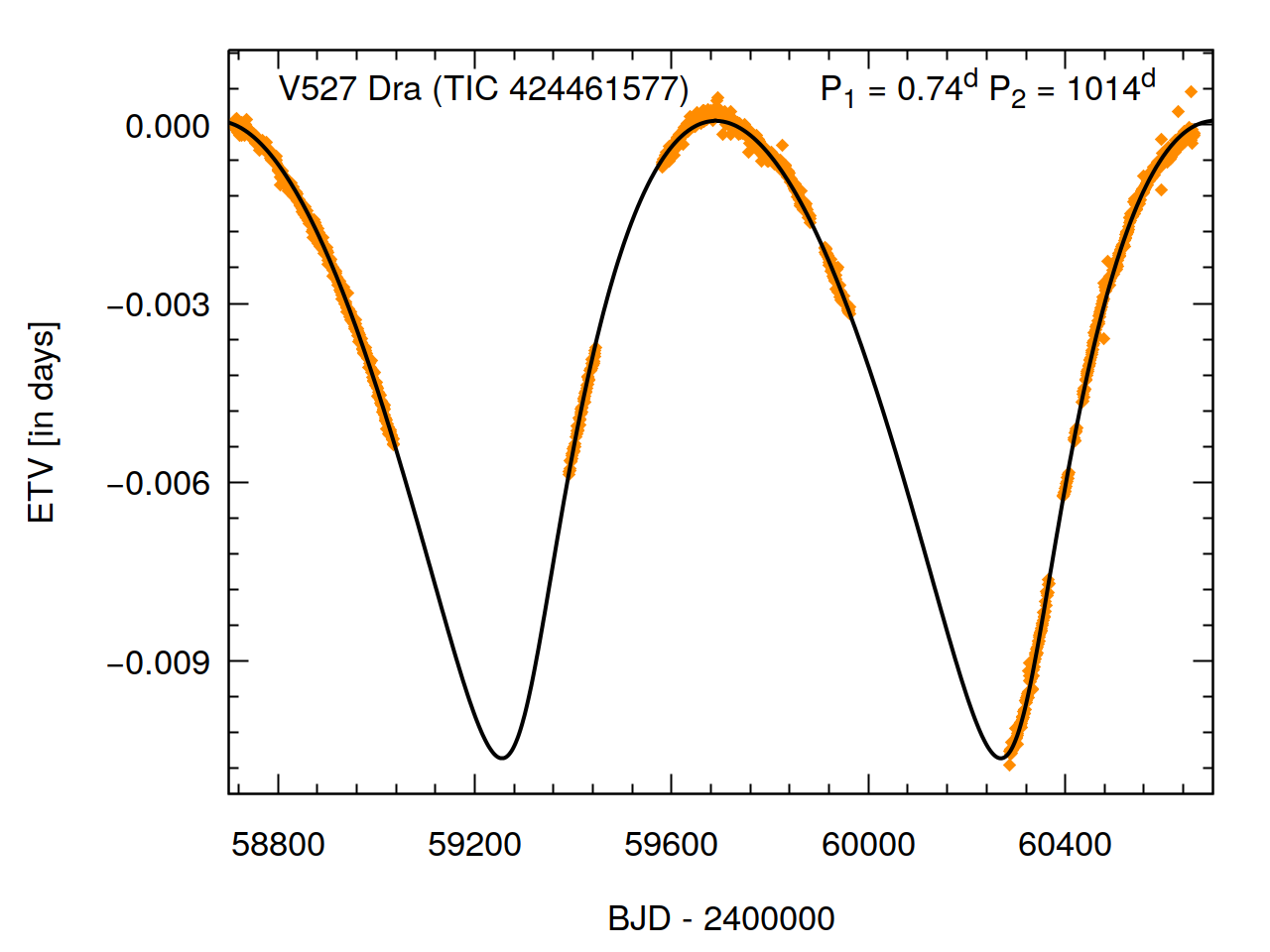}
\includegraphics[width=0.44\textwidth]{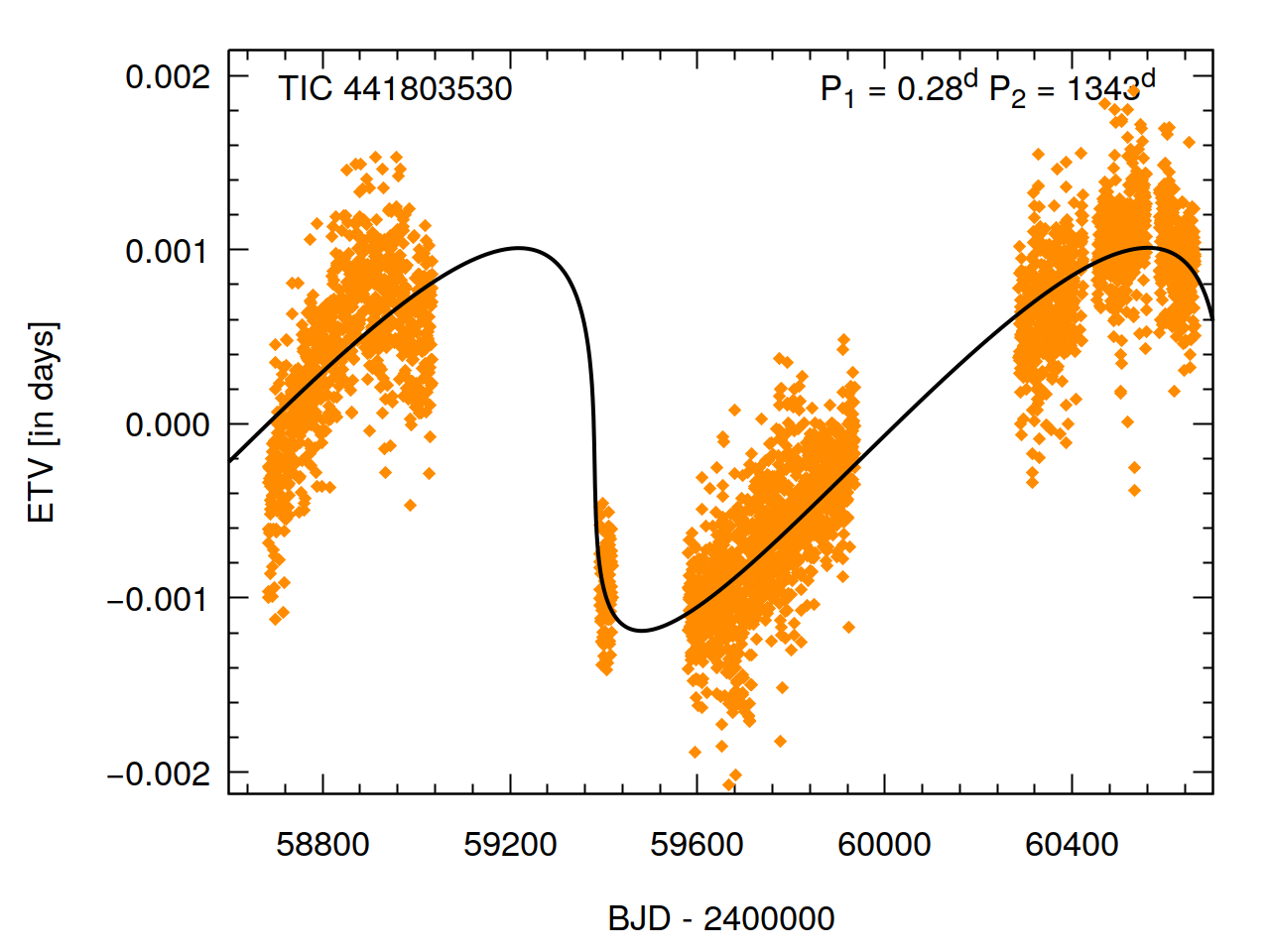}
\end{adjustwidth}
\caption{The averaged ETV points together with the pure LTTE solutions for the last 13 of 28 systems which are ranked into Group $L_2$. For further details, see Table~\ref{Tab:Orbelem_LTTE2}.}
\label{Fig:ETVs_L2b}
\end{figure}

\begin{figure}[H]
\begin{adjustwidth}{-\extralength}{0cm}
\centering
\includegraphics[width=0.43\textwidth]{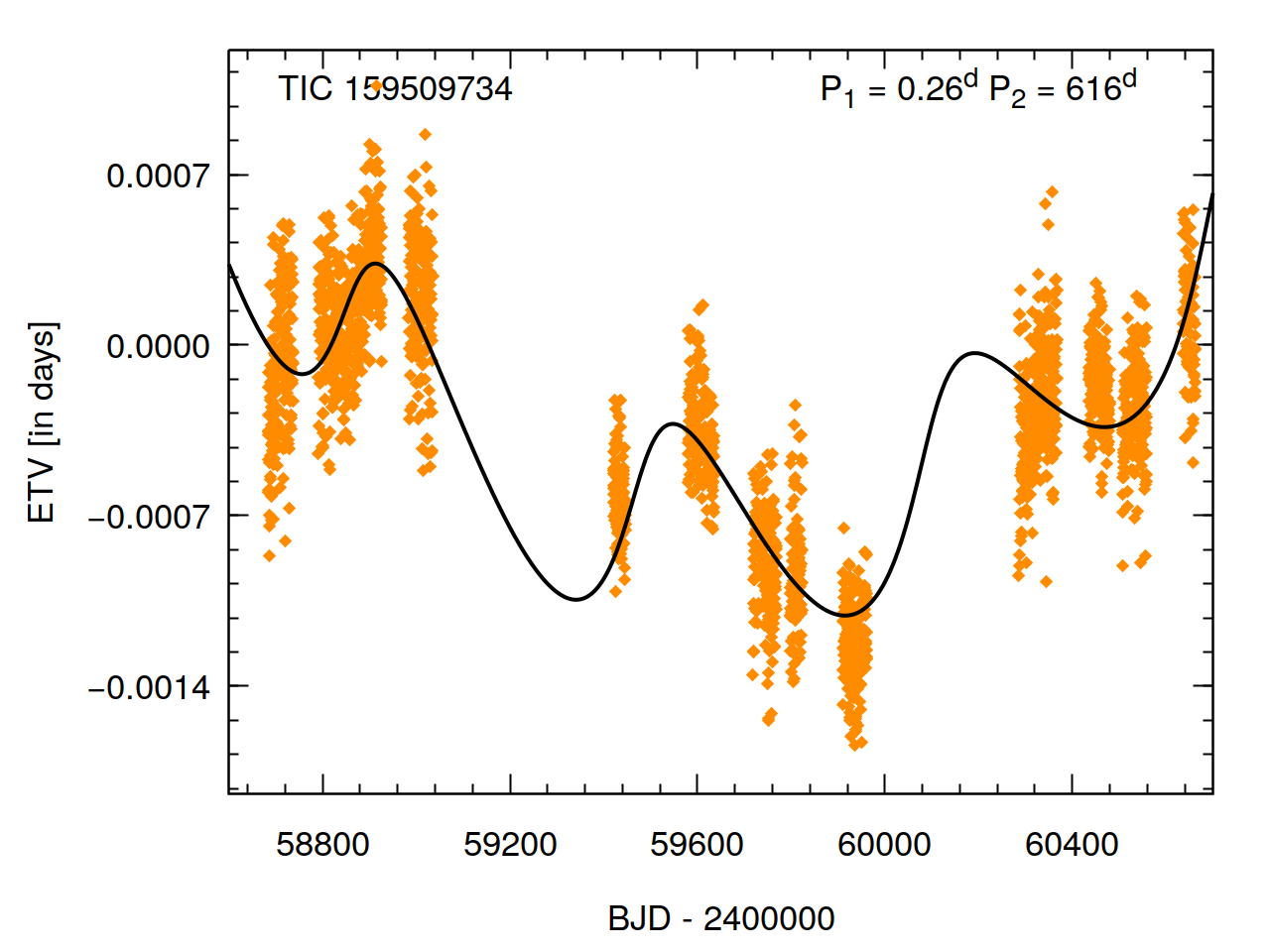}\includegraphics[width=0.43\textwidth]{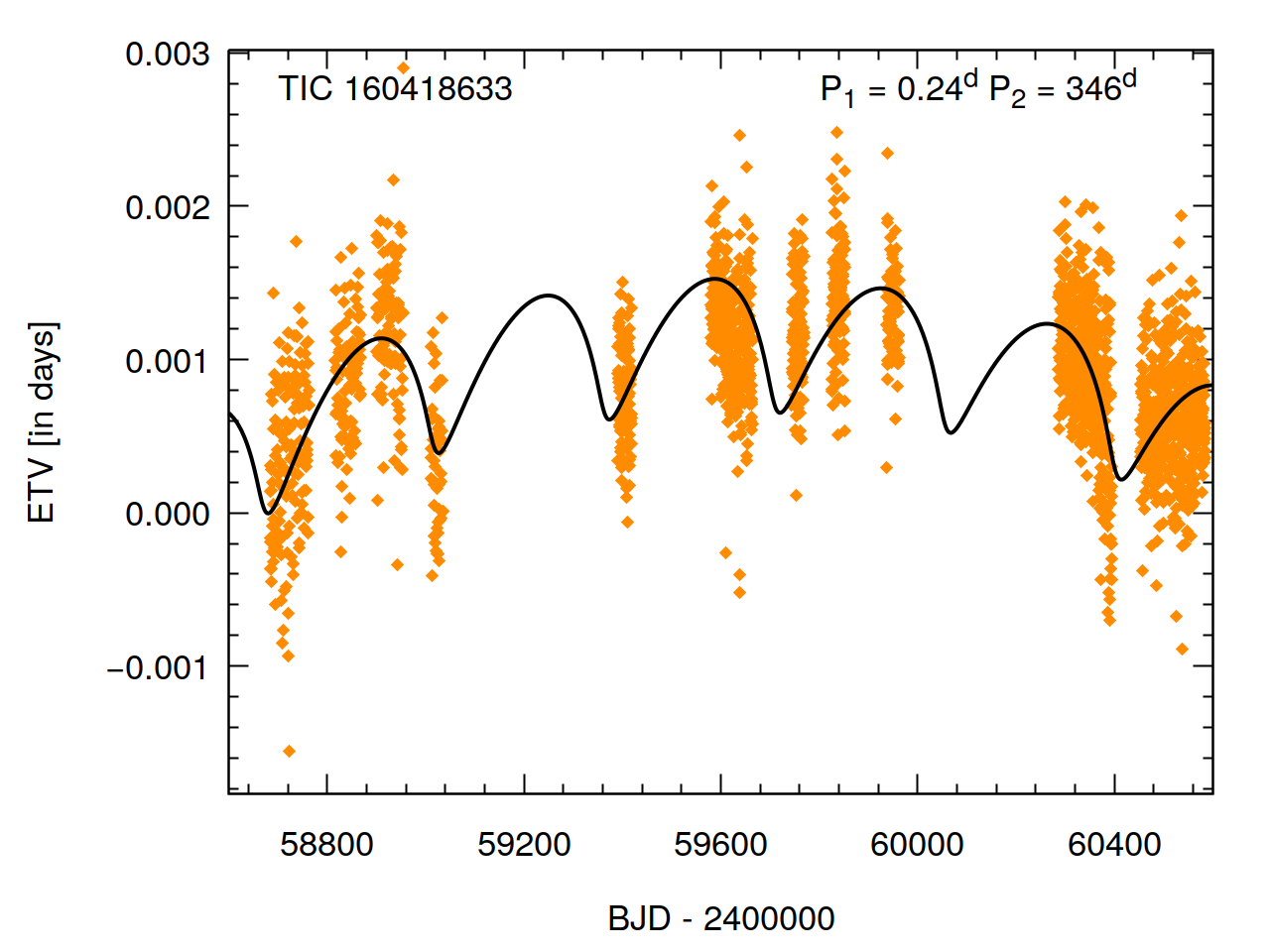}\includegraphics[width=0.43\textwidth]{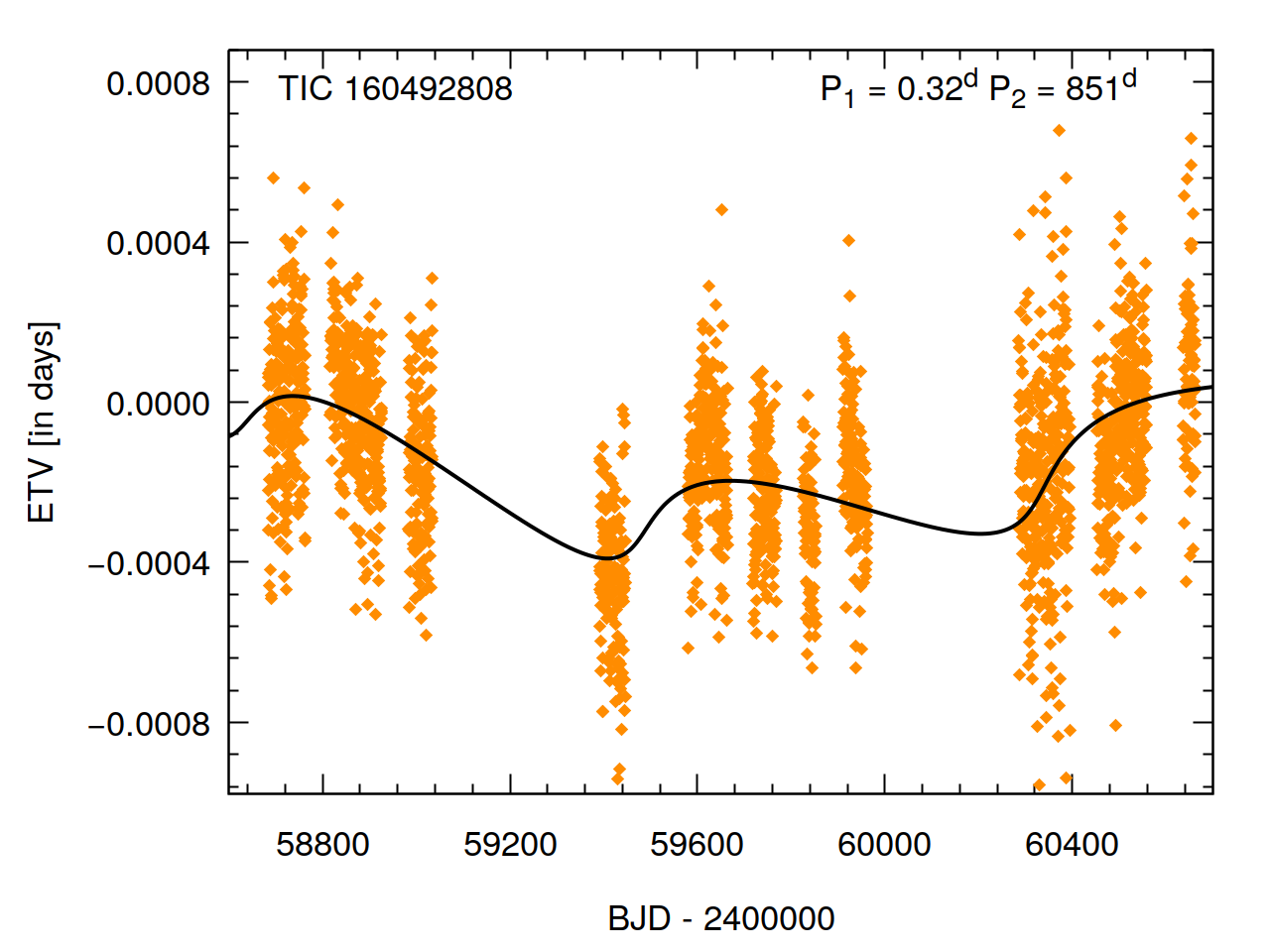}
\includegraphics[width=0.43\textwidth]{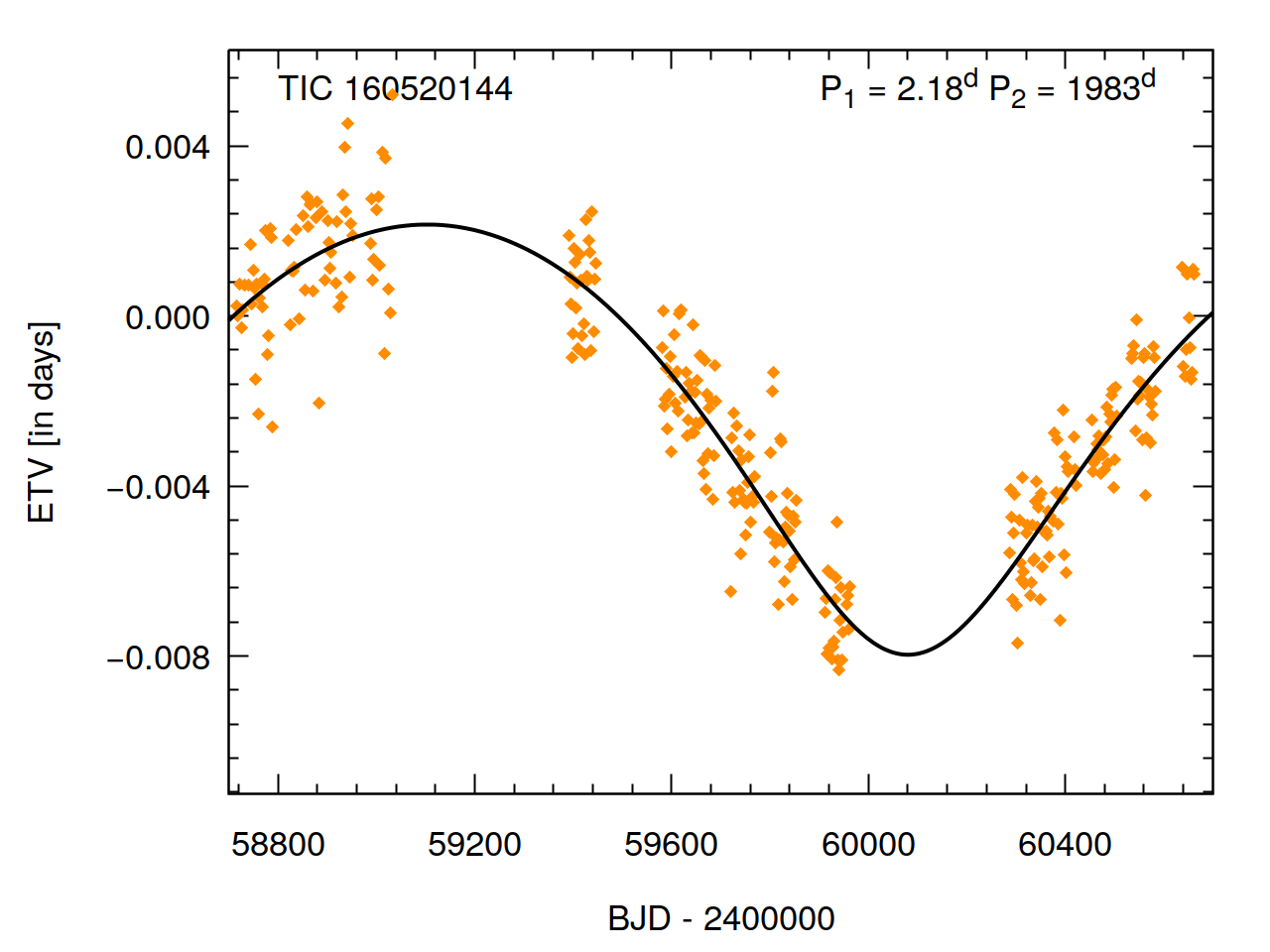}\includegraphics[width=0.43\textwidth]{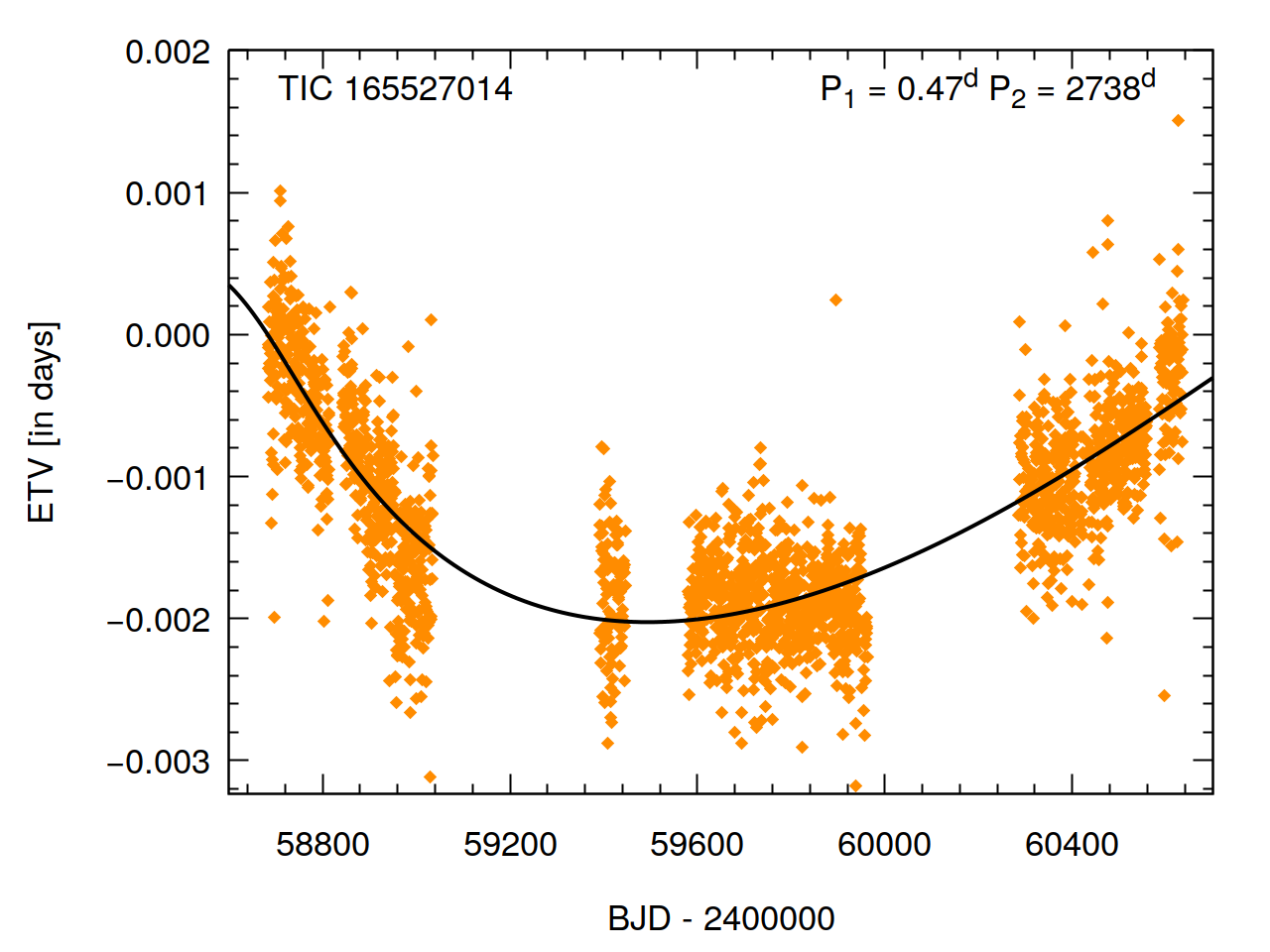}\includegraphics[width=0.43\textwidth]{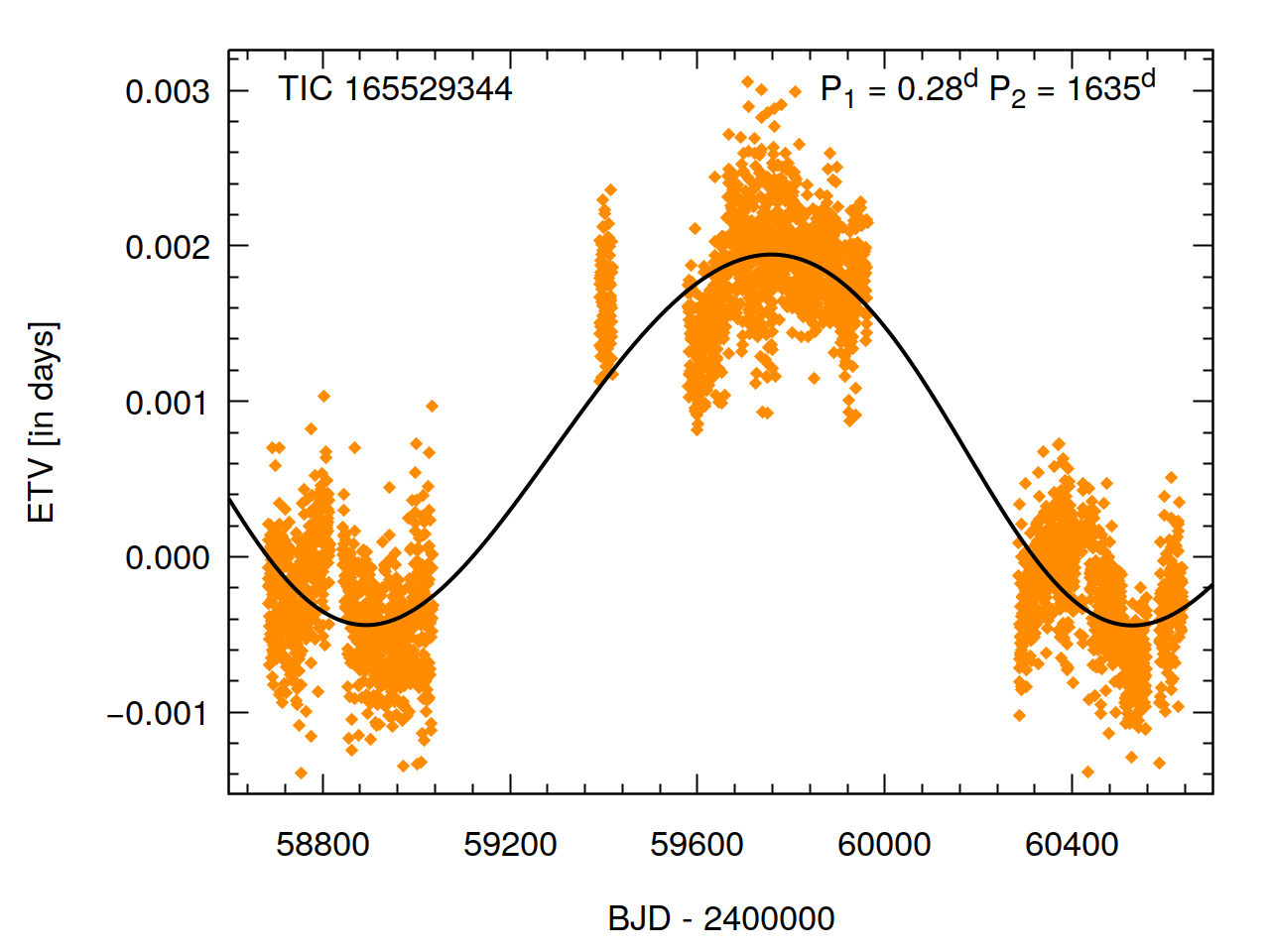}
\includegraphics[width=0.43\textwidth]{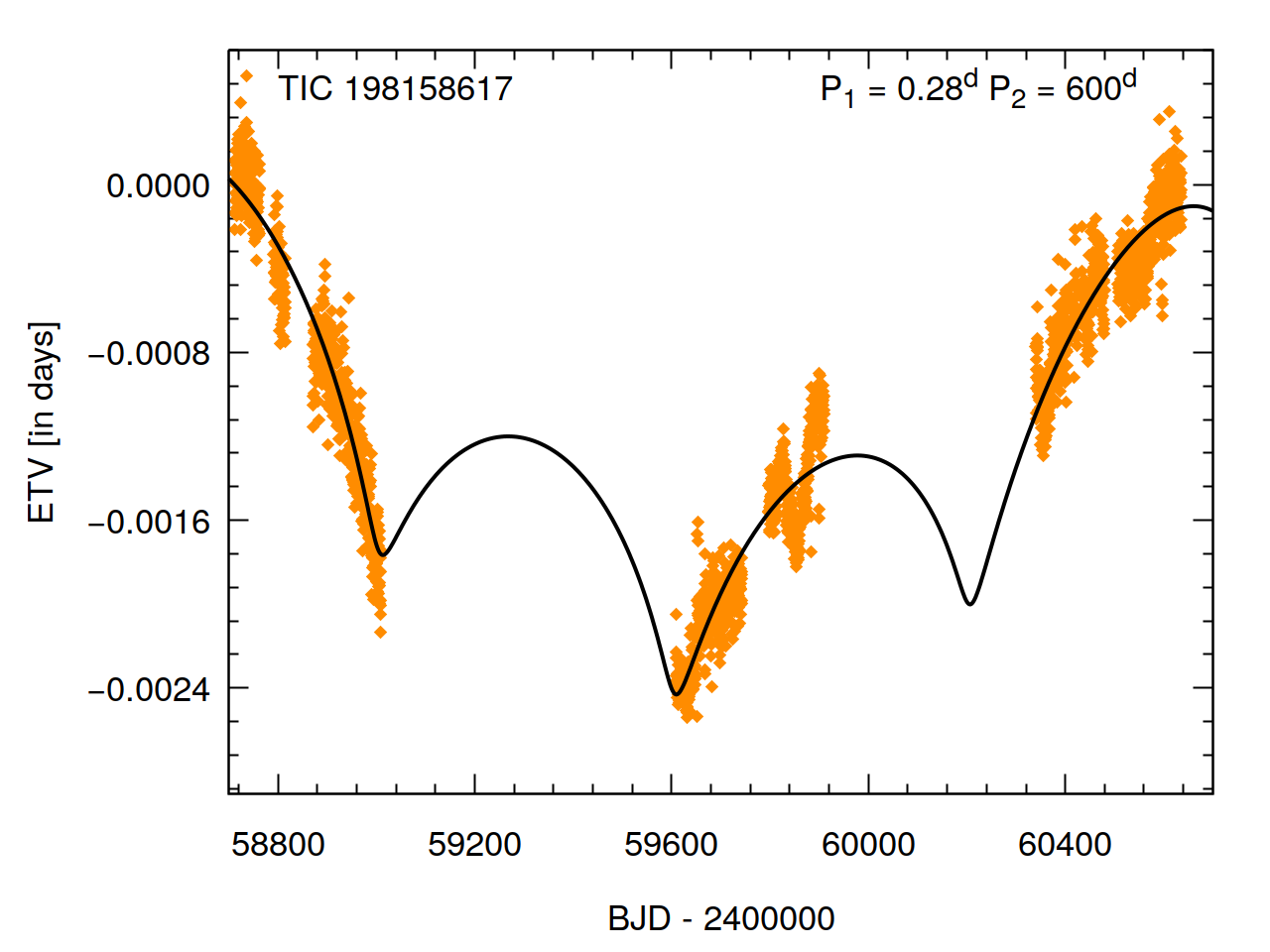}\includegraphics[width=0.43\textwidth]{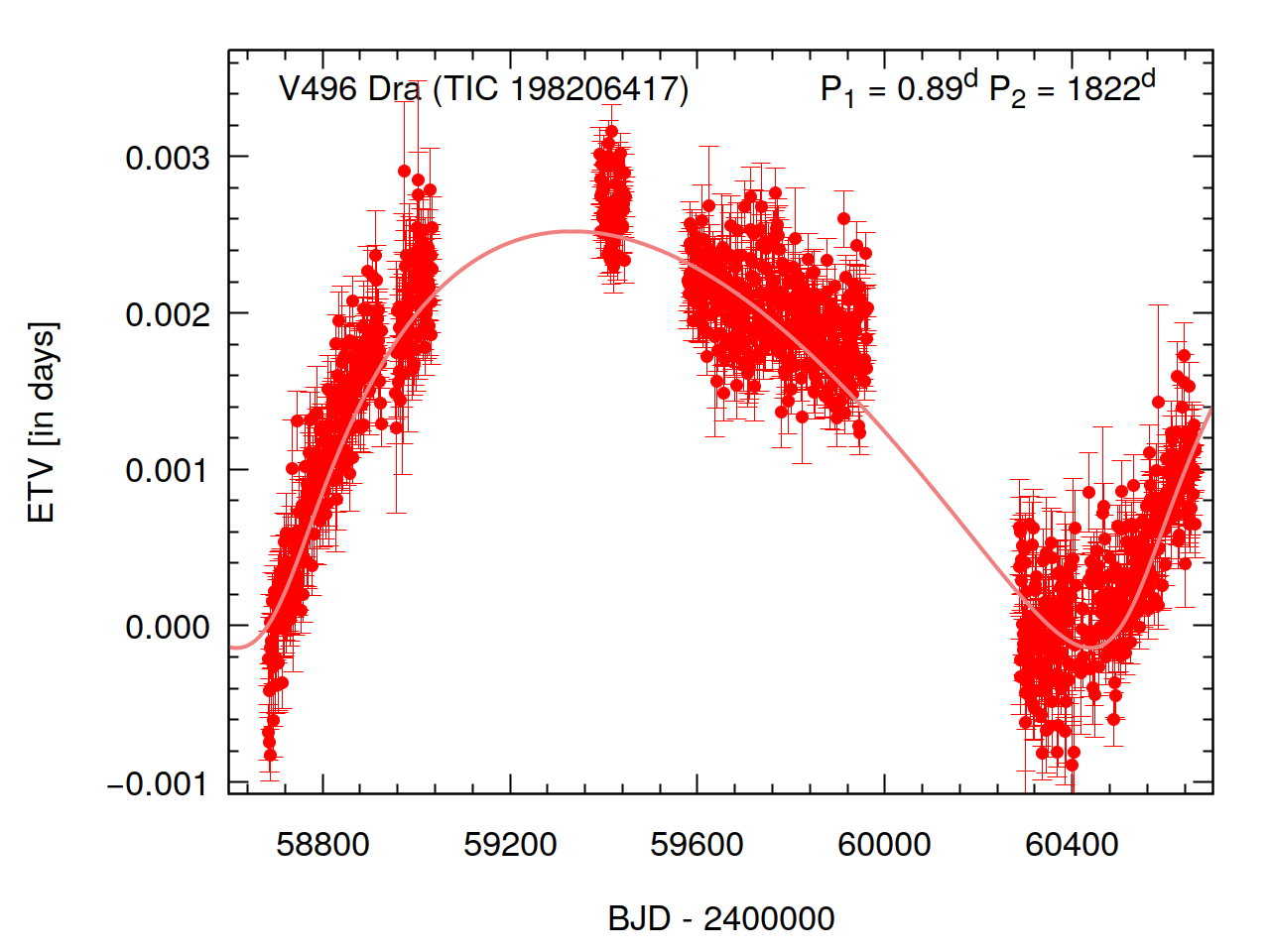}\includegraphics[width=0.43\textwidth]{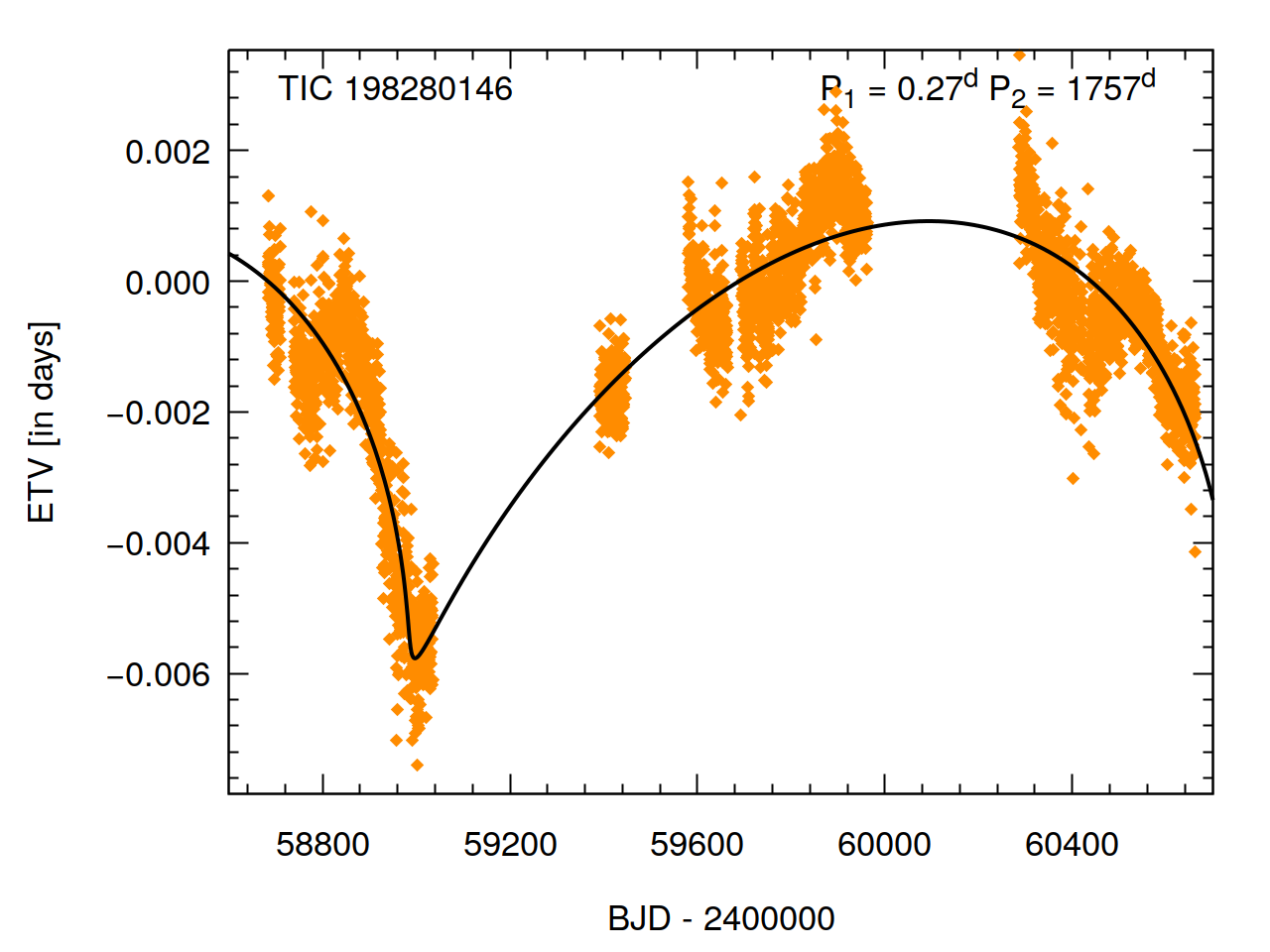}
\includegraphics[width=0.43\textwidth]{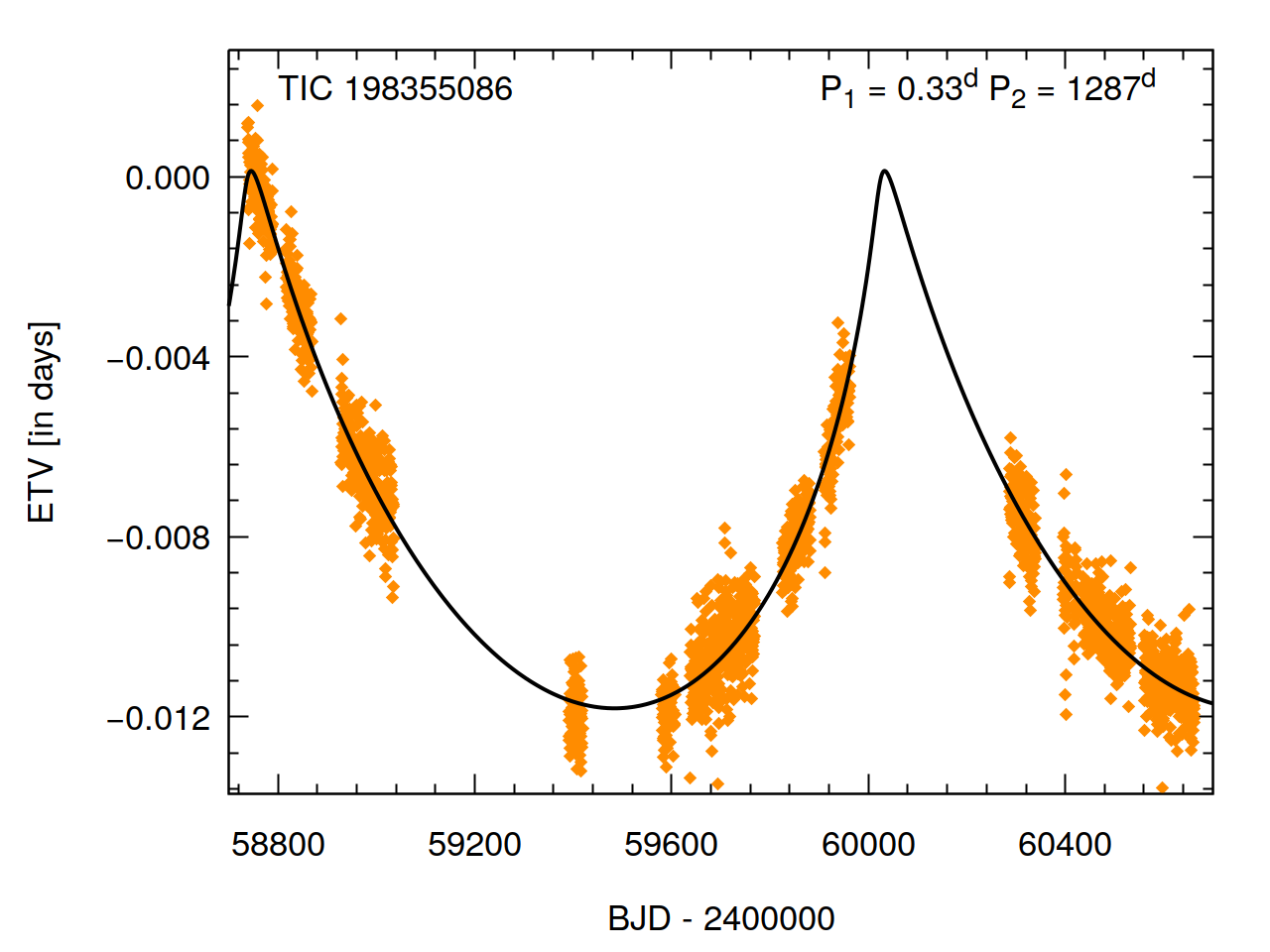}\includegraphics[width=0.43\textwidth]{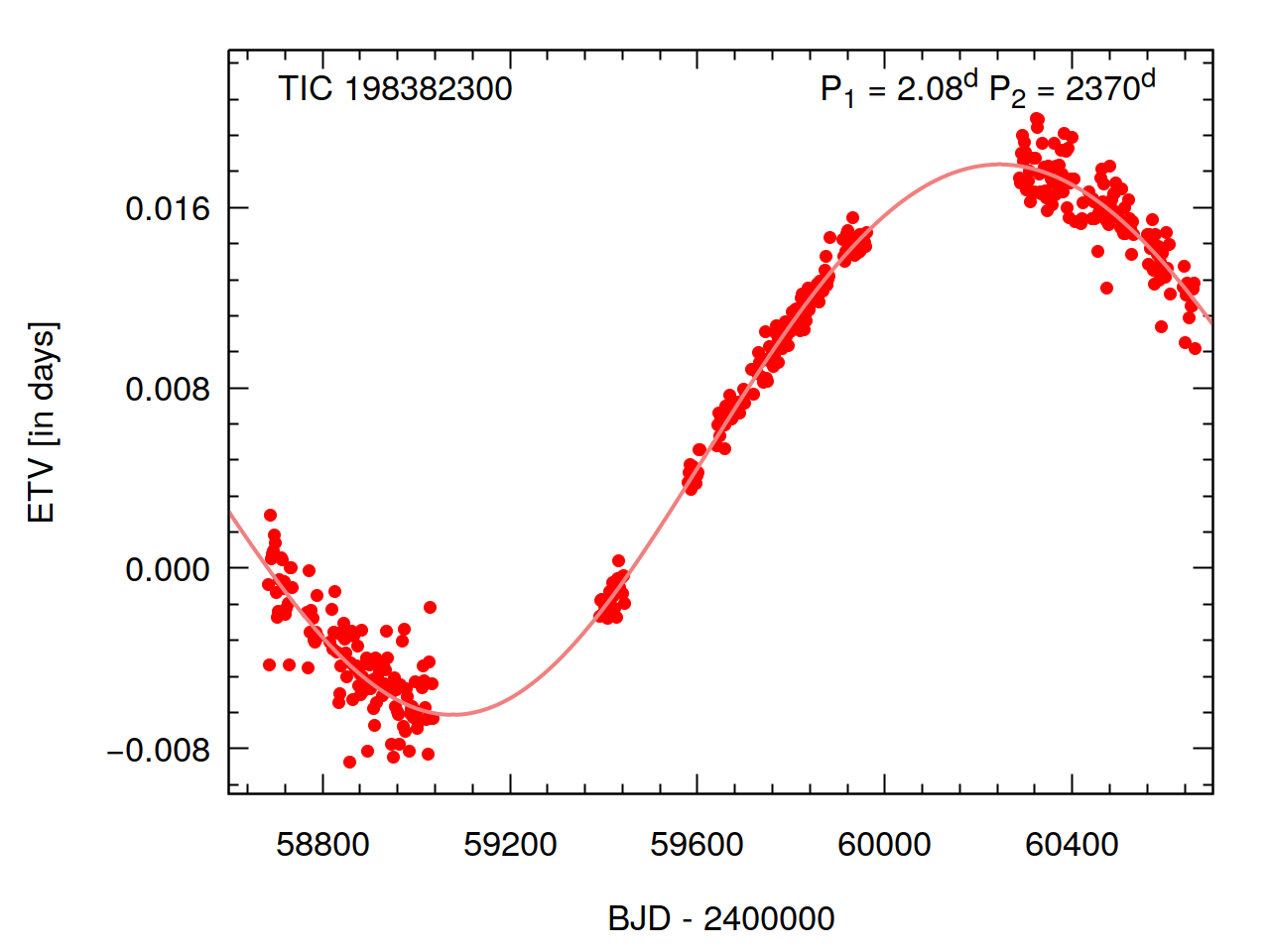}\includegraphics[width=0.43\textwidth]{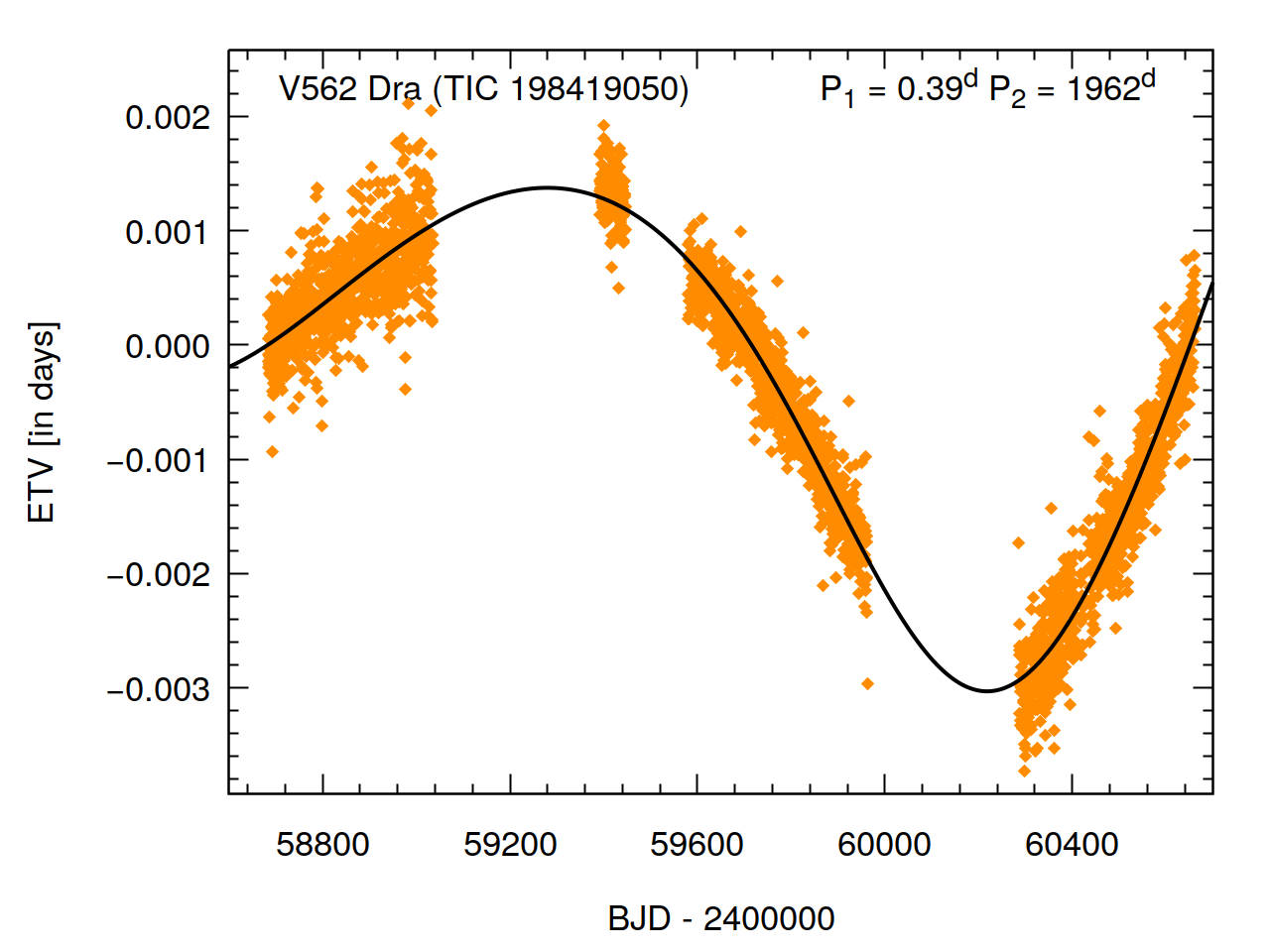}
\includegraphics[width=0.43\textwidth]{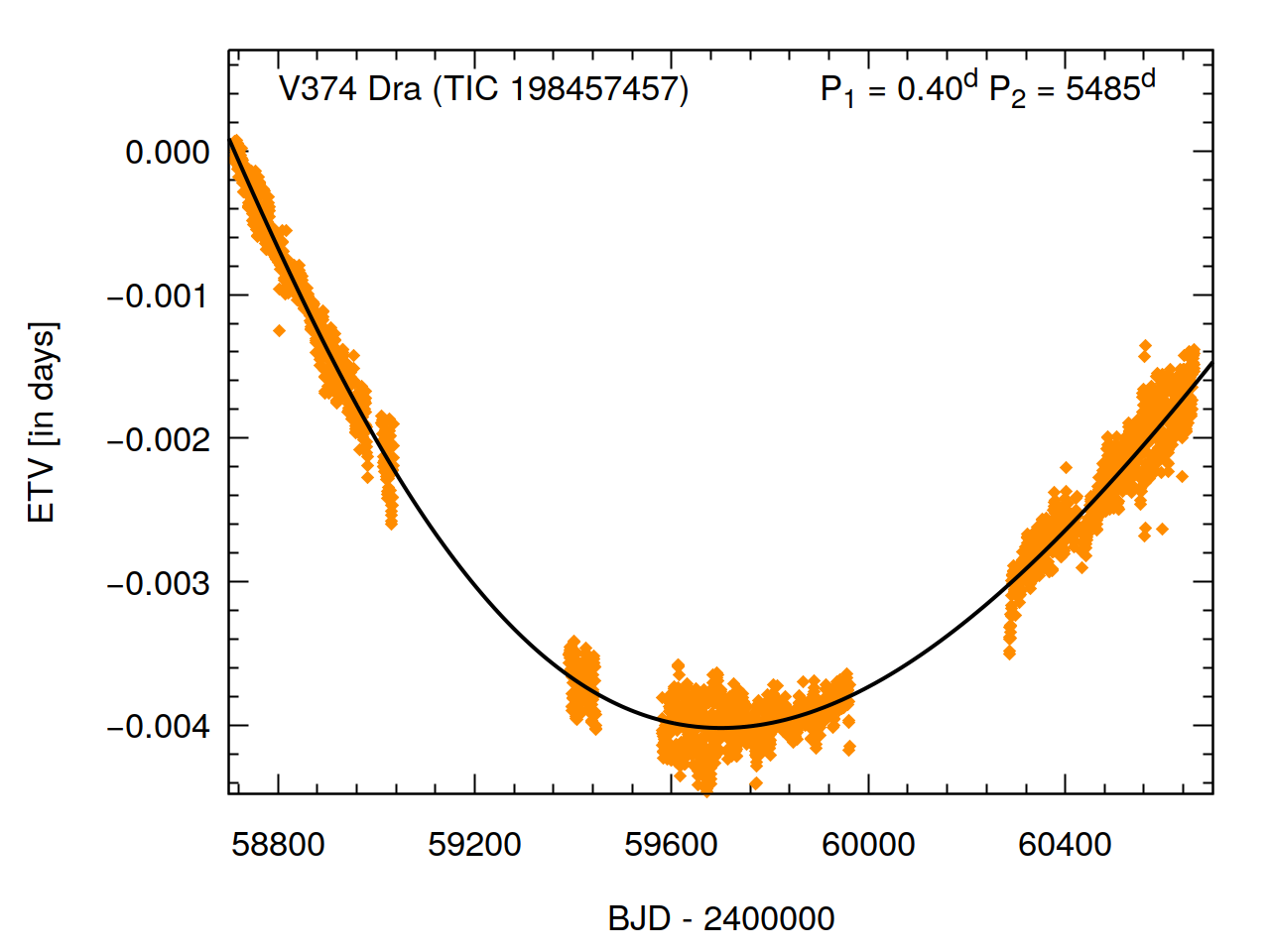}\includegraphics[width=0.43\textwidth]{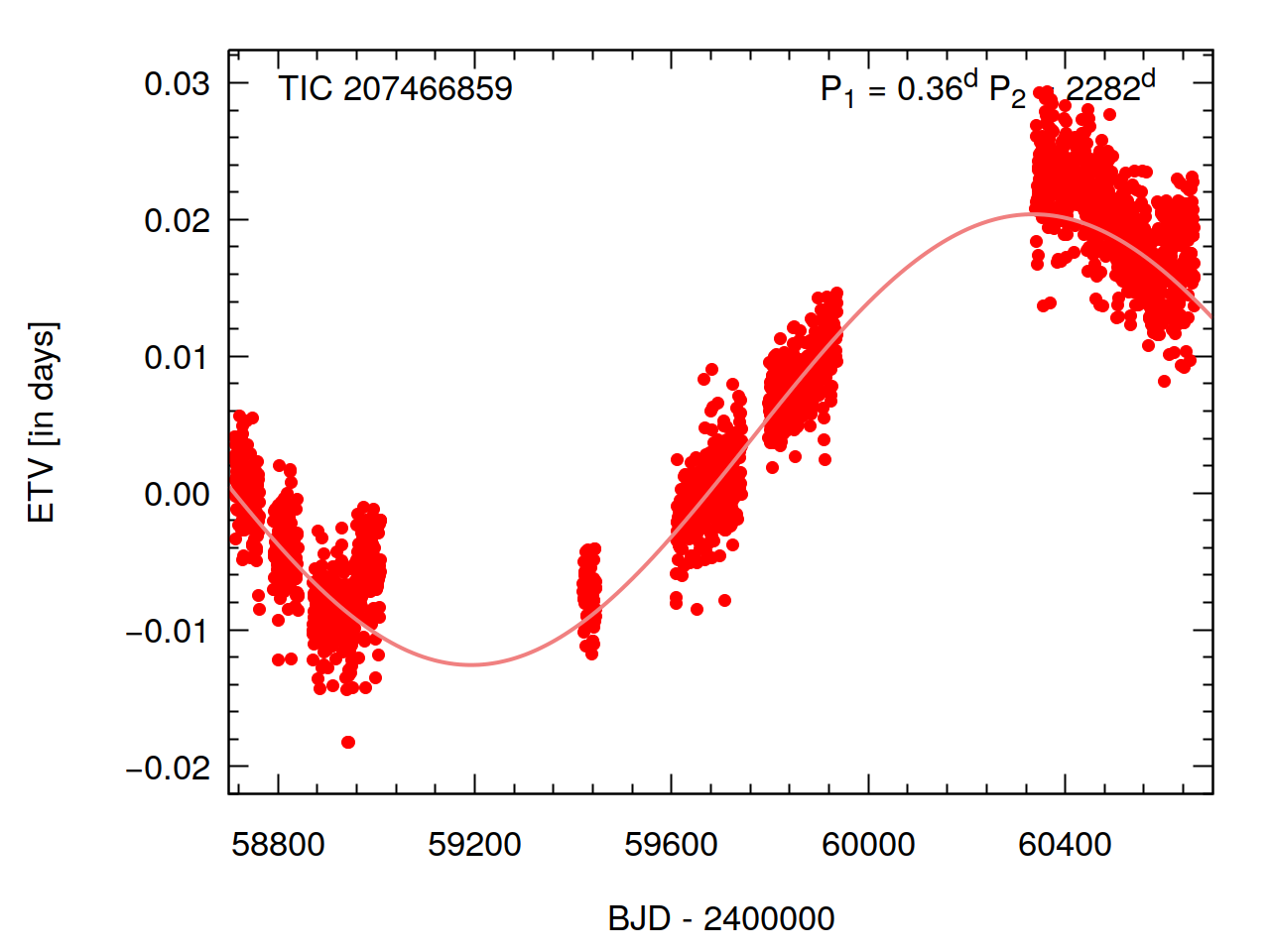}\includegraphics[width=0.43\textwidth]{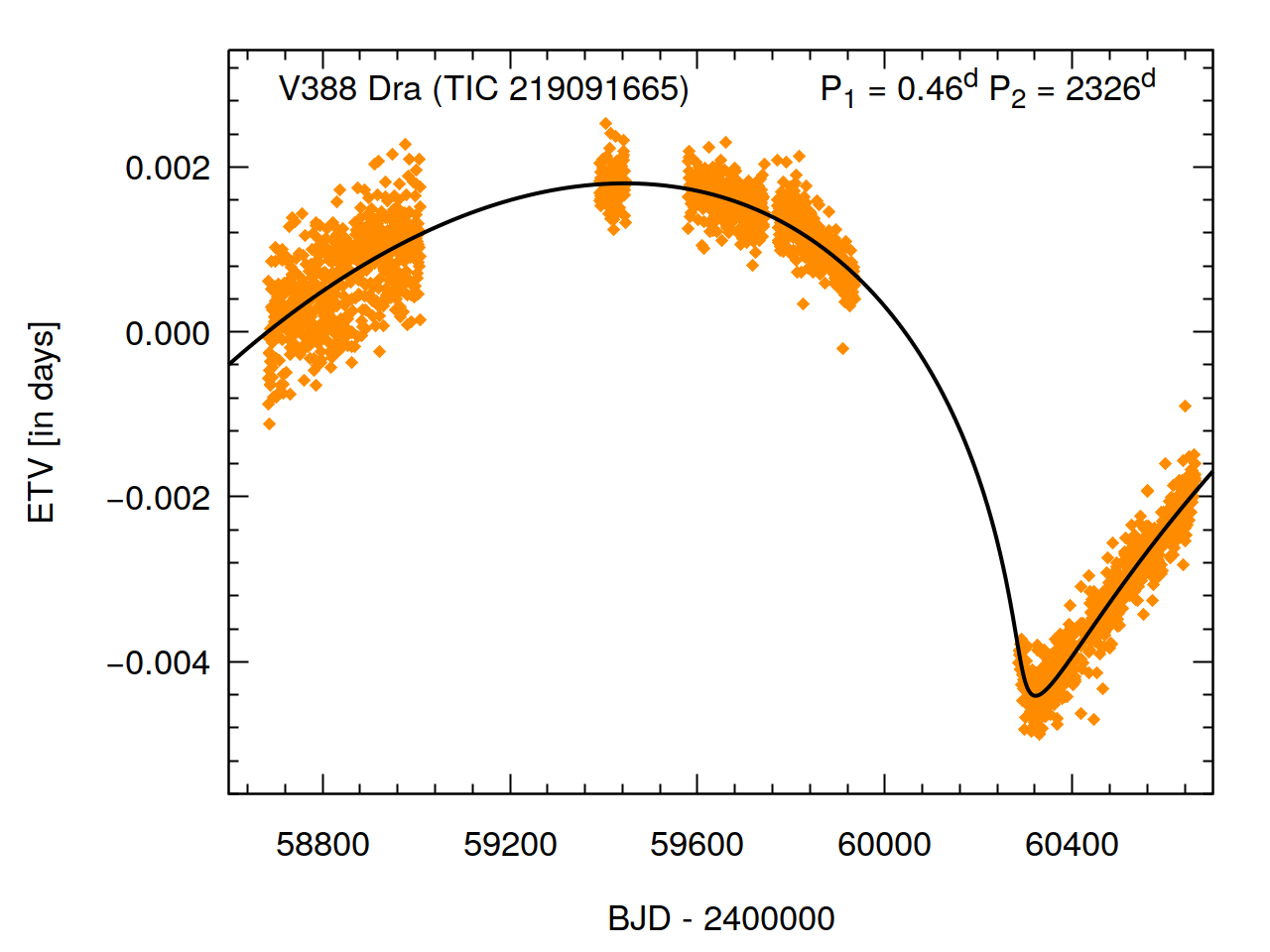}
\end{adjustwidth}
\caption{The first 15 ETVs of those third body candidates which are ranked into the less certain (group $L_3$) pure LTTE third body solutions. Orange diamonds, red circles, and the smooth curves, as well as the error bars, have the same meanings as were explained in the former figure captions. For further details, see Table~\ref{Tab:Orbelem_LTTE3}.}
\label{Fig:ETVs_L3a}
\end{figure}

\begin{figure}[H]
\begin{adjustwidth}{-\extralength}{0cm}
\centering
\includegraphics[width=0.43\textwidth]{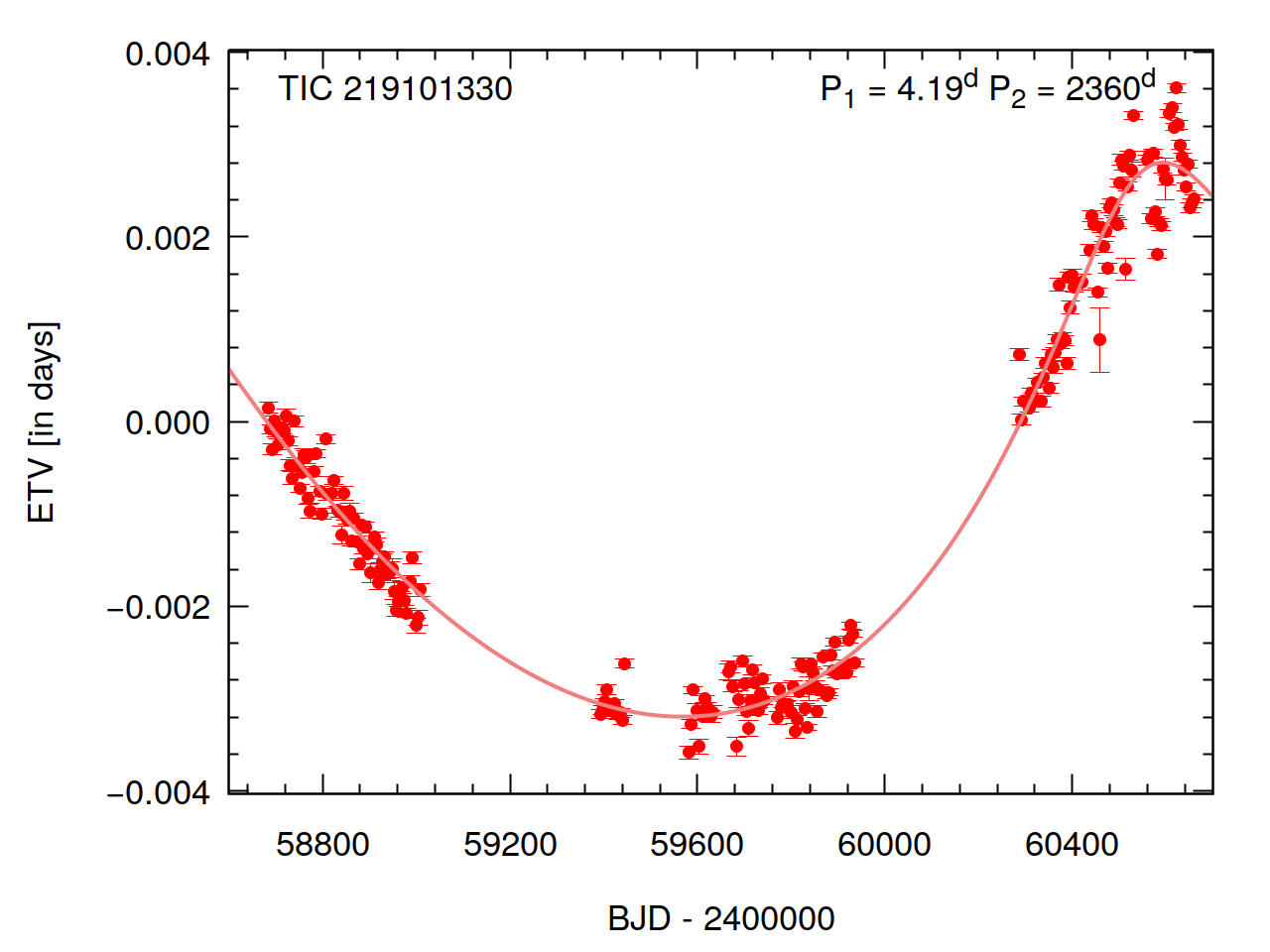}\includegraphics[width=0.43\textwidth]{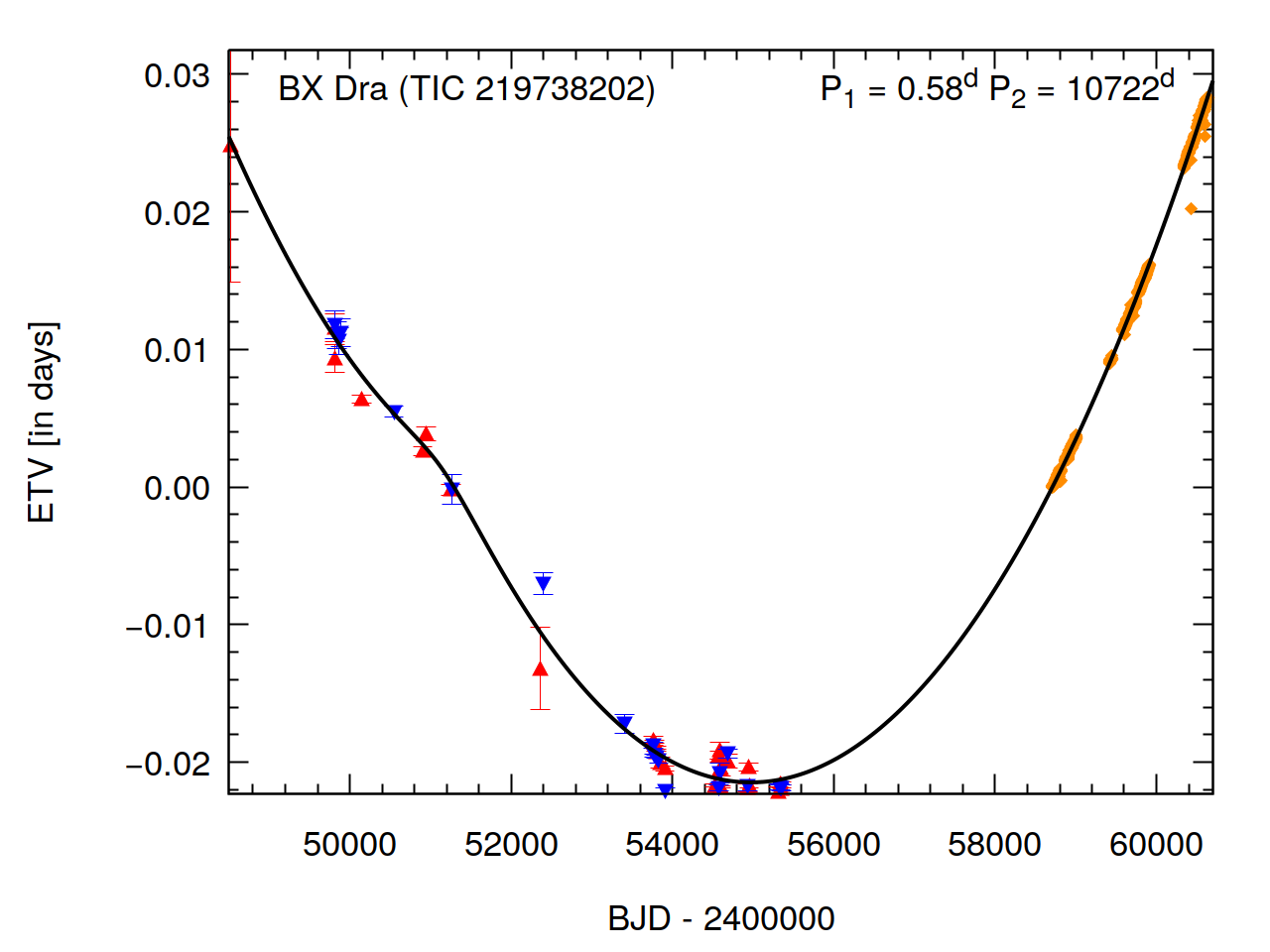}\includegraphics[width=0.43\textwidth]{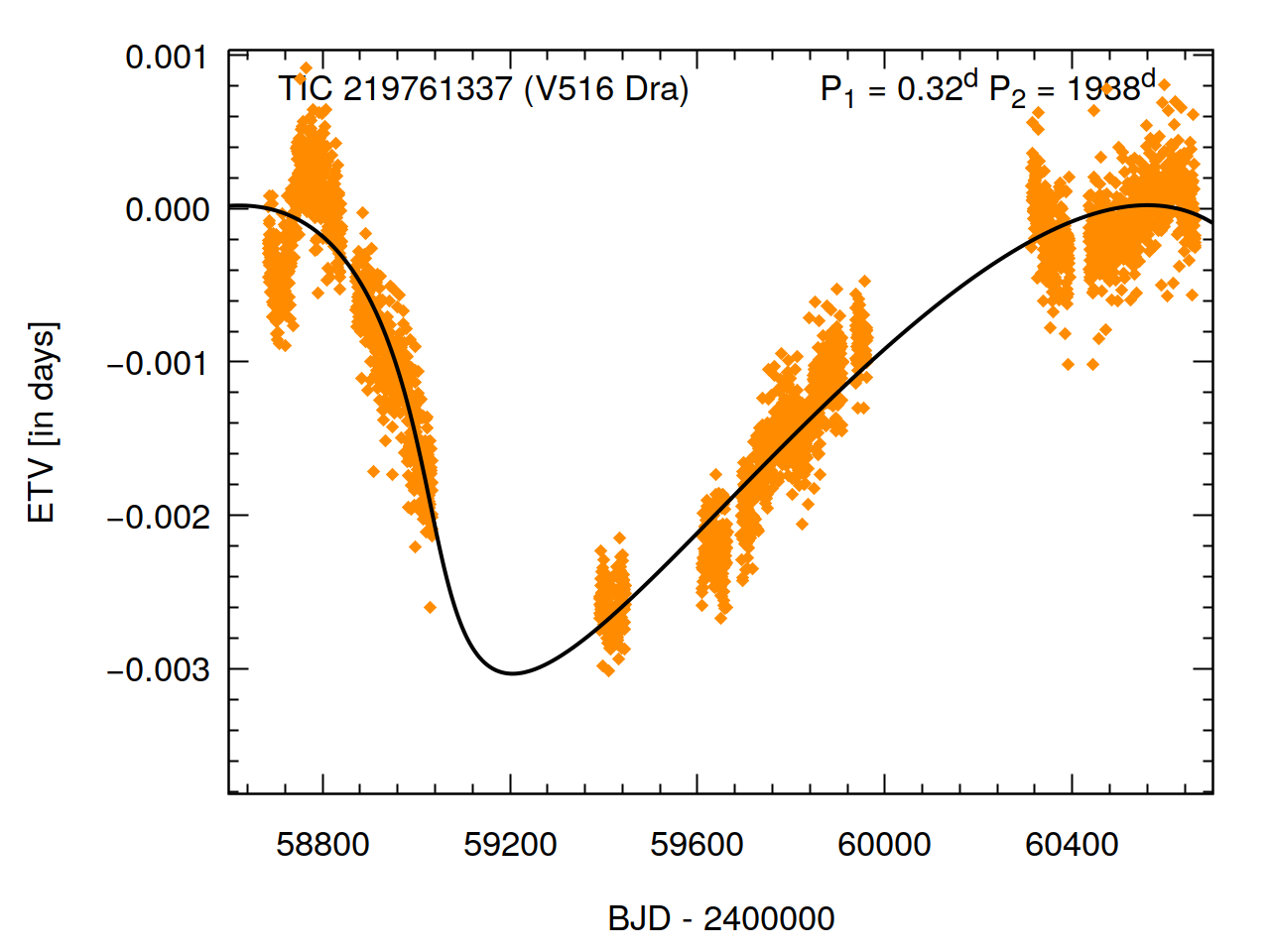}
\includegraphics[width=0.43\textwidth]{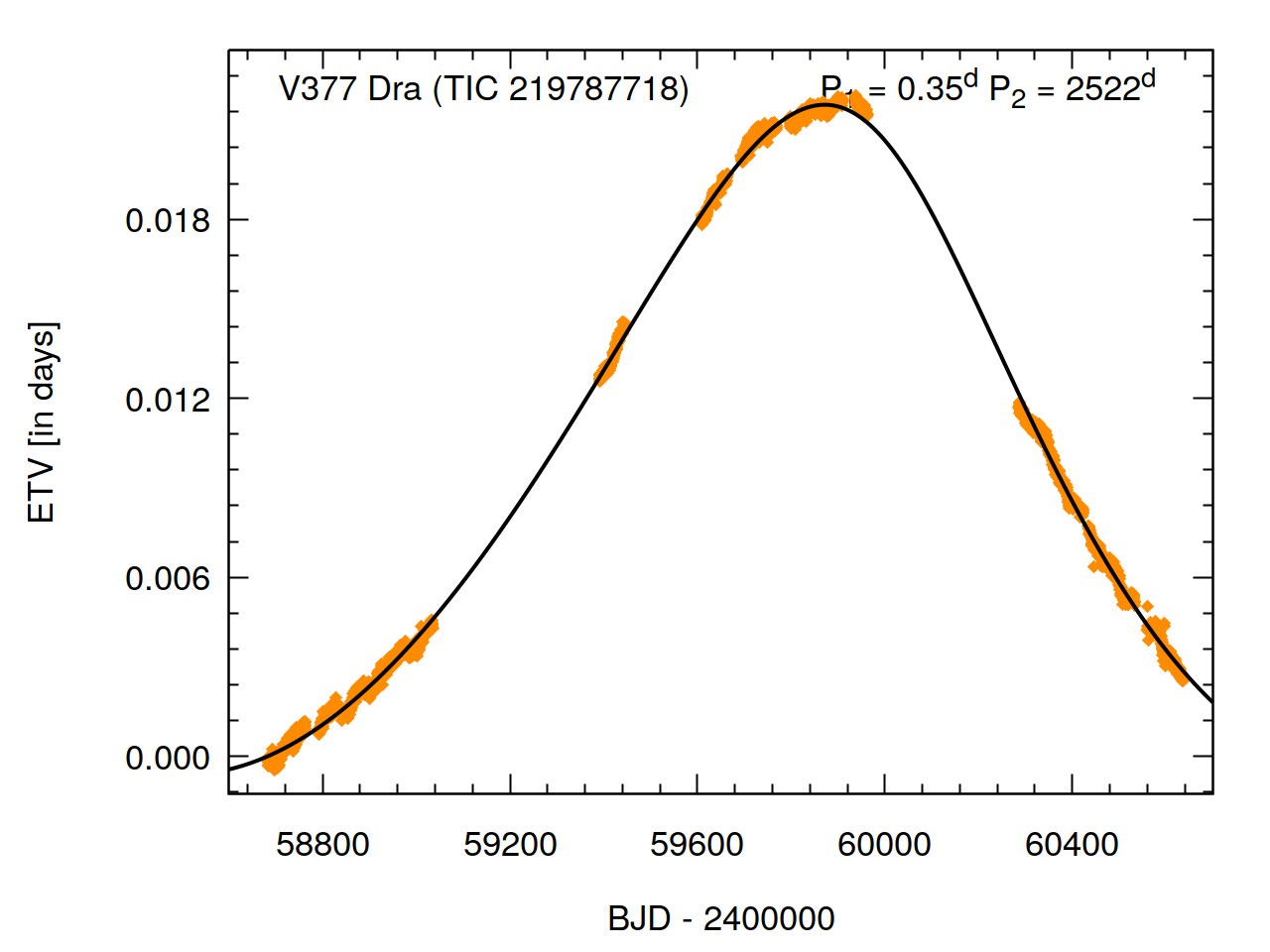}\includegraphics[width=0.43\textwidth]{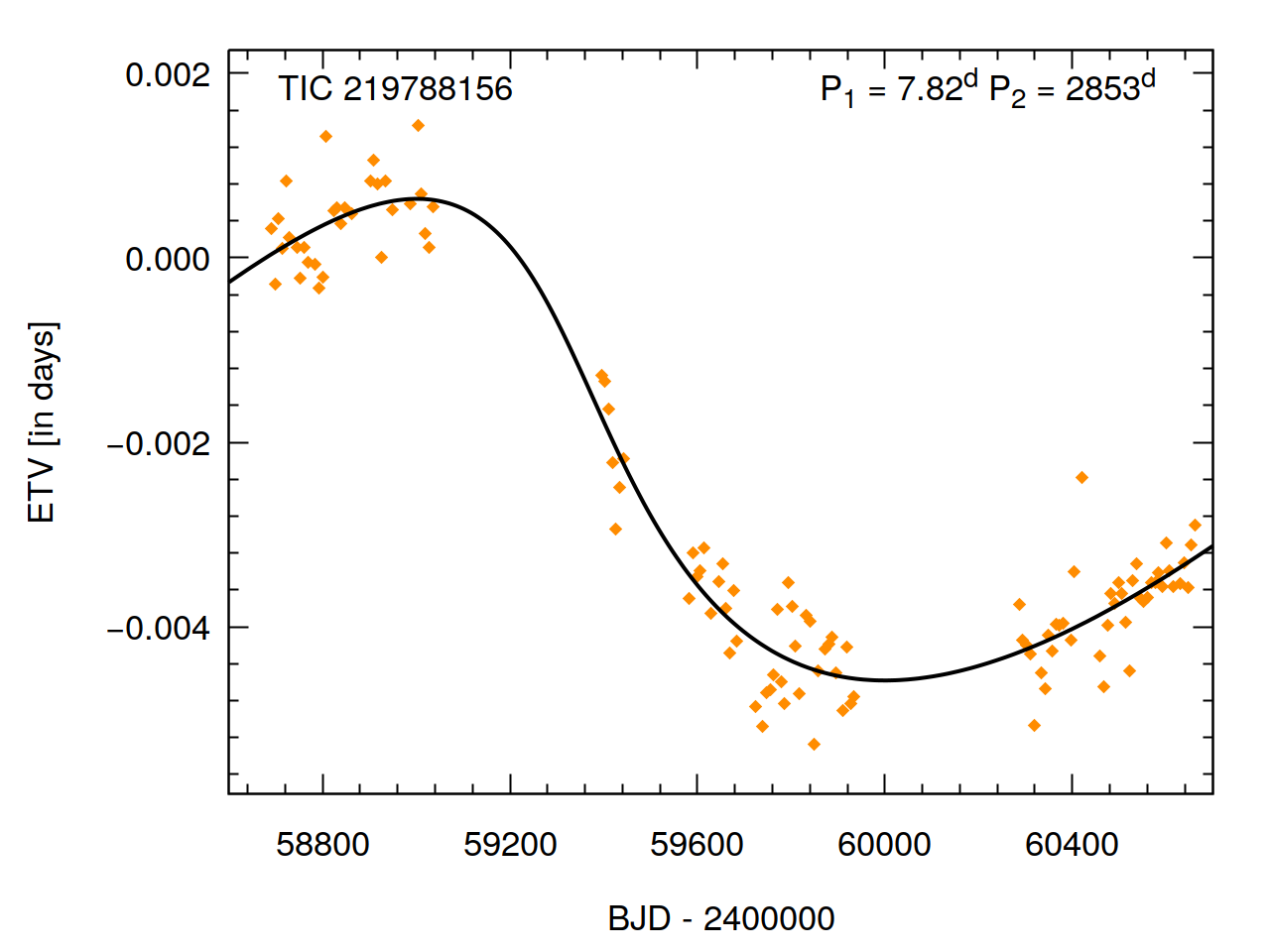}\includegraphics[width=0.43\textwidth]{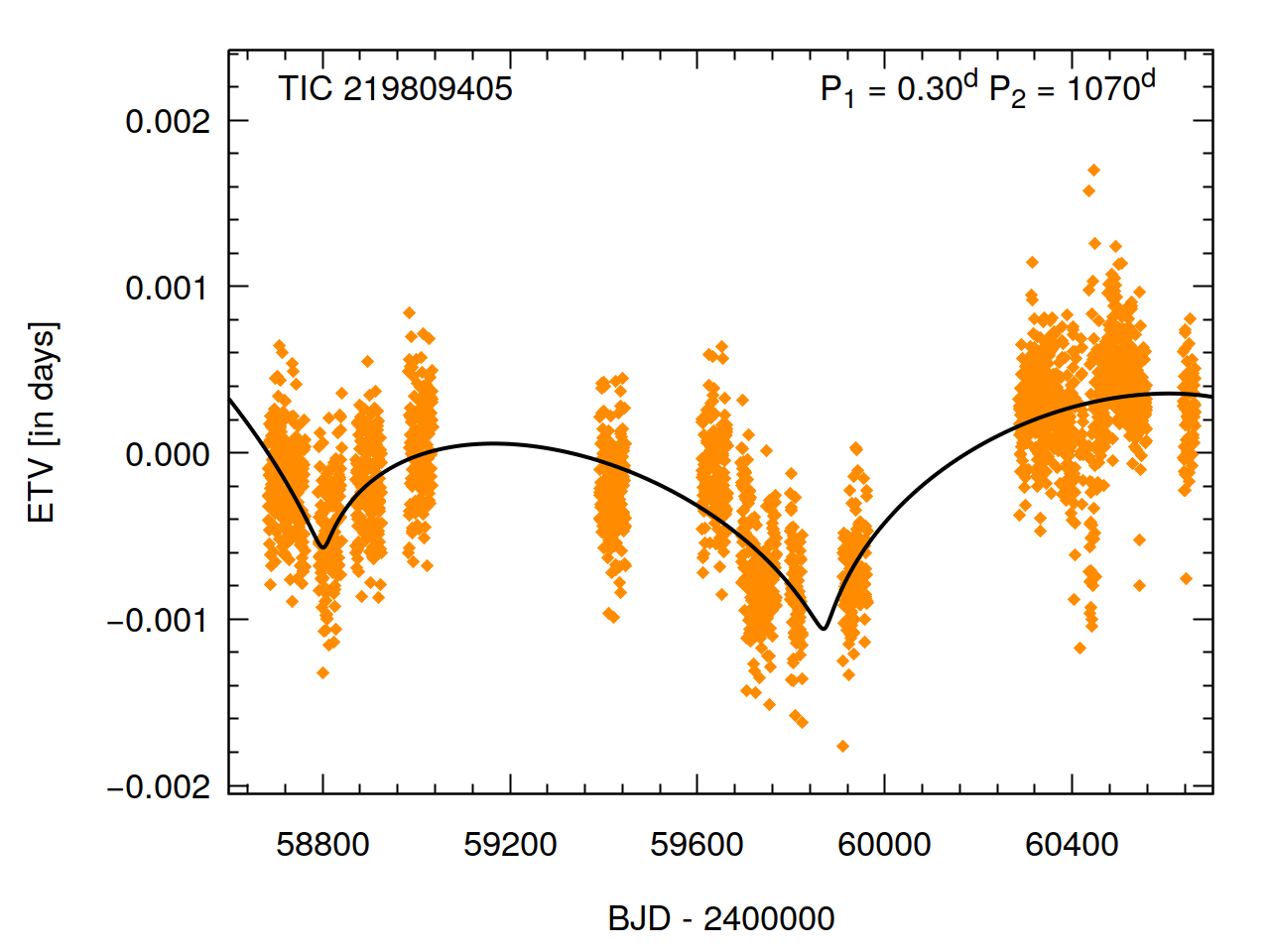}
\includegraphics[width=0.43\textwidth]{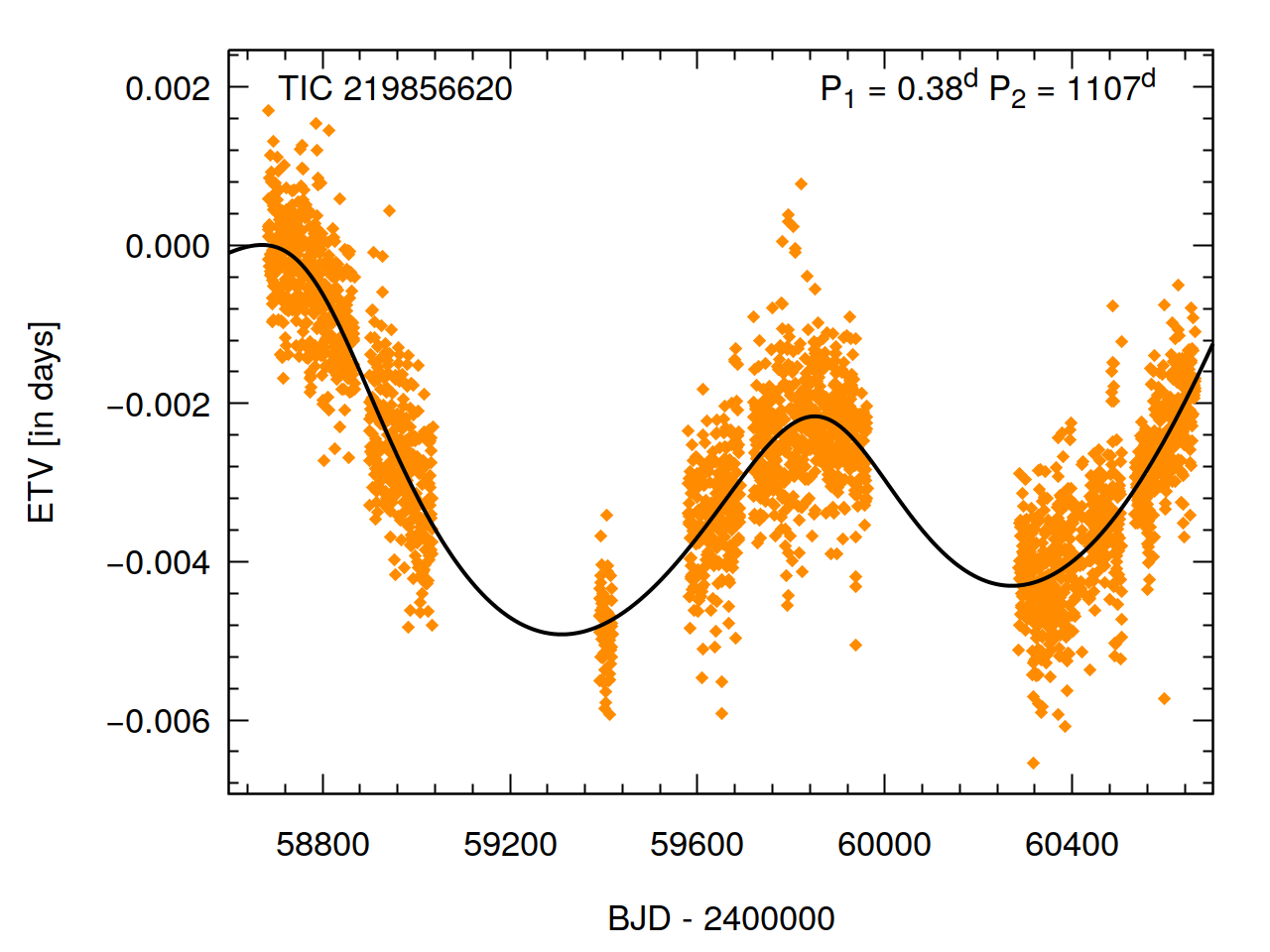}\includegraphics[width=0.43\textwidth]{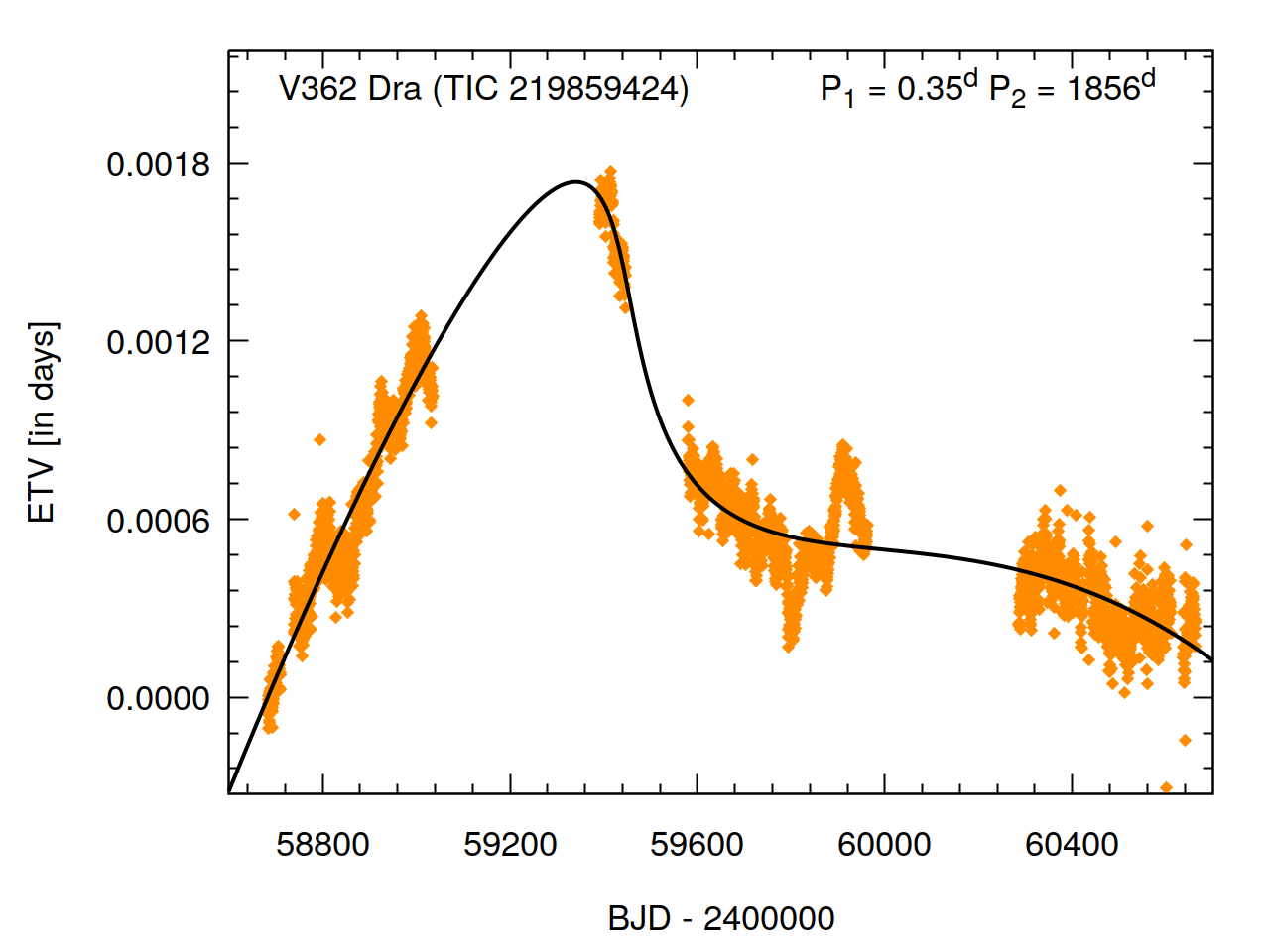}\includegraphics[width=0.43\textwidth]{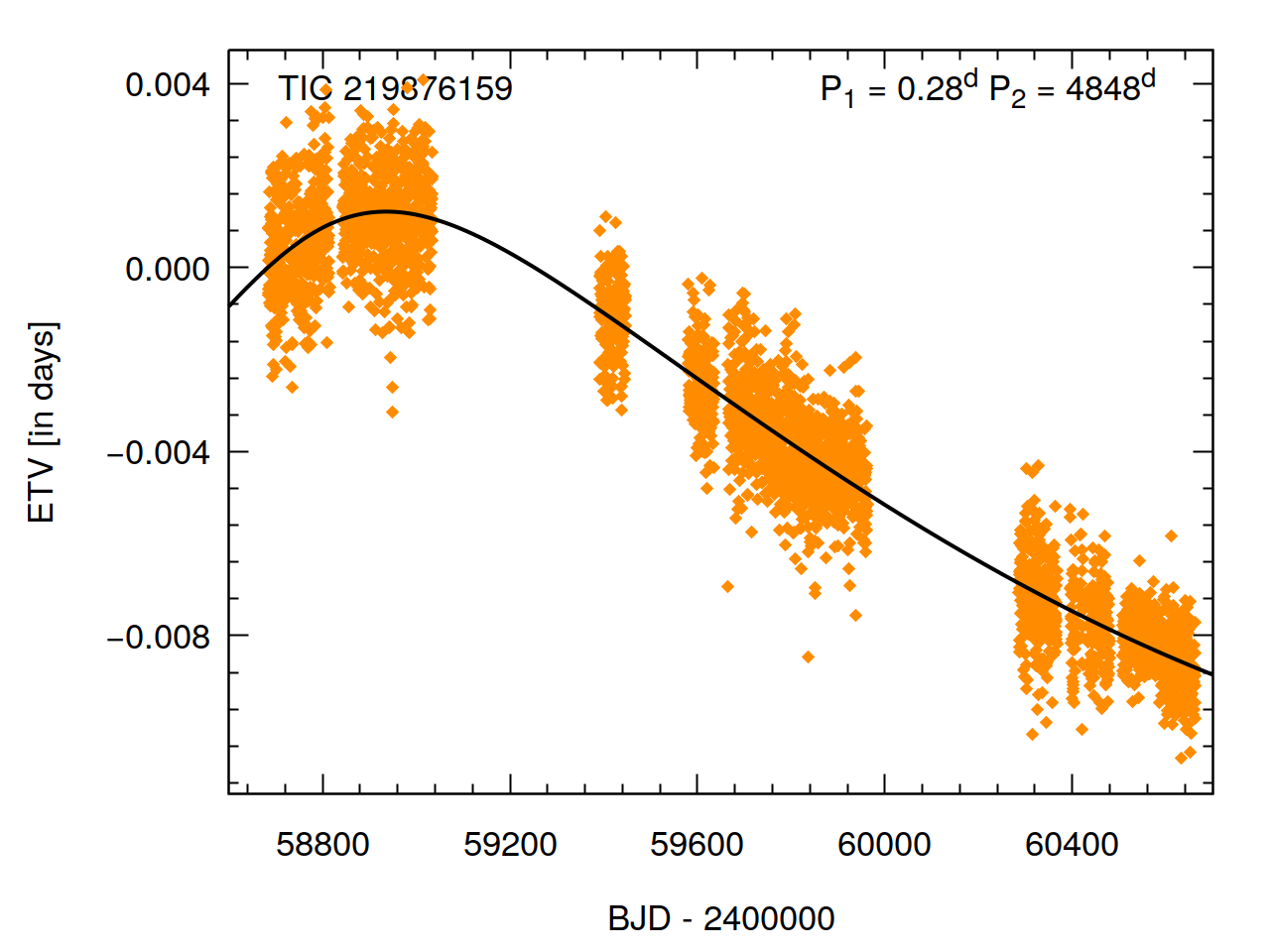}
\includegraphics[width=0.43\textwidth]{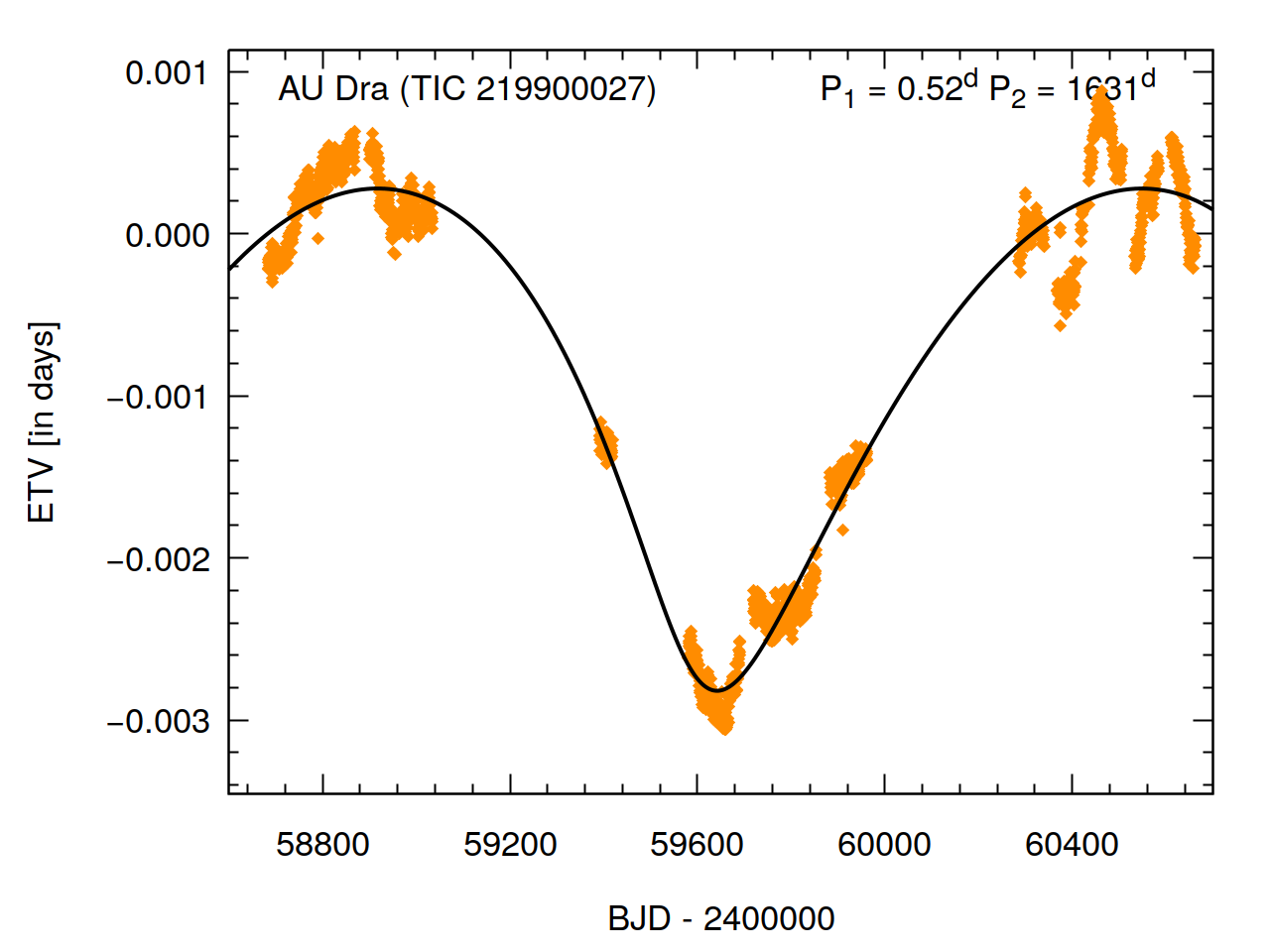}\includegraphics[width=0.43\textwidth]{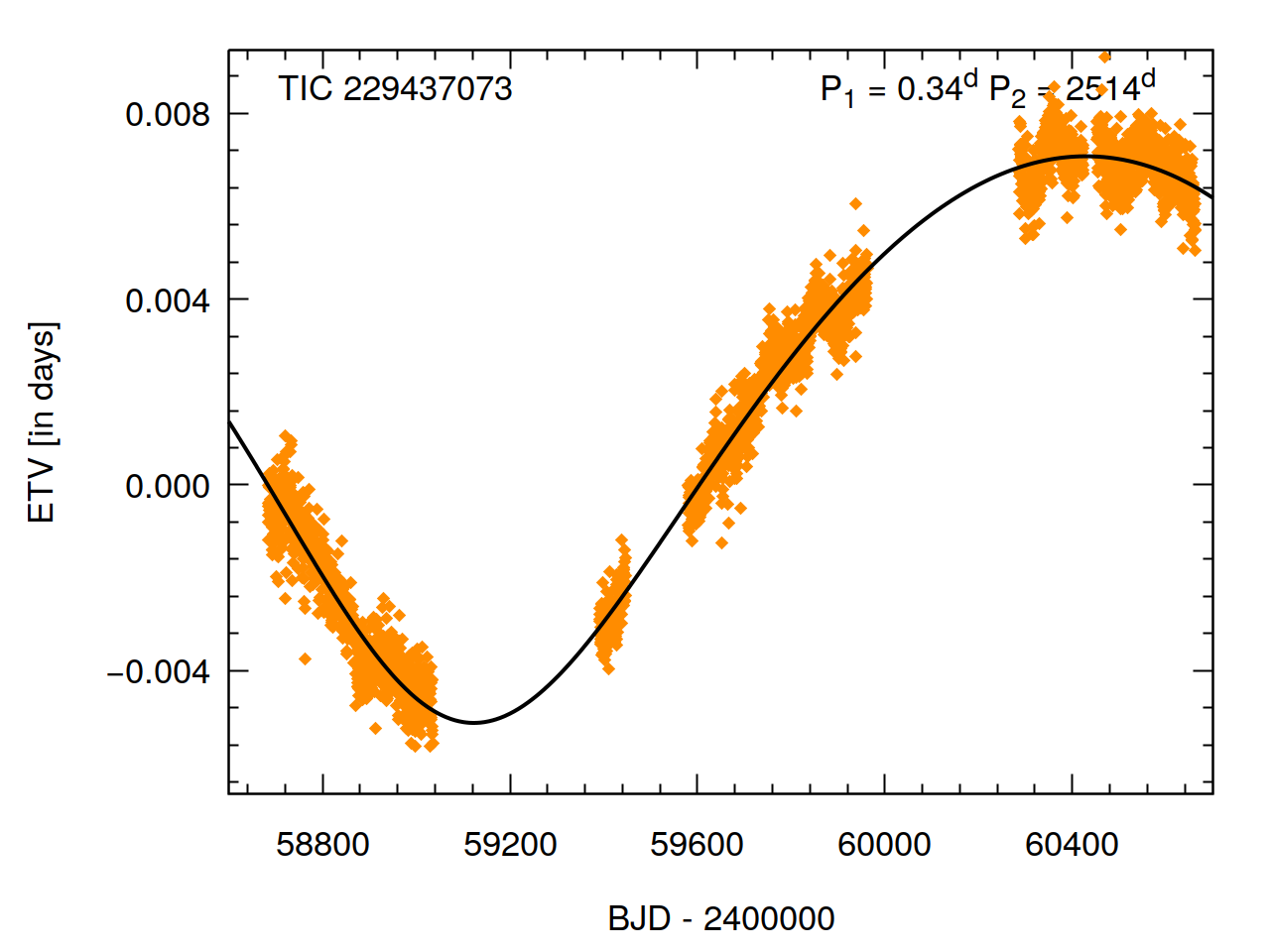}\includegraphics[width=0.43\textwidth]{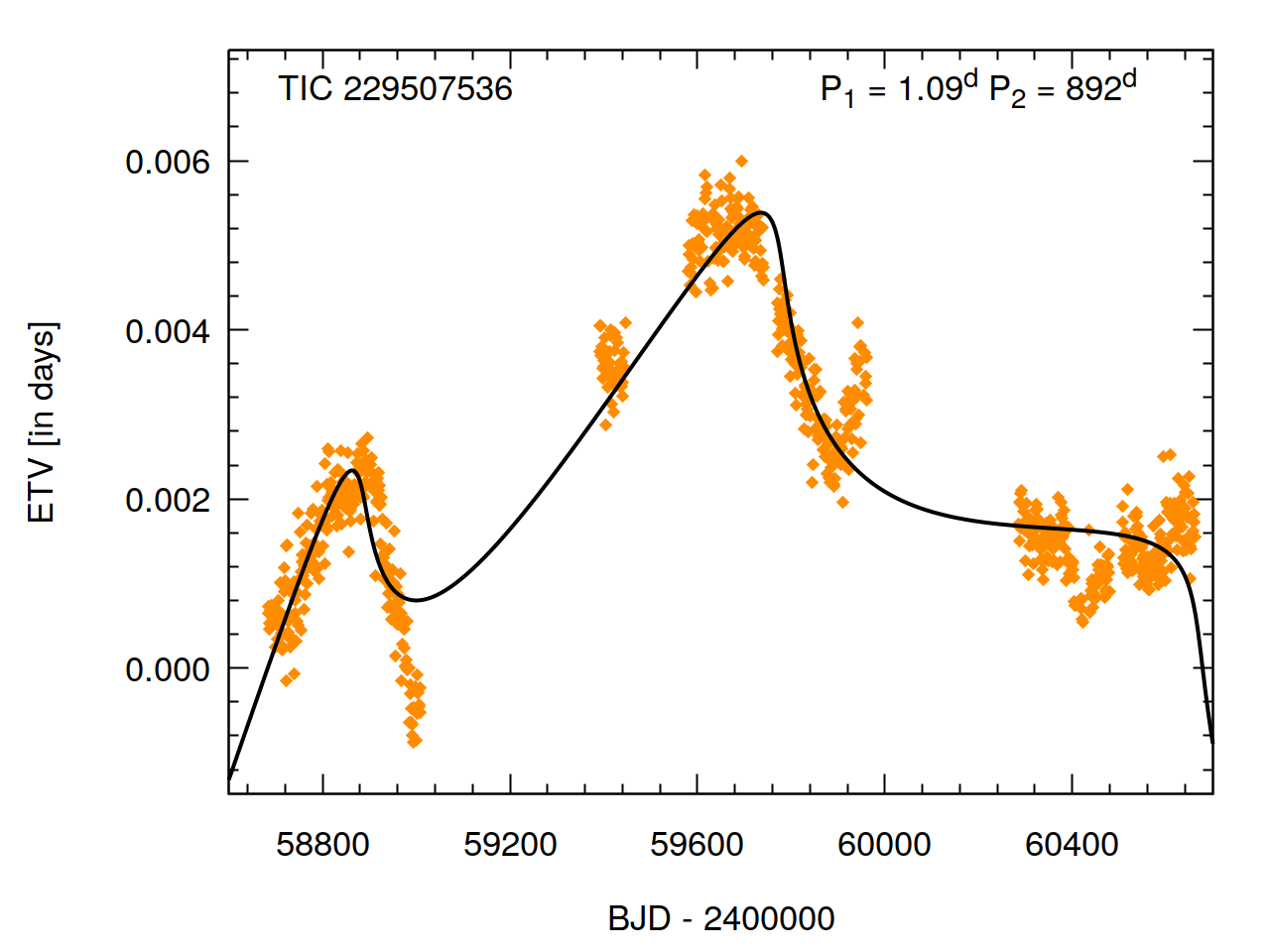}
\includegraphics[width=0.43\textwidth]{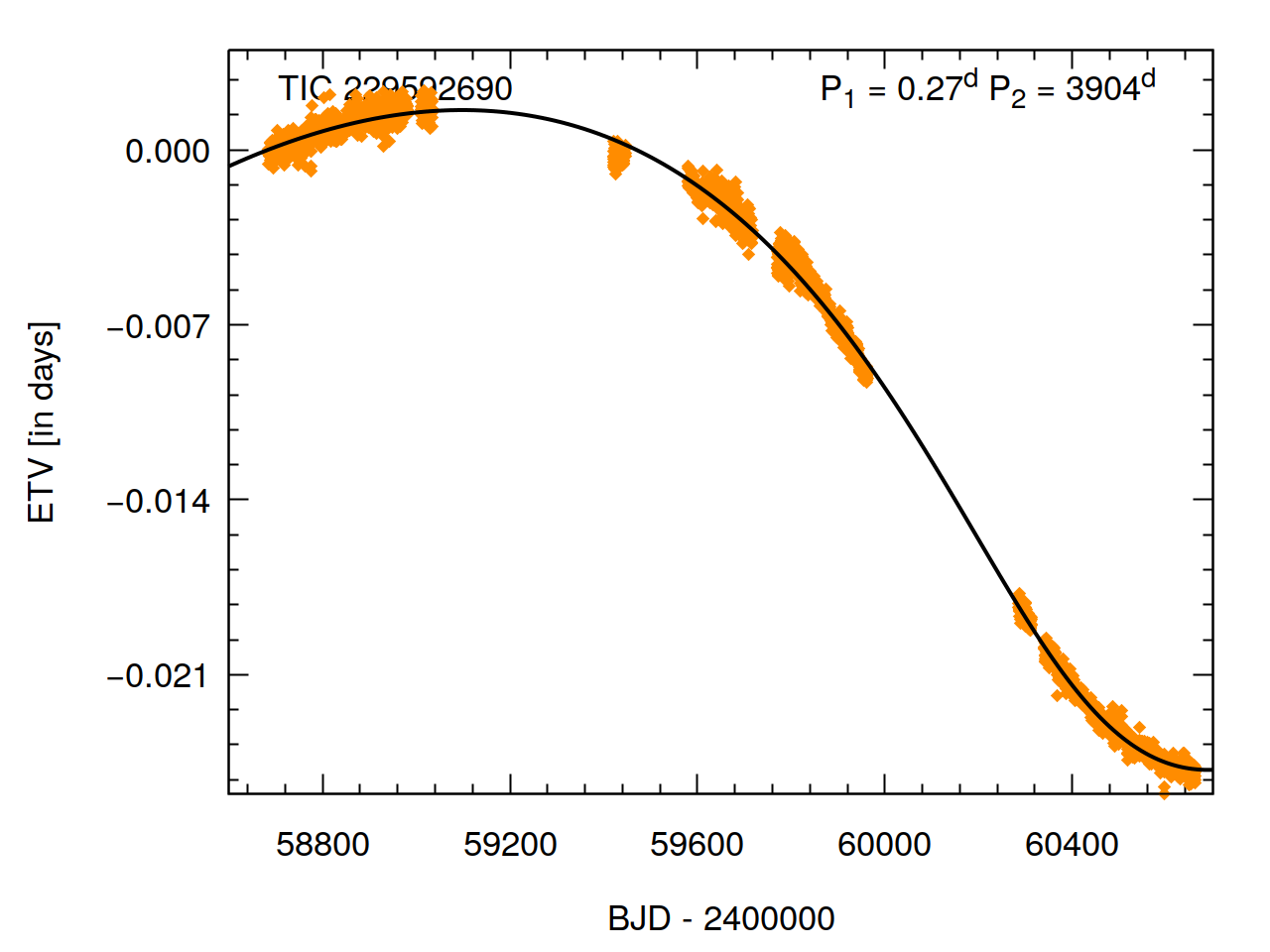}\includegraphics[width=0.43\textwidth]{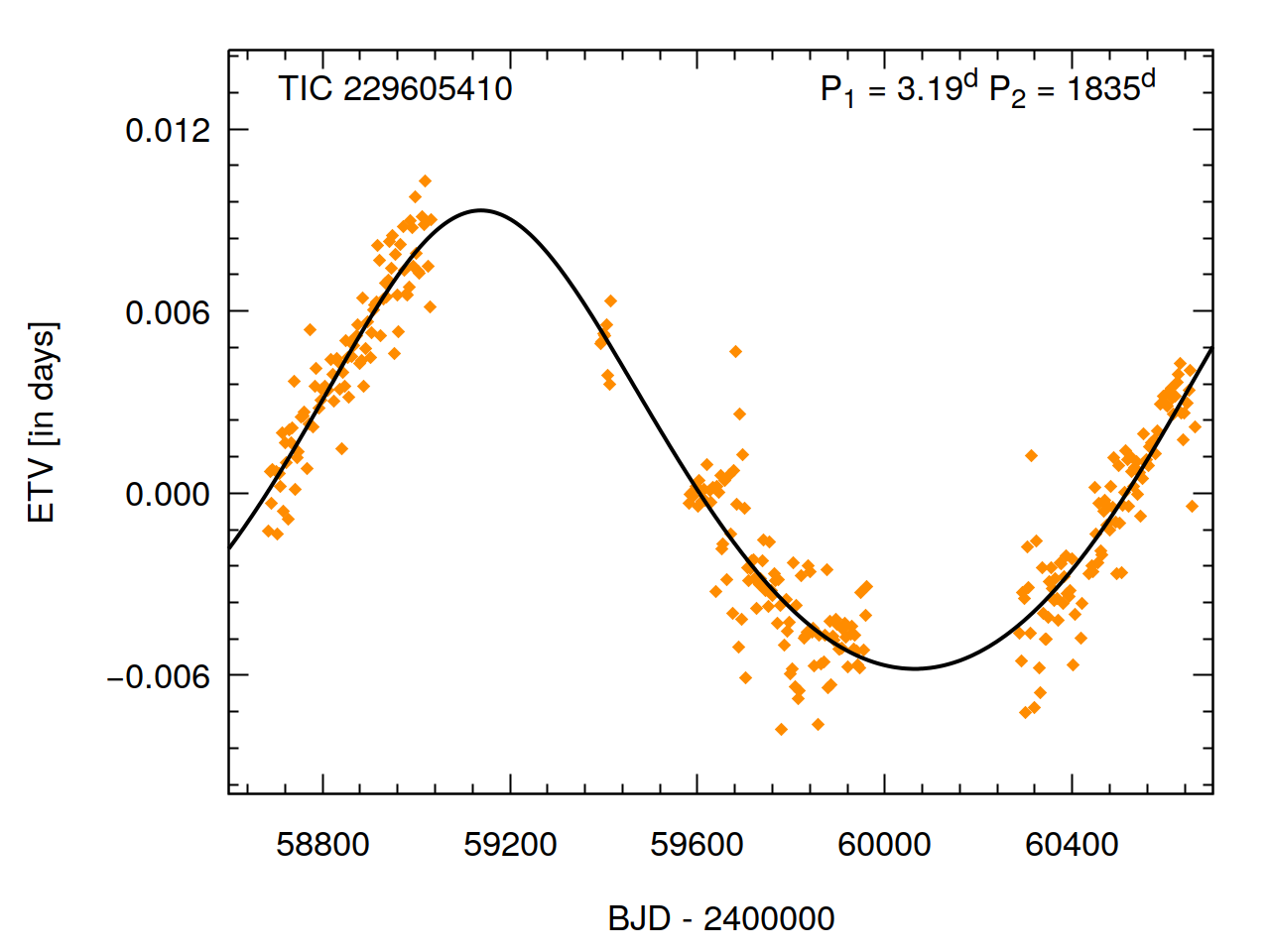}\includegraphics[width=0.43\textwidth]{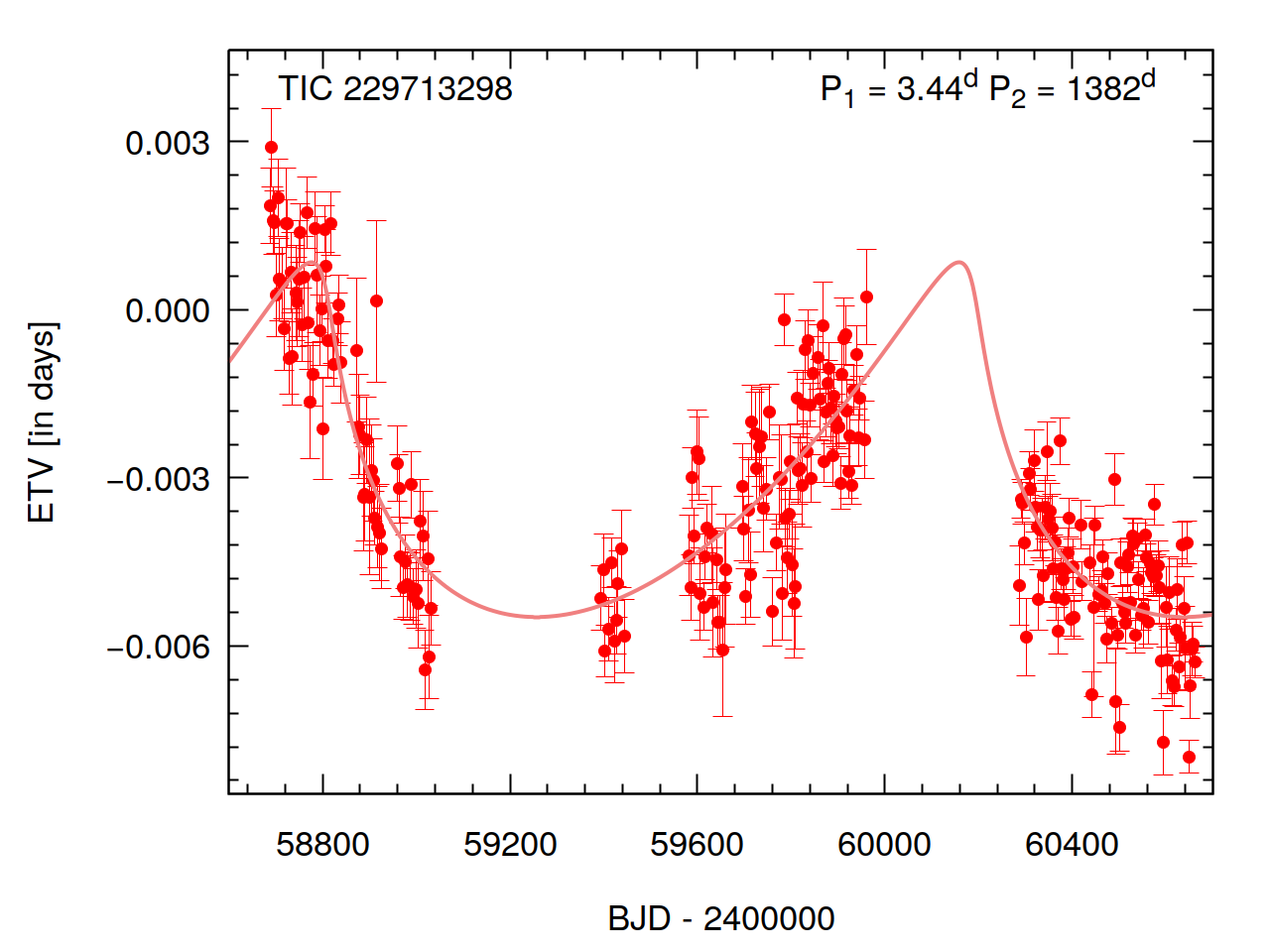}
\end{adjustwidth}
\caption{ETVs of the second 15 such systems which are classified into Group $L_3$. We call attention to the case of TIC 219738202, where earlier ground-based mid-minima times are also utilized and denoted by red upward and blue downward triangles for primary and secondary minima, respectively. The meanings of all the symbols are the same as in the former figures. See Table~\ref{Tab:Orbelem_LTTE3} for further details.}
\label{Fig:ETVs_L3b}
\end{figure}

\begin{figure}[H]
\begin{adjustwidth}{-\extralength}{0cm}
\centering
\includegraphics[width=0.43\textwidth]{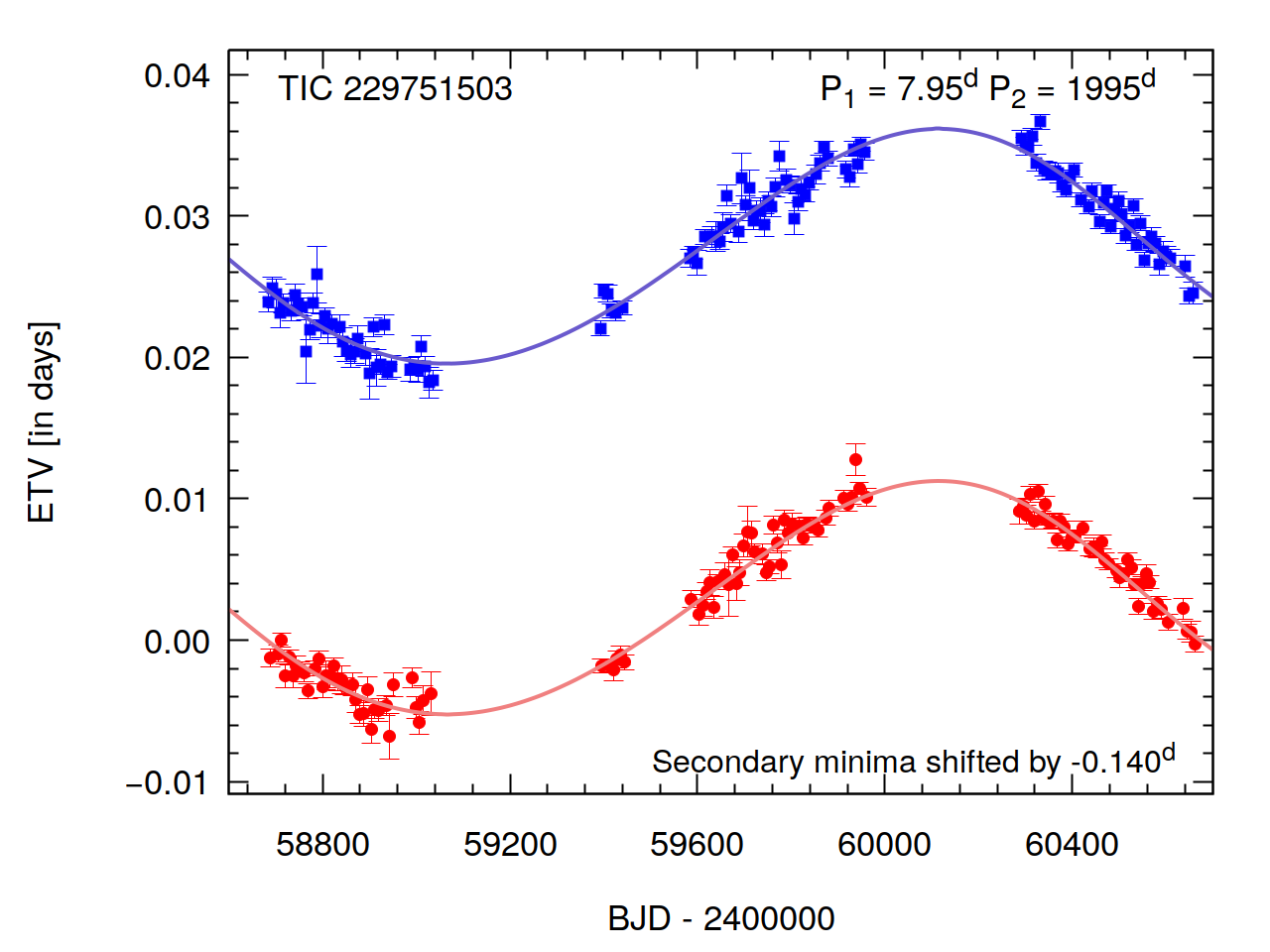}\includegraphics[width=0.43\textwidth]{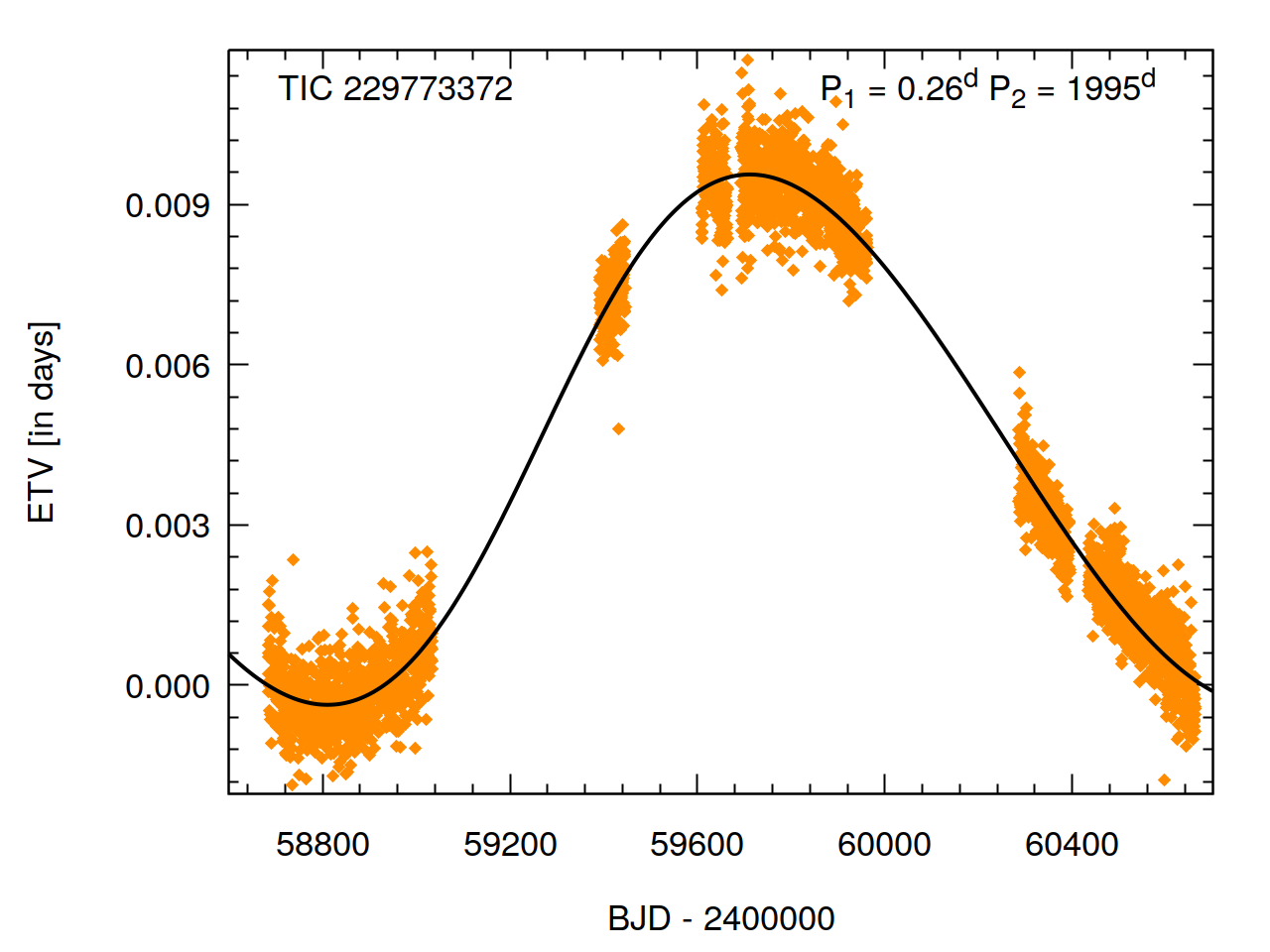}\includegraphics[width=0.43\textwidth]{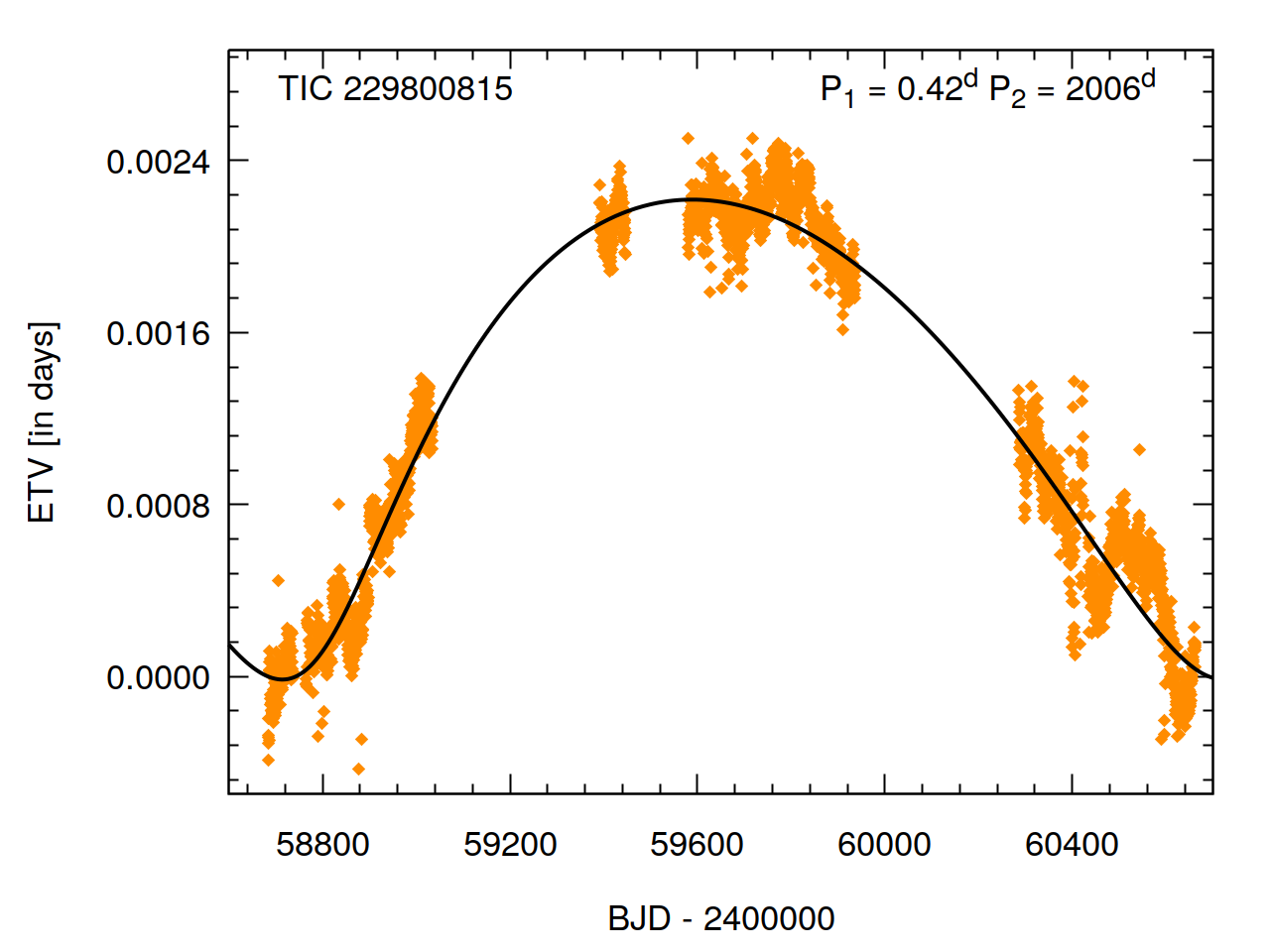}
\includegraphics[width=0.43\textwidth]{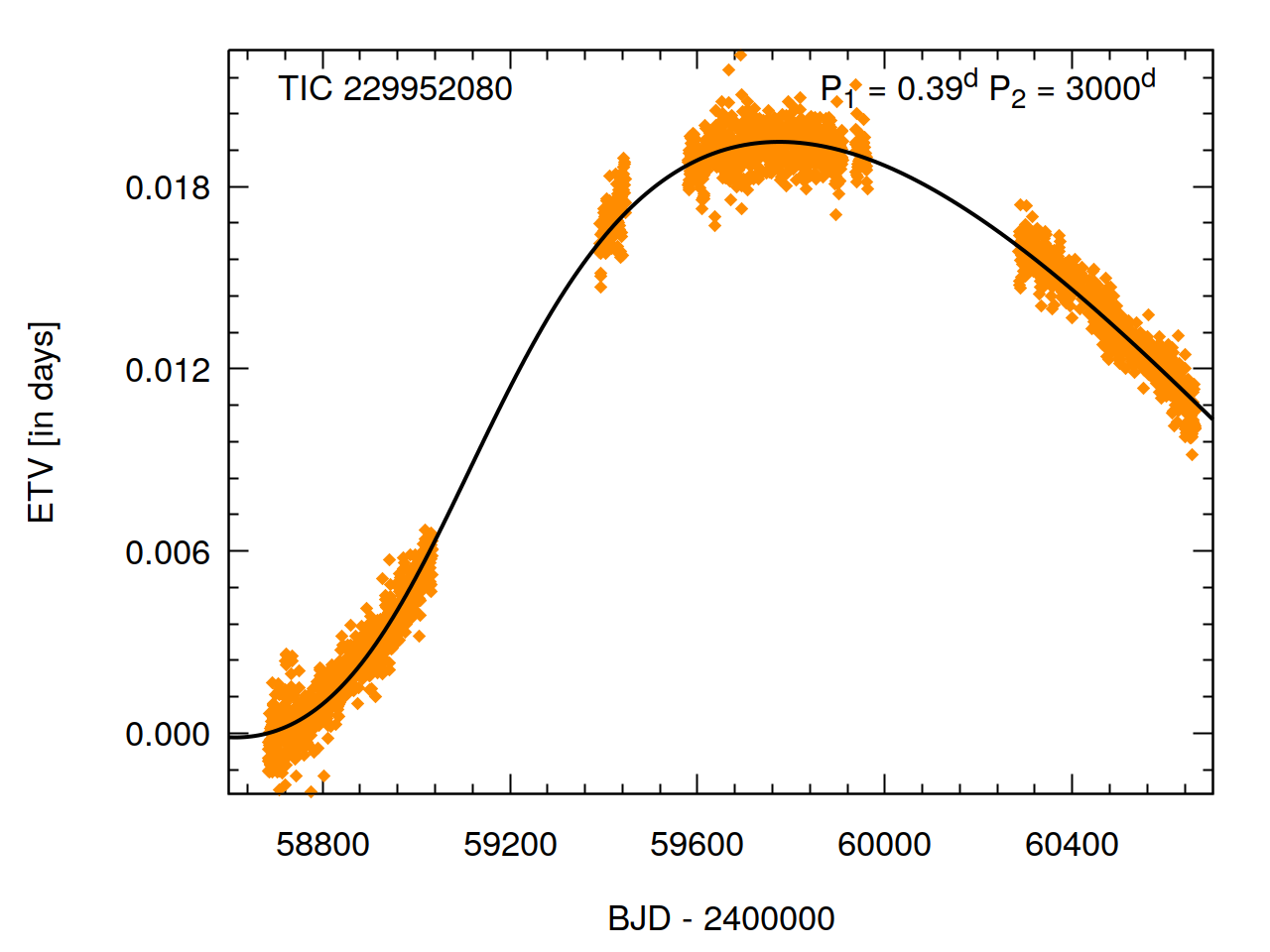}\includegraphics[width=0.43\textwidth]{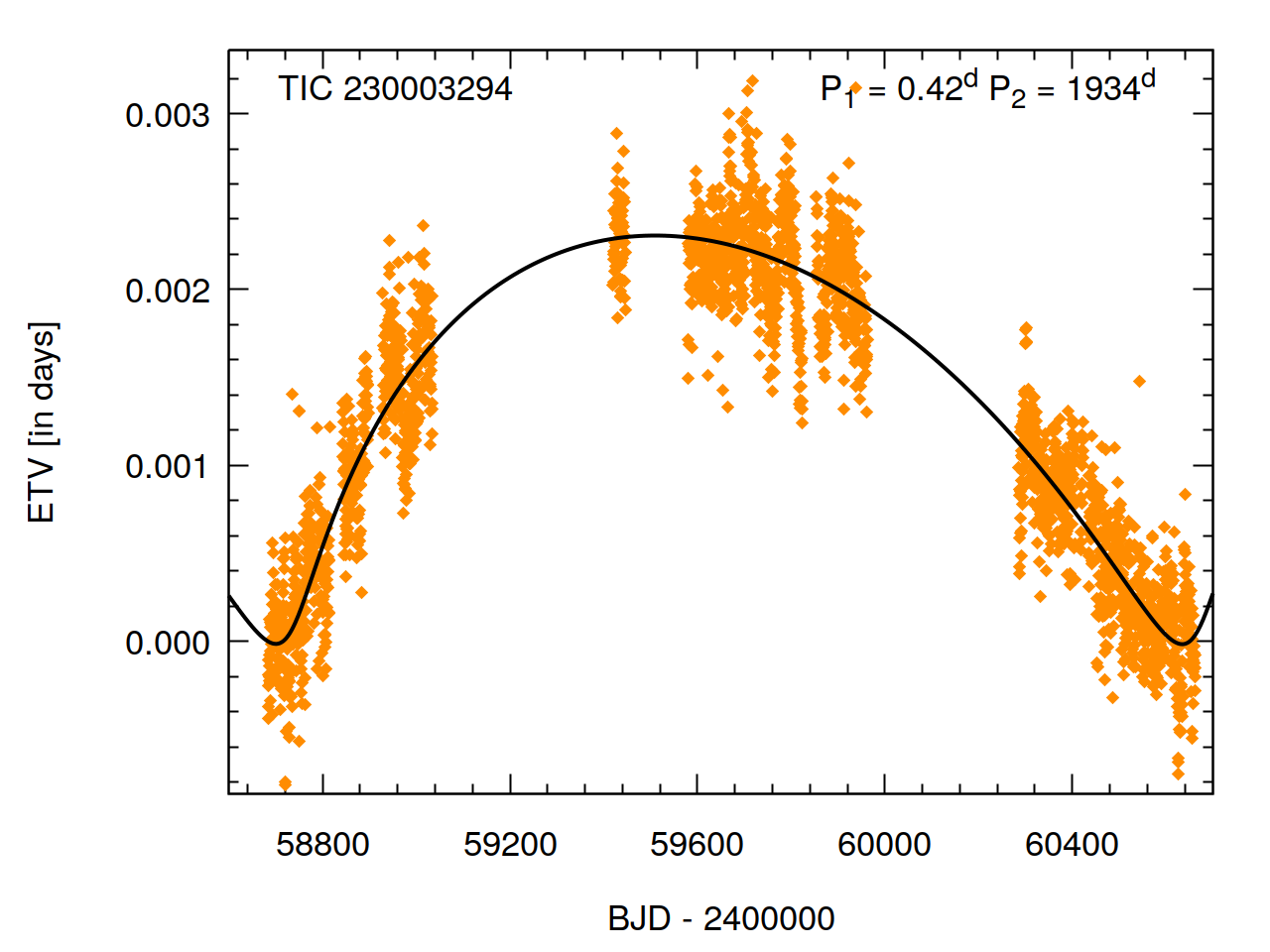}\includegraphics[width=0.43\textwidth]{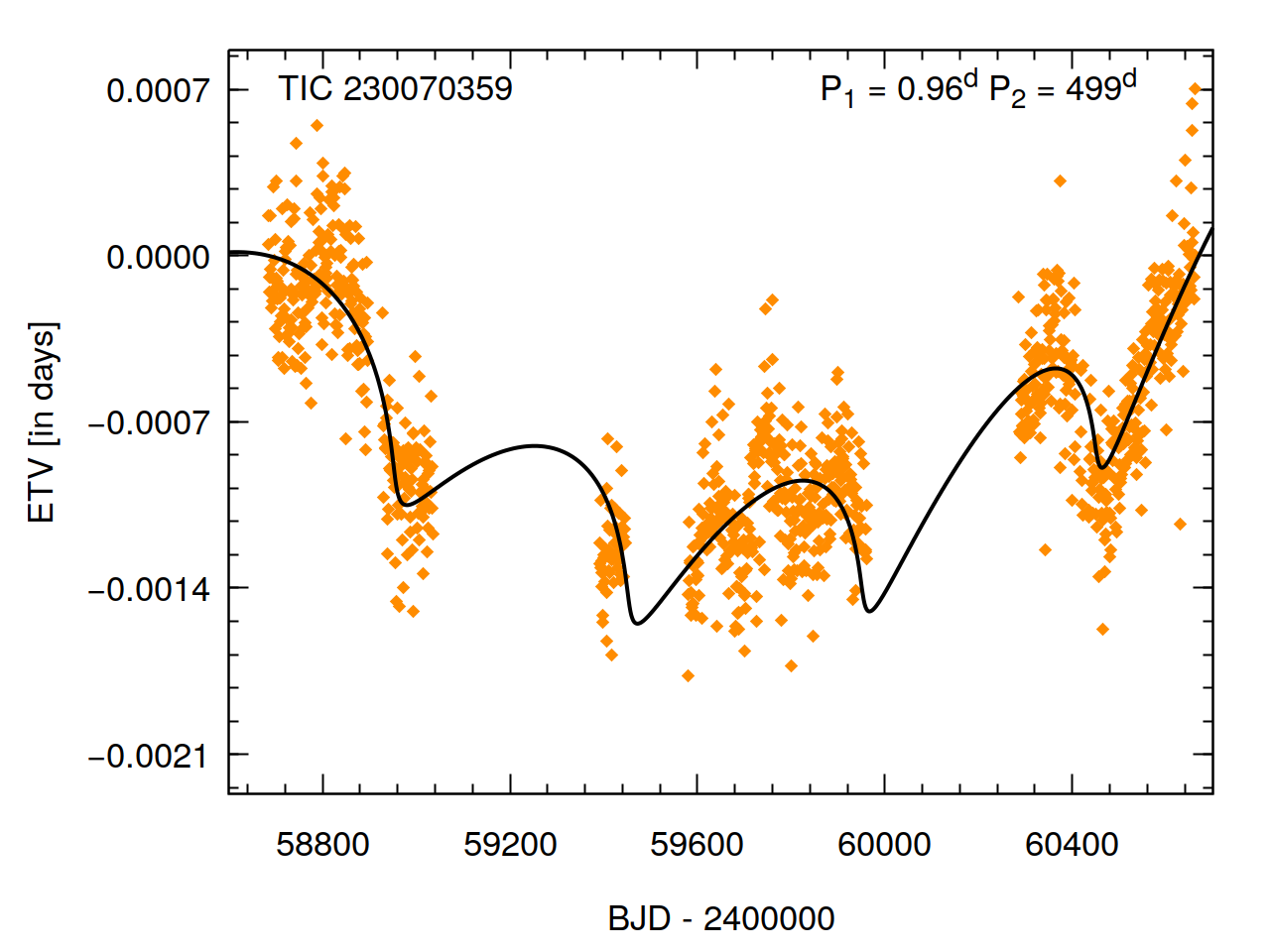}
\includegraphics[width=0.43\textwidth]{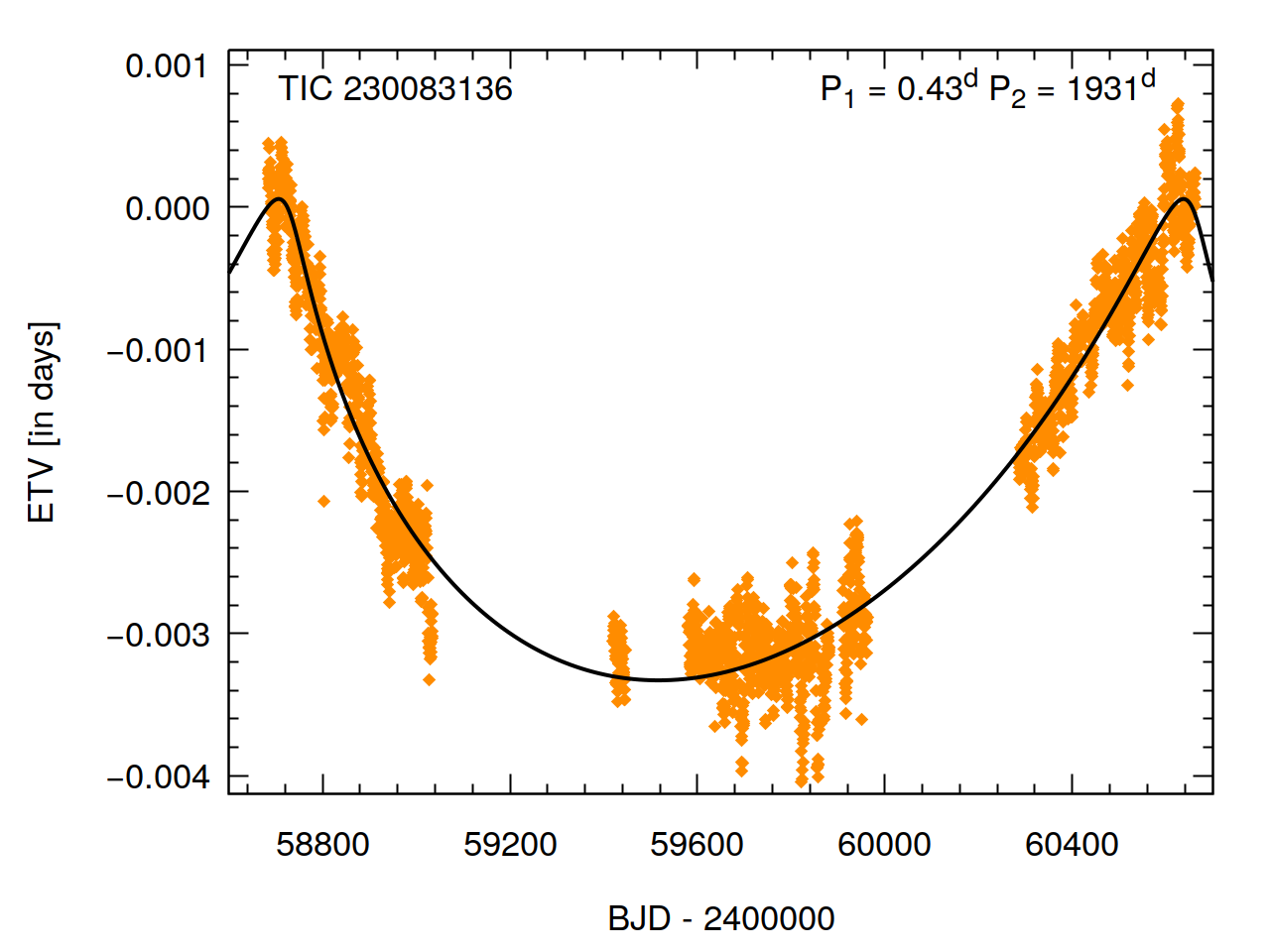}\includegraphics[width=0.43\textwidth]{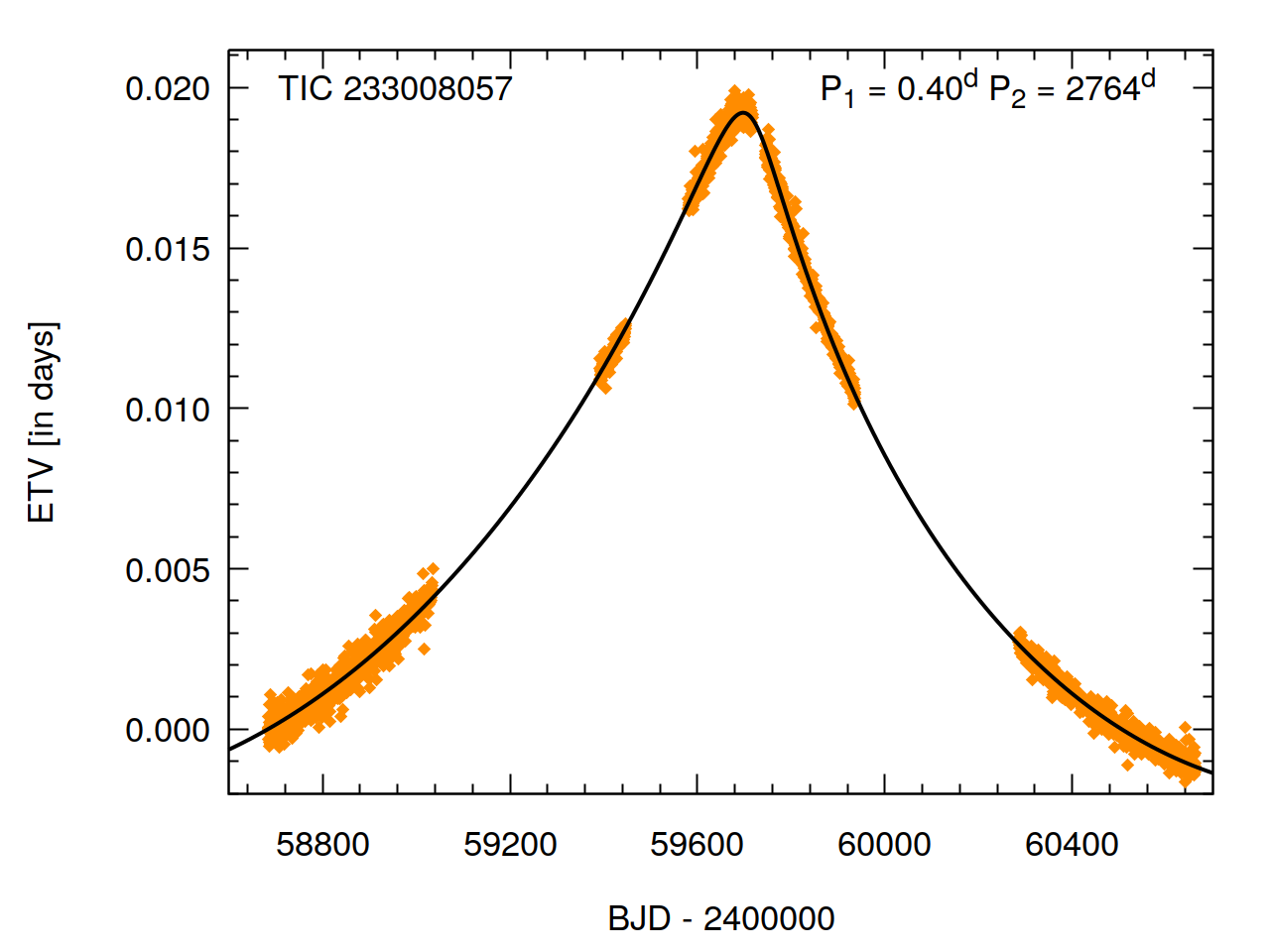}\includegraphics[width=0.43\textwidth]{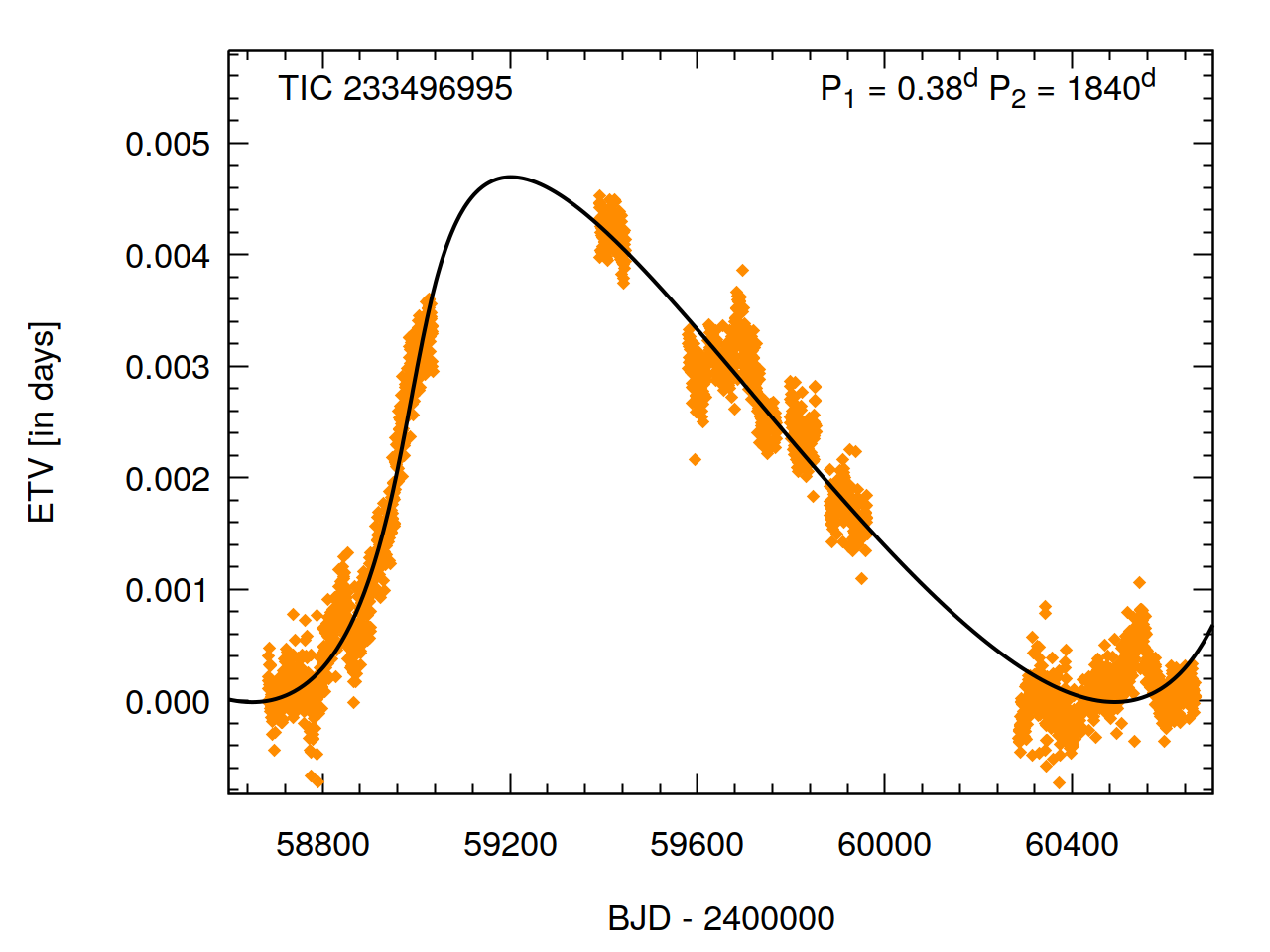}
\includegraphics[width=0.43\textwidth]{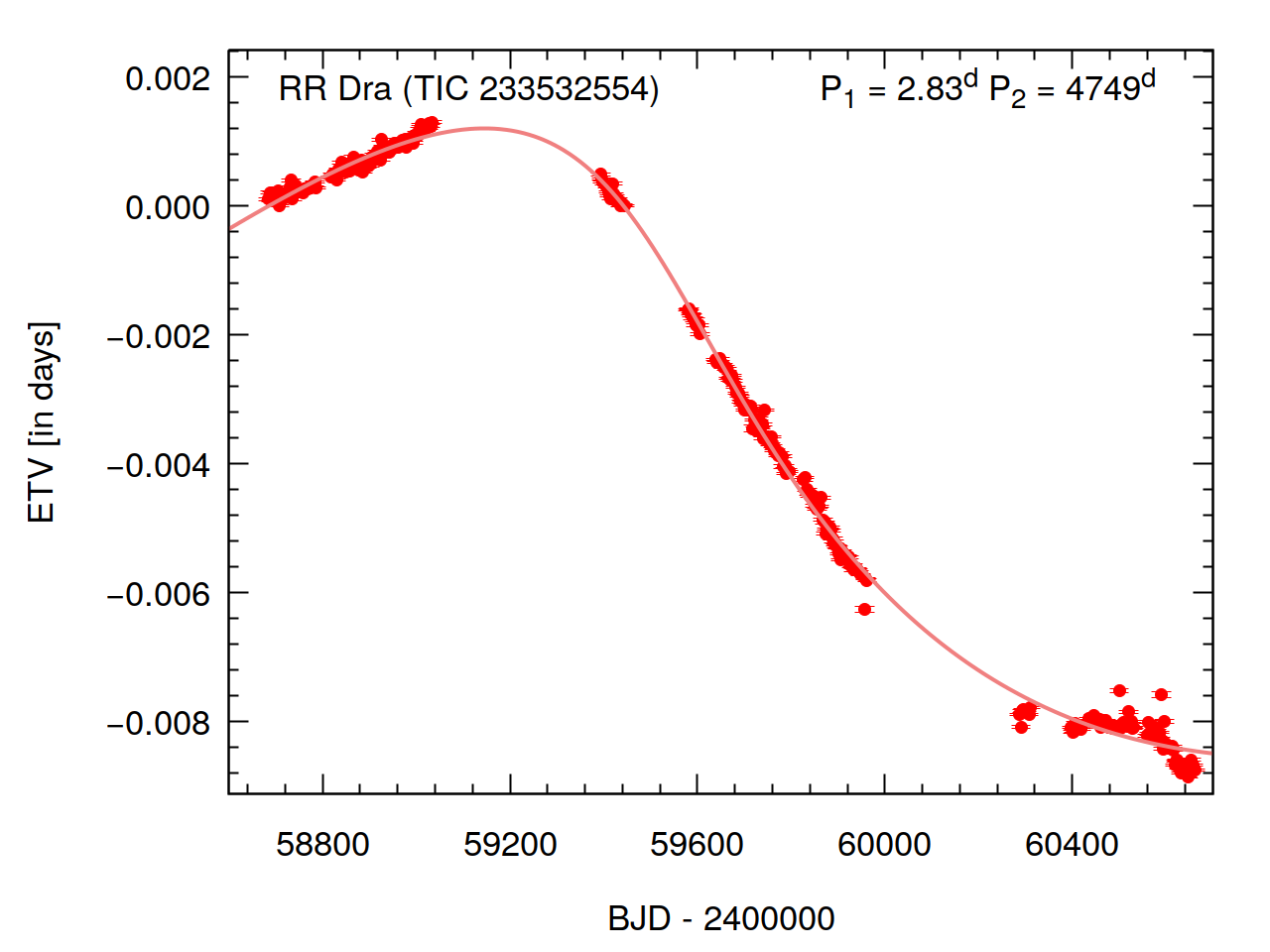}\includegraphics[width=0.43\textwidth]{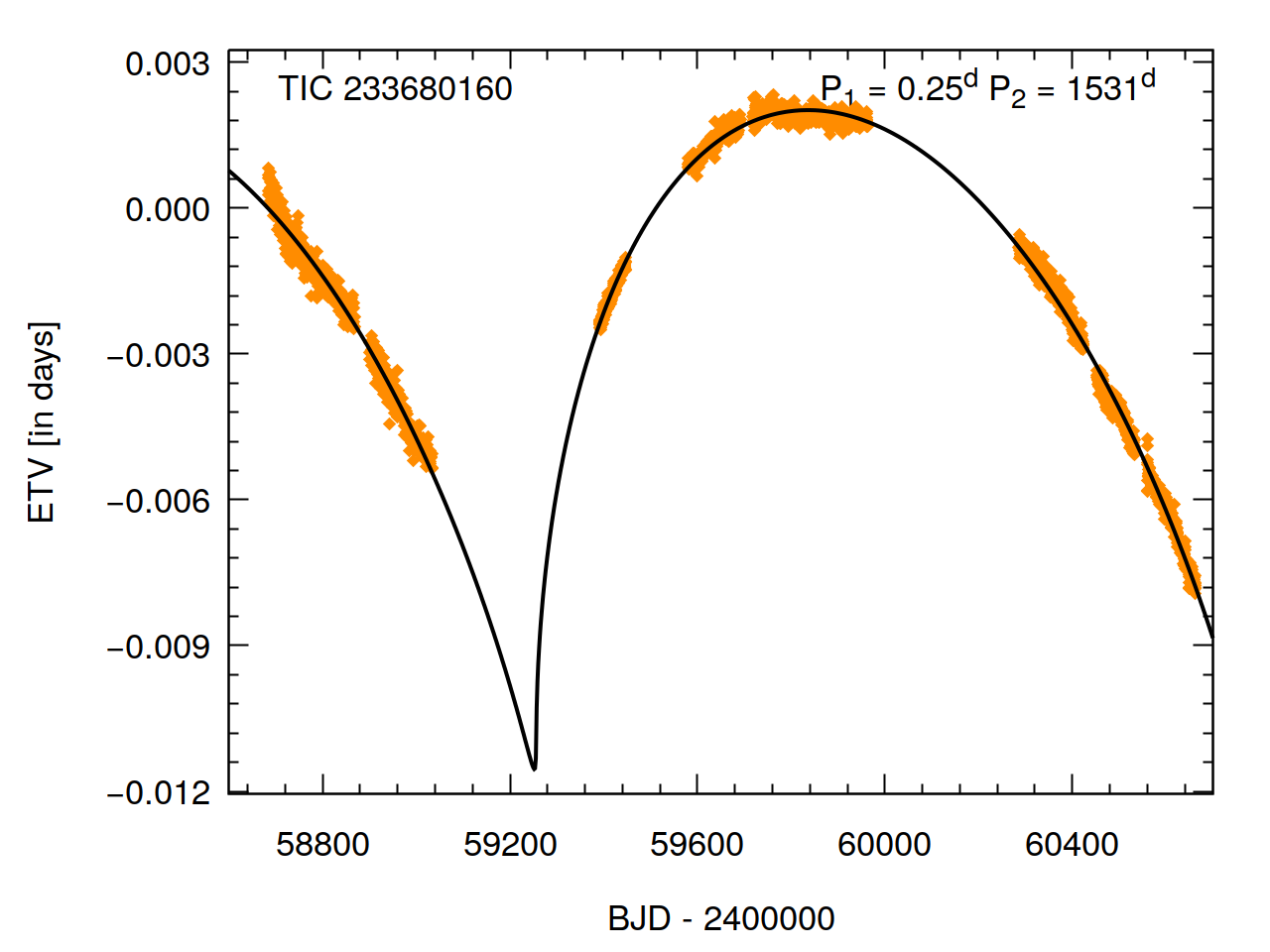}\includegraphics[width=0.43\textwidth]{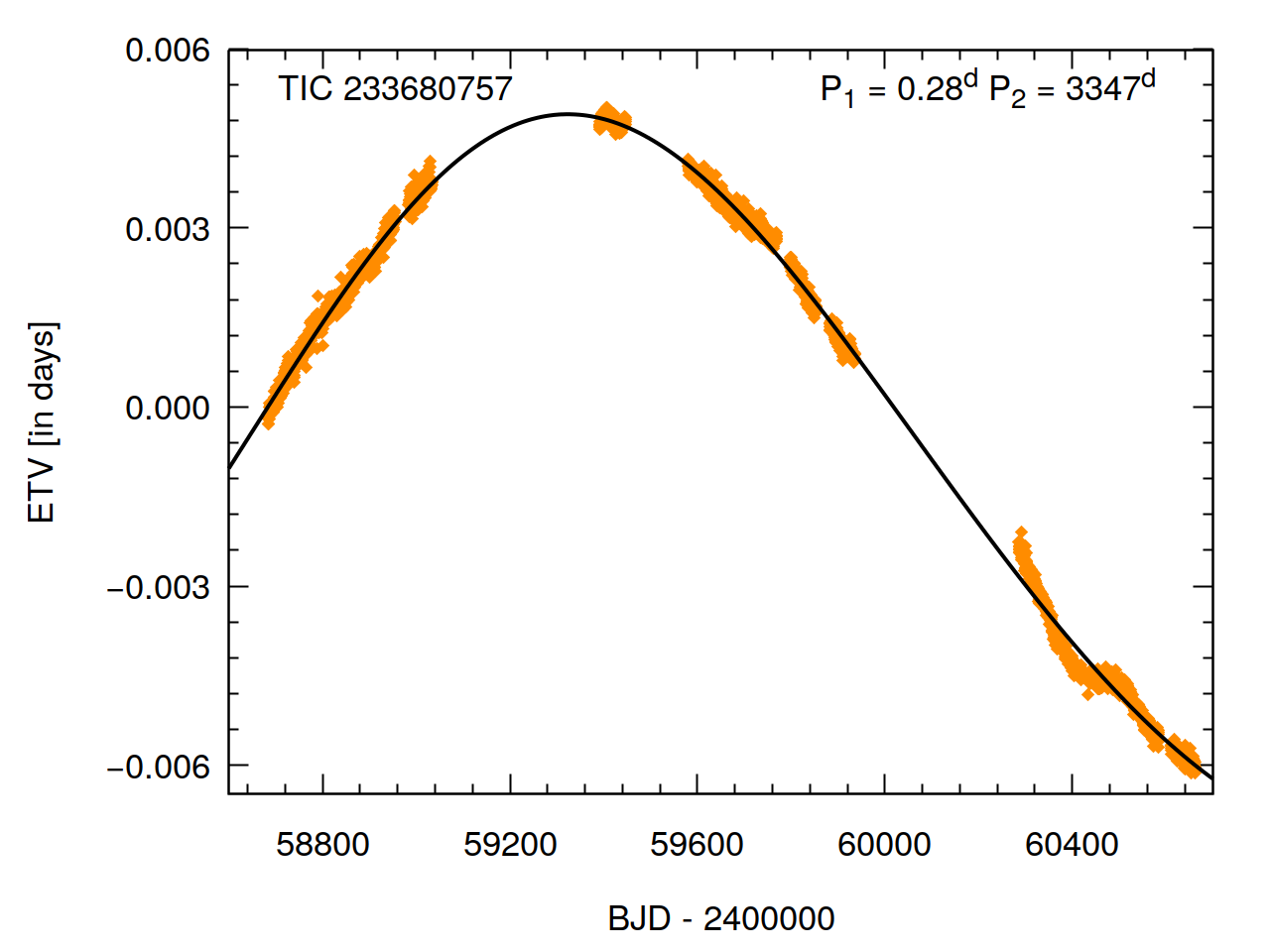}
\includegraphics[width=0.43\textwidth]{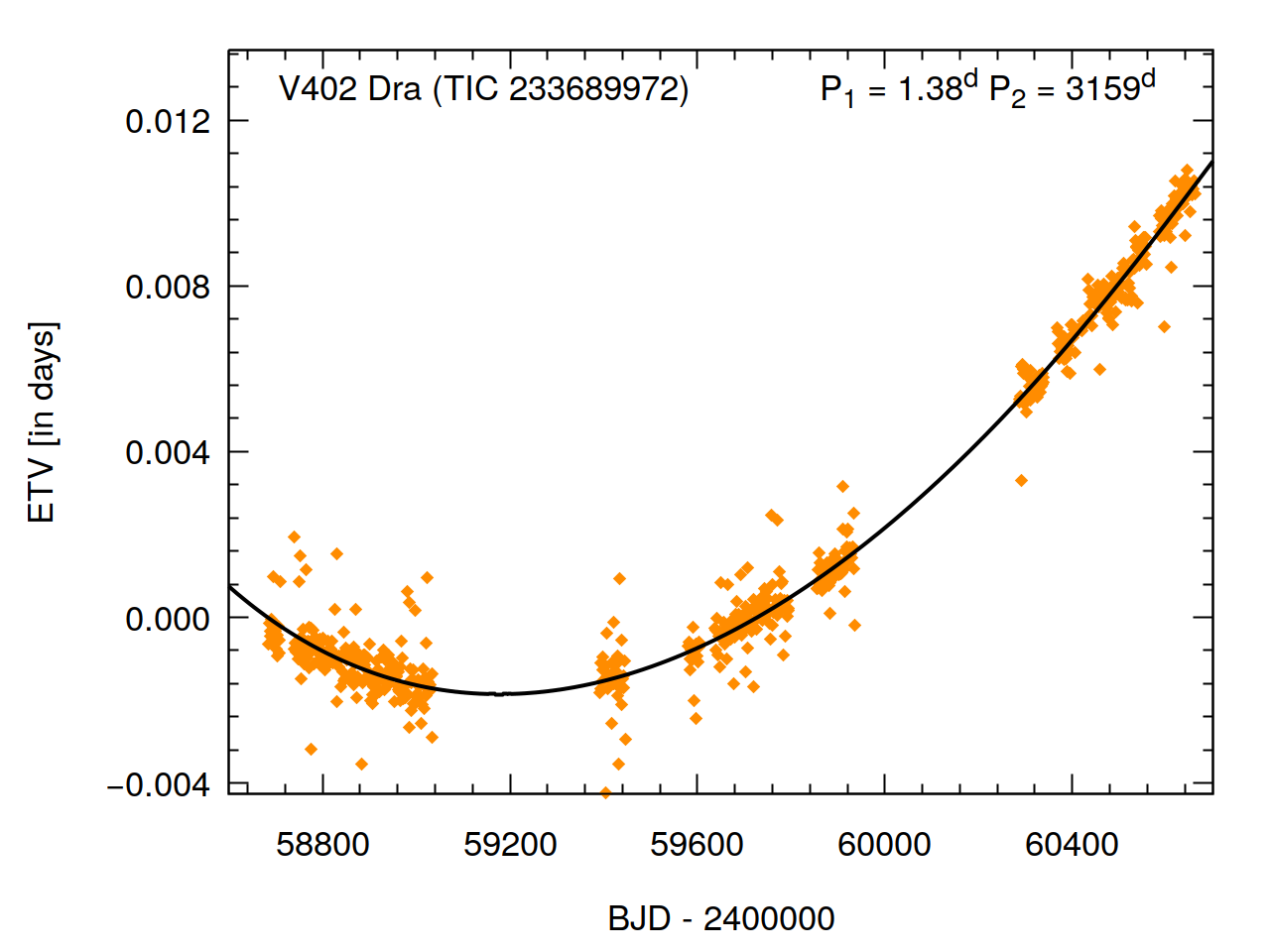}\includegraphics[width=0.43\textwidth]{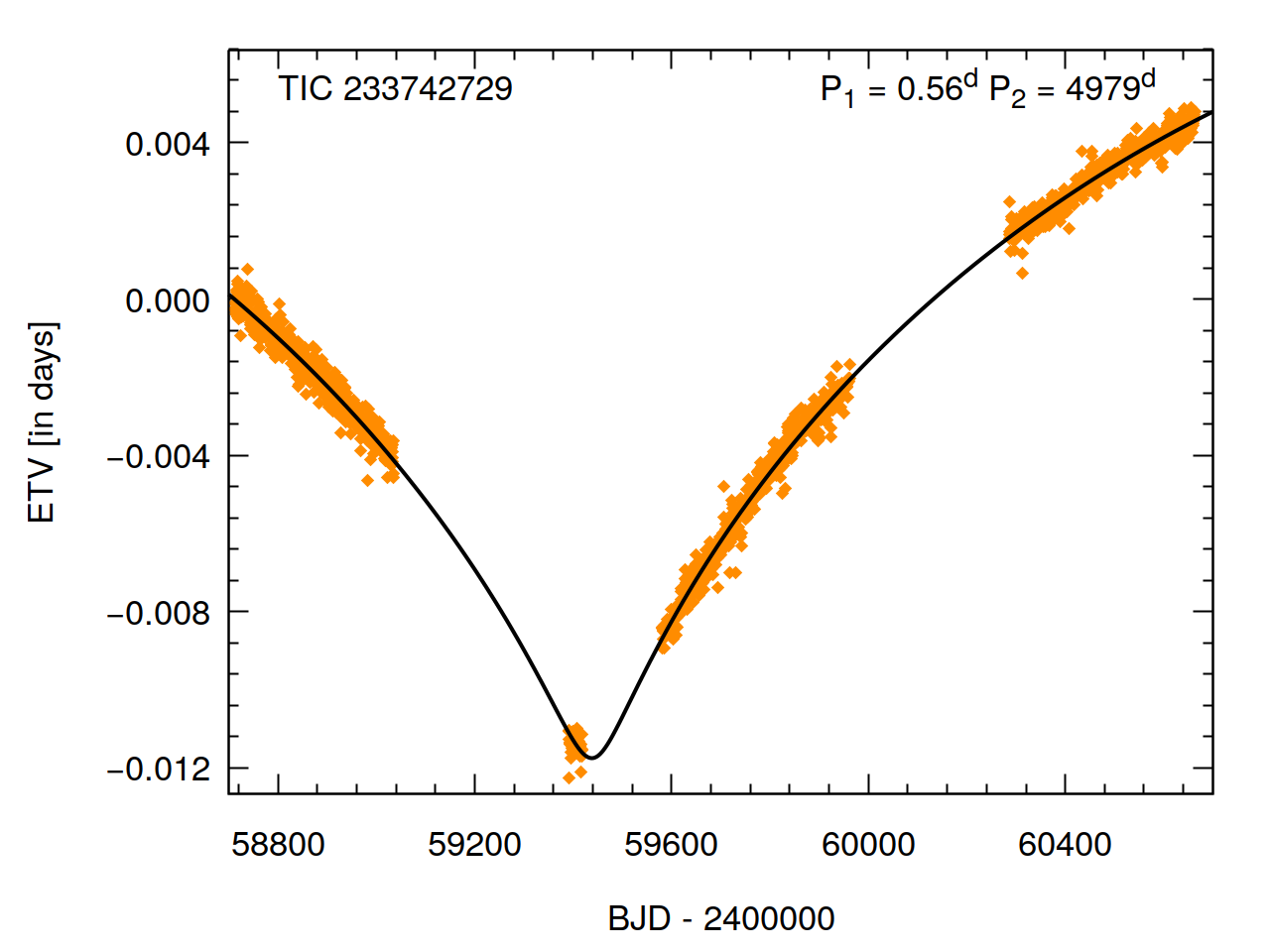}\includegraphics[width=0.43\textwidth]{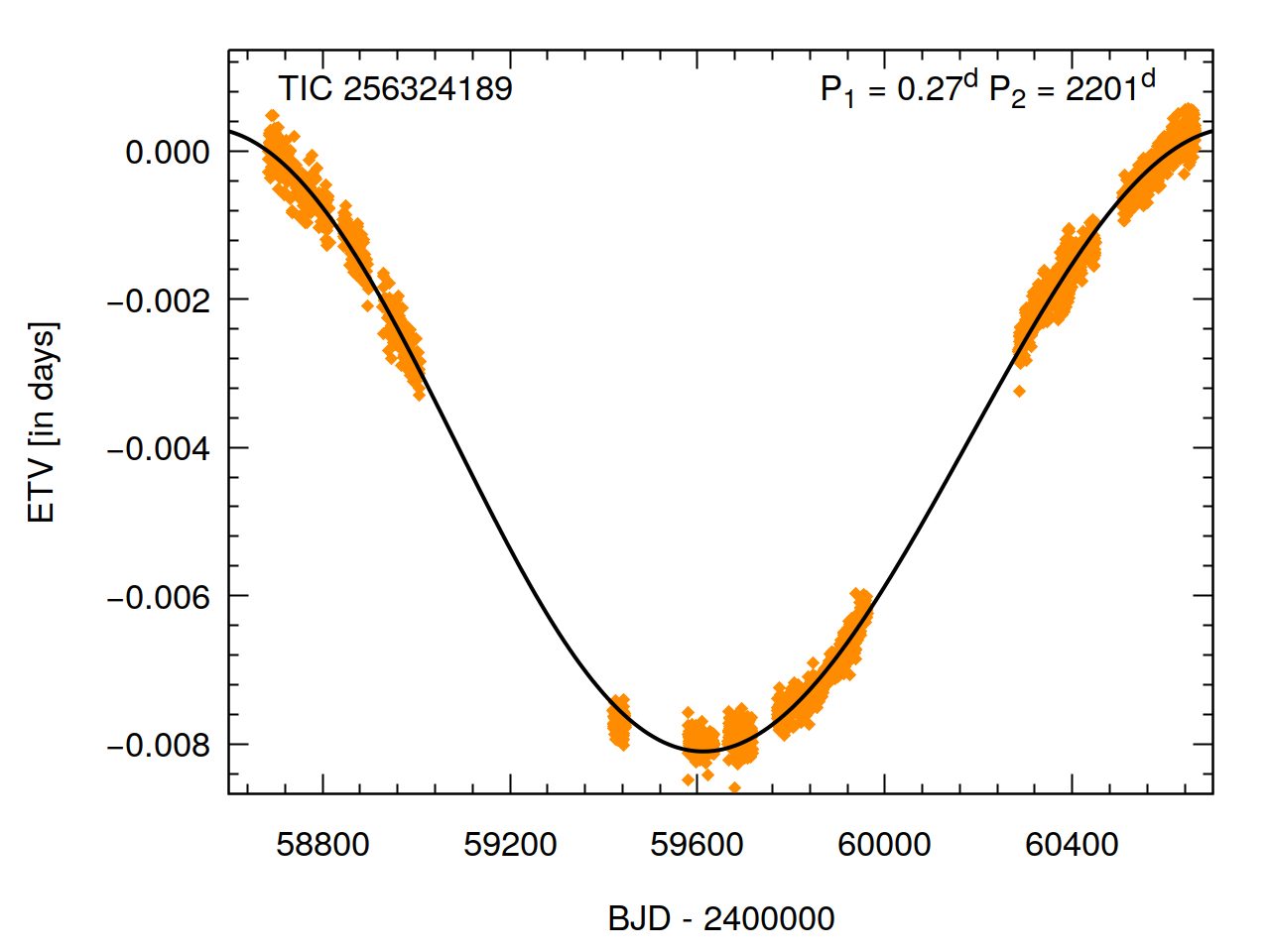}
\end{adjustwidth}
\caption{ETVs of the third set of 15 such systems that are classified into Group $L_3$. In the case of TIC 229751503 (uppermost, left panel), the separate primary and secondary TESS--ETV curves (red circles and blue boxes, respectively) show directly that the EB is eccentric, as the offset of the secondary curve relative to the primary one is proportional to $e_1\cos\omega_1$. The meaning of all the other symbols on the remaining 14 ETV panels are the same as were described formerly. See Table~\ref{Tab:Orbelem_LTTE3} for further details.}
\label{Fig:ETVs_L3c}
\end{figure}

\begin{figure}[H]
\begin{adjustwidth}{-\extralength}{0cm}
\centering
\includegraphics[width=0.43\textwidth]{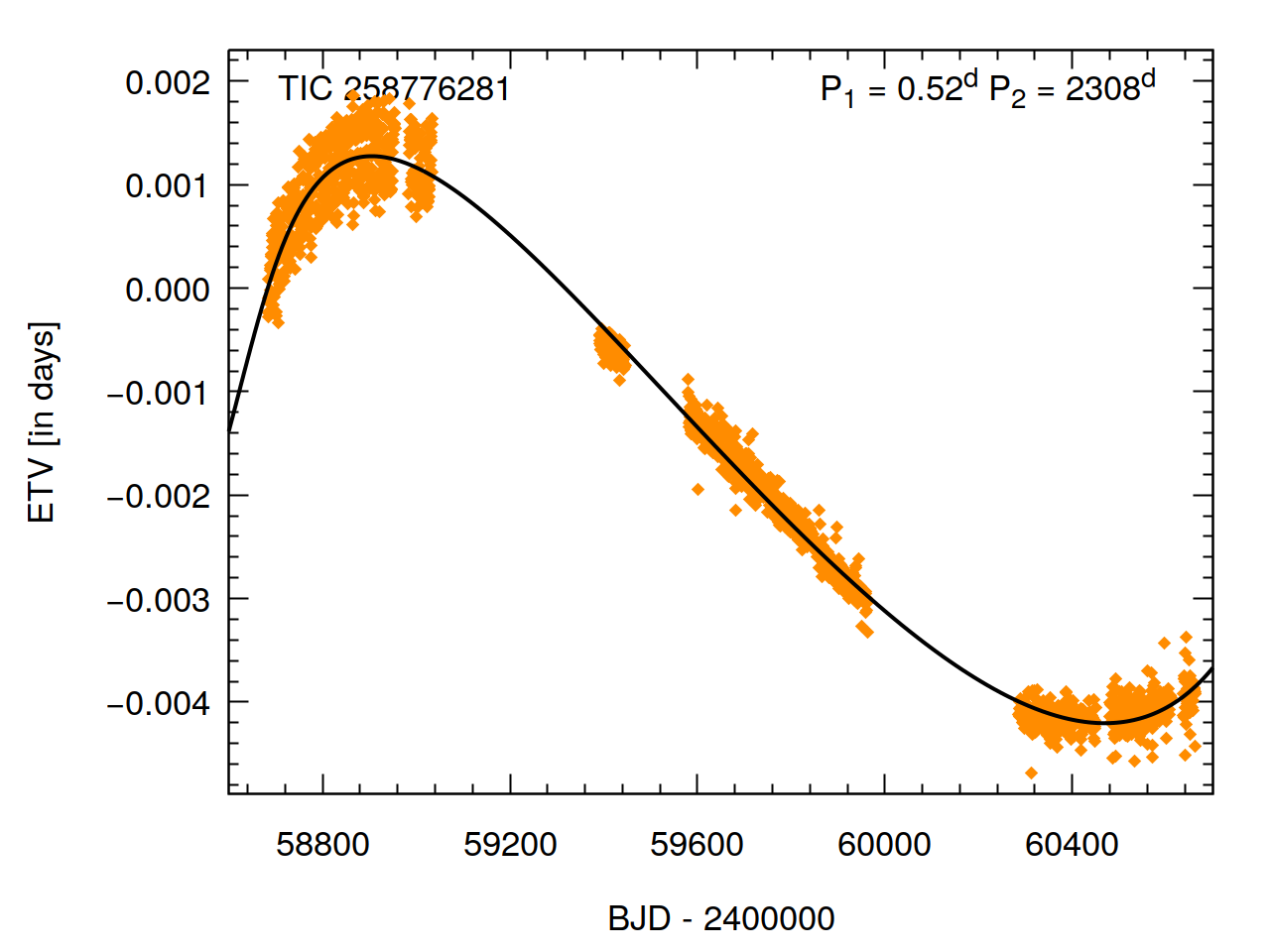}\includegraphics[width=0.43\textwidth]{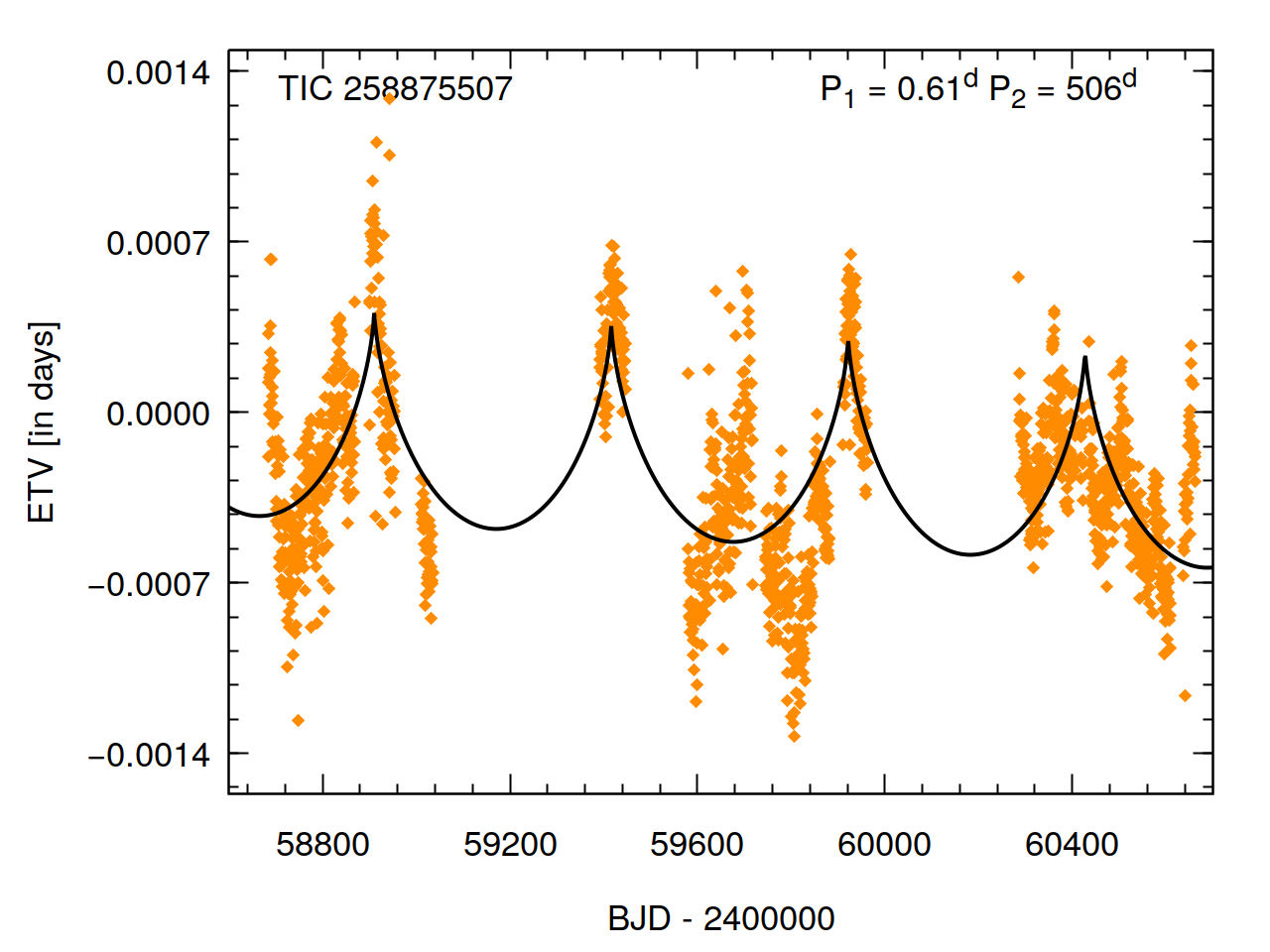}\includegraphics[width=0.43\textwidth]{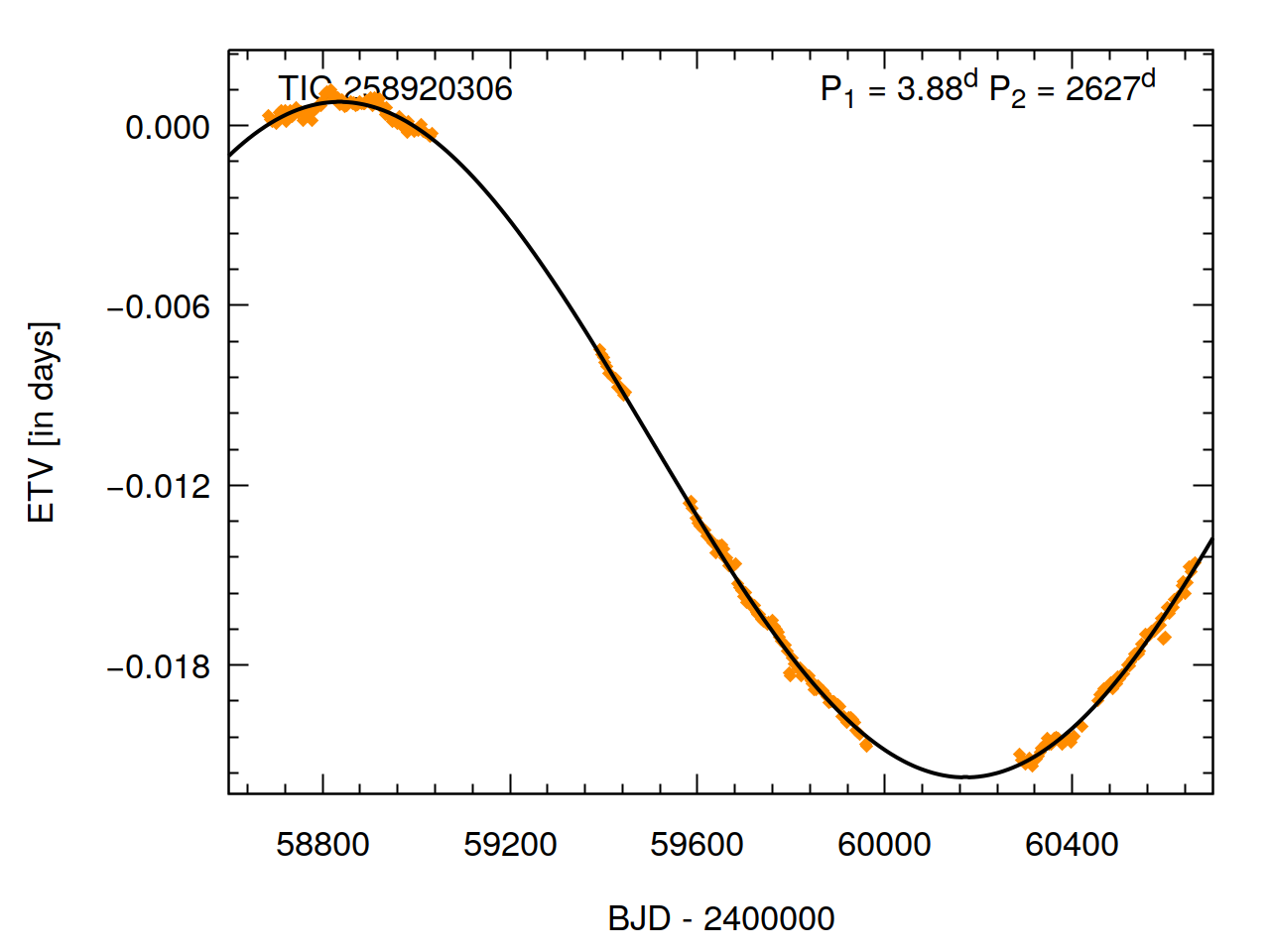}
\includegraphics[width=0.43\textwidth]{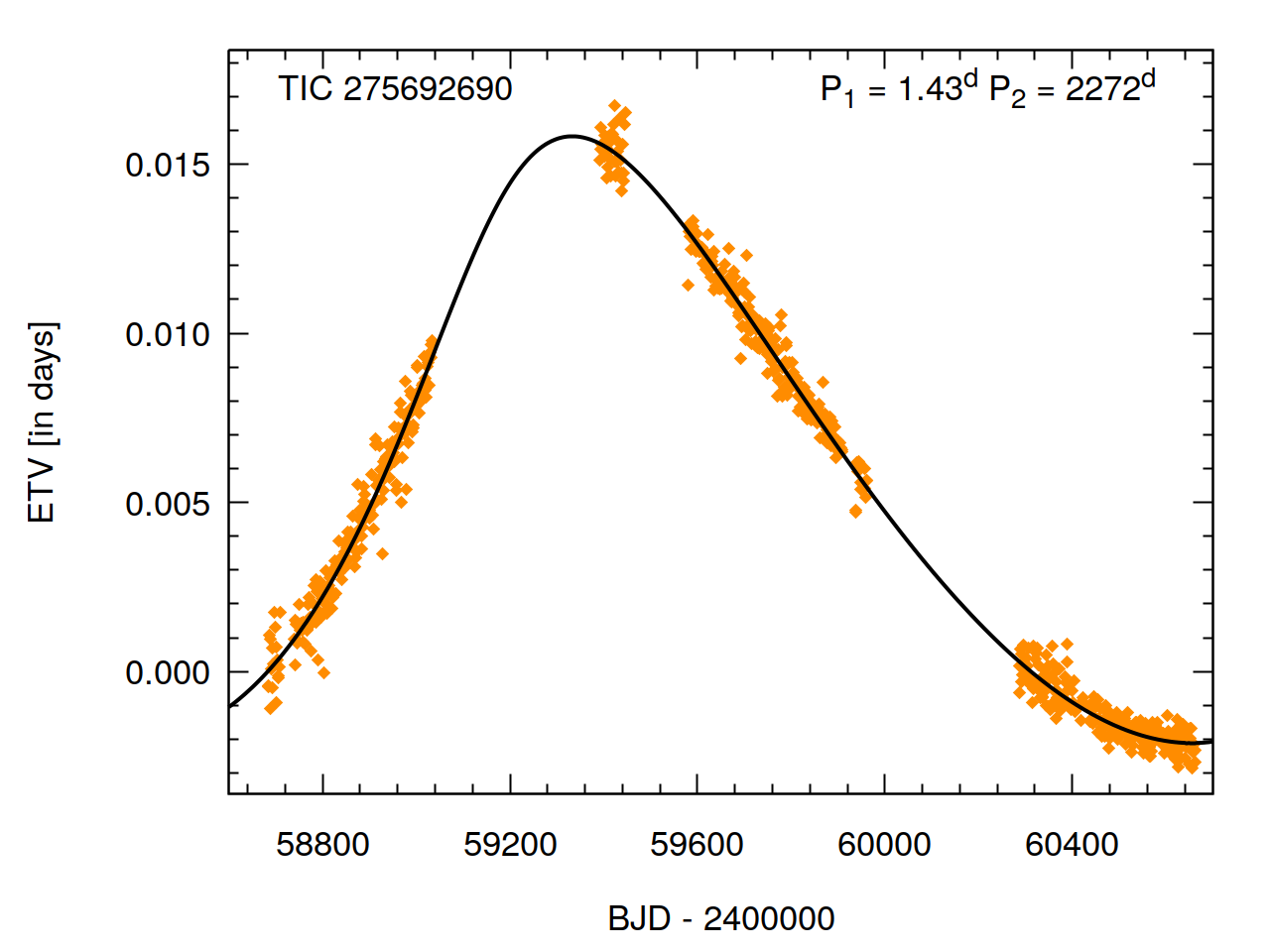}\includegraphics[width=0.43\textwidth]{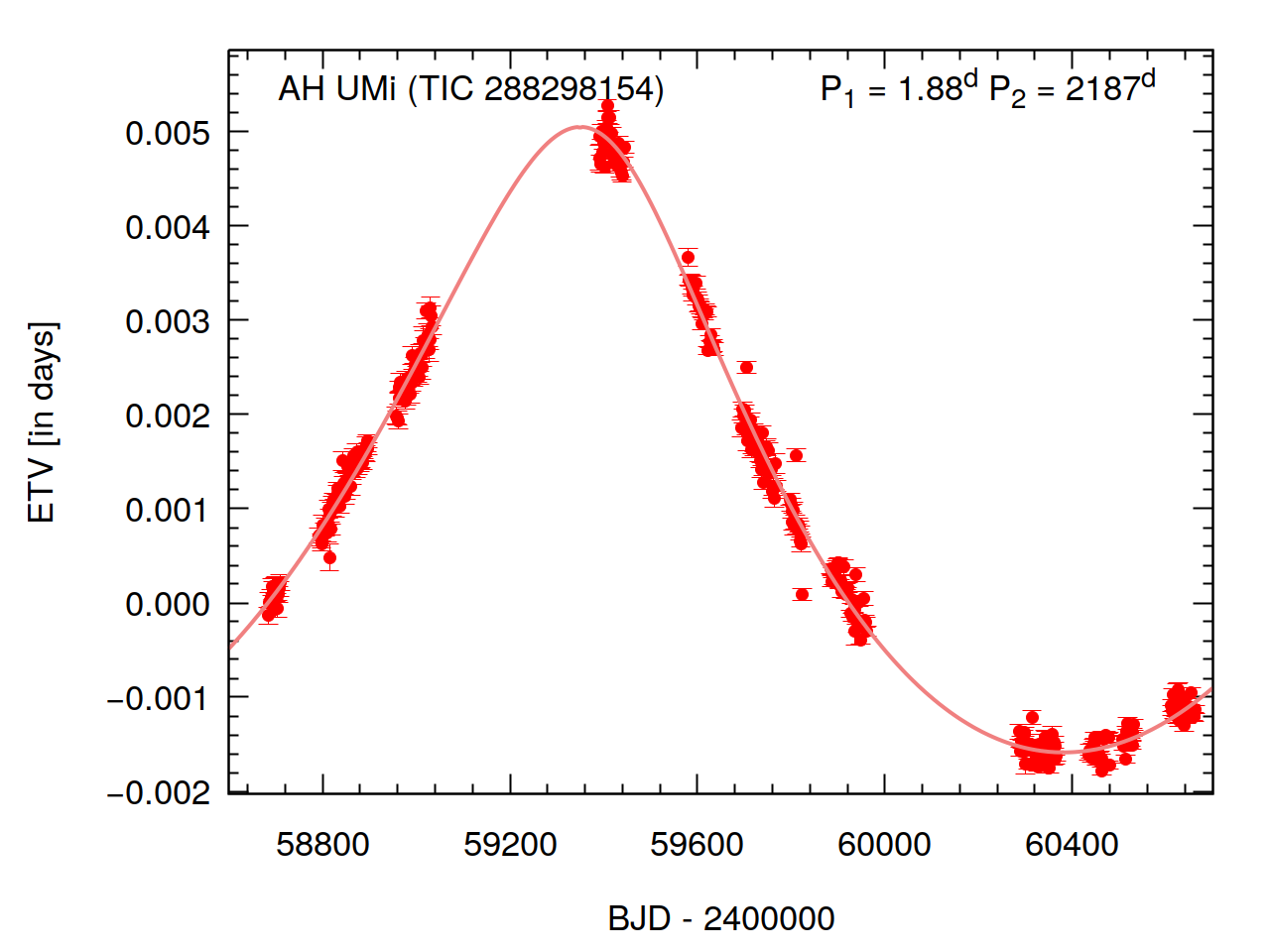}\includegraphics[width=0.43\textwidth]{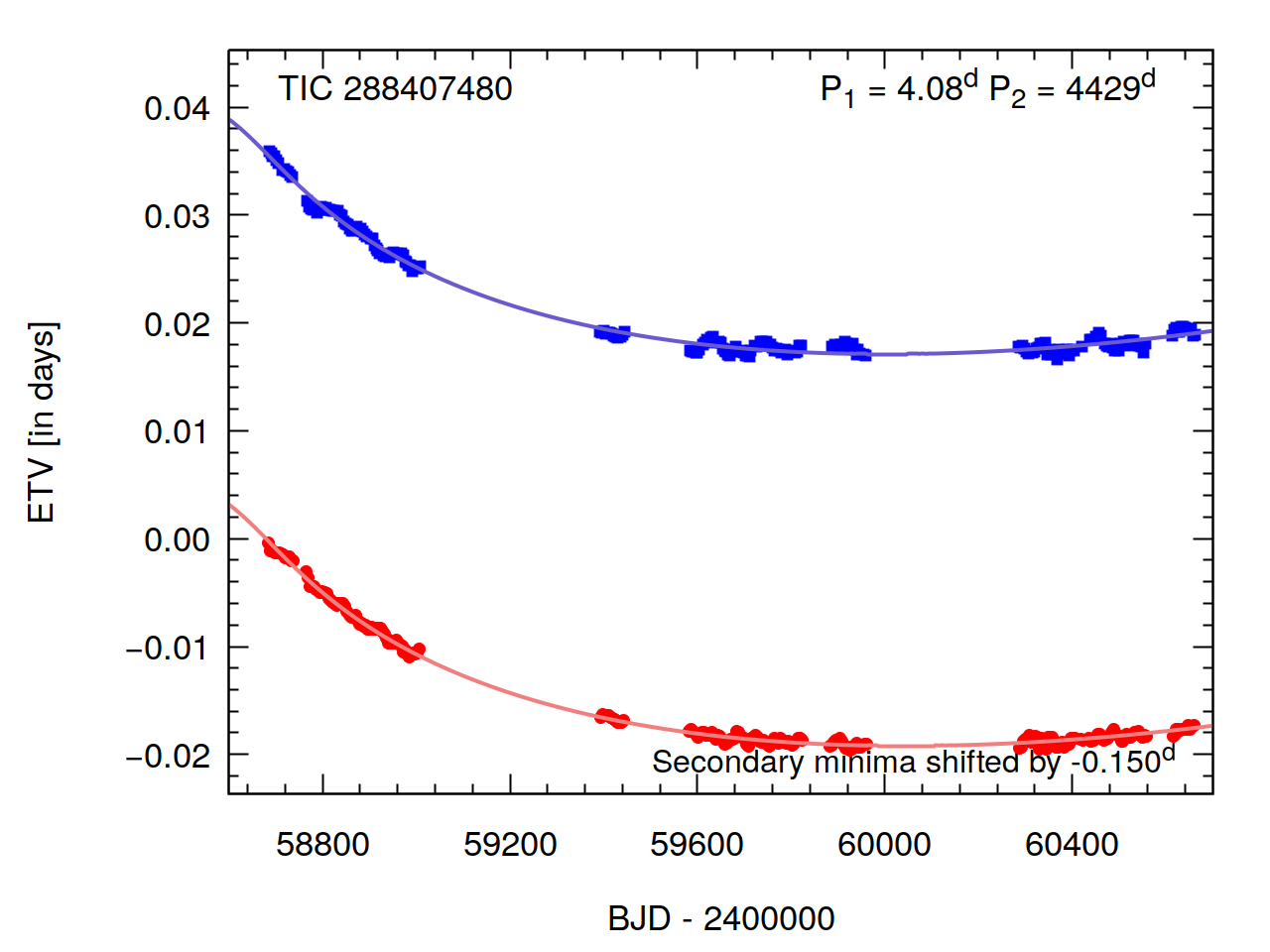}
\includegraphics[width=0.43\textwidth]{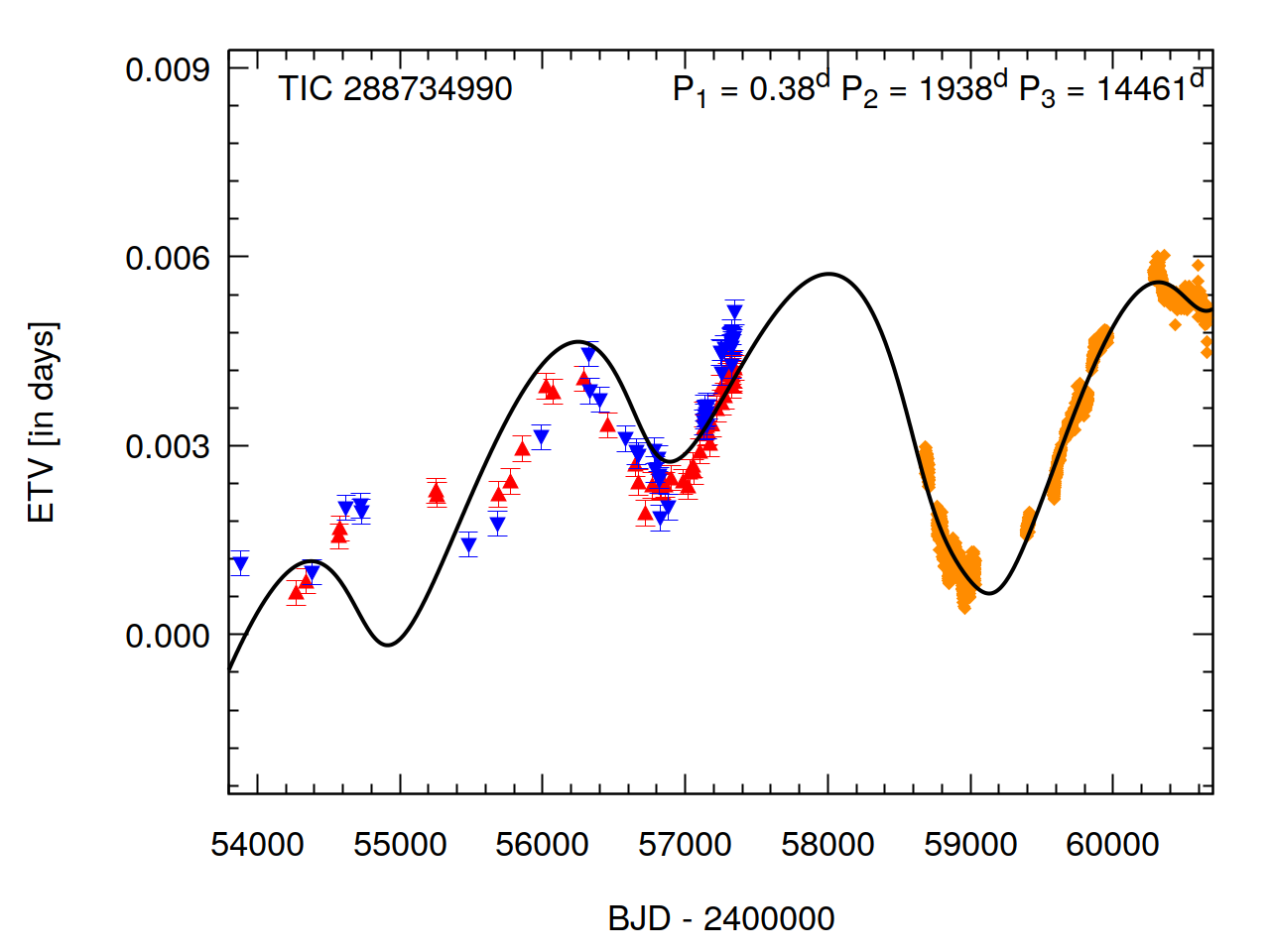}\includegraphics[width=0.43\textwidth]{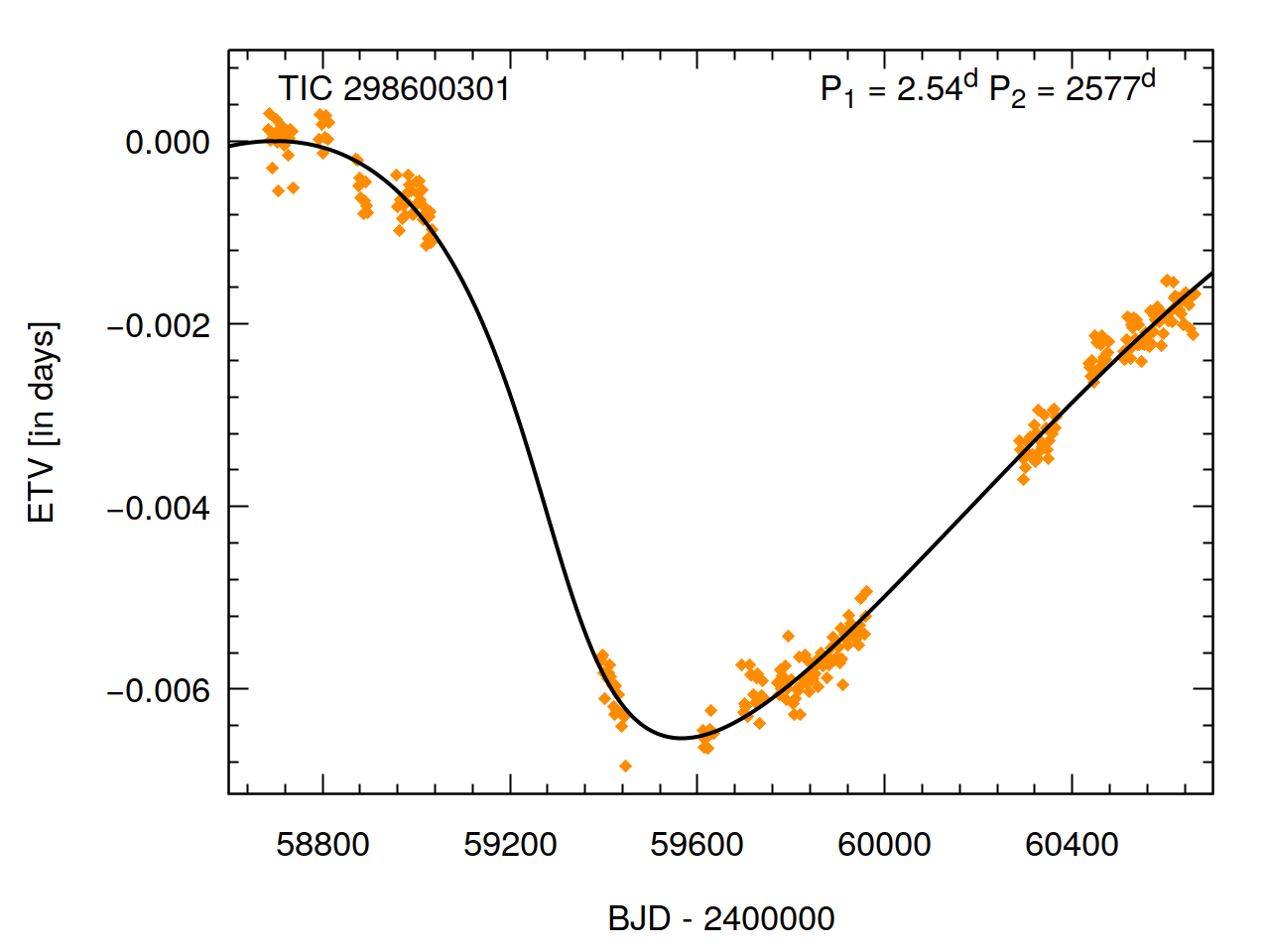}\includegraphics[width=0.43\textwidth]{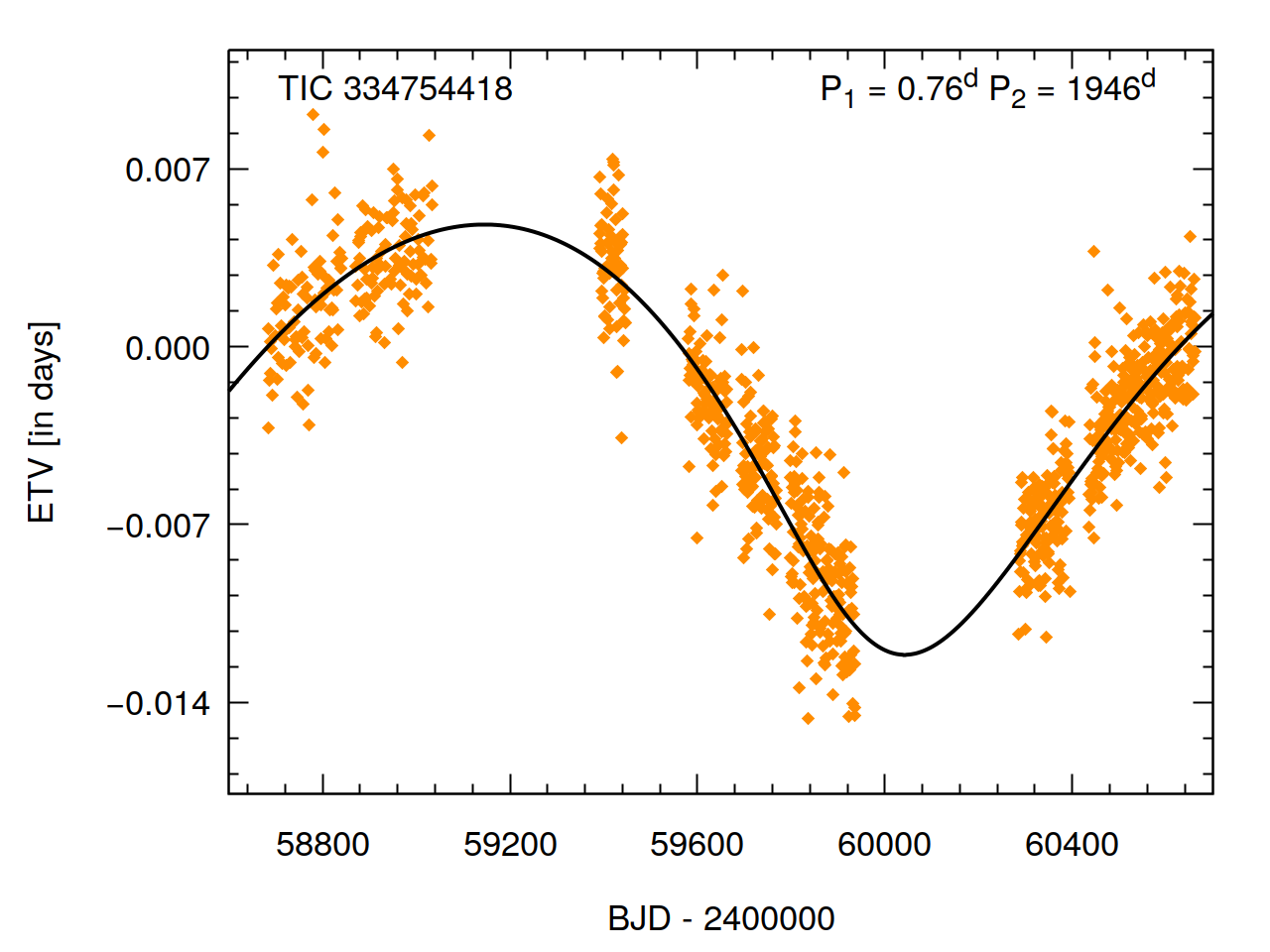}
\includegraphics[width=0.43\textwidth]{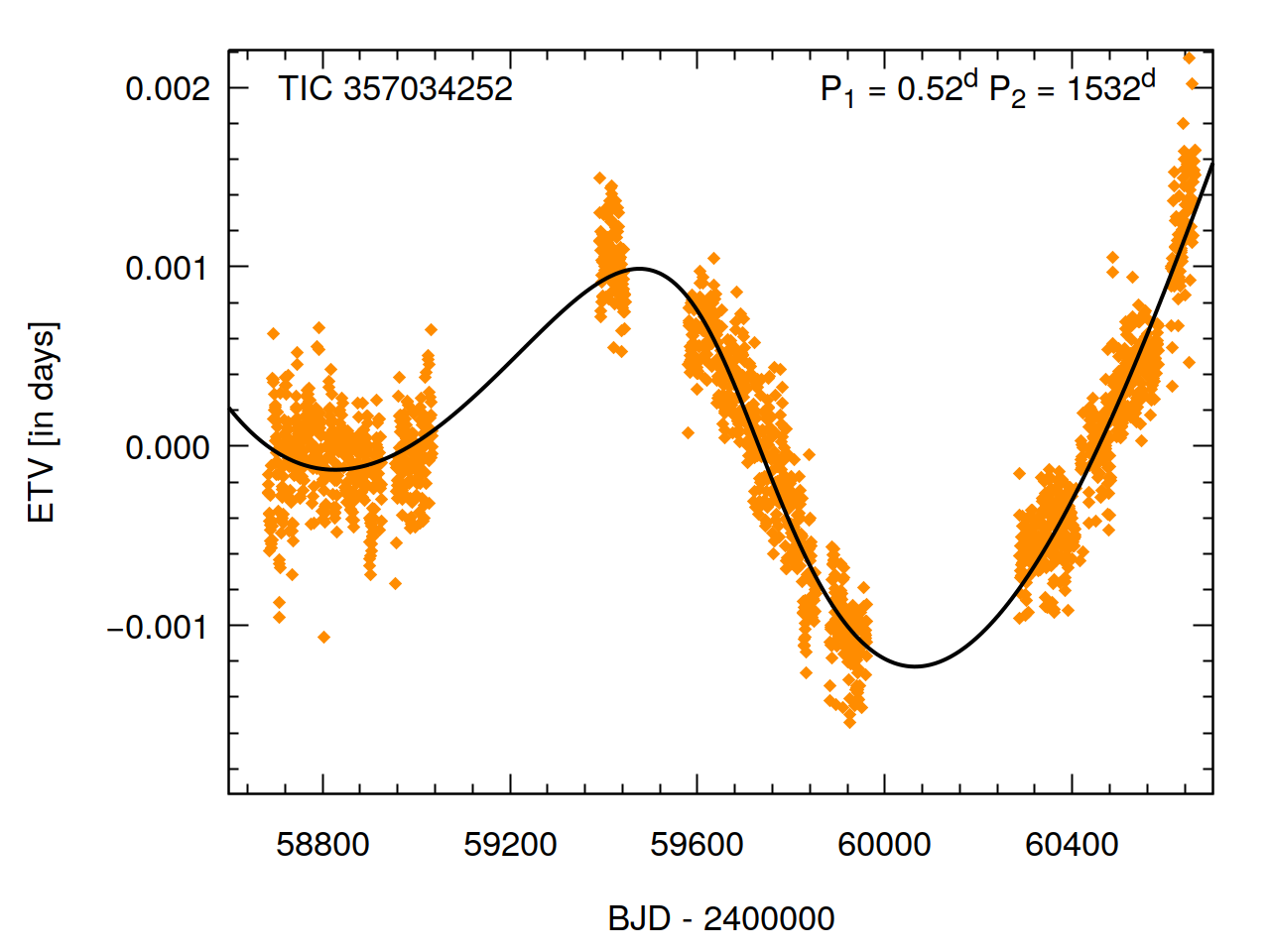}\includegraphics[width=0.43\textwidth]{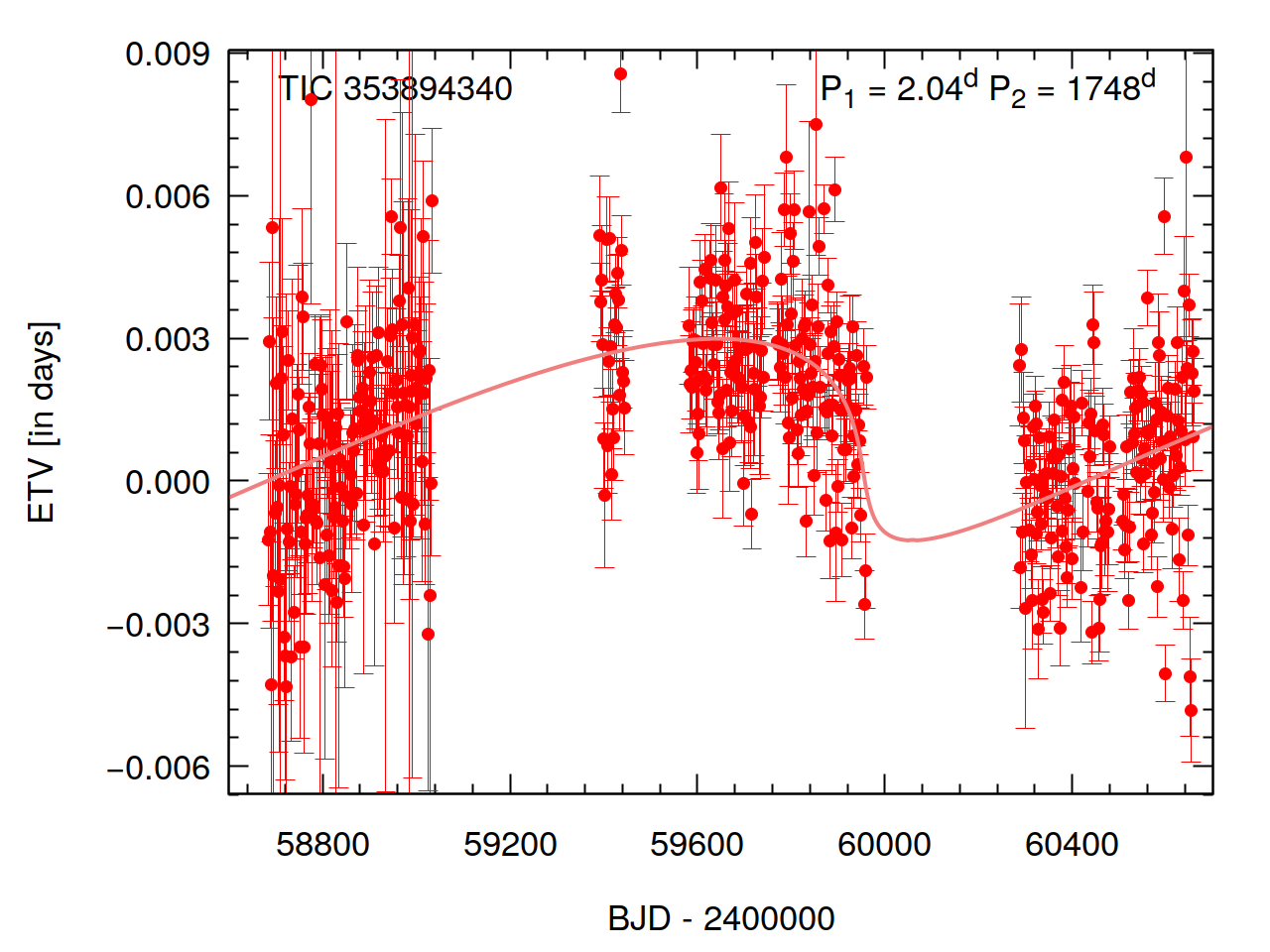}\includegraphics[width=0.43\textwidth]{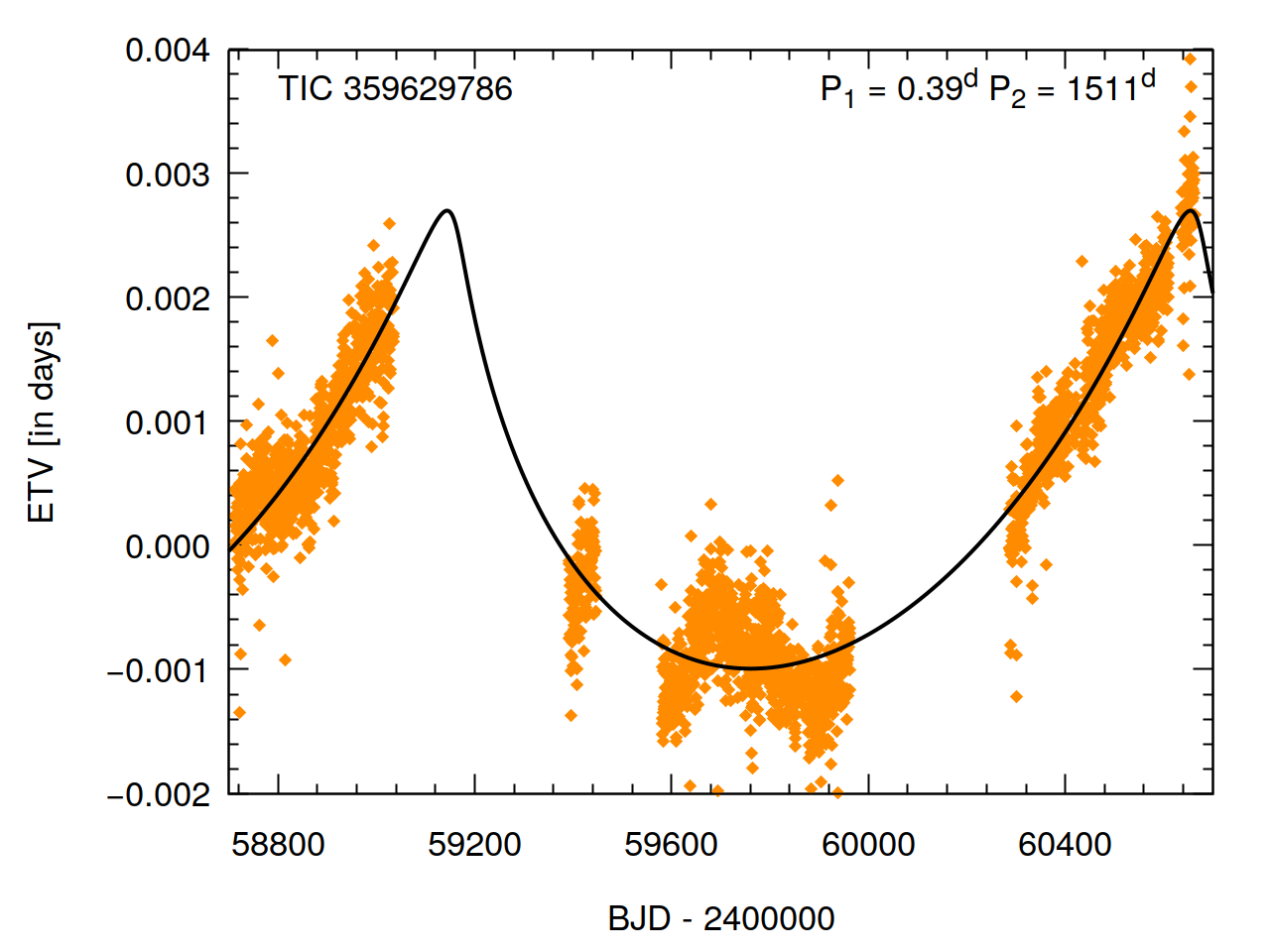}
\includegraphics[width=0.43\textwidth]{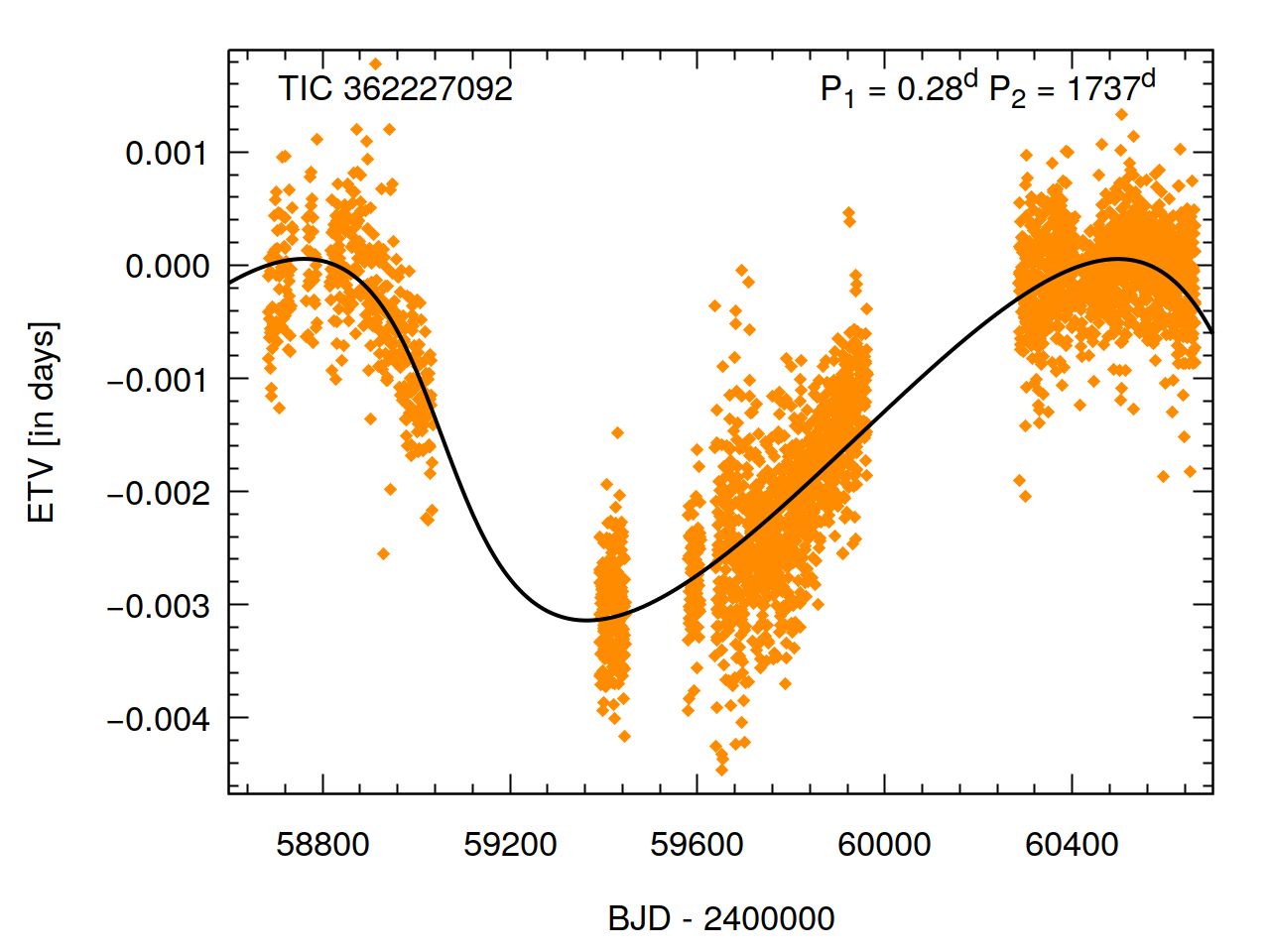}\includegraphics[width=0.43\textwidth]{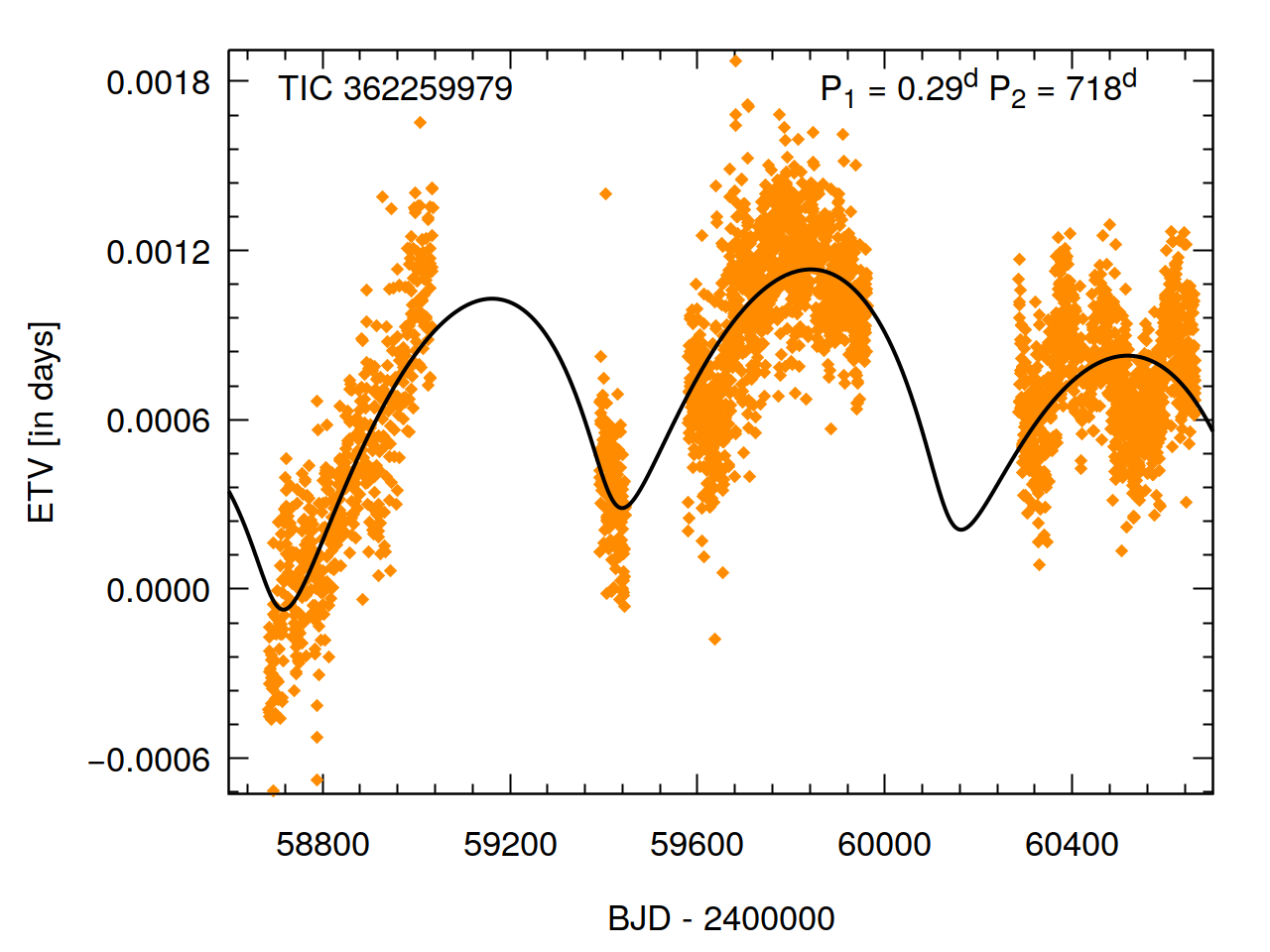}\includegraphics[width=0.43\textwidth]{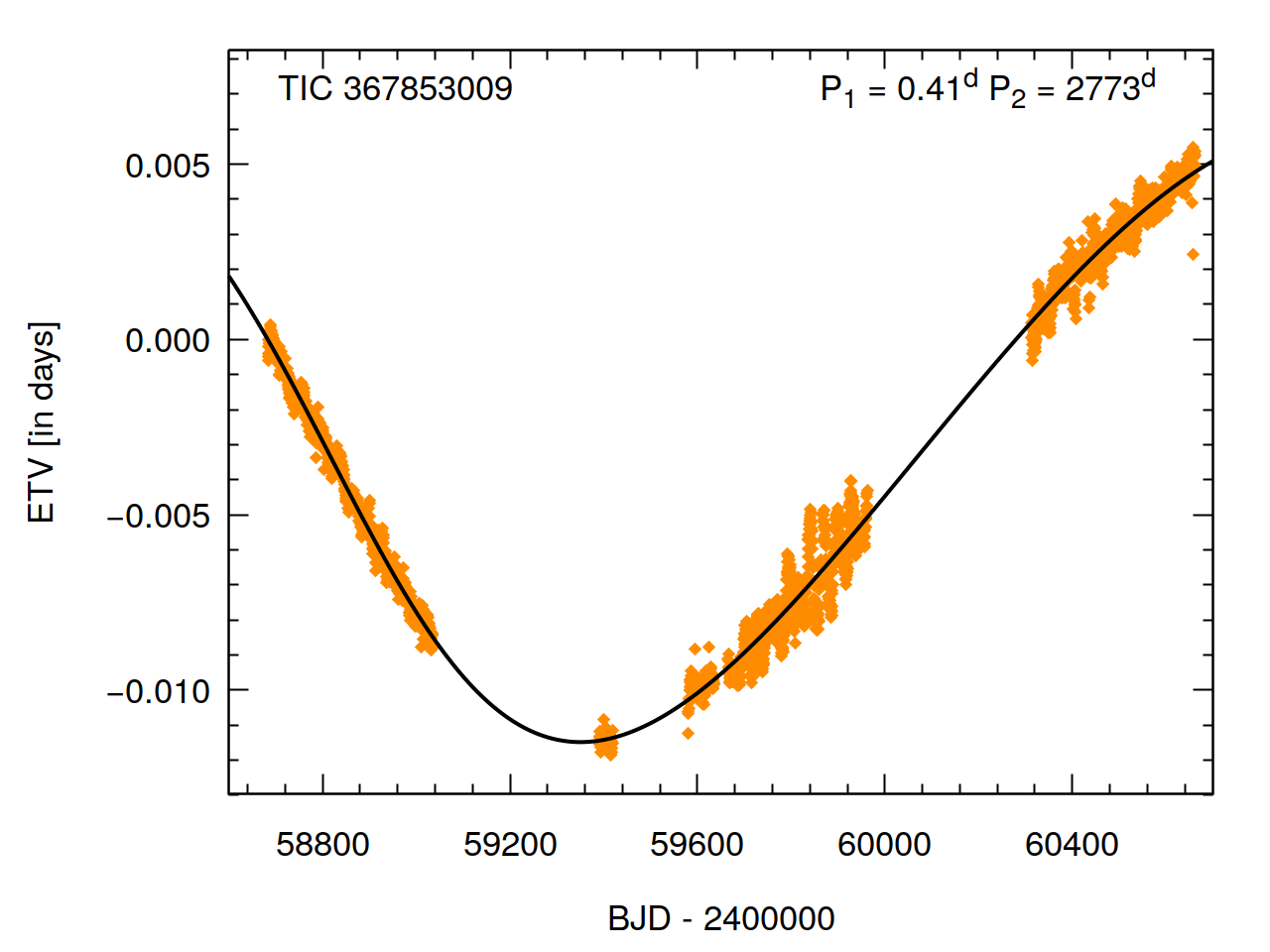}
\end{adjustwidth}
\caption{ETVs of the fourth set of 15 such systems which are classified into Group $L_3$. We note that in the case of TIC 288407480 (third row, left panel), a fourth body LTTE solution is also included. The meaning of all the symbols is the same, as were described formerly. See Table~\ref{Tab:Orbelem_LTTE3} for further details.}
\label{Fig:ETVs_L3d}
\end{figure}

\begin{figure}[H]
\begin{adjustwidth}{-\extralength}{0cm}
\centering

\includegraphics[width=0.43\textwidth]{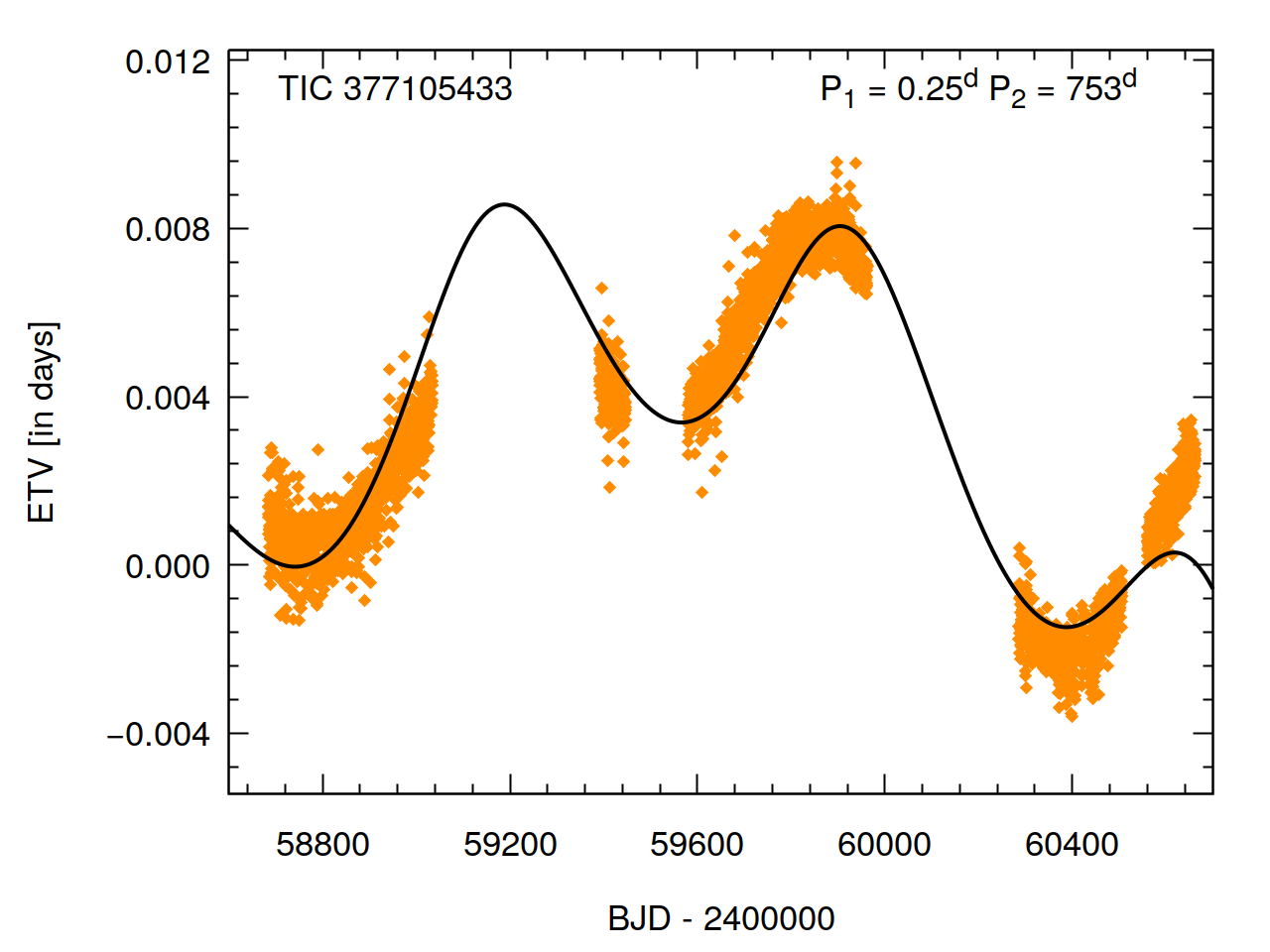}\includegraphics[width=0.43\textwidth]{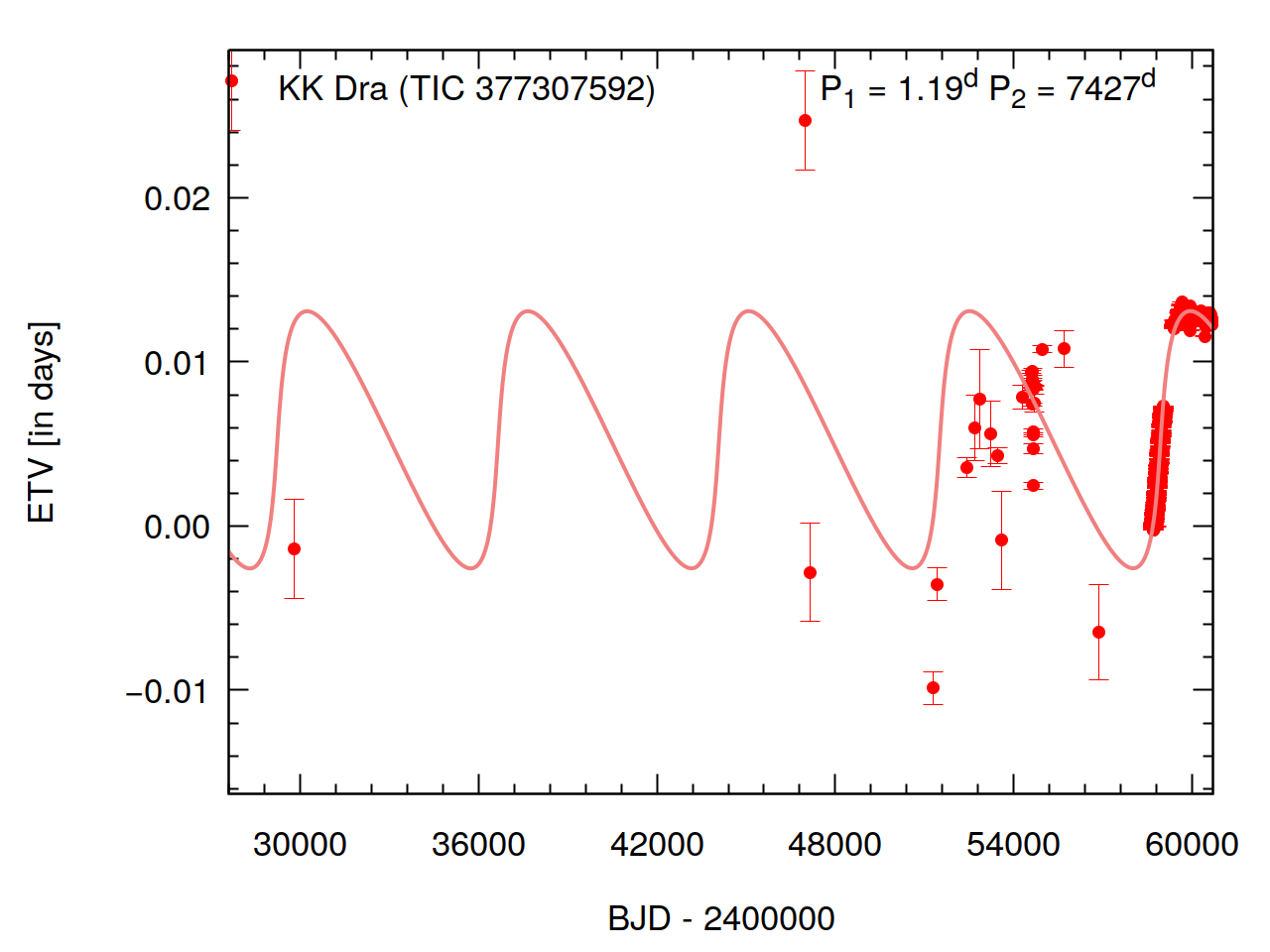}\includegraphics[width=0.43\textwidth]{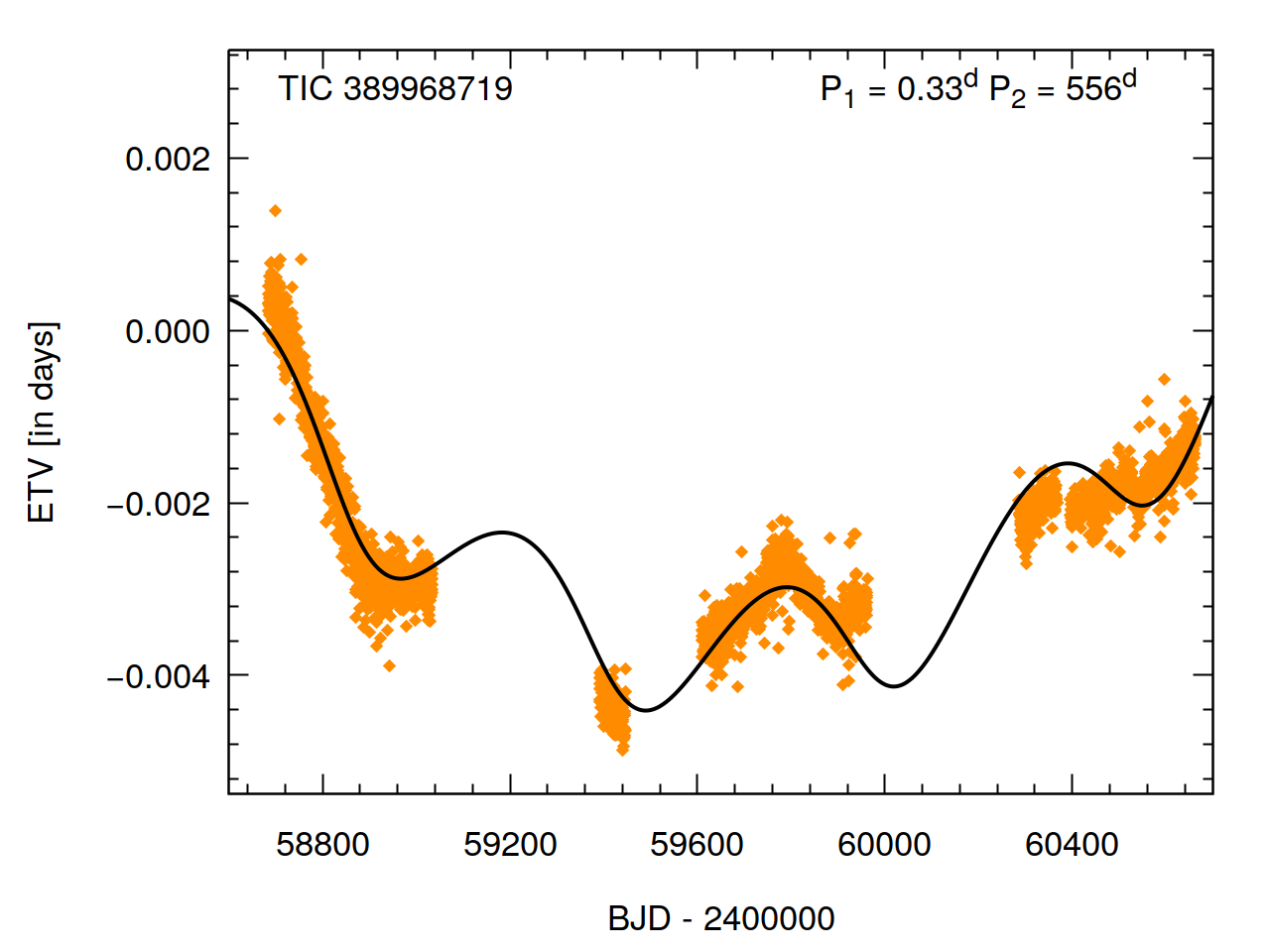}
\includegraphics[width=0.43\textwidth]{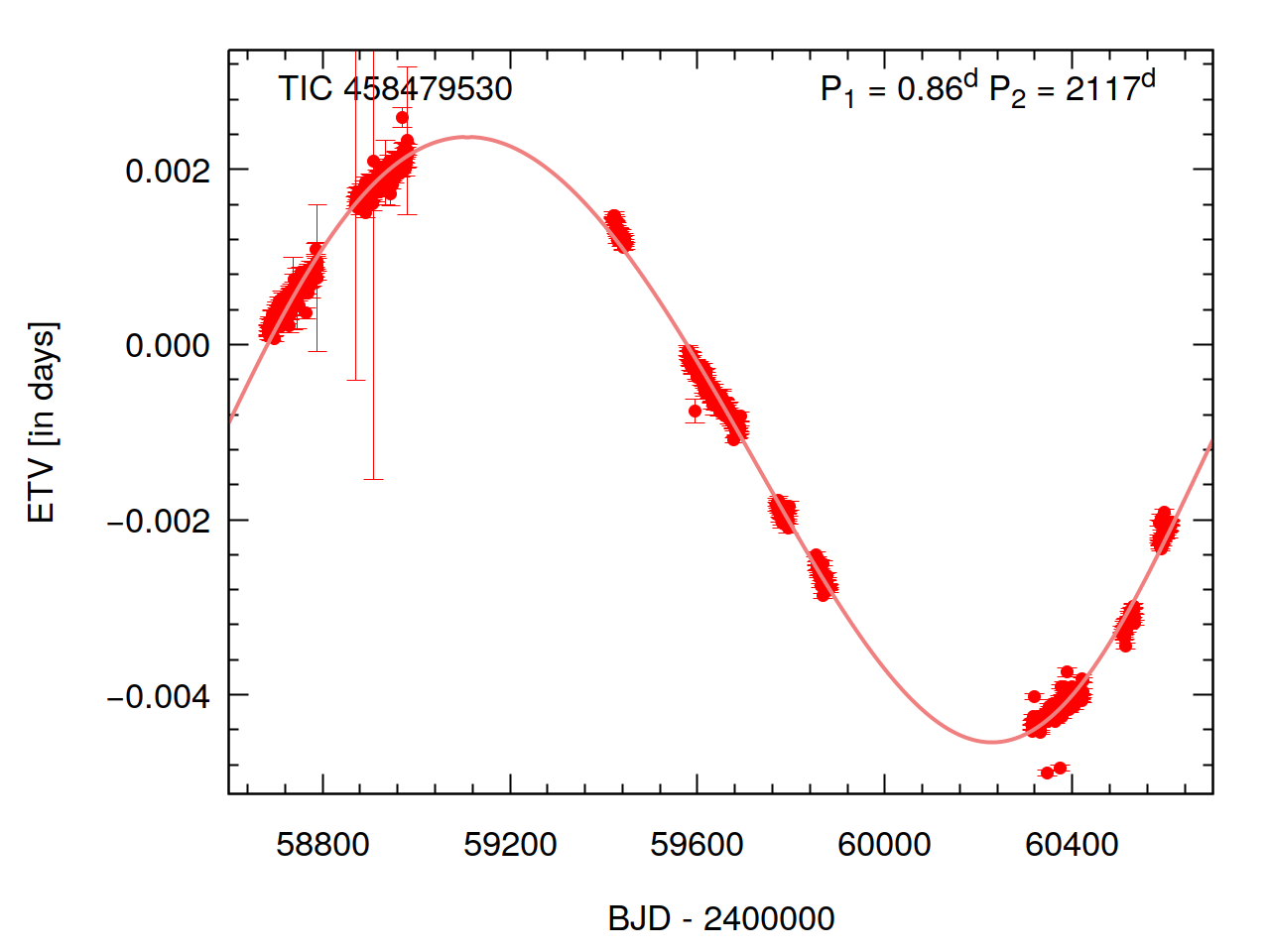}
\end{adjustwidth}
\caption{ETV points together with the pure LTTE solutions for the last 4 of 64 systems which are ranked into the most uncertain Group $L_3$. For further details, see Table~\ref{Tab:Orbelem_LTTE3}.}
\label{Fig:ETVs_L3e}
\end{figure}

\begin{figure}[H]
\begin{adjustwidth}{-\extralength}{0cm}
\centering
\includegraphics[width=0.40\textwidth]{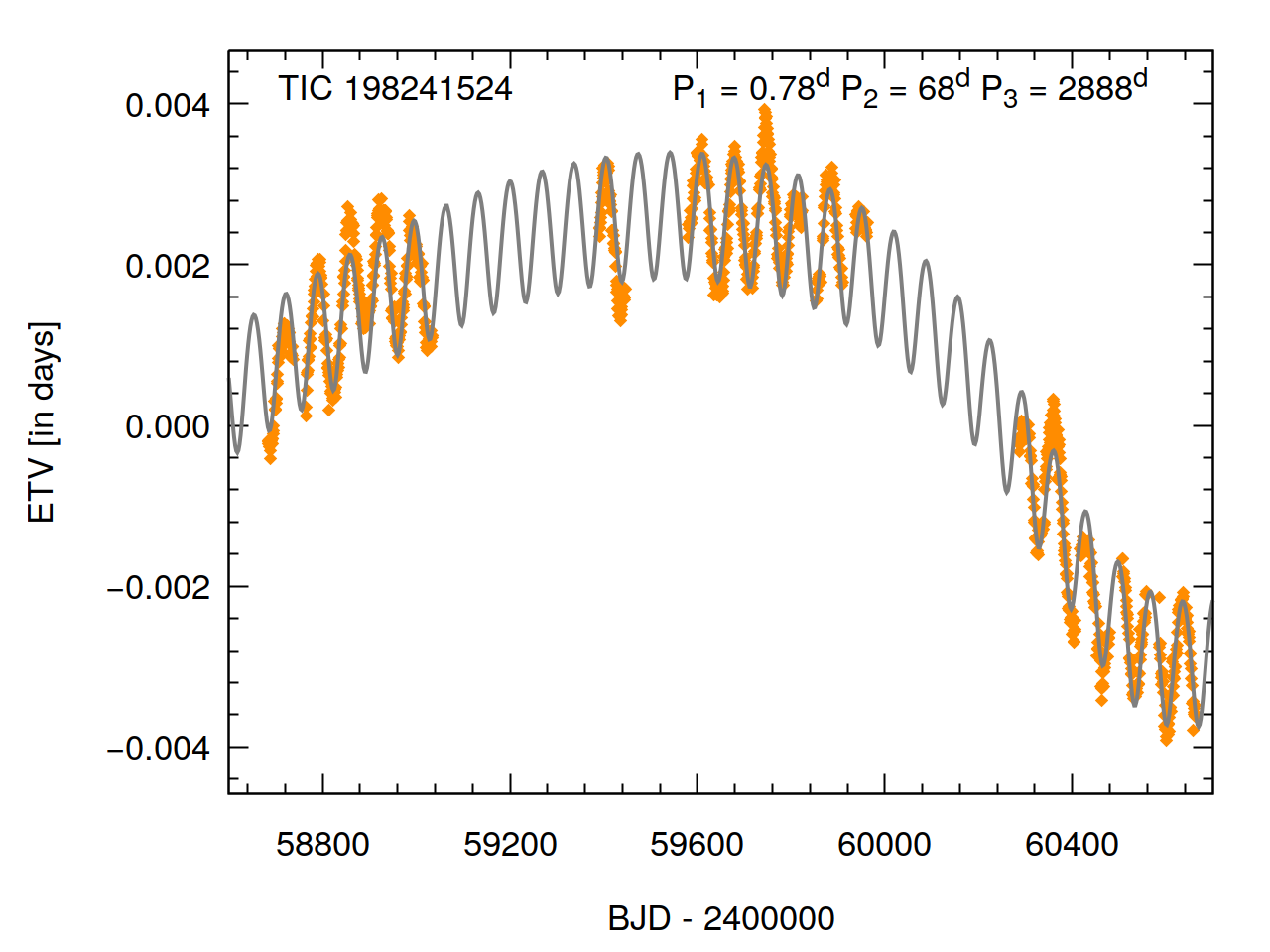}\includegraphics[width=0.40\textwidth]{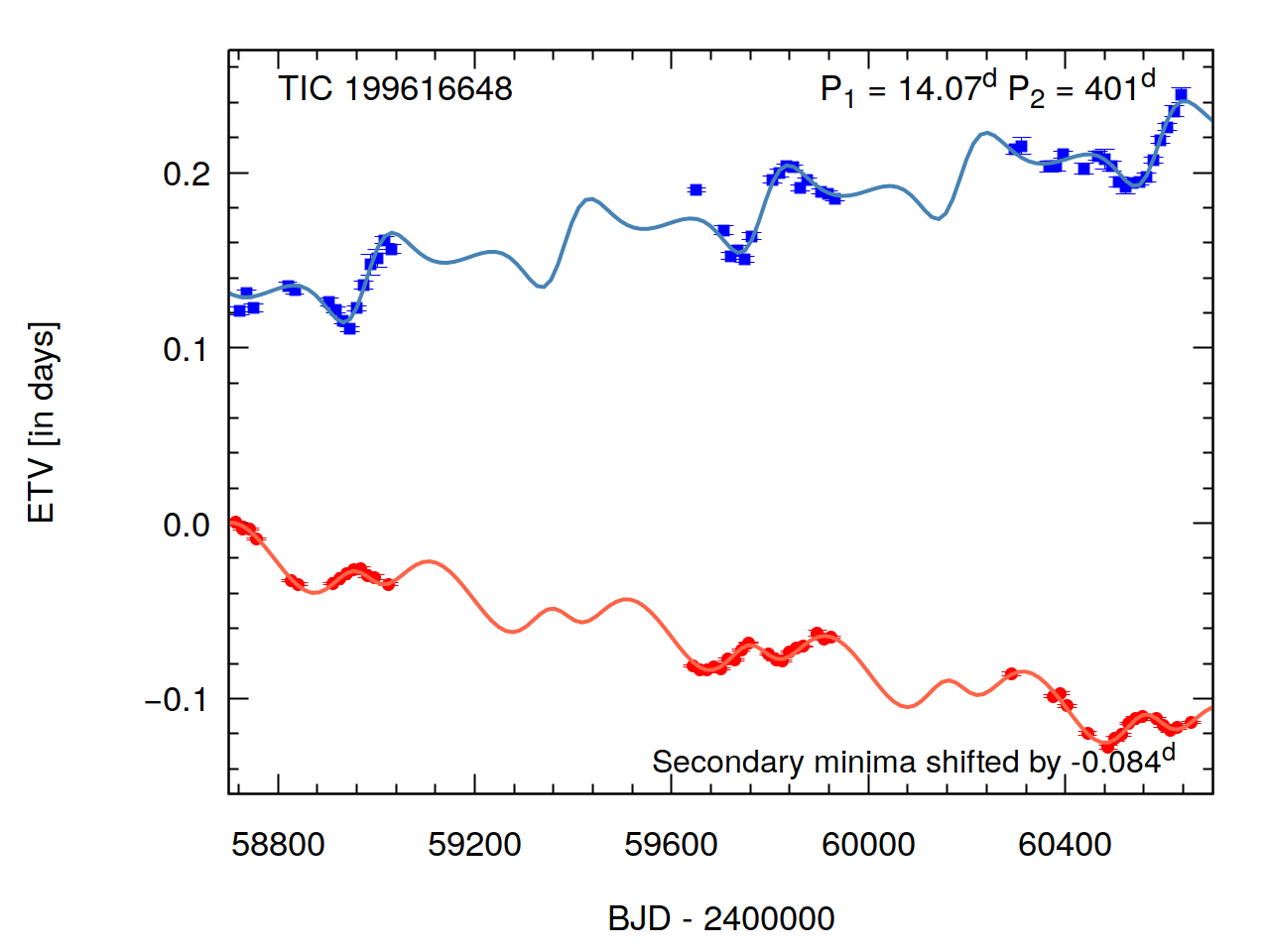}\includegraphics[width=0.40\textwidth]{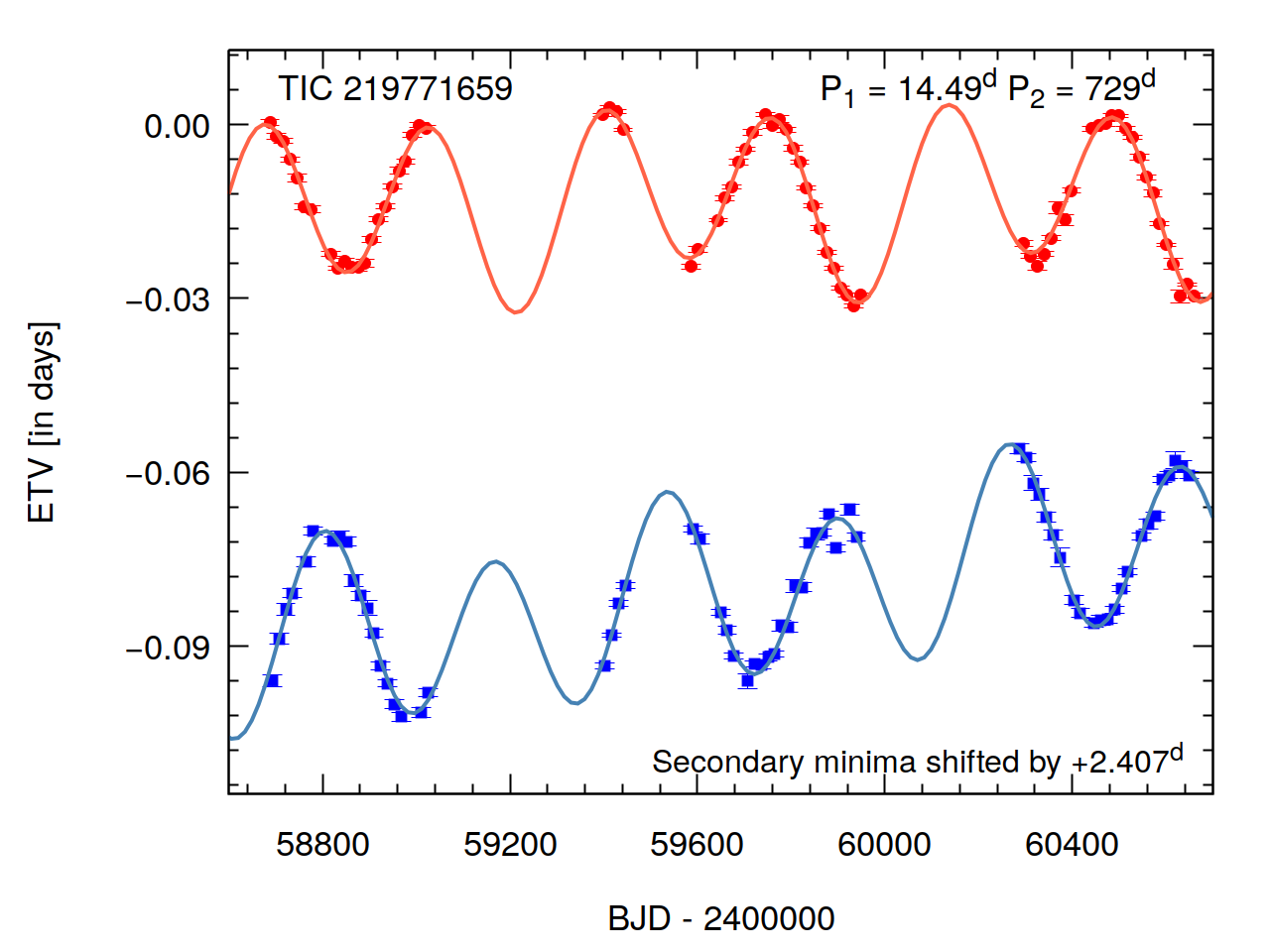}
\includegraphics[width=0.40\textwidth]{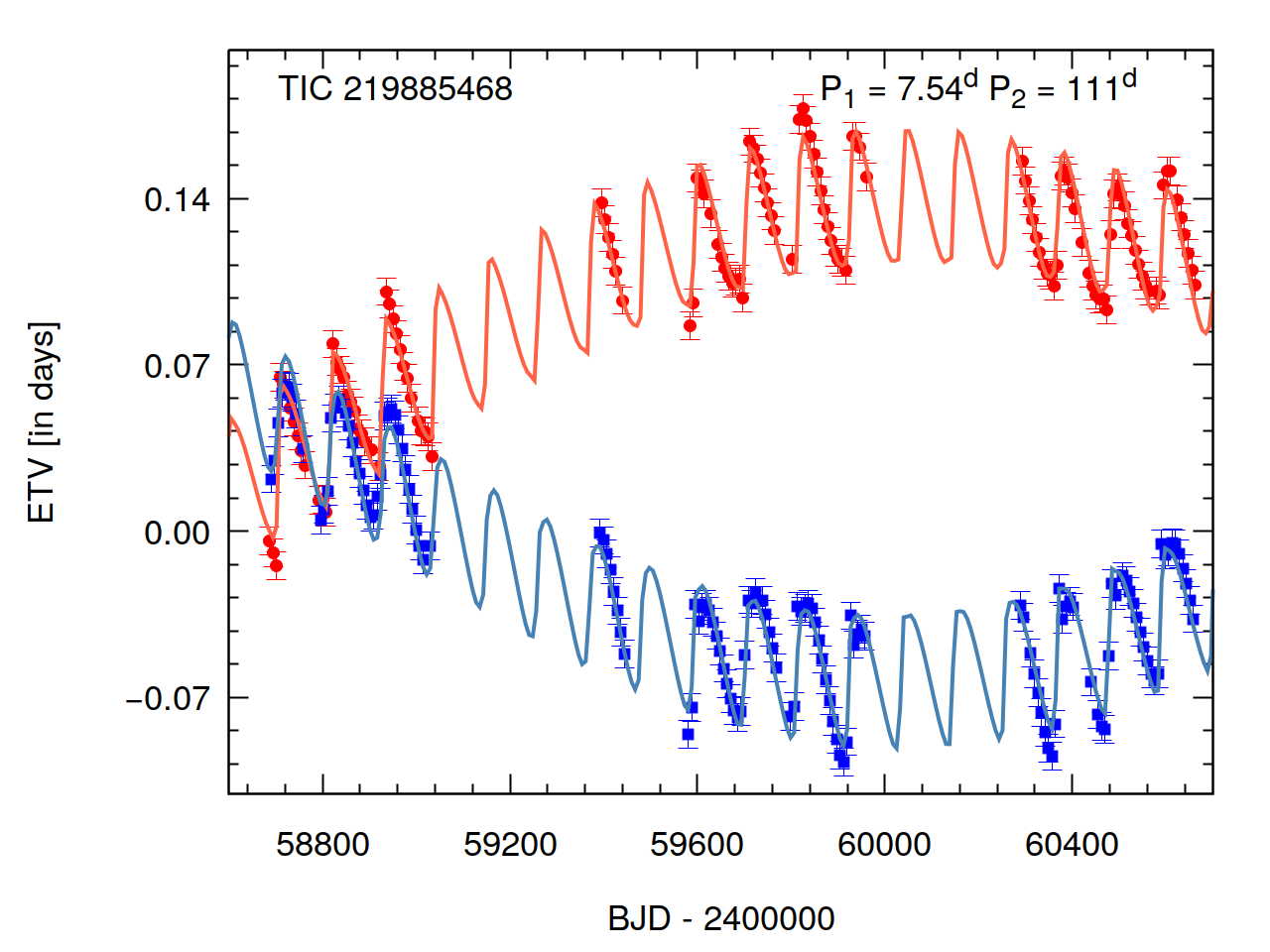}\includegraphics[width=0.40\textwidth]{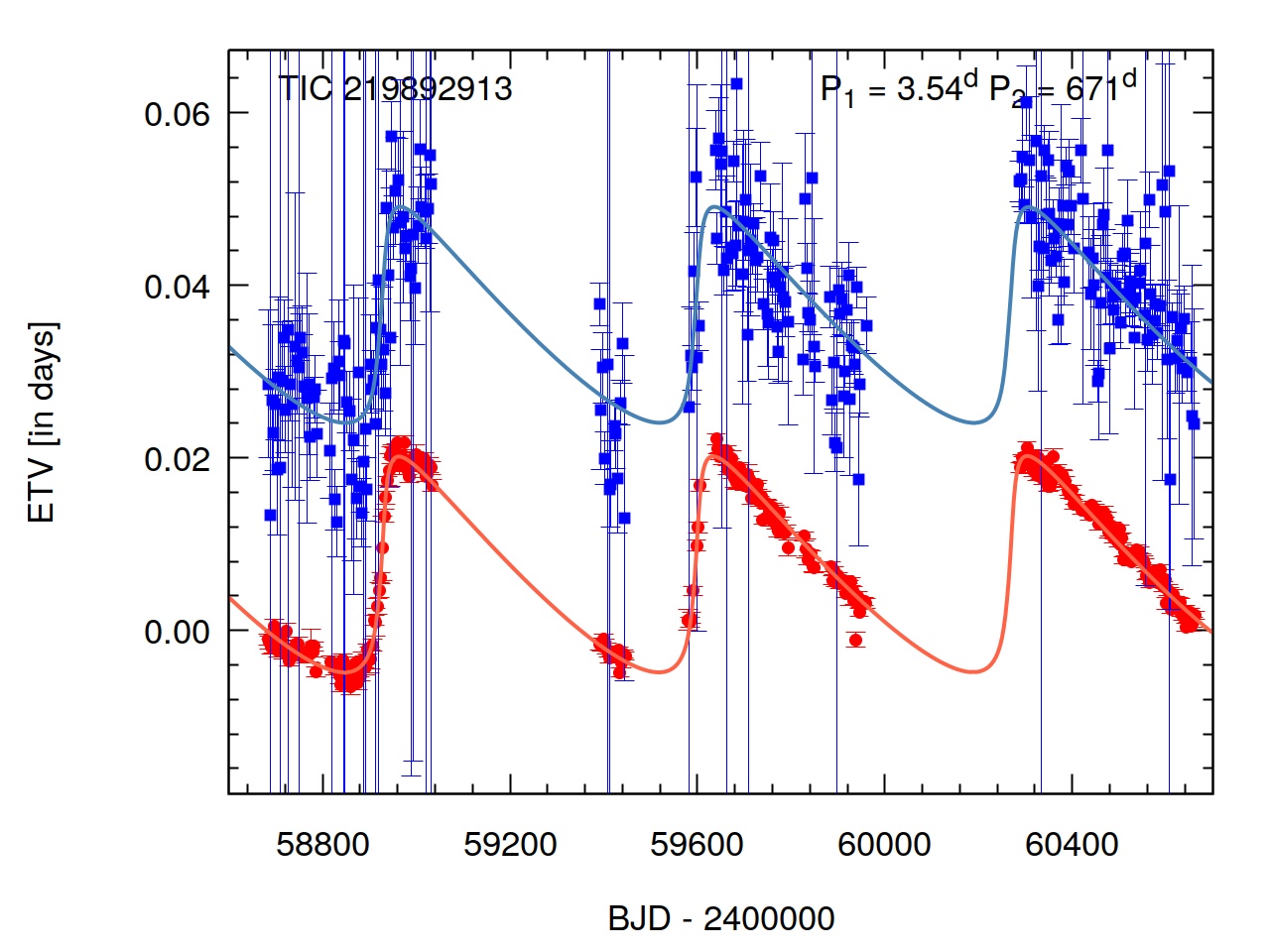}\includegraphics[width=0.40\textwidth]{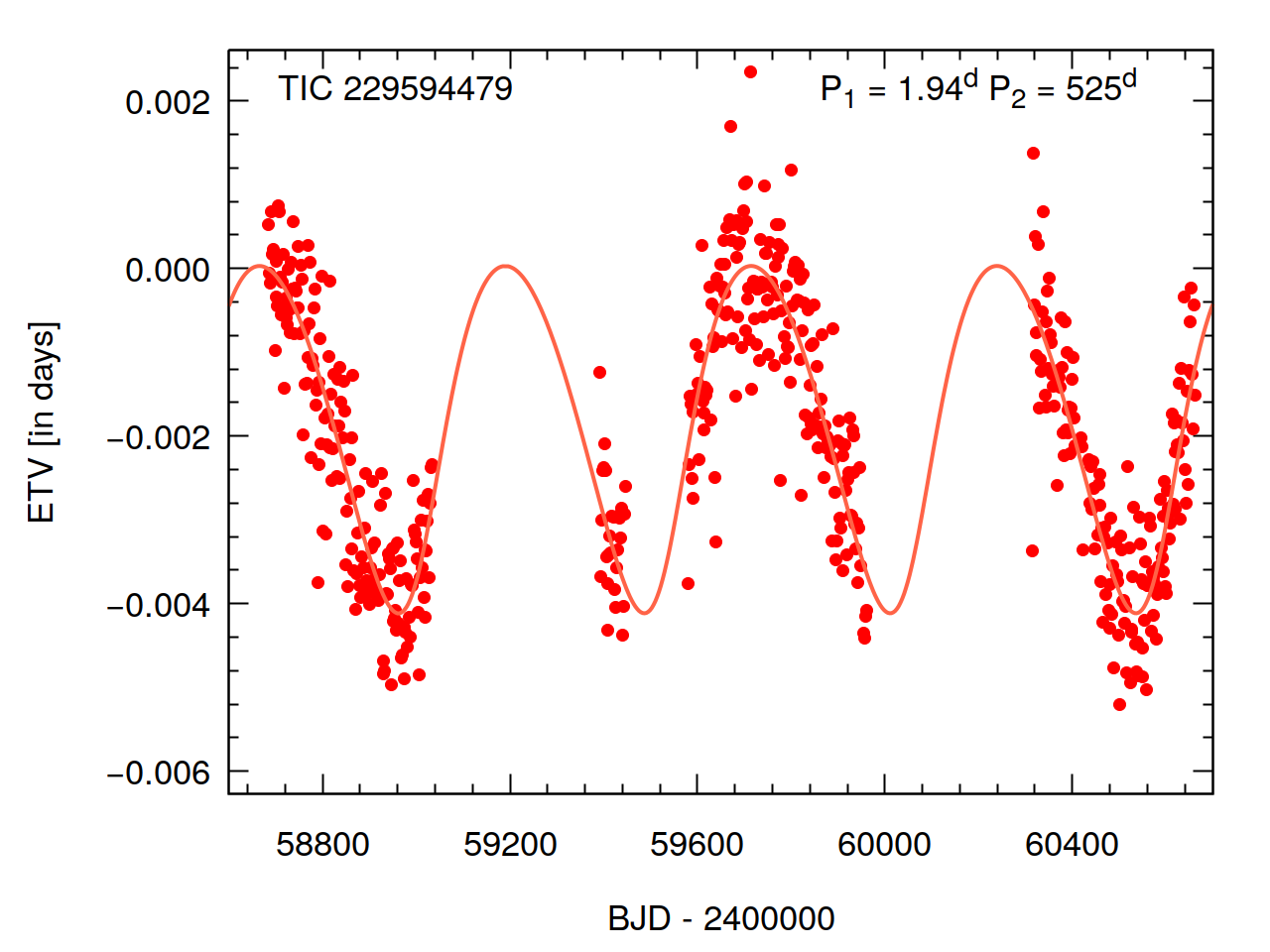}
\includegraphics[width=0.40\textwidth]{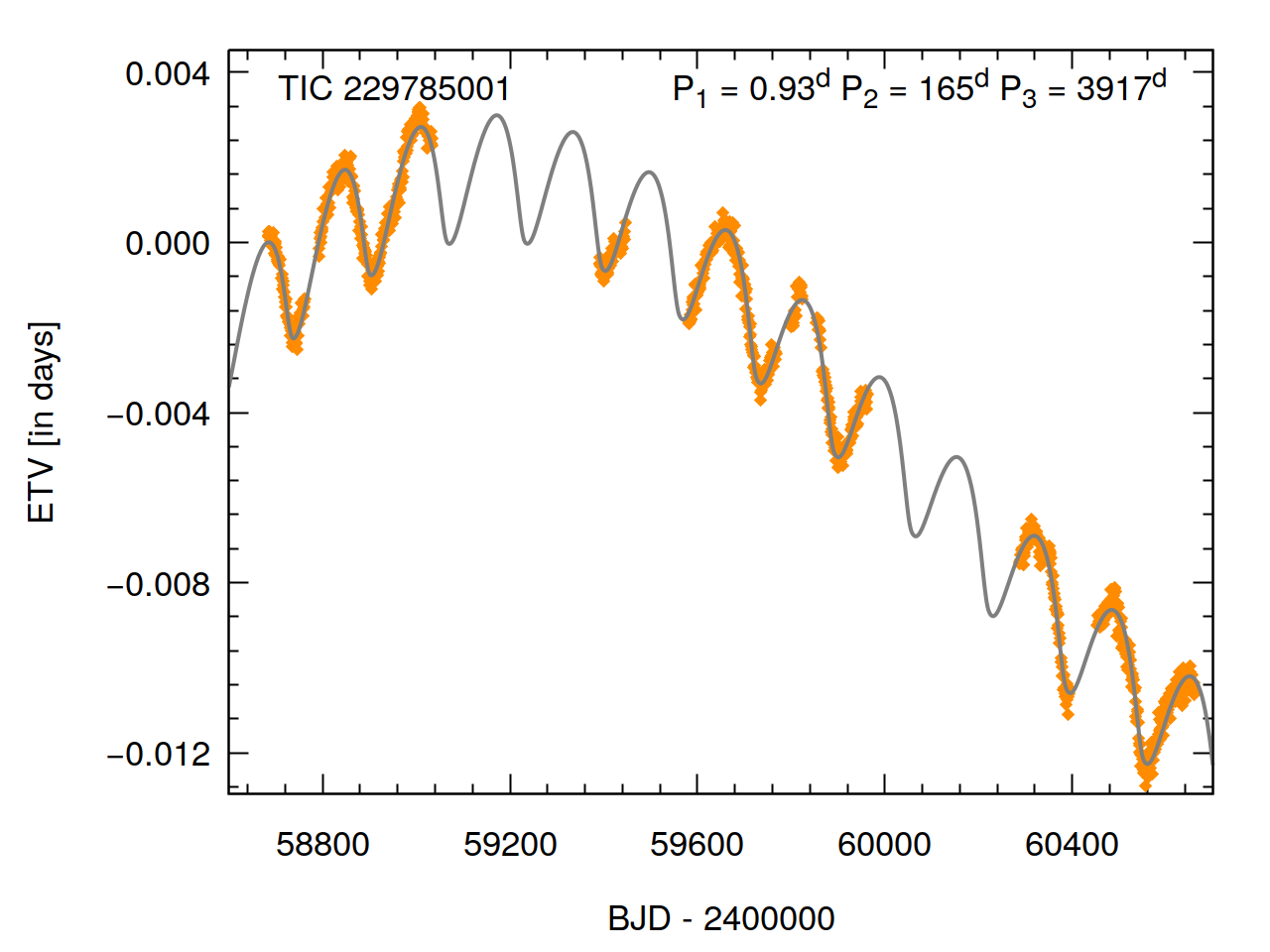}\includegraphics[width=0.40\textwidth]{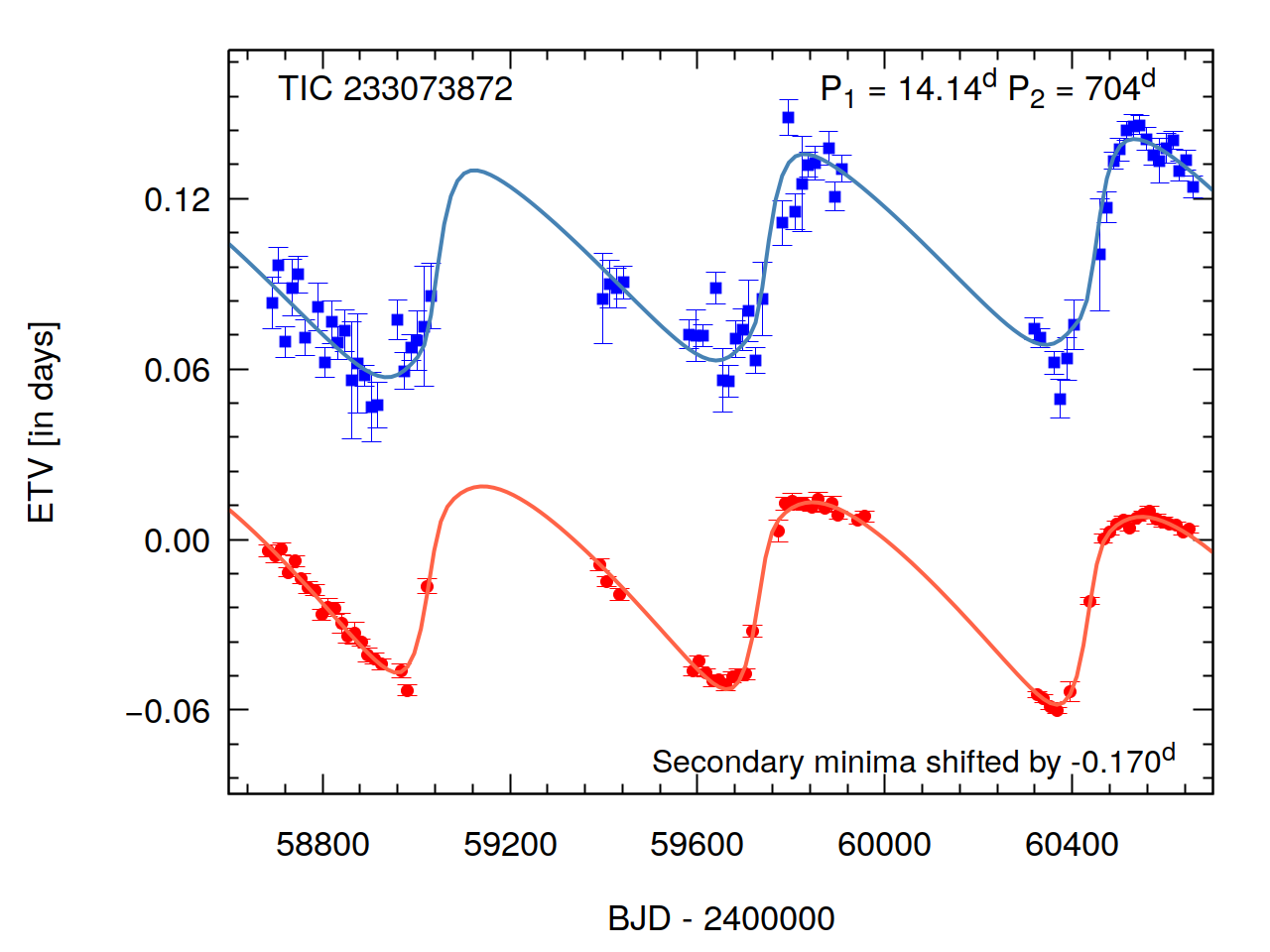}\includegraphics[width=0.40\textwidth]{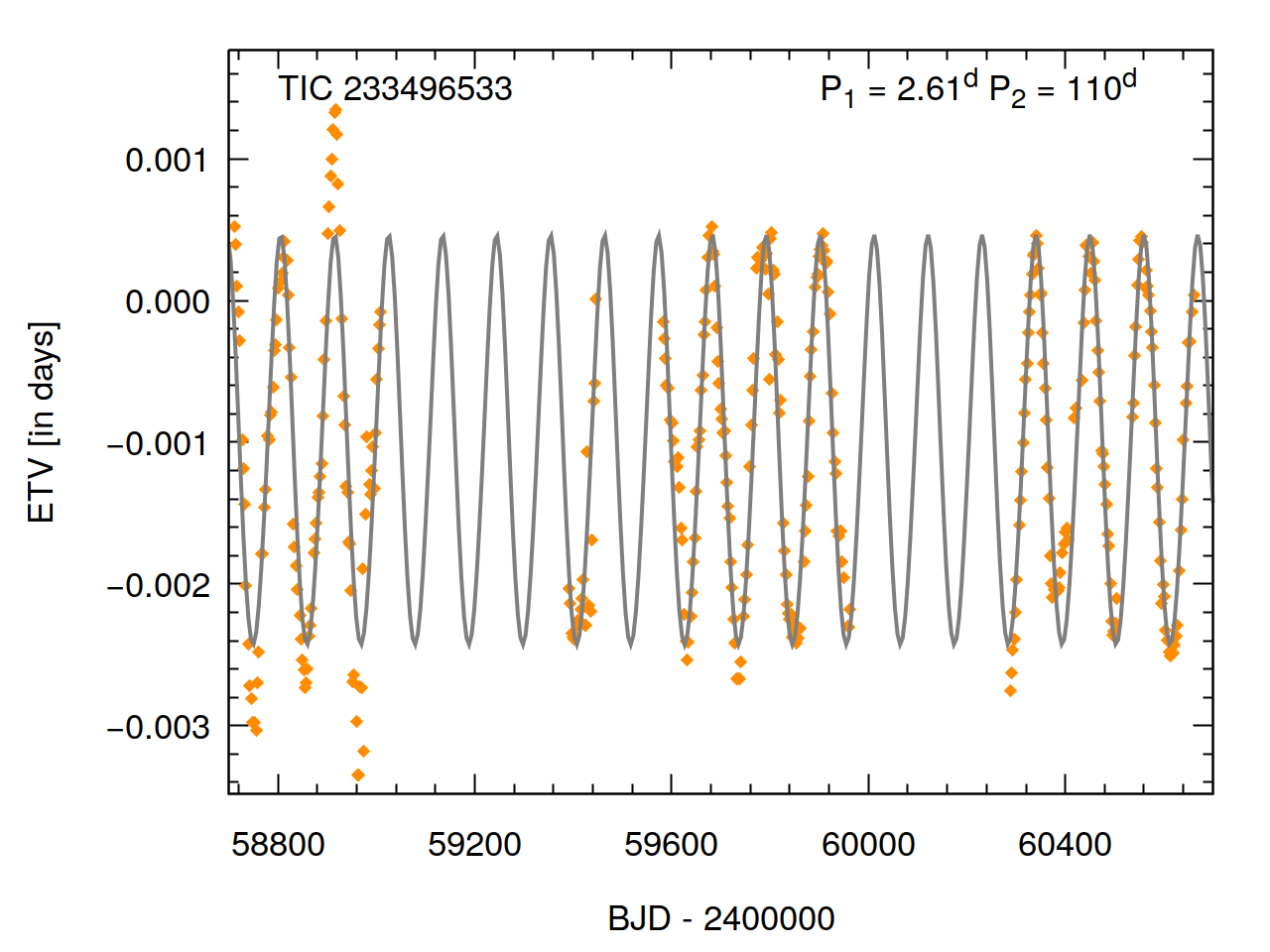}
\includegraphics[width=0.40\textwidth]{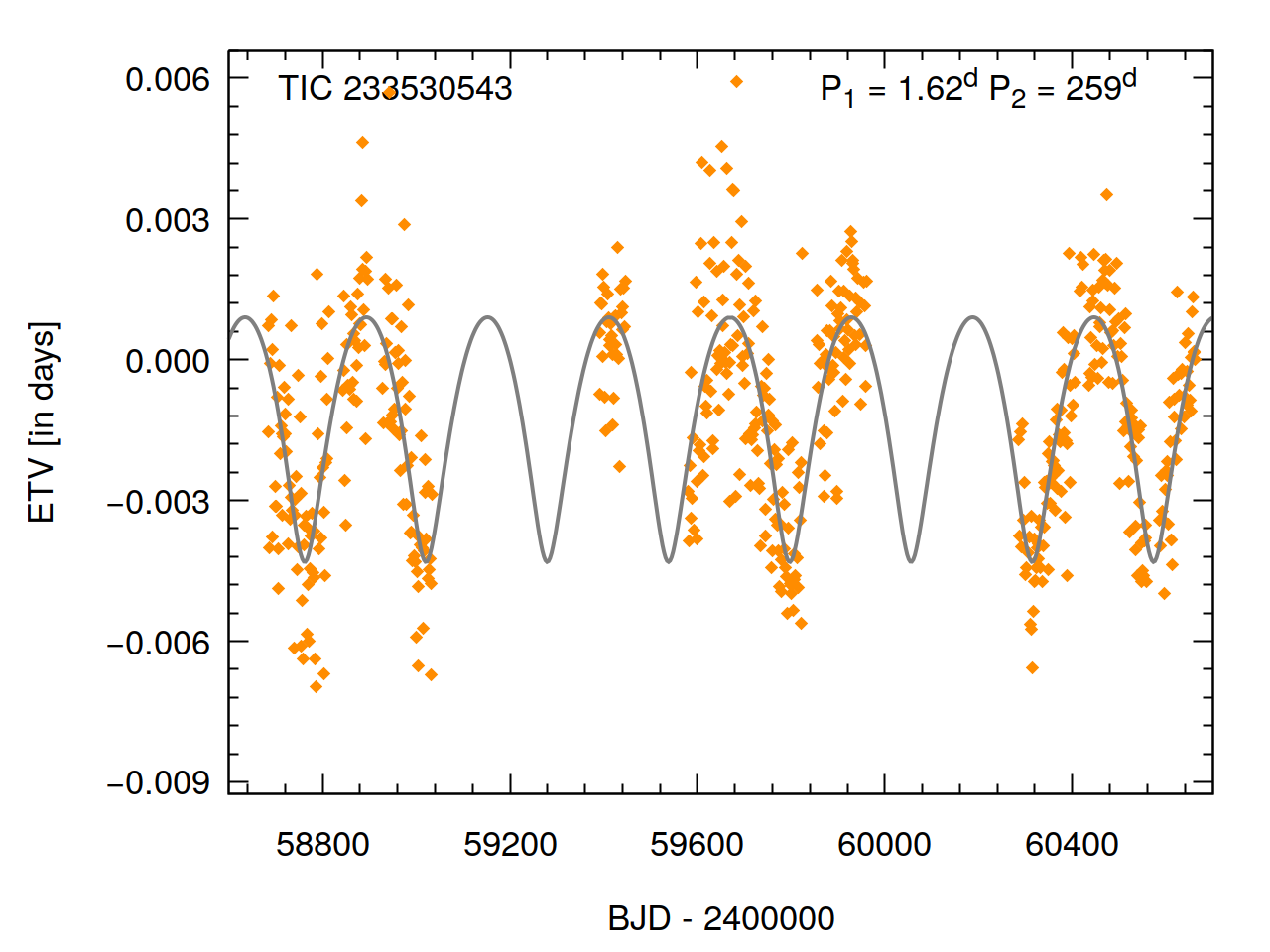}\includegraphics[width=0.40\textwidth]{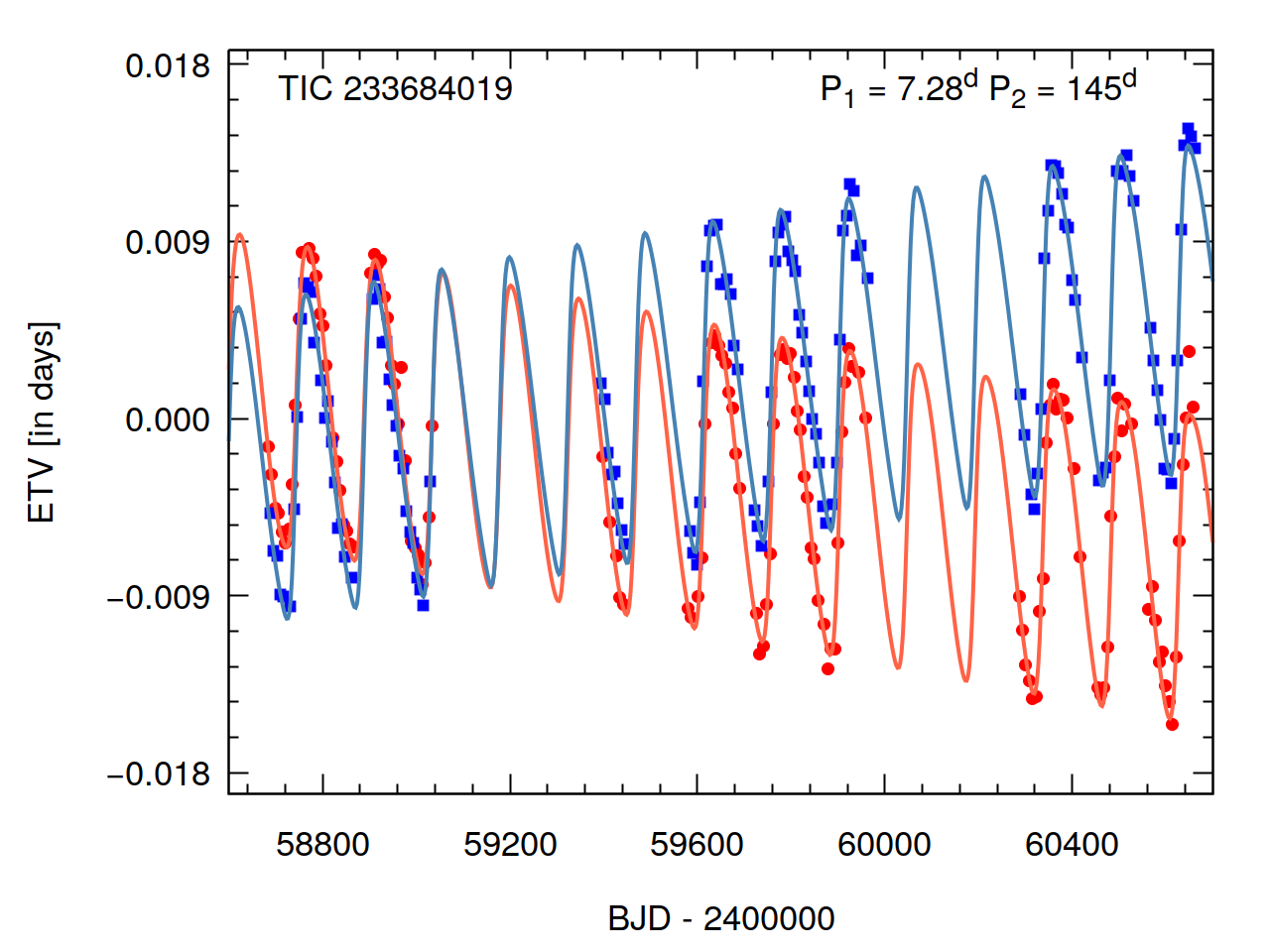}\includegraphics[width=0.40\textwidth]{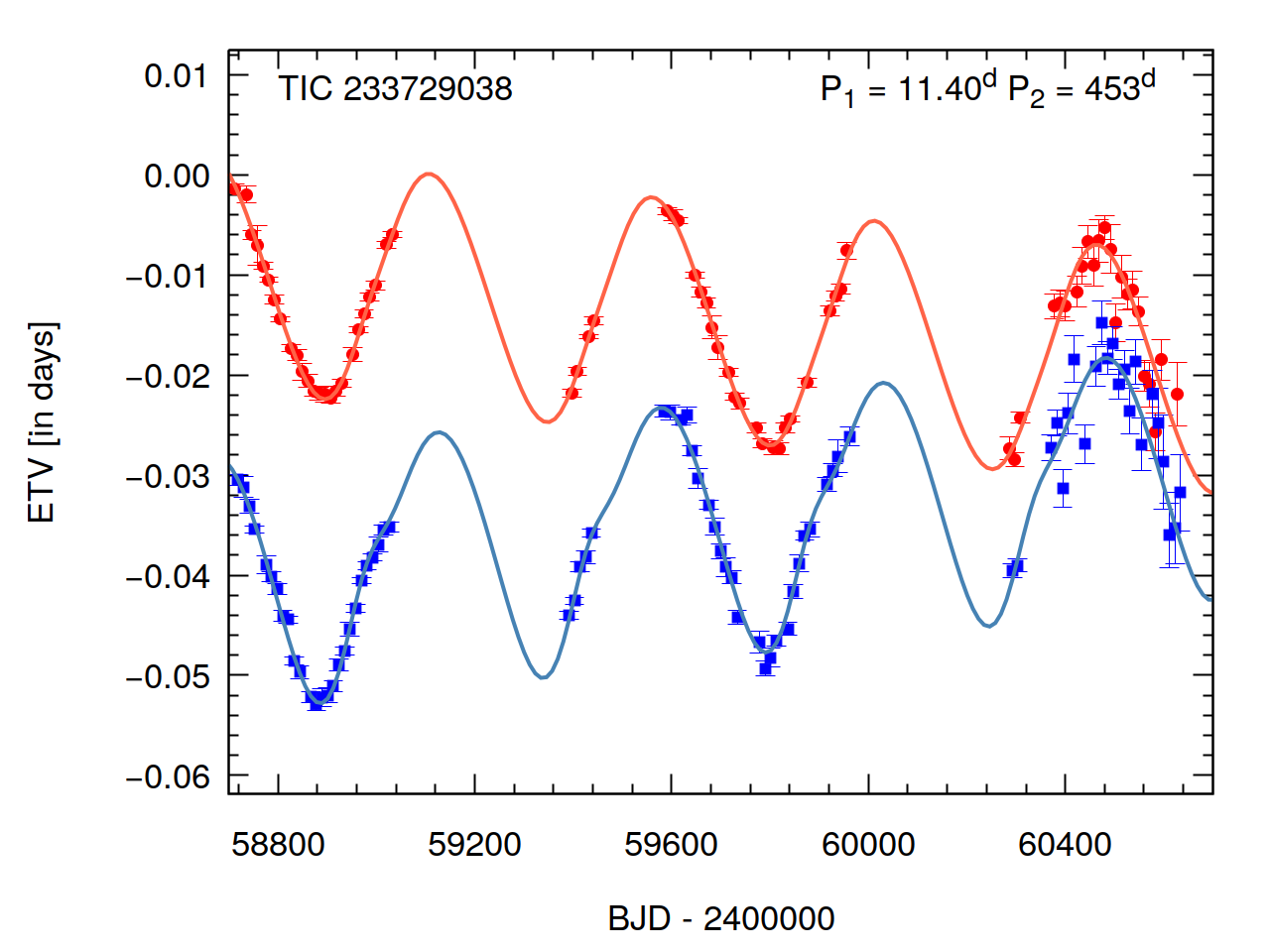}
\includegraphics[width=0.40\textwidth]{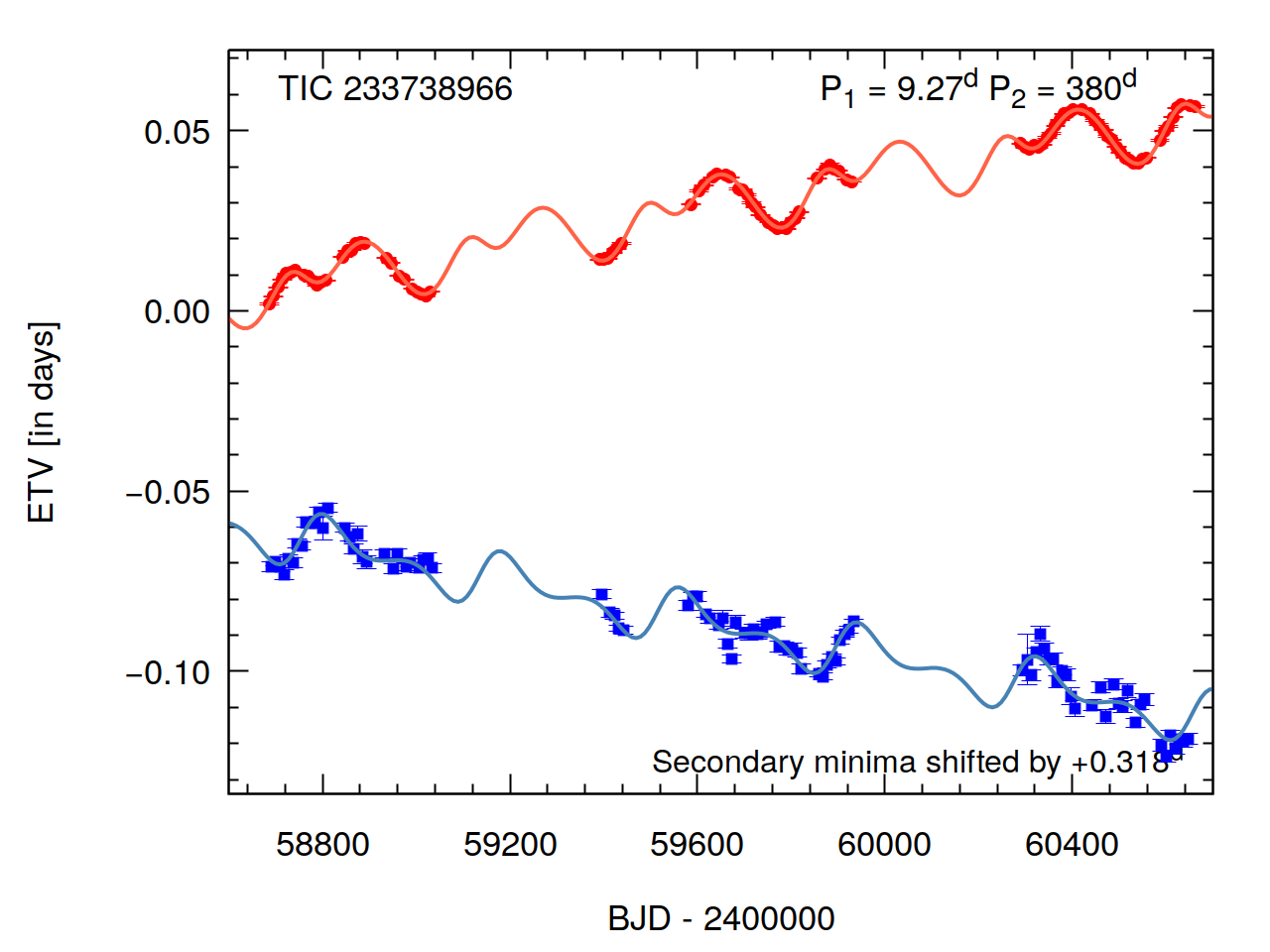}\includegraphics[width=0.40\textwidth]{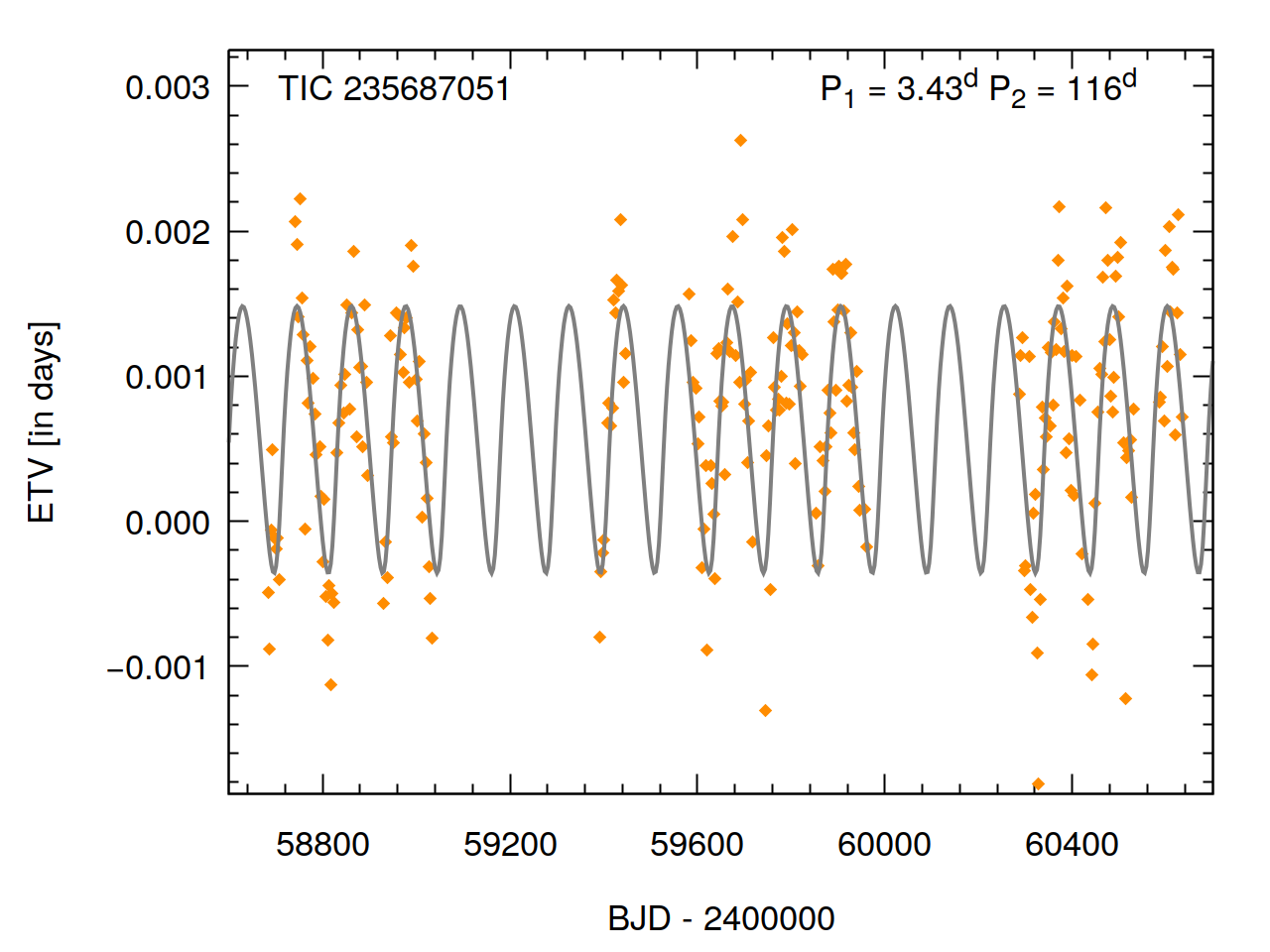}\includegraphics[width=0.40\textwidth]{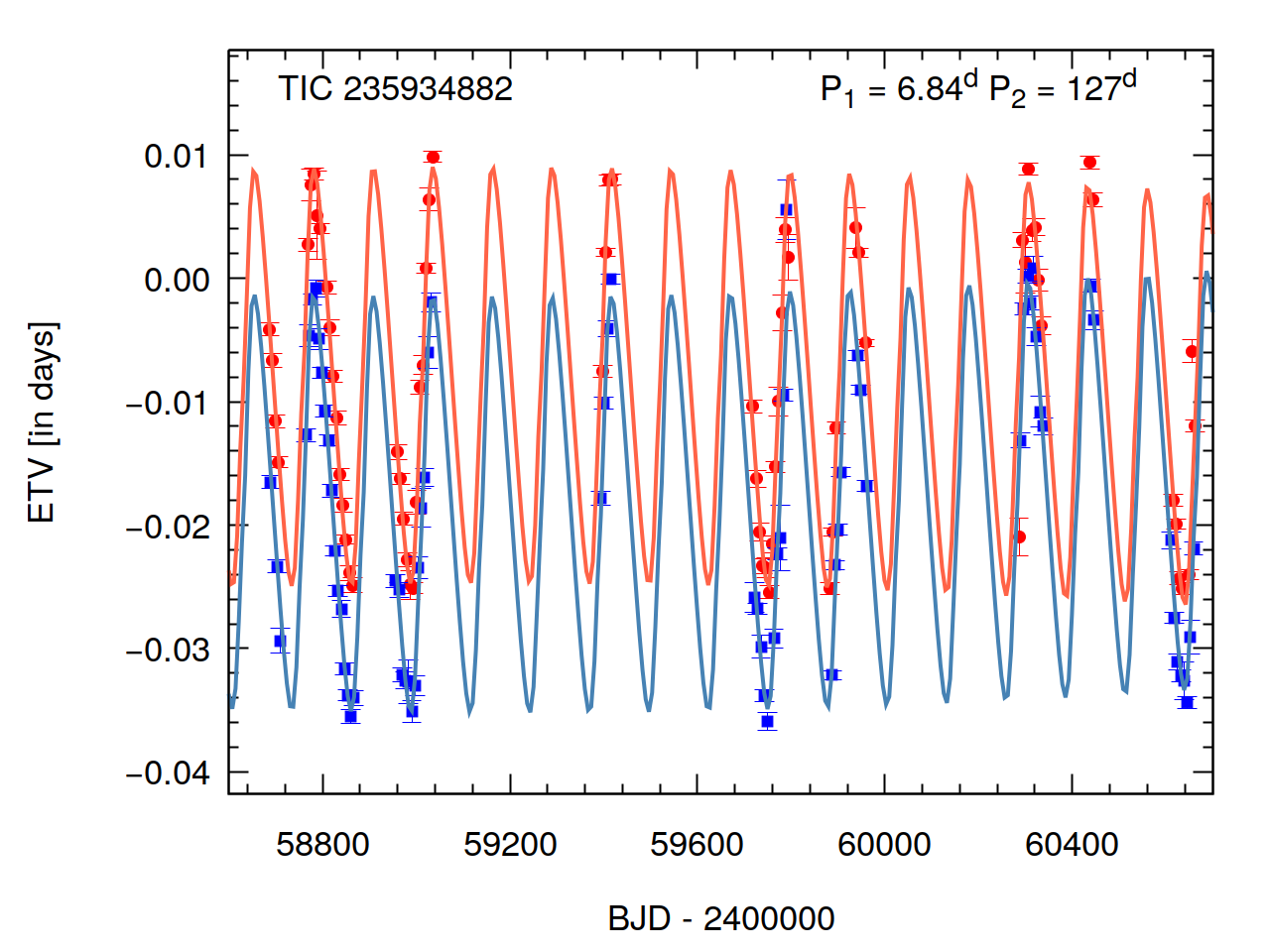}
\end{adjustwidth}
\caption{ETV plots with certain (group $D_1$) LTTE + DE third body solutions. For circular systems with similarly accurate primary and secondary ETV curves, we use the averaged ETVs (orange diamonds) for fitting the analytic solutions. For eccentric systems, however, both the primary (red) and secondary (blue) ETV curves are used simultaneously in the fitting process. In addition to the LTTE + DE third body solutions,  fourth body pure LTTE solutions are also calculated in the case of TICs 198241524 and 229785001 (left panels in the first and third rows, respectively). For further details, see {Tables~\ref{Tab:Orbelemdyn1} and \ref{Tab:AMEparam}}.}
\label{Fig:ETVs_D1a}
\end{figure}

\begin{figure}[H]
\begin{adjustwidth}{-\extralength}{0cm}
\centering
\includegraphics[width=0.35\textwidth]{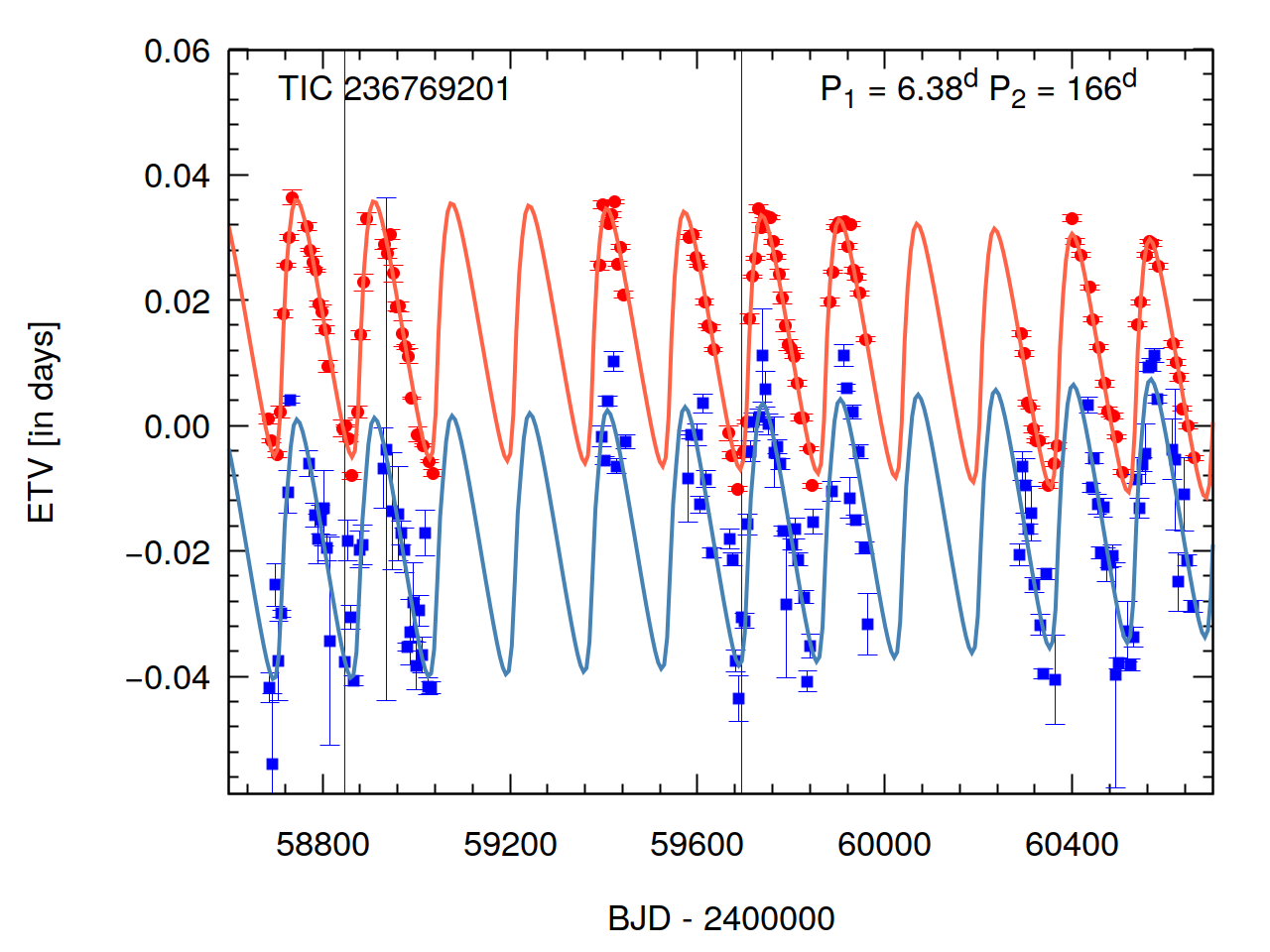}\includegraphics[width=0.35\textwidth]{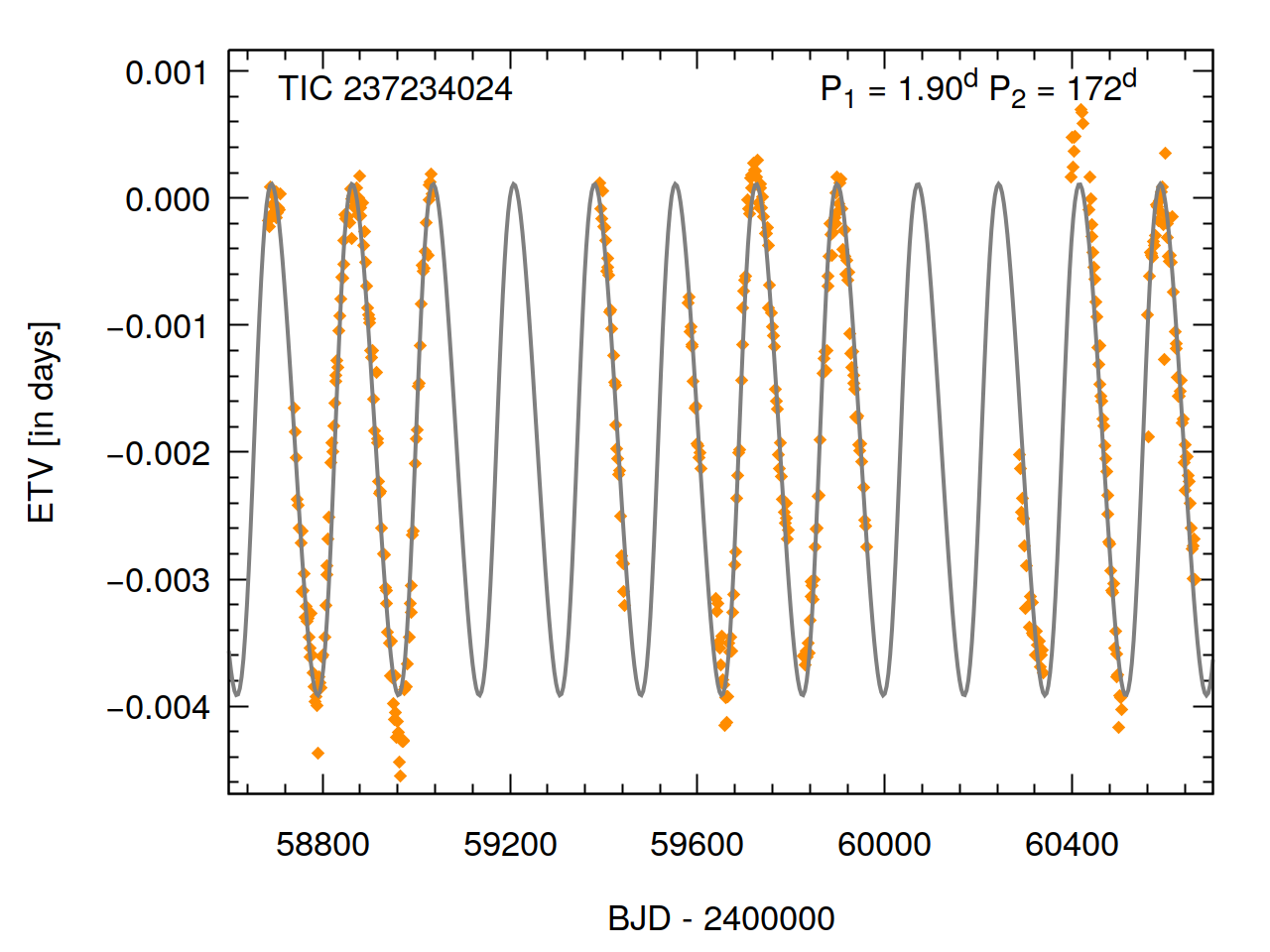}\includegraphics[width=0.35\textwidth]{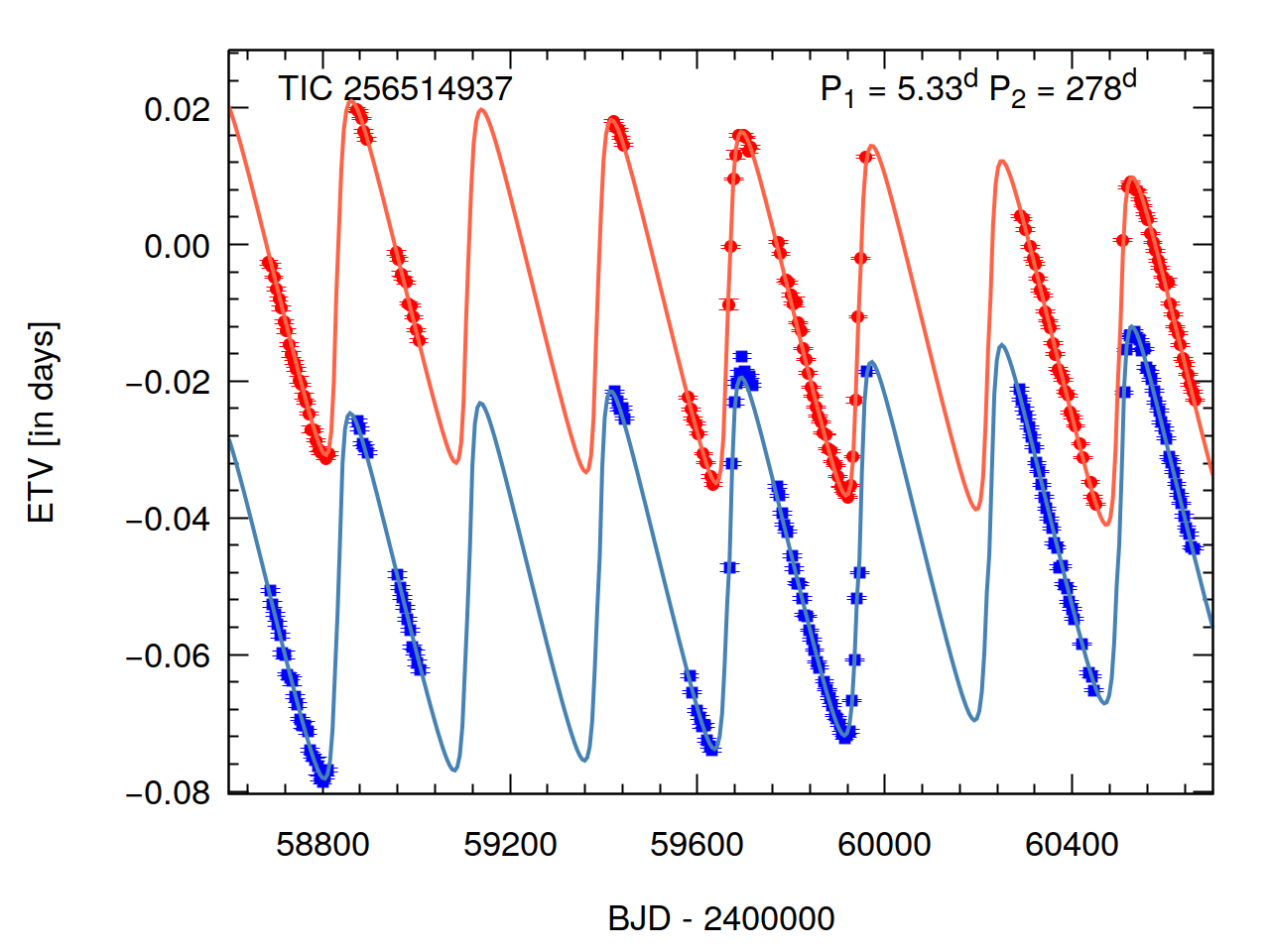}
\includegraphics[width=0.35\textwidth]{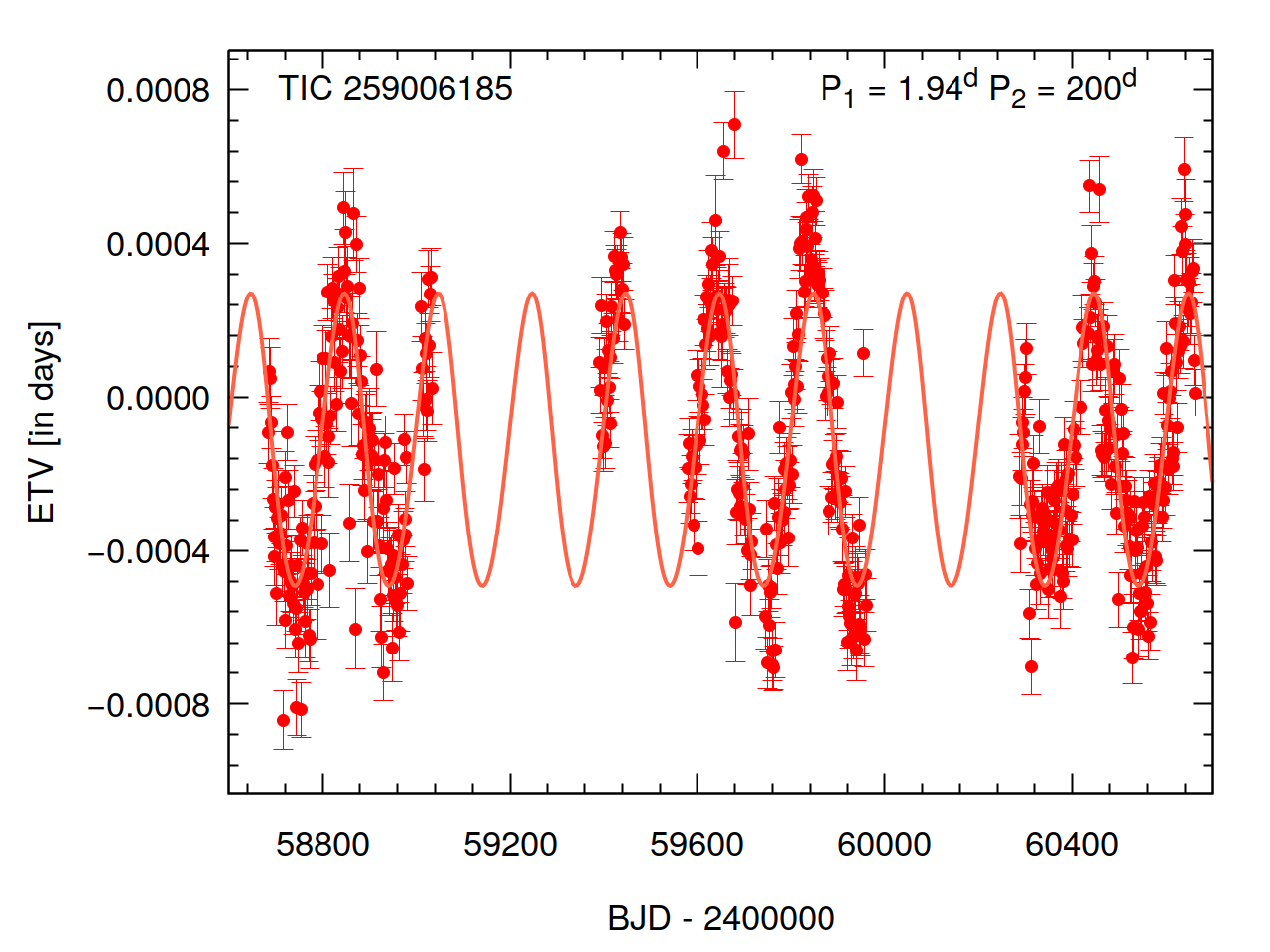}\includegraphics[width=0.35\textwidth]{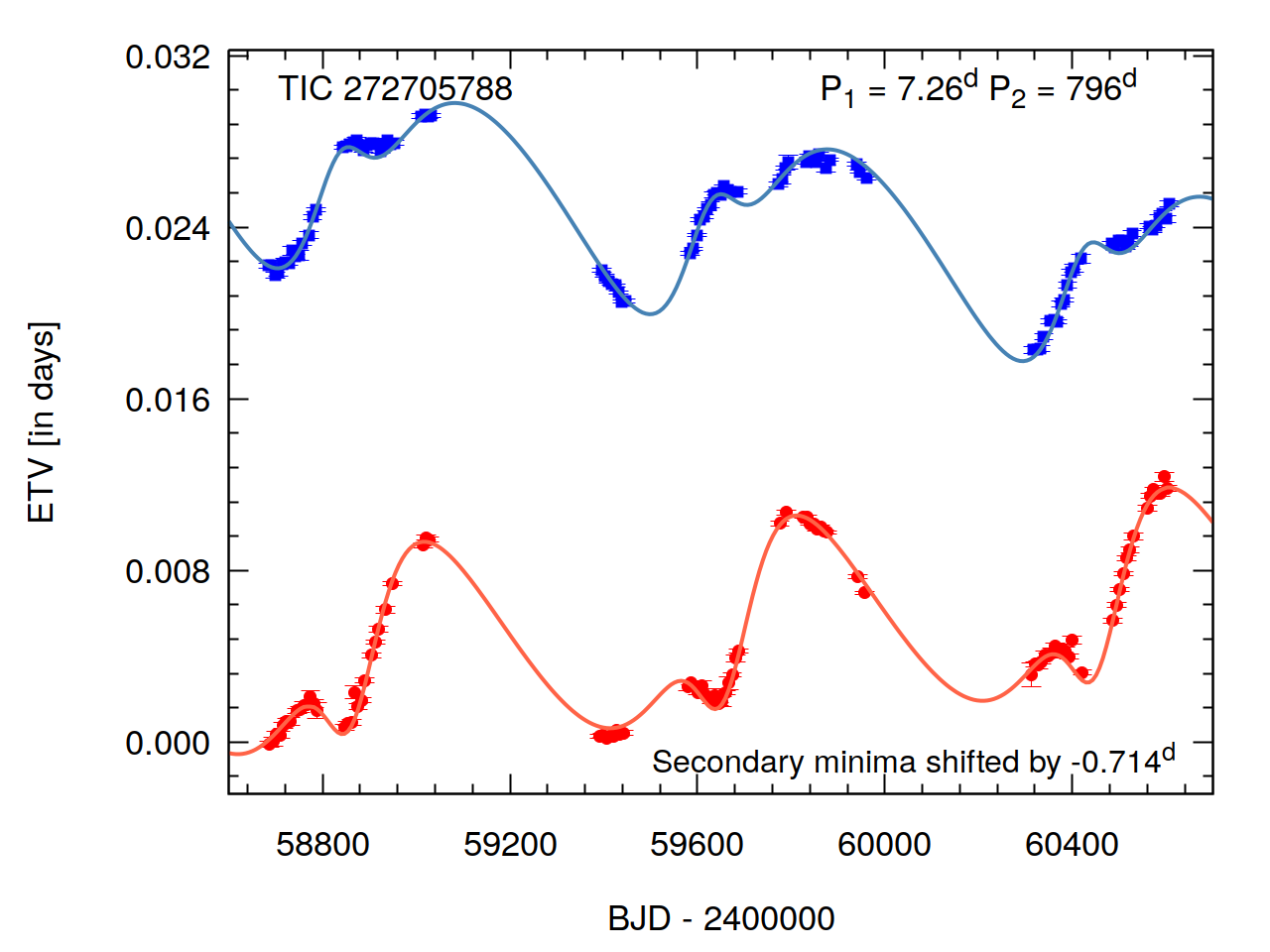}\includegraphics[width=0.35\textwidth]{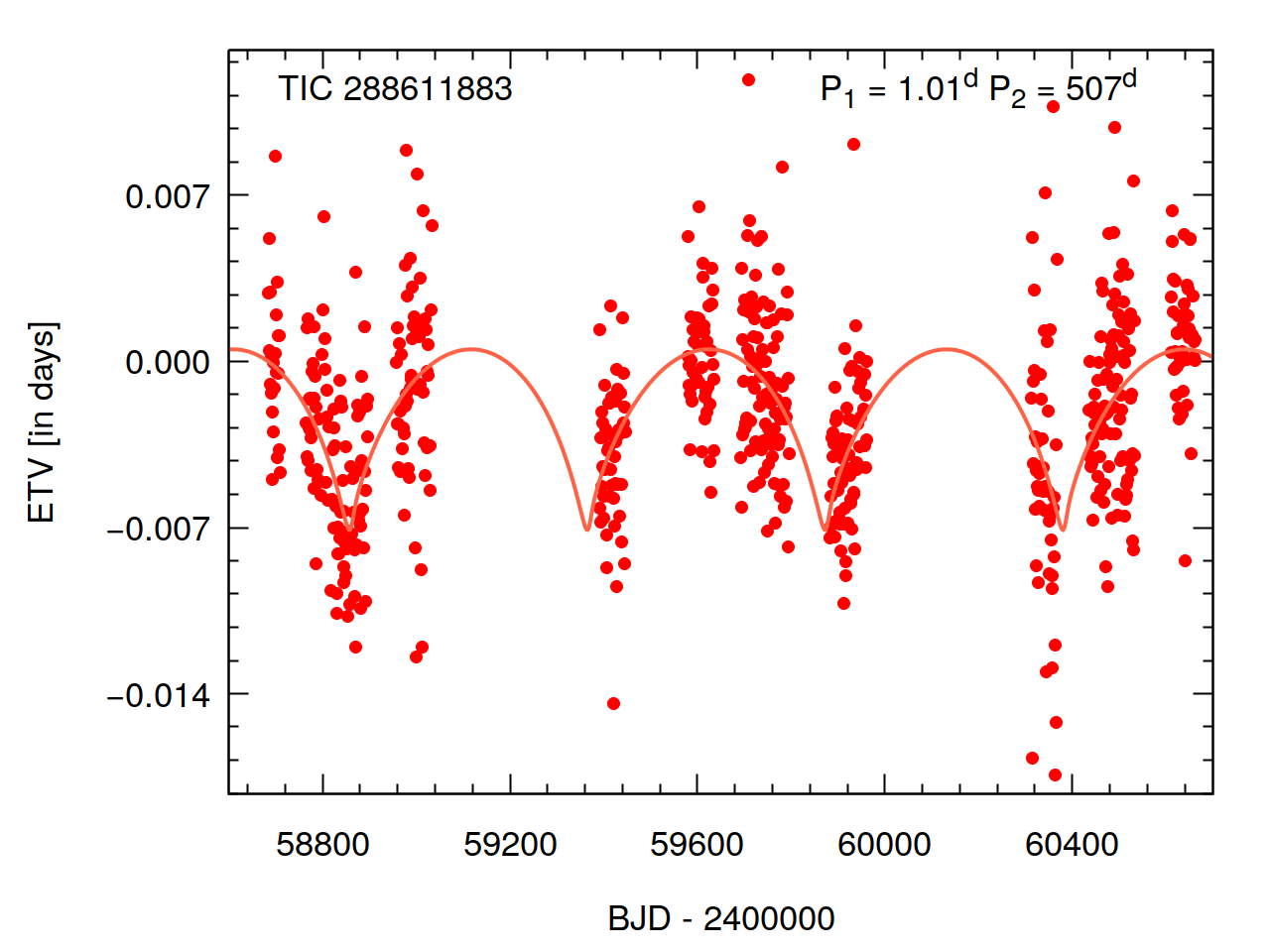}
\includegraphics[width=0.35\textwidth]{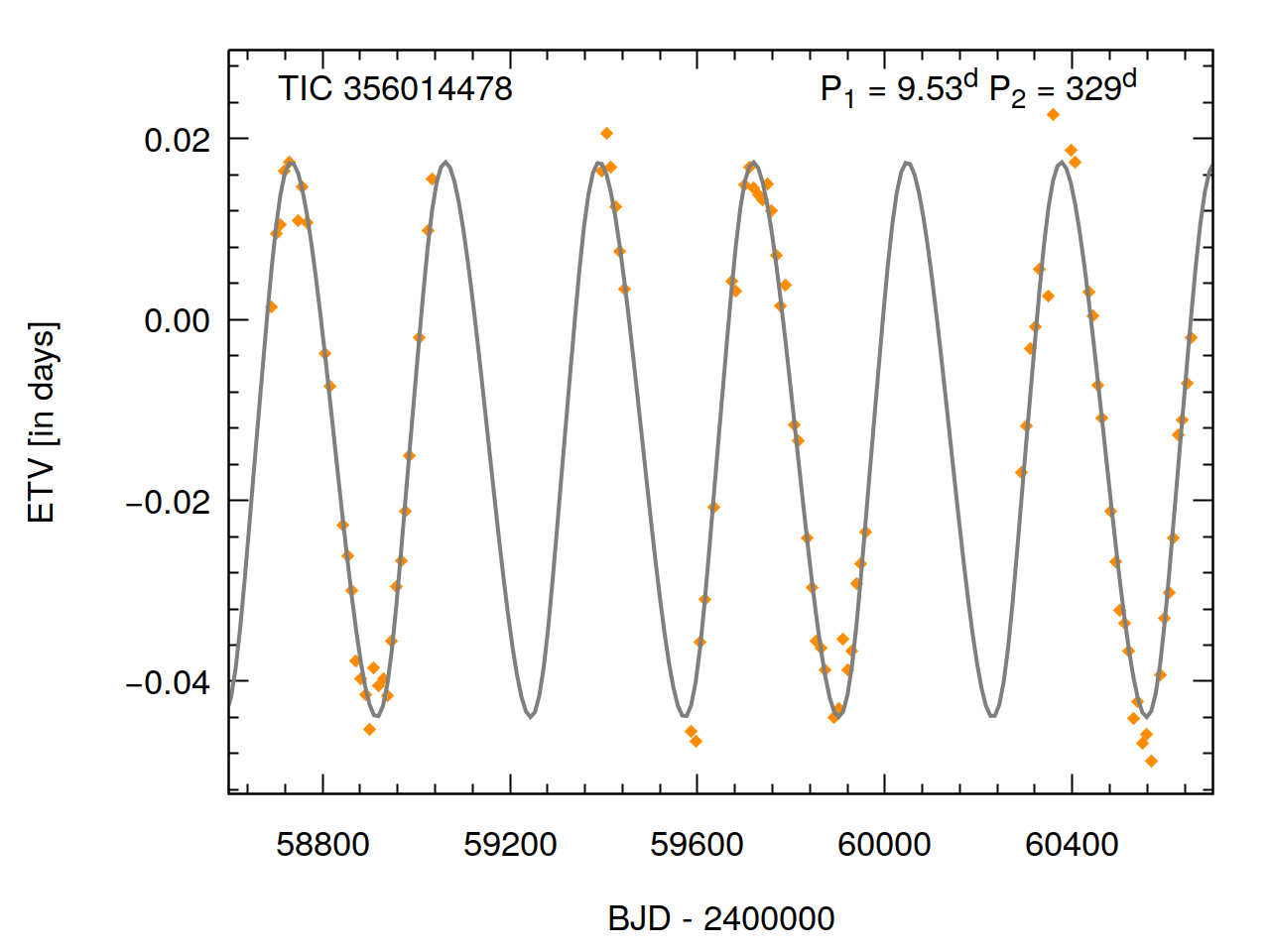}\includegraphics[width=0.35\textwidth]{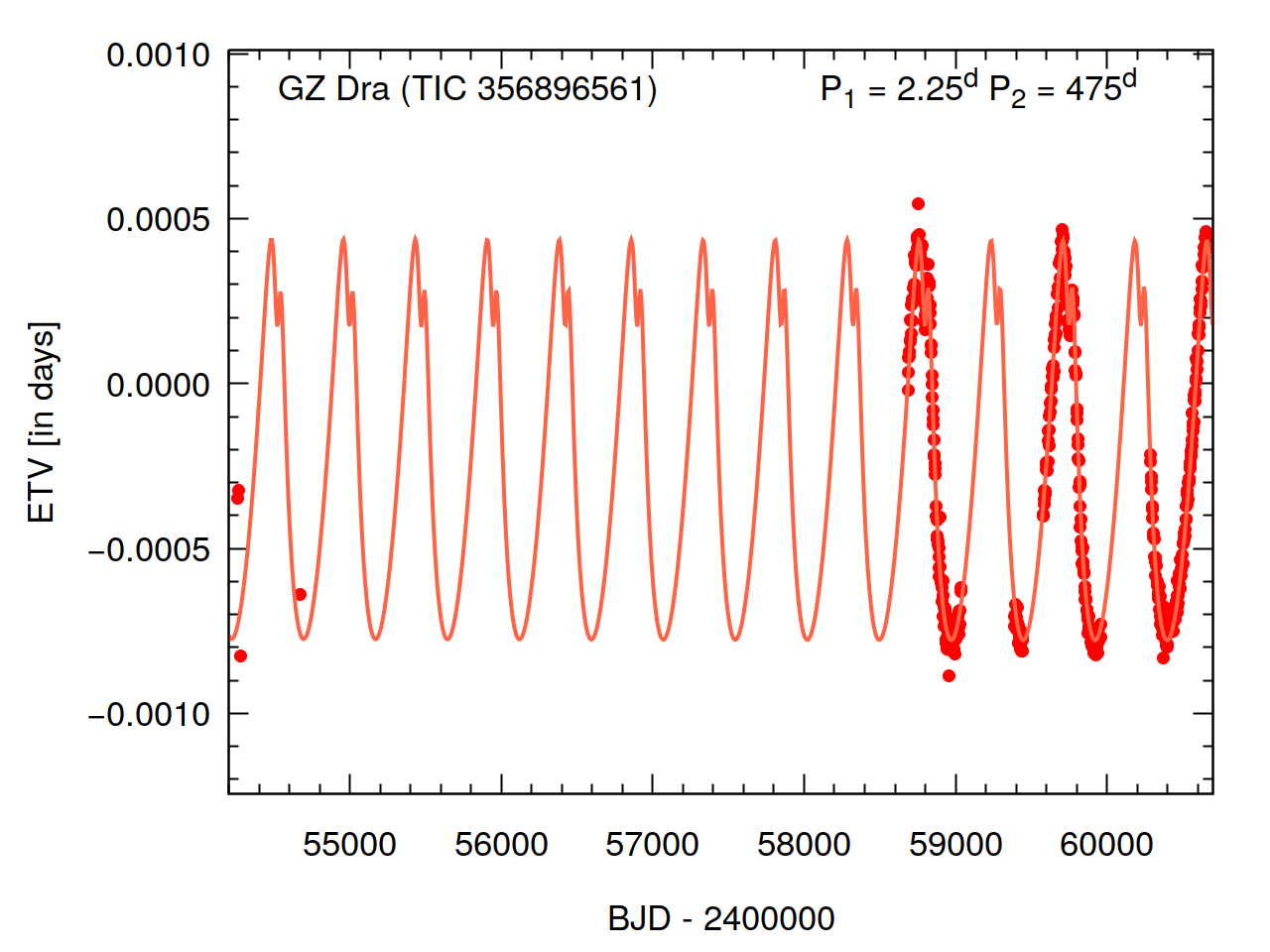}\includegraphics[width=0.35\textwidth]{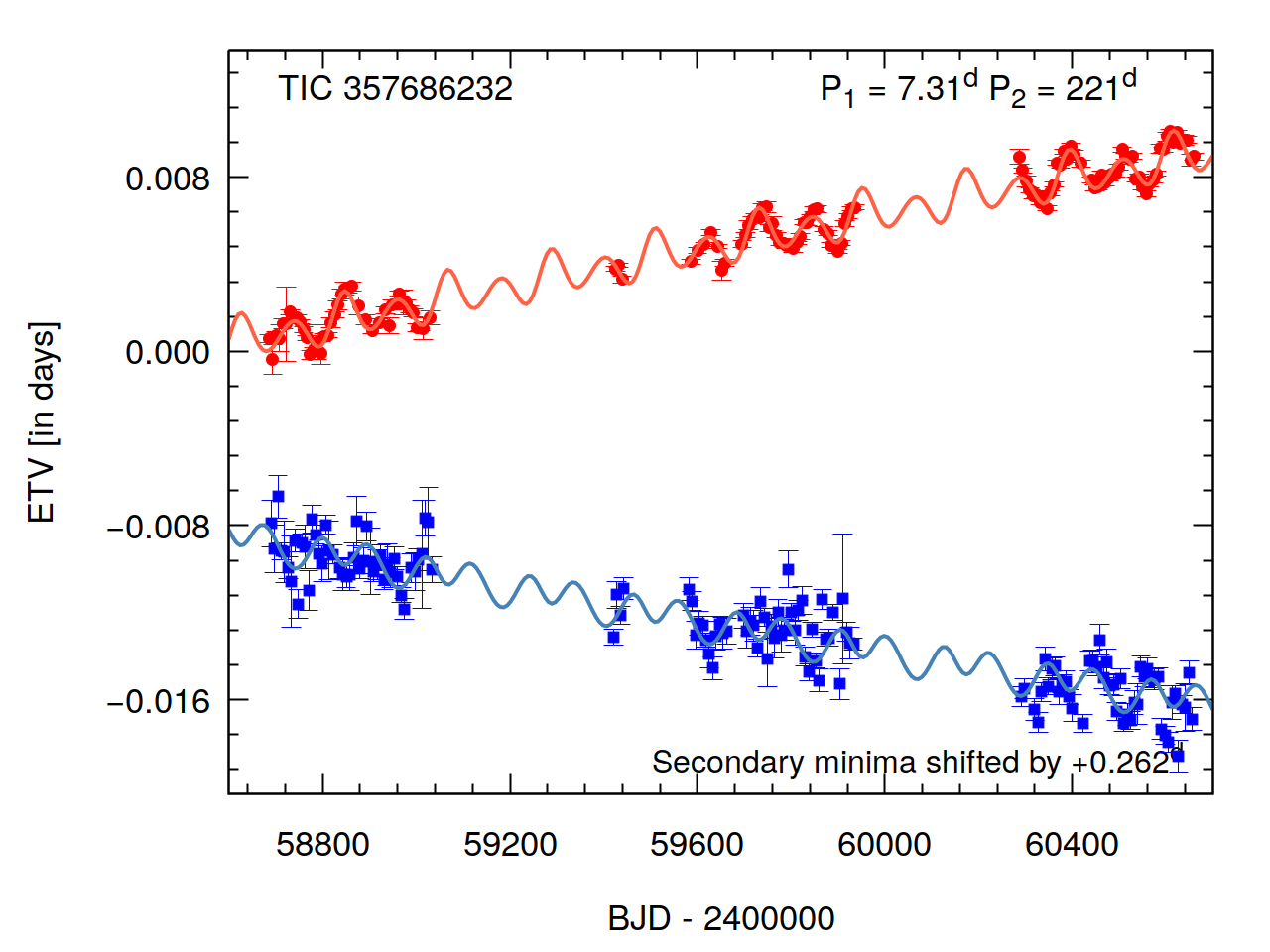}
\includegraphics[width=0.35\textwidth]{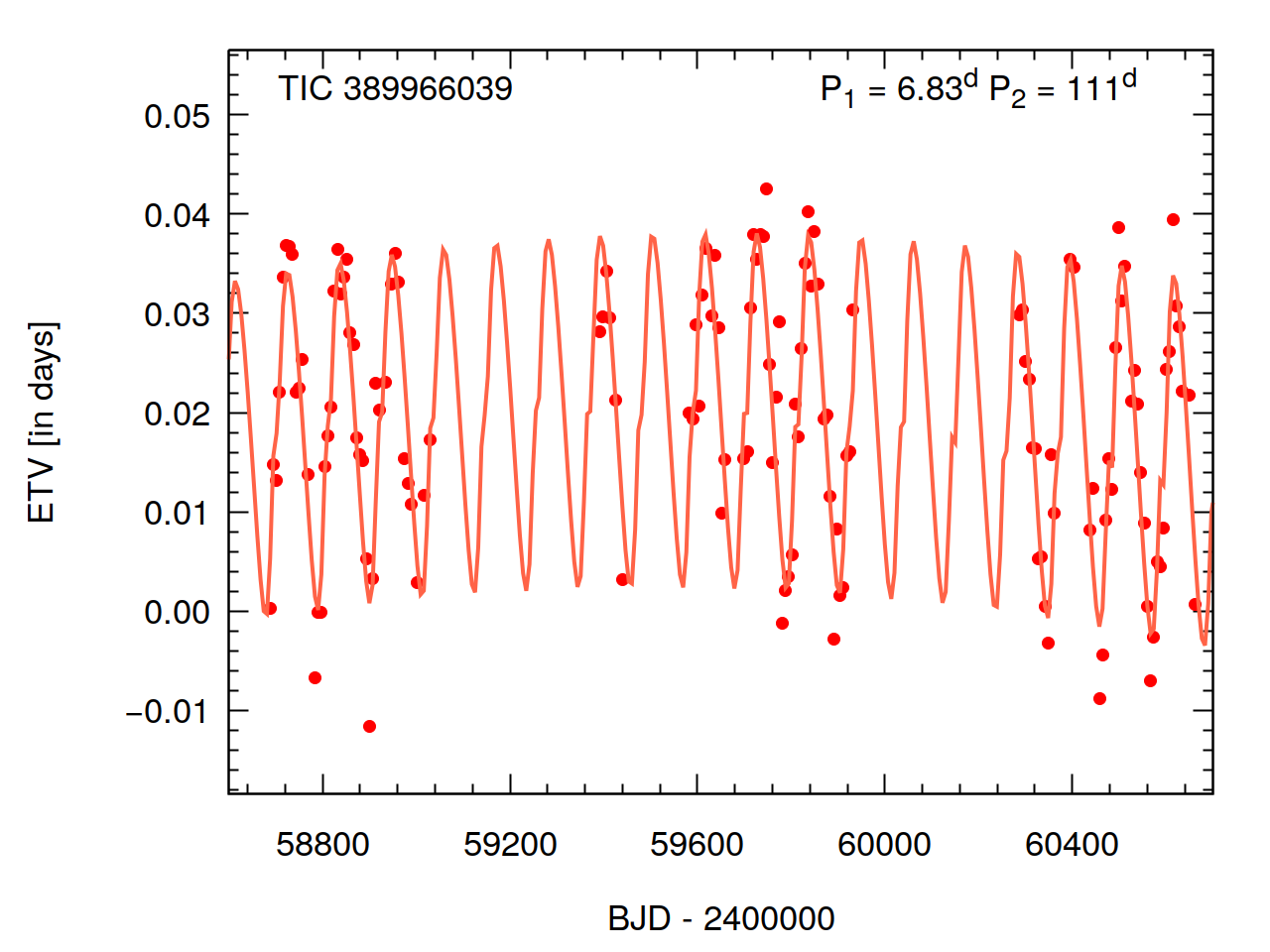}\includegraphics[width=0.35\textwidth]{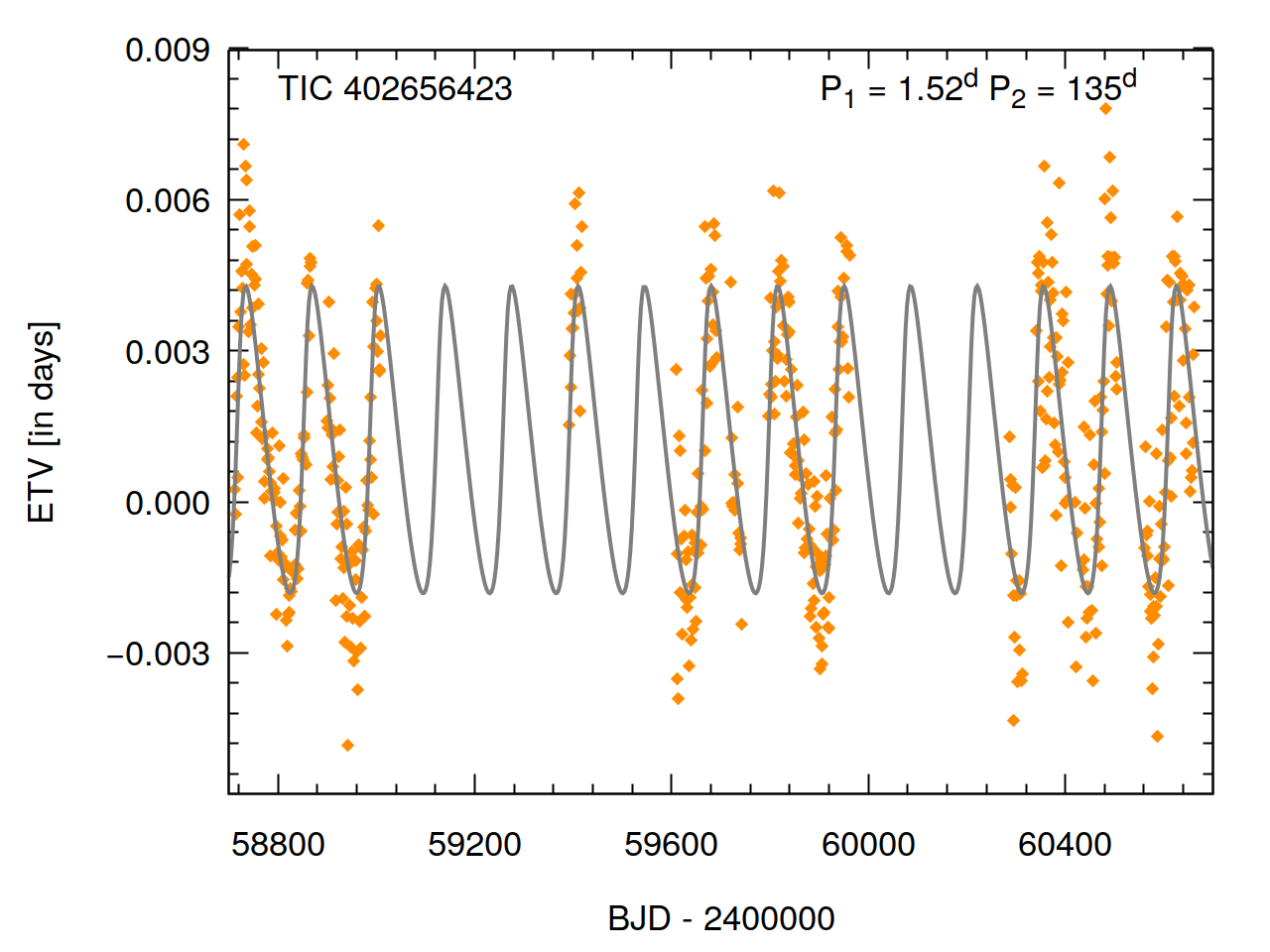}\includegraphics[width=0.35\textwidth]{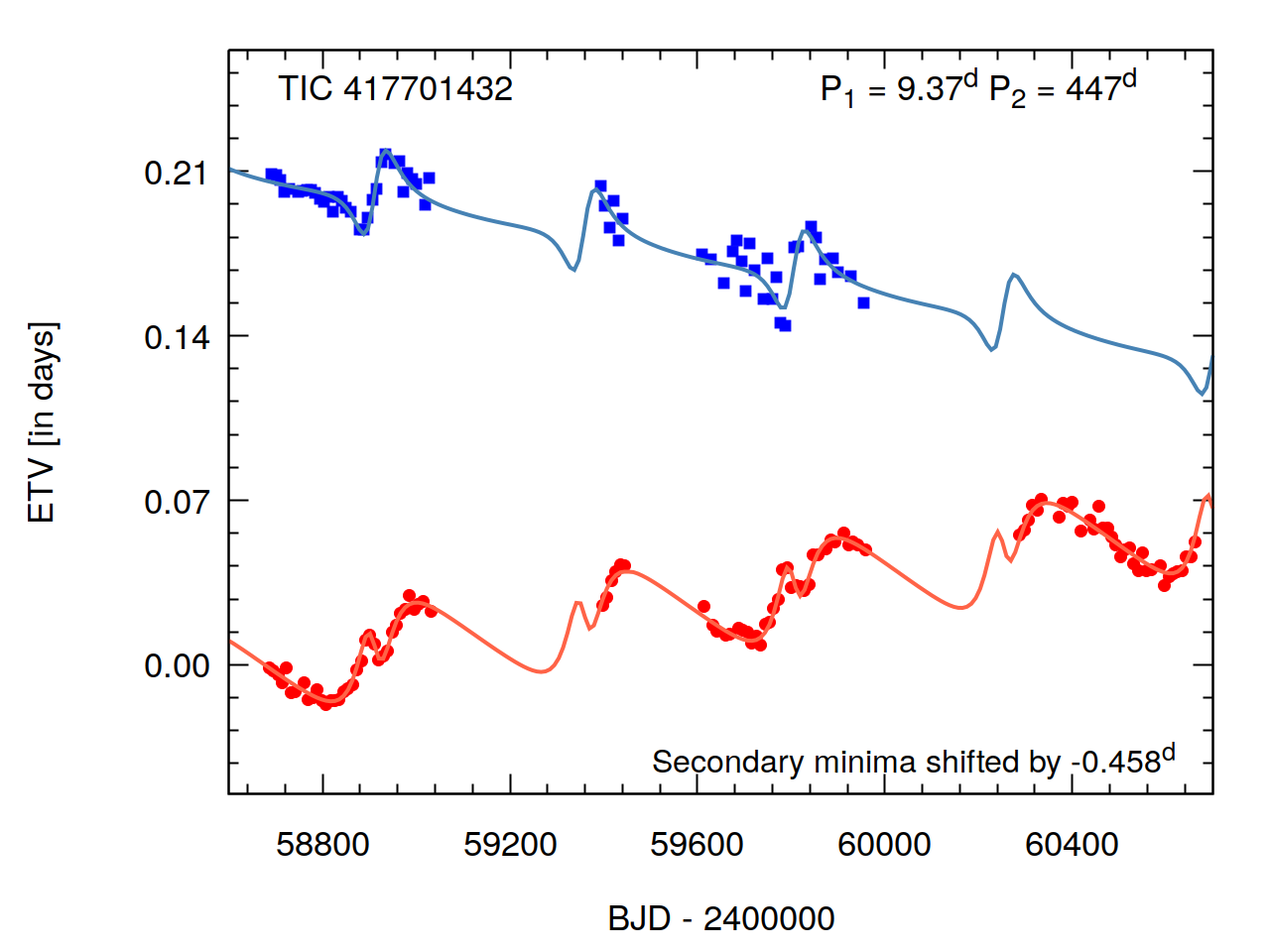}
\includegraphics[width=0.35\textwidth]{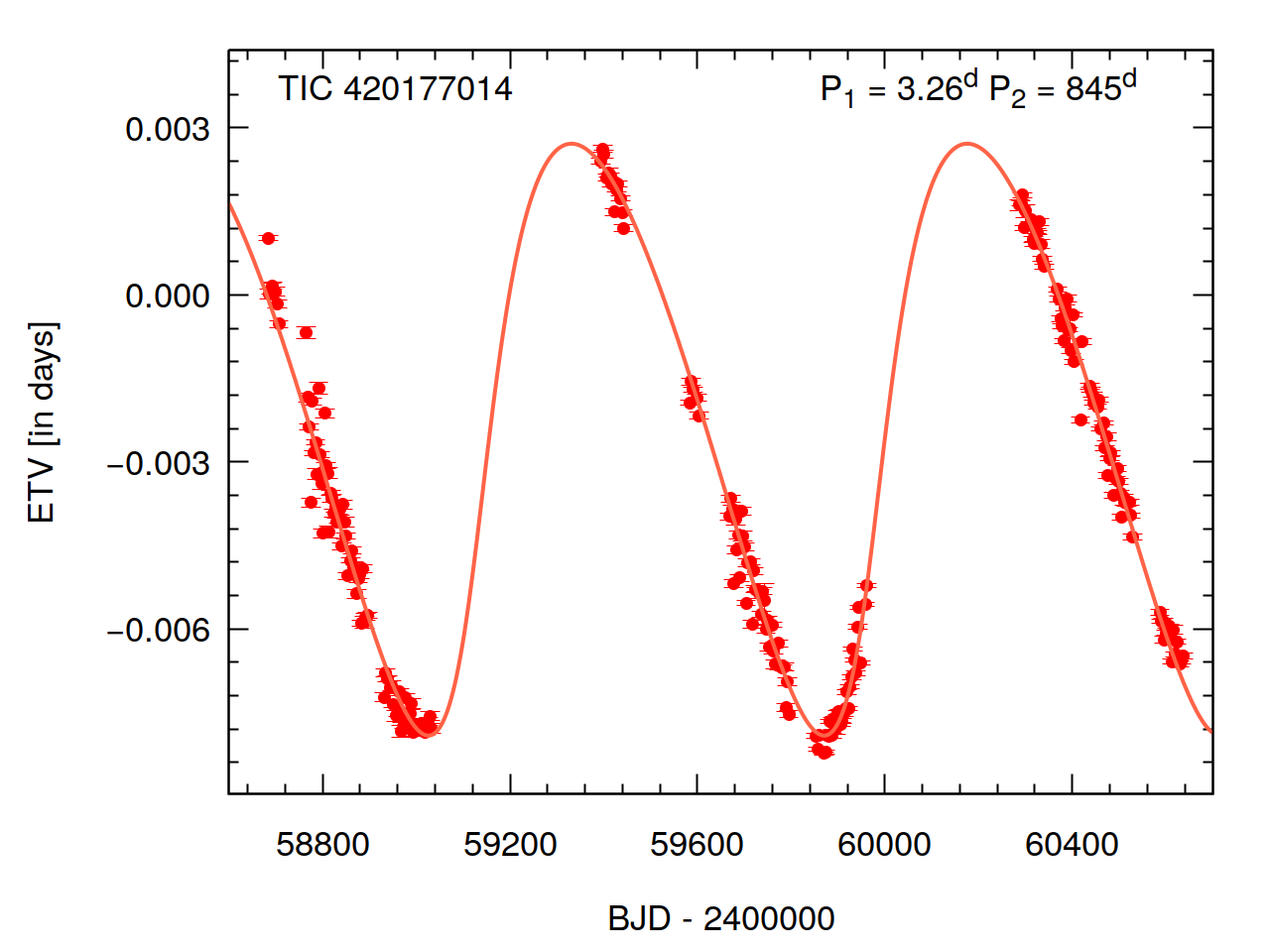}\includegraphics[width=0.35\textwidth]{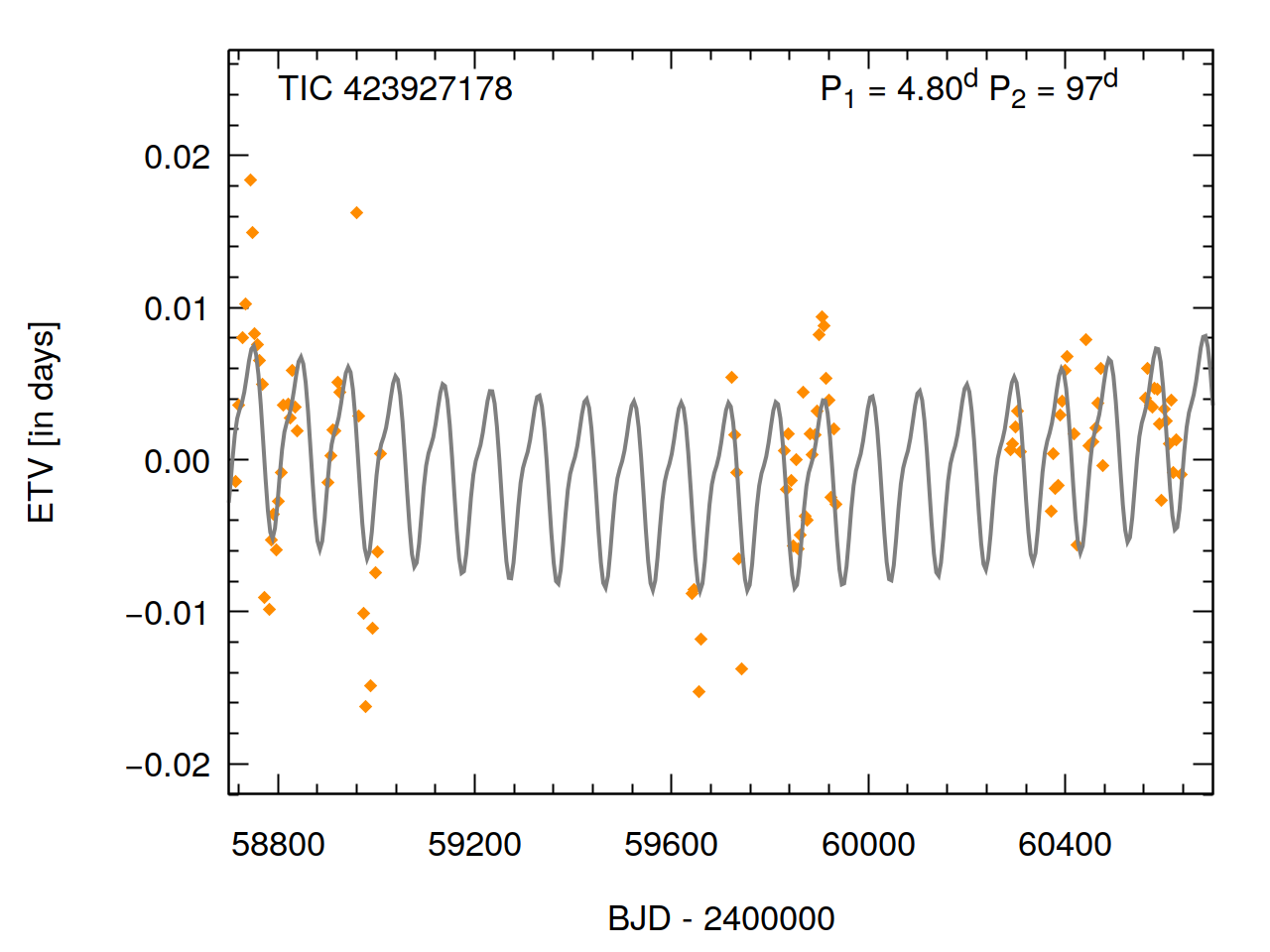}\includegraphics[width=0.35\textwidth]{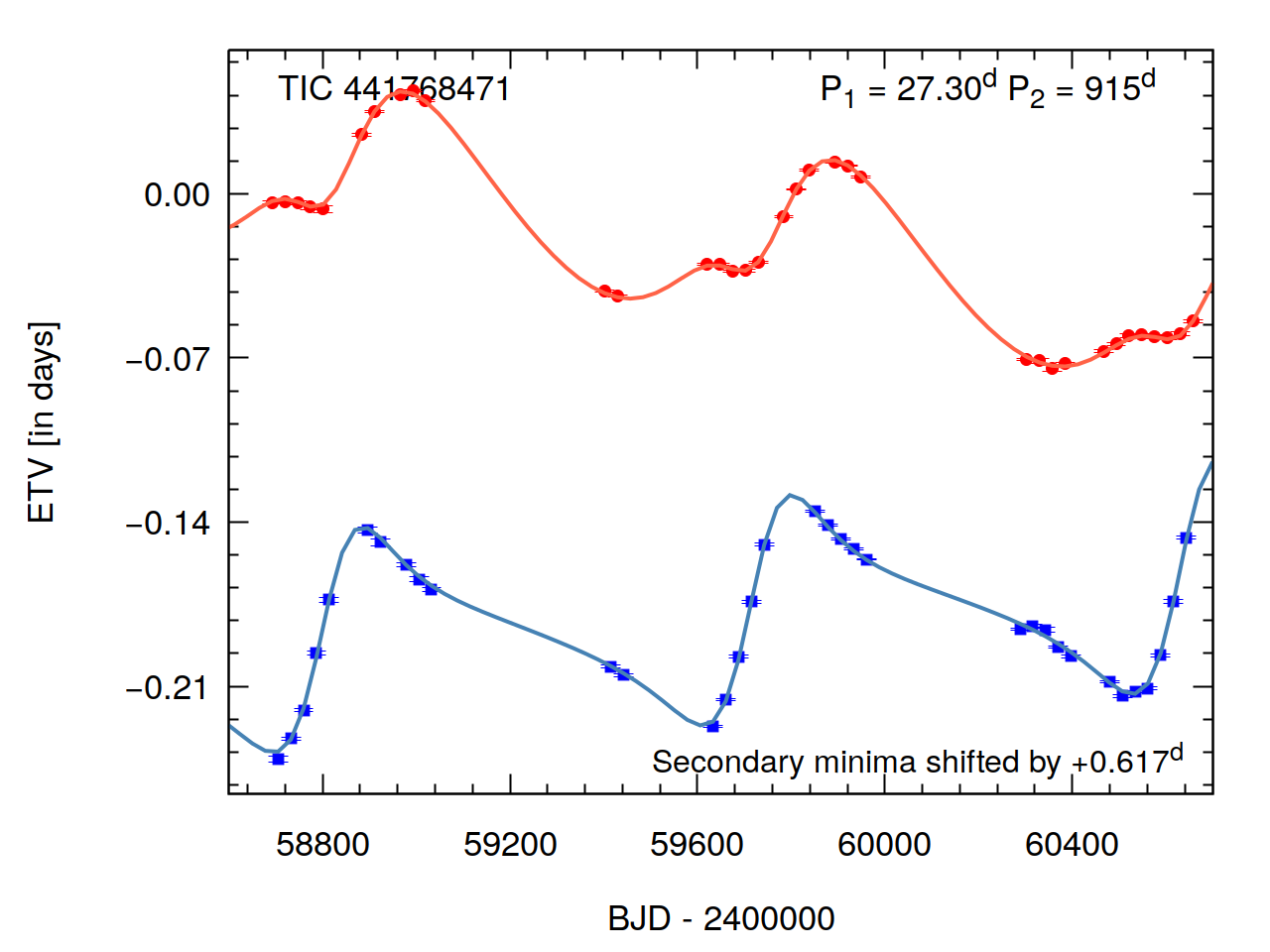}
\includegraphics[width=0.35\textwidth]{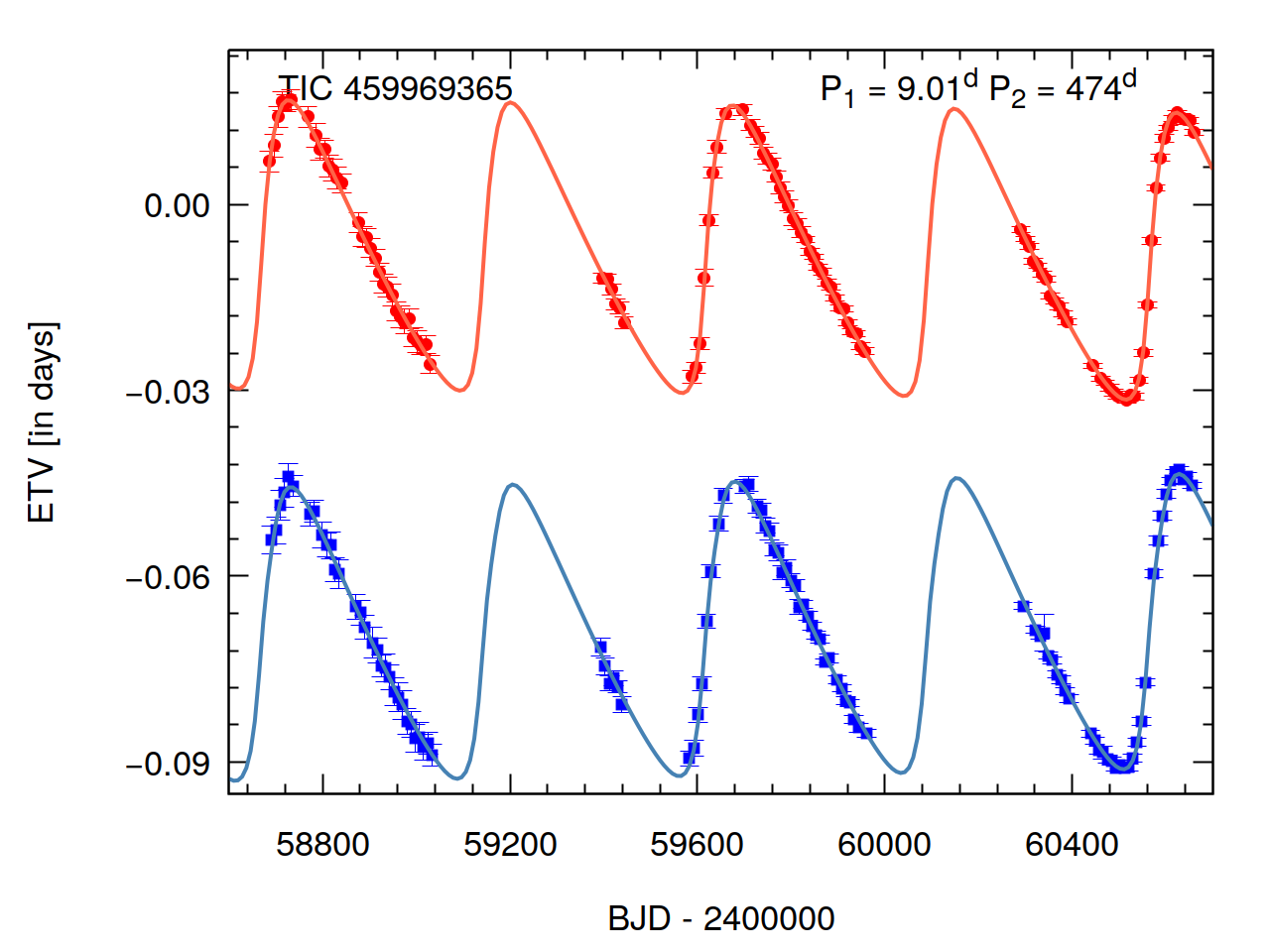}
\end{adjustwidth}
\caption{The last 16 of the 31 certain LTTE + DE solution (group $D_1$) third body candidates. The meaning of each symbol is described in the former figure captions. See  { Tables~\ref{Tab:Orbelemdyn1} and \ref{Tab:AMEparam}} for further details.}
\label{Fig:ETVs_D1b}
\end{figure}

Naturally, any period variations are hidden on the right side of the equation above, that is, in the term $\Delta$. We are searching for $\Delta$ in the following form:
\begin{equation}
\Delta=\sum_{\mathrm{i}=0}^{3}c_{\mathrm{i}}E^{\mathrm{i}}+[\Delta_{\mathrm{LTTE}}+\Delta_{\mathrm{DE}}+\Delta_{\mathrm{AM}}]_{0}^{E}.
\label{Eq:Delta}
\end{equation}
The first component, that is, the $\Sigma$ on the r.h.s.,~may describe a cubic polynomial. We say `may' because the use of the second- (quadratic) and third- (cubic) order terms is only optional and, therefore, for many of the actual ETV curves they are not used. Oppositely, the zeroth- and first-order terms were (and must be) used in all cases. Here, the zeroth-order term (that is, $c_0$) represents a simple vertical shift along the $Y$ axis. When there are no  period variations, it comes simply from the fact that the mid-time of the zeroth eclipse (similar to all the other eclipses) can be calculated only with some uncertainties, while in the case of the presence of any other kinds of period variations, this term simply counterbalances any other constant vertical shifts, which may occur from the mathematical model. Moreover, the first-order term, that is, $c_1\times E$, naturally gives any simple correction to the previously used `average' period $P$. When there are no other non-zero values on the r.h.s., this gives directly the slope of the ETV curve and, therefore, describes the correction to the `average' period $P$, which arises simply from the fact that, again, this period cannot be obtained without measurement uncertainties. Moreover, as far as only (quasi-)periodic period variations (see below) are present (therefore, there are no quadratic or cubic polynomial terms), this $c_1$-term helps to maintain the `average' slope of the ETV curve near zero (that is, the time axis of the cyclic ETV should be approximately horizontal). Considering the second- and third-order polynomial terms, their presence naturally results in varying local slopes and, hence, observed period variations. It can be readily shown that the second-order, i.e., quadratic, term describes such a period variation which is constant during one eclipsing period. More exactly, it is easy to show that $c_2=\Delta P/2$, where $\Delta P\approx P\dot P$, that is, the (constant) period variation during two consecutive primary eclipses\endnote{It can also be shown easily that in the case when $\dot P=\mathrm{const}$, that is, the systemic period varies constantly in time but not between two consecutive primary eclipses, the ETV curve can be described with an exponential curve instead of a pure parabola; however, the first- and second-order terms of the Taylor series of that exponential give back the very same parabola.}. Naturally, when the timescale of an uneven period variation is much longer than the time span of the observations, even a non-constant period variation can be linearized and, hence, can be approximated by a parabola. Hence, the usual interpretation of a parabolic ETV curve is the local manifestation of some very long timescale phenomena, such as mass transfer, mass loss, some kinds of magnetic effects, or even the existence of some strictly periodic but very long-term effect, e.g., due to a very distant third body. Similarly, a cubic coefficient in theory may represent such period variations for which the rate varies linearly in time (or, more precisely, in between two eclipses), but, mathematically, cubic polynomials can be used to describe even long-period periodic signals as well. Later, in Figures~\ref{Fig:ETVs_L1a}--\ref{Fig:ETVs_D1b} we will show examples of ETVs, for which mathematical descriptions were possible only with the use of quadratic or cubic polynomials.

Regarding the second to fourth terms on the r.h.s. of Equation~(\ref{Eq:Delta}), these are periodic terms which, in the current work, describe such effects that are caused by a third, more distant stellar (or planetary) component. Amongst them,
\begin{eqnarray}
\Delta_\mathrm{LTTE}&=&-\frac{a_\mathrm{AB}\sin i_2}{c}\frac{\left(1-e_2^2\right)\sin(v_2+\omega_2)}{(1+e_2\cos v_2)}, \nonumber \\
&=&-\mathcal{A}_\mathrm{LTTE}\sin\left(\mathcal{E}_2+\phi_2\right)+\frac{\mathcal{A}_\mathrm{LTTE}}{\sqrt{1-e_2^2\cos^2\omega_2}}e_2\sin\omega_2,
\label{Eq:LTTEdef}
\end{eqnarray}
is the well-known and widely used light-travel time effect (LTTE) component. Here, $i_2$, $e_2$, and $\omega_2$ are the inclination, eccentricity, and argument of pericenter of the outer orbit (usually denoted with the index `2'), $v_2$ and $\mathcal{E}_2$ are the true and eccentric anomalies of this same orbit, $c$ is the speed of light, and $a_\mathrm{AB}$ is the semi-major axis of the orbit of the center of mass of the inner EB components (that is, components A and B) around the center of mass of the triple system. Finally, note that the negative sign on the r.h.s. comes from the fact that, in contrast to the usual usage of the formula above, where the argument of the pericenter of the orbit of the center of mass of the inner EB around the center of mass of the triple (that is, $\omega_\mathrm{AB}$) is applied, we calculate the orbit with $\omega_2$, which differs from $\omega_\mathrm{AB}$ by $180^\circ$. Moreover, from the second row, where the eccentric anomaly was used instead of the true anomaly, one can see that the LTTE curve can be represented with a pure sine curve, of which the amplitude is
\begin{eqnarray}
\mathcal{A}_\mathrm{LTTE}&=&\frac{m_\mathrm{C}}{m_\mathrm{ABC}}\frac{a_2\sin i_2}{c}\sqrt{1-e_2^2\cos^2\omega_2}  \\
&=&\left(\frac{G}{4\pi^2}\right)^{1/3}f^{1/3}(m_\mathrm{C})\frac{P_2^{2/3}}{c}\sqrt{1-e_2^2\cos^2\omega_2}, \nonumber
\label{Eq:A_LTTE}
\end{eqnarray}
and, as one can infer from the second row, by the analogy of the mass function used for single-lined spectroscopic binaries (SB1), it is usual to introduce the same mass function as
\begin{equation}
f(m_\mathrm{C})=\frac{m_\mathrm{C}^3\sin^3i_2}{m_\mathrm{ABC}^2}.
\end{equation}
Note that it can also be very easily shown that the phase of the LTTE curve, $\phi_2$ is equal to
\begin{equation}
\tan\phi_2=\tan\left(\frac{\sin\omega_2}{\sqrt{1-e_2^2}\cos\omega_2}\right).
\end{equation}
Moreover, the very last term of the second row of Equation~(\ref{Eq:LTTEdef}) is constant (at least as far as the outer orbital elements remain constant) and, therefore, mathematically it yields only an extra contribution to $c_0$, that is, to the zero-point shift.

LTTE naturally can be put into the category of `apparent' period variations, as this is a simple geometric effect (due to the finite speed of light), caused by the unperturbed two-body (or Keplerian) revolution of the inner EB around the center of mass of the triple star. Due to this revolution, the distance of the EB from the observer varies periodically, assuming that the motion does not happen exactly in the plane of the sky (that is, in the tangential plane of the line of sight of the target). As the LTTE depends on the varying distances projected onto the radial axis $z$, that is, on $\Delta z$, there is nothing surprising about the fact that a correct LTTE ETV solution is equivalent to that which can be inferred from the radial velocity ($\dot z$) RV-solution of an SB1 system. (Naturally, there is the difference that in the case of an SB1 solution, one can find the orbital parameters and projected masses for the spectroscopic binary, while here, one can obtain analogous parameters for the outer orbit.) Consequently, we can determine the minimum mass for the tertiary component (that is, for component C), if we have some a priori assumption about the mass of the inner binary, that is, of the EB. We solve for this minimum mass in an exactly similar way as in our predecessor paper of \citet{borkovitsetal25} on the {\textit Kepler} triple star candidates, reobserved with TESS. In that paper, and, similarly, in the current work, if there are any available values for $m_\mathrm{AB}$, e.g., such as those obtained from former RV studies of double-lined spectroscopic binaries (SB2), we use those masses and cite the former works in the appropriate tables. When such pre-determined EB masses are unavailable for overcontact (that is, W UMa-type) EBs, we follow the empirical mass--period relations of \citet{gazeasstepien08}, while for other EBs, we simply, but reasonably, assume that $m_\mathrm{AB}=2.0\,\mathrm{M}_\odot$.

The next term, $\Delta_\mathrm{DE}$, represents those period variations which are due to the gravitational or dynamical interactions between the inner and outer stars in tight triple stellar systems. By the expression `tight,' we usually mean such triple or multiple systems where, due to the proximity of the third component to the inner binary members, the motion of neither the inner nor the outer stars can be considered to be pure, practically unperturbed Keplerian motions.  In this case, there are additional periodic (or at least quasi-periodic) features induced on the ETV curves due to the mutual gravitational perturbations---and these can be detected even on rather short timescales. 

Naturally, in the case of a bound third star, gravitational perturbations are always present. However, the largest amplitude, so-called `apse-node timescale' perturbations (see, e.g., the short review in \citet{borkovits22} and further references therein), have a characteristic timescale of $P_\mathrm{apse-node}\propto P_2^2/P_1$ and, hence, this quickly tends to centuries or millennia with growing outer periods, as well as with growing $P_2/P_1$ ratios. Additionally, the shorter timescale perturbations---of which the periods are in the orders of $P_2$ or $P_1$, and their amplitudes are related to $P_1/P_2$ or $(P_1/P_2)^2$---become quickly very small and, hence, undetectable with an increasing ratio of $P_2/P_1$.

\vspace{-12pt}

\begin{table}[H]
\tablesize{\fontsize{7.0}{7.0}\selectfont} 
\caption{Triple system candidates with certain or very likely LTTE solutions.} 
\label{Tab:Orbelem_LTTE1}  
\begin{adjustwidth}{-\extralength}{0cm}
\centering

\begin{tabularx}{\fulllength}{lccccccccccc}

\toprule
\textbf{TIC No.} & \boldmath{$P_1$} & \boldmath{$\Delta P_1$} & \boldmath{$P_2$} & \boldmath{$a_\mathrm{AB}\sin i_2$} & \boldmath{$e_2$} & \boldmath{$\omega_2$} & \boldmath{$\tau_2$} & \boldmath{$f(m_\mathrm{C})$} & \boldmath{$(m_\mathrm{C})_\mathrm{min}$} & \boldmath{$\frac{{\cal{A}}_\mathrm{dyn}}{{\cal{A}}_\mathrm{LTTE}}$} & \boldmath{$m_\mathrm{AB}$}\\

        & \textbf{(day)} &\boldmath{$\times10^{-10}$} \textbf{(d/c)}&\textbf{(day)}&\textbf{(R}\boldmath{$_\odot$}\textbf{)}  &       &   \textbf{(deg)}    &   \textbf{(MBJD)} & \textbf{(M}\boldmath{$_\odot$}\textbf{})       & \textbf{(M}\boldmath{$_\odot$}\textbf{)}            & &  \textbf{(M}\boldmath{$_\odot$}\textbf{)}    \\
\midrule
159465833     & 1.8528488 (9)& $-$ & 702 (11)& 205 (35) & 0.4 (2) &168 (18)&58,695 (37) & 0.24 (12)  & 1.39 & 0.09 & 2: \\
160518449     &2.25604664 (2)& $-$ &926.8 (5)&103.8 (4) &0.250 (7)&92.7 (9)& 58,760 (2) & 0.0175 (2) & 0.47 & 0.05 & 2: \\
165550395 $^a$ &0.32253914 (7)&$-23.7$ (6)&220.9 (1)&63.7 (7)&0.21 (1)&312 (6)&58,708 (4) & 0.071 (2)  & 0.65 & 0.01 & 1.30\\
198280388     &0.488399772 (5)&$-$ &626.8 (3)& 95.5 (3) &0.148 (5)&  0 (2) & 58,677 (2) & 0.0297 (2) & 0.58 & 0.002& 2: \\
229412530     &0.36690400 (3)&2.5 (1)&529 (1) & 110 (4)  &0.79 (2) & 24 (1) & 58,585 (3) & 0.064 (6)  & 0.71 & 0.06 &1.66 *\\
229711743     &0.30085000 (3)&2.21 (8)&403.4 (3)&108.1 (9)&0.42 (1)&340.0 (8)&58,640 (2) & 0.104 (3)  & 0.81 & 0.008&1.46 *\\
229751802     &0.516255787 (3)&$-$ &216.96 (6)& 111 (2) & 0.49 (2)& 191 (1)&58,593.9 (7)& 0.39 (2)   & 1.77 & 0.08 & 2: \\
              &             &     &6156 (910)& 143 (36)& 0.39 (9)&341 (40)&59,473 (734)& 0.0010 (7) & 0.22 &      & 3: \\
229762991     &0.33846177 (2)&$-$2.20 (7)&417.9 (3)&75.8 (5)&0.11 (1)&274 (7)&58,562 (8) & 0.0334 (7) & 0.53 & 0.002&1.57 *\\
229771231     &0.82095349 (4)&6.5 (3)&359.1 (2)&85.5 (7) & 0.50 (1)&146 (1) & 58,853 (1) & 0.065 (2)  & 0.80 & 0.09 & 2: \\
229787617     &0.41265209 (1)& $-$ &454.7 (7)&  80 (1)  & 0.11 (3)&138 (16)& 58,732 (20)& 0.033 (2)  & 0.57 & 0.006&1.79 *\\ 
229789945     &0.36334574 (4)&3.4 (1)& 720 (3)&  58 (1)  & 0.25 (4)&135 (9) & 58,911 (17)& 0.0050 (3) & 0.26 & 0.002&1.65 *\\
229972643     &0.87325549 (8)&$-7.5$ (6)&763 (2)&64.7 (8)& 0.31 (2)&351 (5) & 58,907 (11)& 0.0062 (2) & 0.32 & 0.01 & 2: \\
230394086     &0.34662012 (3)&$-4.0$ (1)&773.1 (9)&140 (1)&0.35 (1)&316 (2) & 58,921 (5) & 0.061 (1)  & 0.68 & 0.003&1.60 *\\
232606864 $^b$ &0.29430458 (7)&12.8 (5)&522.6 (3)&94.9 (5)&0.466 (7)& 48 (1) & 58,790 (2) & 0.0420 (7) & 0.55 & 0.006&1.44 *\\ 
232610209     &0.37794075 (4)&6.2 (1)&410.7 (3)&100.3 (9)& 0.12 (2)&128 (9) & 58,530 (10)& 0.080 (2)  & 0.79 & 0.002&1.69 *\\
232632890     & 1.3480688 (3)& $-$ & 547 (3)  & 165 (6) & 0.26 (6)&151 (14)& 58,489 (23)& 0.20 (2)   & 1.30 & 0.04 & 2: \\  
233050384     &0.552454646 (5)&$-$ &194.04 (4)&82.7 (3) &0.059 (7)&316 (7) & 58,613 (4) & 0.201 (2)  & 1.36 & 0.008&2.17 *\\ 
233070708     &0.98221030 (8)& $-$ & 421 (1) & 123 (3)  &0.25 (4) & 53 (9) & 58,561 (11)& 0.142 (9)  & 1.11 & 0.03 & 2: \\ 
233130175     &1.4940859 (1) & $-$ &189.9 (3)& 64 (2)   &0.17 (7) &275 (26)& 58,691 (14)&  0.10 (1)  & 0.94 & 0.19 & 2: \\
233574062     &0.29953854 (1)&1.42 (4)&348.5 (1)&86.1 (3)&0.179 (8)& 65 (3) & 58,685 (3) & 0.0704 (8) & 0.69 & 0.003&1.46 *\\
233611409 $^c$ &0.696175248 (9)&$-$ & 410.7 (2)& 88.6 (5)&0.25 (1) & 12 (2) & 58,619 (2) & 0.055 (1)  & 0.86 & 0.02 &2.53 *\\
236760823     &1.11377207 (1)& $-$ &274.35 (8)& 68.8 (3)&0.010 (8)& 36 (45)& 58,752 (35)& 0.0579 (7) & 0.76 & 0.11 & 2: \\
237202126     &0.305522858 (6)&$-$ & 534.6 (3)&173.1 (6)&0.049 (7)& 198 (9)& 58,943 (13)& 0.243 (3)  & 1.20 &0.0004&1.47 *\\
258351350 $^d$ &0.772936146 (9)&$-$ & 384.6 (1)&120.4 (4)&0.295 (6)& 196 (1)& 58,832 (1) & 0.158 (2)  & 1.58 & 0.03 & 3.43\\  
259038777     &0.43327127 (1)& $-$ &540.3 (8)& 75 (1)   & 0.31 (3)& 178 (5)& 58,476 (7) & 0.019 (1)  & 0.47 & 0.006&1.85 *\\
264218477     &0.30855580 (1)&2.35 (3)&109.23 (2)&44.9 (4)&0.38 (1)& 298 (2)&58,694.5 (7)& 0.102 (3)  & 0.85 & 0.06 &1.60 *\\
288301339     & 1.8842079 (2)& $-$ & 610 (4) & 80 (3)   & 0.12 (8)&157 (38)& 58,843 (65)& 0.018 (2)  & 0.48 & 0.03 & 2: \\
288430645 $^e$ & 0.7915109 (2)& $-$ & 259 (1) & 96 (6)   &0.16 (11)&193 (43)& 58,624 (32)&  0.18 (3)  & 1.23 & 0.03 & 2: \\
288510753     &2.01985712 (6)& $-$ & 868 (2) & 112.0 (7)& 0.30 (2)& 267 (2)& 59,055 (6) & 0.0250 (5) & 0.55 & 0.05 & 2: \\
320324245     &0.35113328 (1)&4.53 (4)&530.1 (2)&111.9 (3)&0.151 (4)& 20 (2)& 58,879 (3) & 0.0650 (5) & 0.70 & 0.002&1.61 *\\ 
357596894     &0.46468888 (4)& $-$ & 728 (3) & 171 (8)  & 0.38 (5)& 198 (4)& 58,540 (9) & 0.13 (2)   & 0.90 & 0.006& 1.5: \\
394089883     &0.26693075 (2)& $-$ & 916 (3) & 201 (7)  & 0.27 (5)& 197 (5)& 58,408 (16)& 0.13 (1)   & 0.86 & 0.001&1.35 *\\
441738417 $^e$ &0.67658255 (5)&6.8 (3)&471.4 (3)&109 (1)  & 0.49 (1)&  12 (1)& 58,815 (2) & 0.078 (2)  & 0.86 & 0.04 & 2:\\ 
441788515     & 1.4548757 (1)& $-$ & 938 (3) & 175 (4)  & 0.29 (2)&  67 (6)& 58,636 (17)& 0.082 (6)  & 0.88 & 0.02 & 2: \\
441806190     &0.270233943 (5)&$-$ & 880 (1) & 117 (1)  & 0.71 (1)& 271 (1)& 58,590 (4) & 0.0278 (7) & 0.45 & 0.007&1.36 *\\
\bottomrule
\end{tabularx}
\end{adjustwidth}
\noindent{\footnotesize{\textbf{Notes}. In the last column of the current and the forthcoming two tables, the $m_\mathrm{AB}$ values denoted with * and $:$ refer to estimated values with the use of either the formulae of \citet{gazeasstepien08} or our own reasonable estimations, respectively. Values without such additional characters are taken from the literature and referenced in this and the forthcoming Table notes. $^a$: Cubic ephemeris: $c_3=1.29(3)\times10^{-13}\,\mathrm{d/c}^2$; $^b$: V504 Dra---cubic ephemeris: $c_3=-3.1(2)\times10^{-14}\,\mathrm{d/c}^2$; $^c$: V411 Dra; $^d$: HZ Dra---$m_\mathrm{AB}$ was taken from \citep{liakosniarchos17}; $^e$: third body eclipses.}}
 
\end{table}

\begin{table}[H]
\tablesize{\fontsize{7.2}{7.2}\selectfont} 
\caption{Triple system candidates with less certain but likely LTTE solutions.} 
\label{Tab:Orbelem_LTTE2}  

\begin{adjustwidth}{-\extralength}{0cm}
\centering
\begin{tabularx}{\fulllength}{lccccccccccc}

\toprule
\textbf{TIC No.} & \boldmath{$P_1$} & \boldmath{$\Delta P_1$} & \boldmath{$P_2$} & \boldmath{$a_\mathrm{AB}\sin i_2$} & \boldmath{$e_2$} & \boldmath{$\omega_2$} & \boldmath{$\tau_2$} & \boldmath{$f(m_\mathrm{C})$} & \boldmath{$(m_\mathrm{C})_\mathrm{min}$} & \boldmath{$\frac{{\cal{A}}_\mathrm{dyn}}{{\cal{A}}_\mathrm{LTTE}}$} & \boldmath{$m_\mathrm{AB}$}\\
        & \textbf{(day)} &\boldmath{$\times10^{-10}$} \textbf{(d/c)}&\textbf{(day)}&\textbf{(R}\boldmath{$_\odot$}\textbf{)}  &       &   \textbf{(deg)}    &   \textbf{(MBJD)} & \textbf{(M}\boldmath{$_\odot$}\textbf{)}       & \textbf{(M}\boldmath{$_\odot$}\textbf{)}            & &  \textbf{(M}\boldmath{$_\odot$}\textbf{)}    \\
\midrule
159398028     &0.480077629 (9)&$-$ & 675 (2) & 26.1 (7) & 0.27 (4)& 296 (9)&58,860 (17) &0.00052 (4) & 0.13 & 0.004& 2: \\
165500662     &0.36683666 (2)& $-$ &1704 (12)& 63 (2)   & 0.47 (3)& 156 (3)& 58,609 (16)& 0.00114 (9)& 0.16 & 0.001&1.66 *\\
199632809     & 1.5980432 (2)& $-$ &1282 (9) & 152 (2)  & 0.33 (2)& 165 (6)& 58,498 (22)& 0.029 (1)  & 0.58 & 0.02 & 2: \\
219111461     & 0.6303173 (1)& $-$ & 626 (2) & 207 (7)  & 0.32 (4)& 235 (7)& 58,812 (13)& 0.31 (3)   & 1.57 & 0.01 & 2: \\
229500406     &0.38340528 (4)& $-$ & 1497 (9)& 235 (2)  & 0.19 (2)& 131 (5)& 58,155 (21)& 0.078 (2)  & 0.79 &0.0008&1.71 *\\
229651225     &0.34981342 (2)& $-$ &1638 (10)& 177 (1)  & 0.29 (1)& 203 (2)& 57,904 (11)& 0.0277 (6) & 0.49 &0.0007&1.61 *\\ 
229799471     &0.39832986 (3)& $-$ & 1742 (8)& 328 (1)  &0.240 (9)& 222 (2)& 57,826 (10)& 0.155 (3)  & 1.07 &0.0007&1.75 *\\
229910746     &2.50633079 (7)& $-$ &1791 (17)& 79.0 (5) & 0.20 (2)&  68 (3)& 57,922 (20)& 0.00206 (5)& 0.22 & 0.03 & 2: \\
230007820 $^a$ & 5.253226 (1) & $-$ &3030 (95)& 413 (11) & 0.62 (1)& 197 (2)& 60,169 (65)& 0.10 (1)   & 0.97 & 0.11 & 2: \\
230116482     & 1.9288578 (6)& $-$ & 983 (13)& 182 (21) &0.53 (10)&243 (17)& 59,022 (51)& 0.083 (30) & 0.89 & 0.10 & 2: \\
230387571     &1.27996134 (7)& $-$ &1557 (23)& 45 (1)   & 0.37 (4)&  88 (7)& 58,447 (33)& 0.00051 (5)& 0.13 & 0.01 & 2: \\ 
230393824     &0.50549559 (2)& $-$ &1005 (2) & 140 (1)  & 0.24 (1)& 121 (2)& 58,673 (7) & 0.0361 (8) & 0.63 & 0.003& 2: \\
233056681     &0.60112683 (1)& $-$ &1589 (14)& 33.0 (5) & 0.14 (3)&132 (16)& 59,165 (72)&0.000191 (9)& 0.09 & 0.002& 2: \\
233657543     &0.238725164 (3)&$-$ & 818 (2) & 80 (2)   & 0.74 (2)& 153 (1)& 58,610 (5) & 0.0103 (9) & 0.29 & 0.01 &1.26 *\\
233719825     & 2.0509408 (2)& $-$ &1384 (4) & 285 (2)  & 0.31 (1)&  95 (3)& 58,722 (11)& 0.162 (4)  & 1.18 & 0.03 & 2: \\
237201620     &0.44299702 (4)& $-$ &1729 (3) & 593 (1)  &0.246 (2)&119.3 (7)&59,444 (4) & 0.935 (7)  & 2.69 &0.0007&1.88 *\\
237287886 $^b$ & 1.6832976 (5)& $-$ &1884 (81)& 197 (7)  & 0.70 (1)& 3.9 (2)& 57,887 (57)& 0.029 (4)  & 0.58 & 0.04 & 2.0: \\
237308045     &0.34650003 (2)&$-8.45$ (7)&919 (1)&114 (2)& 0.14 (1)& 160 (7)& 58,450 (19)& 0.024 (1)  & 0.47 &0.0007&1.60 *\\
259168350     & 1.4770131 (3)& $-$ &1322 (14)& 243 (4)  & 0.34 (4)& 182 (7)& 58,814 (28)& 0.109 (6)  & 0.99 & 0.02 & 2: \\
259264759     &0.52044443 (2)& $-$ & 1134 (2)& 152.3 (8)& 0.25 (1)& 288 (2)& 58,648 (7) & 0.0368 (6) & 0.65 & 0.002&2.09 *\\
284800646     &0.263959800 (5)&$-$ & 1403 (6)& 45.3 (4) & 0.29 (1)& 335 (3)& 57,995 (13)& 0.00063 (2)& 0.11 &0.0006&1.34 *\\
288515928     &0.29021808 (7)& $-$ &1624 (24)& 200 (3)  &0.302 (8)& 332 (2)& 59,449 (20)& 0.041 (2)  & 0.54 &0.0005&1.43 *\\
353990339     &0.268682550 (9)&$-$ & 1493 (7)& 100 (1)  & 0.28 (1)& 355 (2)& 59,380 (12)& 0.0060 (2) & 0.25 &0.0005&1.36 *\\
377054541     &0.32556131 (2)& $-$ & 1770 (4)& 213.3 (8)&0.455 (5)&65.5 (5)& 58,880 (3) & 0.0415 (5) & 0.57 &0.0007&1.54 *\\    
377251803 $^c$ &0.571369255 (8)&$-$ & 1437 (5)& 61.9 (4) & 0.17 (1)& 300 (5)& 58,280 (20)& 0.00154 (3)& 0.28 & 0.002& 3.5: \\
389966320     & 0.3039899 (1)& $-$ & 1597 (5)& 196 (2)  & 0.39 (1)& 335 (2)& 59,175 (12)&  0.039 (1) & 0.54 &0.0007&1.47 *\\ 
424461577 $^d$ &0.744836056 (9)&$-$ &1013.7 (6)&205.6 (8)&0.480 (4)&121.3 (7)& 58,284 (2)& 0.113 (1)  & 1.22 & 0.007& 2.77 \\
441803530     &0.275245056 (5)&$-$ & 1343 (9)& 159 (26) & 0.97 (1)& 2.1 (6)& 58,038 (9) &  0.030 (15)& 0.47 & 0.13 &1.38 *\\
\bottomrule
\end{tabularx}
\end{adjustwidth}
\noindent{\footnotesize{\textbf{Notes}. {$^a$: SWASP data used; $^b$: V589 Dra; $^c$: DG Dra; $^d$: V527 Dra---$m_\mathrm{AB}$ was taken from \citep{cekietal24}}}.}

\end{table}

\begin{table}[H]
\tablesize{\fontsize{6.8}{6.8}\selectfont} 
\caption{Triple system candidates with very uncertain, questionable LTTE solutions.} 
\label{Tab:Orbelem_LTTE3}  
\begin{adjustwidth}{-\extralength}{0cm}
\centering
\begin{tabularx}{\fulllength}{lccccccccccc}
\toprule
\textbf{TIC No.} & \boldmath{$P_1$} & \boldmath{$\Delta P_1$} & \boldmath{$P_2$} & \boldmath{$a_\mathrm{AB}\sin i_2$} & \boldmath{$e_2$} & \boldmath{$\omega_2$} & \boldmath{$\tau_2$} & \boldmath{$f(m_\mathrm{C})$} & \boldmath{$(m_\mathrm{C})_\mathrm{min}$} & \boldmath{$\frac{{\cal{A}}_\mathrm{dyn}}{{\cal{A}}_\mathrm{LTTE}}$} & \boldmath{$m_\mathrm{AB}$}\\
        & \textbf{(day)} &\boldmath{$\times10^{-10}$} \textbf{(d/c)}&\textbf{(day)}&\textbf{(R}\boldmath{$_\odot$}\textbf{)}  &       &   \textbf{(deg)}    &   \textbf{(MBJD)} & \textbf{(M}\boldmath{$_\odot$}\textbf{)}       & \textbf{(M}\boldmath{$_\odot$}\textbf{)}            & &  \textbf{(M}\boldmath{$_\odot$}\textbf{)}    \\
\midrule
159509734     &0.26237697 (1)&1.66 (3)&616 (2)& 17.0 (4) & 0.41 (4)&202 (9) & 58,860 (16)&0.00017 (1) & 0.07 & 0.003& 1.34 *\\
160418633     &0.24114966 (1)&$-0.84 (4)$&346 (1)&18 (2) &0.64 (10)& 49 (7) & 58,670 (9) & 0.0006 (2) & 0.11 & 0.02 & 1.27 * \\
160492808     &0.315173143 (9)&0.52 (3)&851 (12)&4.8 (5) & 0.6 (1) &162 (11)& 58,630 (33)&0.000021 (6)& 0.02 & 0.007& 1.50 * \\
160520144     & 2.179673 (2) & $-$ &1983 (184)& 188 (15)& 0.34 (7)& 86 (19)&58,084 (151)& 0.023 (7)  & 0.44 & 0.02 & 1.5:\\
165527014     &0.47028357 (9)& $-$ &2738 (208)& 53 (5)  & 0.5 (1) &322 (5) & 58,657 (50)&0.00026 (9) & 0.10 & 0.001& 1.95 *\\ 
165529344     &0.280176824 (4)& $-$& 1635 (8)& 44.6 (3) & 0.11 (3)& 34 (11)& 58,663 (50)&0.000444 (9)& 0.10 &0.0004& 1.39 *\\
198158617     &0.283139944 (9)&2.45 (3)&600 (2)&19.3 (4) & 0.70 (3)& 76 (3) & 59,004 (7) & 0.00027 (2)& 0.08 & 0.01 & 1.40 *\\
198206417 $^a$ & 0.8912545 (1)& $-$ &1822 (24)& 53 (2)   & 0.56 (2)& 132 (2)& 58,707 (22)& 0.00062 (9)& 0.14 & 0.006& 2.0: \\ 
198280146     &0.26745213 (1)& $-$ &1757 (7) & 184 (11) & 0.93 (1)& 38 (2) & 58,986 (17)& 0.027 (5)  & 0.44 & 0.01 & 1.35 *\\
198355086     & 0.33370997 (1)&$-$ &1287 (4) & 243 (5)  & 0.85 (1)& 242 (2)& 58,736 (9) & 0.116 (7)  & 0.89 & 0.006& 1.56 *\\
198382300     &   2.08106 (4)& $-$ &2370 (1110)&454 (370)&0.06 (6)&111 (154)&59,206 (1131)&0.22 (59) & 1.36 & 0.01 & 2.0: \\
198419050 $^b$ & 0.3857223 (1)&6.4 (4)&1962 (110)& 91 (7) & 0.20 (2)& 53 (9) & 58,160 (87)& 0.0026 (7) & 0.21 & 0.07 & 1.71 *\\
198457457 $^c$ &  0.403472 (1)& $-$ &5485 (9278)&203 (392)&0.25 (24)&1 (139)&58,564 (2405)&0.004 (25) & 0.25 &0.0001& 1.77 *\\
207466859     &0.35951260 (5)& $-$ & 2282 (8)& 613 (6)  & 0.01 (1)&71 (137)&59,075 (868)& 0.59 (2)   & 1.98 &0.0003& 1.64 *\\
219091665 $^d$ &0.45703632 (1)& $-$ &2326 (94)& 146 (6)  & 0.86 (2)& 45 (2) & 57,966 (73)& 0.0077 (11)& 0.34 & 0.005&  1.92 *\\
219101330     & 4.193170 (9) & $-$ &2360 (543)&115 (40) & 0.55 (4)&244 (5) &58,165 (432)& 0.0037 (42)& 0.27 & 0.08 & 2.0: \\
219738202 $^e$& 0.5790375 (2)&8.3 (1)&10,722 (419)&170 (8)& 0.64 (3)&267 (4) &61,856 (399)& 0.00057 (9)& 0.17 &0.0003& 2.68 \\
219761337 $^f$ &0.317471154 (4)&$-$ &1938 (22)& 75 (3)   & 0.69 (2)& 18 (1) & 59,044 (11)& 0.0015 (2) & 0.16 & 0.001& 1.51 *\\
219787718 $^g$ &0.35258560 (2)& $-$ &2522 (18)& 422 (2)  &0.321 (4)&294 (1) & 57,462 (14)& 0.158 (3)  & 1.04 &0.0003& 1.62 *\\
219788156     & 7.82452 (5)  & $-$ &2853 (3731)&111 (146)&0.51 (35)&342 (19)&59,333 (1279)&0.002 (11)& 0.22 & 0.20 & 2: \\
219809405     &0.30493647 (1)&1.25 (4)&1070 (6)& 19 (1)  & 0.85 (6)&106 (6) & 58,804 (21)&0.00008 (2) & 0.06 & 0.007& 1.47 *\\
219856620     &0.37941344 (4)&6.9 (2)&1107 (9)& 57.0 (9) & 0.28 (4)&291 (6) & 58,780 (21)& 0.0020 (1) & 0.19 & 0.002& 1.70 *\\
219859424 $^h$ &0.34722279 (4)&$-2.94$ (6)&1856 (219)&36 (4)&0.75 (3)&346 (1)&59,449 (130)& 0.00018 (7)& 0.08 & 0.003& 1.60 *\\
219876159     & 0.281605 (2) & $-$ &4848 (6275)&232 (428)& 0.5 (1)&243 (52)&58,769 (1096)& 0.007 (44)& 0.27 &0.0001& 1.40 *\\
219900027 $^i$ &0.515263400 (5)&$-$ &1631 (12)& 58.8 (4) & 0.49 (1)& 66 (2) & 57,962 (11)& 0.00102 (3)& 0.17 &0.002 & 2.07 *\\
229437073     &0.3378895 (2) & $-$ &2514 (100)&228 (13) & 0.27 (1)& 76 (5) & 59,055 (53)& 0.025 (5)  & 0.47 &0.0003& 1.57 *\\
229507536     &1.0926298 (1) &$-110$ (1)&892 (3)& 73 (4) & 0.78 (3)&334 (2) & 58,888 (7) & 0.0066 (12)& 0.33 & 0.08 & 2: \\
229592690     & 0.2749805 (6)& $-$ &3903 (429)&516 (77) & 0.41 (3)& 42 (2) &60,378 (253)& 0.12 (6)   & 0.84 & 0.001& 1.38 *\\
229605410     & 3.1935553 (8)& $-$ &1835 (165)&281 (35) & 0.23 (6)&266 (18)&59,124 (125)& 0.088 (37) & 0.91 & 0.04 & 2: \\
229713298     & 3.4398317 (6)& $-$ &1382 (21)& 152 (20) & 0.81 (7)&321 (7) & 58,810 (37)& 0.025 (10) & 0.54 & 0.45 & 2: \\
229751503     & 7.95120 (2)  & $-$ &1995 (142)& 309 (34)& 0.11 (6)&319 (24)&58,366 (152)& 0.10 (4)   & 0.95 & 0.20 & 2: \\
229773372     &0.26384197 (2)& $-$ &1995 (21)& 187 (2)  & 0.15 (1)&184 (6) & 59,282 (33)& 0.0220 (7) & 0.41 &0.0003&  1.34 *\\
229800815     &0.41679375 (2)& $-$ &2006 (62)& 43 (1)   & 0.47 (2)& 118 (4)& 58,793 (31)& 0.00026 (3)& 0.10 &0.0009& 1.80 *\\
229952080     &0.38538242 (8)& $-$ &3000 (333)& 391 (3) &0.370 (8)& 165 (2)& 59,059 (46)& 0.089 (20) & 0.83 &0.0003& 1.71 *\\
230003294     &0.416511705 (8)&$-$ &1934 (13)& 46.1 (6) & 0.72 (2)& 119 (2)& 58,733 (11)& 0.00035 (1)& 0.11 & 0.002& 1.80 *\\
230070359     &0.96460609 (4)&20.8 (4)&499 (2)& 16 (2)   & 0.80 (6)& 31 (5) & 58,455 (8) & 0.00022 (8)& 0.10 & 0.51 & 2: \\
230083136     &0.430564929 (8)&$-$ & 1931 (7)&68.1 (6)  & 0.79 (1)&298.8 (9)&58,729 (7) & 0.00113 (3)& 0.17 & 0.003& 1.84 *\\
233008057     &0.40050172 (1)& $-$ &2764 (22)& 403 (2)  &0.771 (3)&285.0 (3)&59,715 (12)& 0.115 (2)  & 0.94 & 0.001& 1.76 *\\
233496995     &0.377089513 (6)&$-$ & 1840 (8)& 114 (1)  &0.648 (8)&191.0 (9)& 58,999 (6)& 0.0058 (2) & 0.28 & 0.002& 1.69 *\\
233532554 $^j$ & 2.831268 (5) & $-$ &4749 (1990)&205 (82)& 0.5 (1) & 328 (6)&59,514 (542)& 0.005 (7)  & 0.33 & 0.01 & 2.29 \\
233680160     &0.247398972 (6)&$-$ & 1531 (2)& 645 (41) &0.987 (2)& 159 (1) &59,256 (8) & 1.5 (3)    & 3.08 & 0.17 & 1.29 *\\
233680757     & 0.2792922 (1)& $-$ &3347 (222)&228 (18) & 0.14 (1)&242 (15)&59,096 (162)& 0.014 (4)  & 0.35 &0.0001& 1.39 *\\
233689972 $^k$ & 1.38073 (3)  & $-$ &3159 (2205)&319 (720)&0.6 (8) &300 (69)&57,936 (1558)& 0.04 (30) & 1.04 & 0.005& 4.02 *\\
233742729     &0.55573999 (8)& $-$ &4979 (362)& 374 (17)& 0.86 (2)& 92 (2) & 59,440 (90)& 0.028 (6)  & 0.60 & 0.002& 2.18 *\\
256324189     &0.265594242 (9)&$-$ &2201 (22)& 156.8 (9)&0.052 (5)& 4 (15) & 59,102 (91)& 0.0107 (3) & 0.31 &0.0002& 1.35 *\\ 
258776281     & 0.5217897 (5)& $-$ &2308 (113)& 124 (39)& 0.58 (5)& 194 (2)& 58,641 (34)& 0.0048 (46)& 0.30 & 0.002& 2.09 *\\
258875507     &0.608811436 (6)&$-$ & 506 (1) & 16.1 (7) & 0.99 (3)& 266 (2)& 58,910 (5) & 0.00022 (3)& 0.09 & 9.94 & 1.8: \\
258920306     & 3.88223 (1)  & $-$ &2627 (217)& 419 (63)& 0.03 (3)&195 (23)&58,300 (186)&  0.14 (7)  & 1.12 & 0.03 & 2: \\  
275692690     & 1.4296891 (3)& $-$ &2272 (49)&  347 (4) & 0.39 (1)& 226 (4)& 59,165 (33)&  0.108 (6) & 0.99 & 0.007& 2: \\
288298154 $^l$ & 1.8761884 (2)& $-$ &2187 (49)&  124 (2) & 0.41 (1)& 284 (3)& 59,396 (35)&  0.0054 (4)& 0.31 & 0.01 & 2: \\
288407480     & 4.08242 (2)  & $-$ &4429 (1129)&534 (158)&0.71 (5)& 320 (8)&58,623 (166)&  0.10 (11) & 0.97 & 0.05 & 2: \\
288734990 $^m$ &0.383928467 (2)&$-$ & 1938 (7)&  53 (3)  & 0.34 (3)& 34 (3) & 54,785 (7) & 0.00052 (8)& 0.11 &0.0008& 1.40\\
              &             &     &14,461 (69)& 227 (8) & 0.81 (2)& 43 (2) & 59,111 (97)& 0.00075 (8)& 0.20 &      & 3: \\
298600301     & 2.5414138 (8)& $-$ &2577 (265)& 143 (12)& 0.57 (6)& 21 (4) & 59,313 (94)&  0.006 (2) & 0.32 & 0.03 & 2: \\  
334754418     & 0.7644722 (8)& $-$ &1946 (104)& 318 (14)& 0.35 (3)& 68 (11)& 58,025 (86)&  0.11 (2)  & 1.00 & 0.002& 2: \\
353894340     & 2.039645 (6) & $-$ &1746 (1230)&161 (85)& 0.9 (1) & 10 (7) &58,207 (944)&  0.02 (4)  & 0.48 & 0.34 & 2: \\
357034252     &0.52131110 (8)&8.1 (7)&1532 (38)&  47 (2) & 0.32 (3)& 325 (4)& 58,131 (44)& 0.00059 (9)& 0.14 & 0.002& 2: \\
359629786     &0.39148668 (2)& $-$ & 1511 (12)& 77 (3)  & 0.81 (3)& 303 (3)& 59,162 (14)& 0.0026 (3) & 0.24 & 0.005& 2: \\
362227092     &0.283993153 (7)&$-$ & 1737 (14)& 68 (2)  & 0.48 (4)& 358 (3)& 59,052 (16)& 0.0014 (1) & 0.15 &0.0007& 1.41 *\\
362259979     &0.29305573 (1)&$-0.72$ (2)&718 (4)&16.1 (6)&0.53 (5)&  61 (5)& 58,699 (13)& 0.00011 (1)& 0.06 & 0.005& 1.44 *\\ 
367853009     & 0.4108690 (2)& $-$ &2773 (123)& 335 (15)& 0.24 (1)&  19 (8)& 58,914 (74)& 0.066 (10) & 0.75 &0.0004& 1.79 *\\
377105433 $^n$ & 0.2514049 (1)&$-8.4$ (7)&753 (3)& 108 (2)& 0.13 (2)& 247 (3)& 58,378 (8) &  0.030 (1) & 0.45 &0.0005& 1.30 *\\
377307592 $^o$ & 1.1928338 (1)& $-$ &7427 (184)& 446 (9) &0.758 (6)&178.5 (4)&58,931 (14)&  0.022 (2) & 0.52 & 0.003& 2: \\
389968719     &0.33178789 (2)&6.80 (5)&556 (1)& 28.2 (3) & 0.17 (2)& 286 (1)& 58,858 (13)& 0.00097 (4)& 0.14 & 0.002& 1.55 *\\ 
458479530     &0.85990211 (2)& $-$ &2117 (80)& 129.2 (9)&0.122 (6)& 139 (4)& 58,366 (26)& 0.0065 (5) & 0.33 & 0.002& 2: \\
\bottomrule
\end{tabularx}
\end{adjustwidth}
\noindent{\footnotesize{{\bf Notes.} {$^a$: V496 Dra; $^b$: V562 Dra; $^c$ V374 Dra; $^d$: V388 Dra; $^e$: BX Dra, mass and former times of minima taken from \citep{parketal13}; $^f$: V516 Dra; $^g$: V377 Dra; $^h$: V362 Dra; $^i$: AU Dra; $^j$: RR Dra. Binary mass taken from \citet{senavcietal22};  $^k$: V402 Dra; $^l$: AH UMi; $^m$: Mass and former times of minima were taken from \citep{wolfetal16,zhangetal18}; $^n$: Cubic coefficient: $c_3=-04(27)\times10^{-15}\,\mathrm{c/d}^2$; $^o$: KK Dra. SWASP and other ground-based minima were used.}}}

\end{table}

Thus, we say that a triple is `tight' when the $P_2/P_1$ ratio is small enough to produce observable perturbations on a timescale of $P_2$. According to previous experience, the upper limit on the `tightness' in most cases is in the vicinity of $50\lesssim P_2/P_1\lesssim100$. The current limit for a specific $P_2/P_1$ ratio also depends on several different parameters, mainly on the eccentricities of the inner and outer orbits, their mutual inclination angle, and, moreover, on the masses (or, more strictly speaking, the mass ratios) of the constituent bodies. Restricting ourselves to the lowest order of these $P_2$-timescale third body perturbations, it was shown in \citet{borkovitsetal16} that the amplitude of these dynamical effects (DEs) on the ETV curve can be estimated as
\begin{equation}
\mathcal{A}_\mathrm{DE}=\frac{1}{2\pi}\frac{m_\mathrm{C}}{m_\mathrm{ABC}}\frac{P_1^2}{P_2}\left(1-e_2^2\right)^{-3/2}.
\label{Eq:A_dyn}
\end{equation}
It was shown, however, in \citet{rappaportetal13} that for nearly coplanar inner and outer orbits (that is, when the sine of the mutual inclination of the orbits is close to zero), a more useful estimation can be obtained with the use of the following expression:
\begin{equation}
\mathcal{A}^{\mathrm{coplanar}}_\mathrm{DE}=\frac{3}{2\pi}\frac{m_\mathrm{C}}{m_\mathrm{ABC}}\frac{P_1^2}{P_2}\left(1-e_2^2\right)^{-3/2}e_2.
\label{Eq:A_dyn^cop}
\end{equation}
This latter expression nicely illustrates that in the case of two circular orbits and a flat configuration (that is, when $\sin i_\mathrm{mut}=0$), the $P_2$ timescale dynamical perturbations disappear, at least to the lowest order, i.e., the so-called quadrupole-order perturbations.

\begin{table}[H]
\tablesize{\fontsize{7.0}{7.0}\selectfont} 
\caption{Orbital elements from combined dynamical and LTTE solutions with outer period shorter than half of the length of the datasets.} 
\label{Tab:Orbelemdyn1}  
\begin{adjustwidth}{-\extralength}{0cm}
\centering
\begin{tabularx}{\fulllength}{lccccccccccc}
\toprule
\textbf{TIC No.} & \boldmath{$P_1$} & \boldmath{$P_2$} & \boldmath{$a_2$} & \boldmath{$e_2$} & \boldmath{$\omega_2$} & \boldmath{$\tau_2$} & \boldmath{$f(m_\mathrm{C})$} & \boldmath{$\frac{m_\mathrm{C}}{m_\mathrm{ABC}}$} & \boldmath{$m_\mathrm{AB}$} & \boldmath{$m_\mathrm{C}$} & \boldmath{$\frac{{\cal{A}}^\mathrm{meas}_\mathrm{dyn}}{{\cal{A}}_\mathrm{LTTE}}$}\\
        & \textbf{(day)} & \textbf{(day)} &\textbf{(R}\boldmath{$_\odot$}\textbf{)}  &       &   \textbf{(deg)}    &   \textbf{(MBJD)} & \textbf{(M}\boldmath{$_\odot$}\textbf{)}       & \textbf{(M}\boldmath{$_\odot$}\textbf{)}            &  \textbf{(M}\boldmath{$_\odot$}\textbf{)}  &   \\
\midrule
198241524 &0.775175037 (3)&68.36 (1) & 105 (1)  & 0.14 (3) &  80 (6) & 58,687 (1) & 0.075 (8)  & 0.29 (1) & 2.39 (10)   & 0.96 (5)  & 0.22\\
          &             &2888 (104) &114 (19) $^a$& 0.54 (4) &  41 (6) & 57,575 (98)& 0.002 (1)  & $-$   & 0.30 $^a$   & 3:
 $^a$   & \\
199616648 &14.074091 (2) & 401.0 (6) & 311 (13) & 0.143 (8)&  89 (4) & 58,572 (6) & 0.11 (2)   & 0.36 (2) & 1.62 (28)   & 0.90 (17) & 7.74\\
219771659 &14.494781 (1) & 728.8 (4) & 495 (10)  &0.000 (2)&  60 (3) & 58,400 (7) & 0.10 (1)   & 0.323 (7)& 2.07 (16)   & 0.99 (8)  & 3.49\\
219885468 & 7.539375 (4) & 110.8 (1) & 131 (5)  & 0.47 (6) &  88 (11)& 58,816 (1) & 0.011 (6)  & 0.17 (4) & 2.04 (27)   & 0.41 (12) & 55 \\ 
219892913 & 3.5444122 (2)& 670.5 (8) & 581 (22) &0.773 (7) & 207 (3) & 58,926 (8) & 0.25 (5)   & 0.35 (1) & 3.81 (58)   & 2.05 (34) & 2.38 \\
229594479 &1.93753692 (9)& 525 (2)   & 358 (28) & 0.29 (7) & 133 (13)& 58,487 (21)& 0.019 (6)  & 0.20 (1) & 1.78 (51)   & 0.45 (14) & 0.12 \\
229785001 $^b$&0.92951067 (1)&165.0 (4)& 204 (4) & 0.51 (10)& 30 (10) & 58,733 (8) & 0.33 (8)   & 0.43 (4) & 2.37       & 1.79 (27) & 0.40\\
          &             & 3917 (874)&307 (120) $^a$&0.2 (2)&212 (124)&58,620 (1369)& 0.025 (27)& $-$     & 0.70 $^a$   & 3: $^a$   & \\
233073872 &14.139792 (4) & 704 (1)   & 571 (16) & 0.51 (1) & 66 (3)  & 59,030 (2) & 0.18 (3)   & 0.33 (1) & 3.38 (37)   & 1.67 (22) & 7.57 \\
233496533 &2.60647771 (7)& 109.63 (7)& 173 (7)  & 0.06 (2) & 229 (20)& 58,679 (6) & 0.08 (3)   & 0.24 (4) & 4.41 (61)   & 1.40 (36) & 0.37 \\
233530543 & 1.6215399 (4)& 258.8 (6) & 269 (10) & 0.39 (9) & 75 (9)  &58,766.9 (1)& 0.19 (6)   & 0.37 (4) & 2.46 (30)   & 1.44 (30) & 0.34 \\
233684019 & 7.2814934 (5)& 145.03 (5)& 143 (7)  & 0.25 (2) & 92 (3)  &58,742.2 (5)& 0.010 (3)  & 0.18 (1) & 1.55 (26)   & 0.33 (6)  & 11.92 \\
233729038 &11.402384 (1) & 453.0 (3) & 383 (32) & 0.16 (2) & 141 (3) & 58,531 (4) & 0.25 (10)  & 0.41 (3) & 2.15 (74)   & 1.52 (55) & 2.21 \\
233738966 & 9.2705778 (2)& 379.9 (3) & 342 (8)  & 0.100 (3)&191.7 (9)&58,754.7 (2)& 0.074 (7)  & 0.278 (3)& 2.69 (23)   & 1.04 (9)  & 2.82 \\
235687051 & 3.4327179 (2)& 116.1 (2) & 171 (13) & 0.20 (6) & 125 (19)& 58,715 (7) & 0.0015 (16)&0.067 (31)& 4.68 (115)  & 0.34 (18) & 2.27 \\ 
235934882 & 6.8382508 (5)& 127.04 (2)& 160.1 (8)& 0.331 (7)& 5.2 (8) &58,631.0 (4)& 0.053 (5)  & 0.25 (1) & 2.56 (2)    & 0.85 (5)  & 17.7 \\
236769201 & 6.3754524 (4)& 165.58 (2)& 202 (2)  & 0.461 (8)& 146 (2) & 58,718 (1) & 0.062 (4)  & 0.249 (6)& 3.05 (9)    & 1.01 (4)  & 15.8 \\
237234024 &1.89581986 (3)& 172.4 (1) & 225 (11) & 0.21 (2) & 180 (5) & 58,654 (2) & 0.077 (29) & 0.25 (4) & 3.88 (70)   & 1.27 (17) & 0.38 \\
256514937 & 5.3289063 (1)& 277.97 (5)& 287 (5)  & 0.660 (3)& 195 (1) &58,832.8 (3)&  0.20 (1)  & 0.376 (3)& 2.56 (18)   & 1.54 (11) & 11.4 \\ 
259006185 &1.939651537 (9)& 200 (1)  & 241 (9)  & 0.12 (3) & 316 (17)& 58,865 (17)& 0.0014 (4) & 0.067 (8)& 4.38 (53)   & 0.31 (5)  & 0.17 \\
272705788 & 7.2566243 (1)&  796 (4)  & 504 (6)  & 0.293 (5)& 168 (1) & 58,845 (7) & 0.048 (2)  & 0.269 (3)& 1.98 (8)    & 0.73 (3)  & 0.73 \\
288611883 & 1.0104128 (2)&  507 (5)  & 332 (15) & 0.78 (7) & 67 (7)  & 58,863 (14)& 0.17 (5)   & 0.45 (5) & 1.06 (16)   & 0.85 (20) & 0.36 \\ 
356014478 & 9.531374 (4) & 328.6 (8) & 458 (38) & 0.19 (2) & 177 (12)& 58,659 (14)& 6.10 (1.91)& 0.80 (4) & 2.41 (49)   & 9.58 (2.87)& 2.17\\
356896561 $^c$&2.253360555 (5)&475.3 (9)&369 (20)& 0.52 (2) & 308 (6) & 58,805 (4) & 0.0007 (3) & 0.09 (2) & 2.73 (49)   & 0.25 (3)  & 0.34 \\
357686232 & 7.3089428 (1)& 220.9 (2) & 204 (3)  & 0.061 (3)& 318 (4) & 58,610 (3) & 0.0010 (1) & 0.076 (4)& 2.15 (10)   & 0.18 (1)  & 3.16 \\
389966039 & 6.832011 (5) & 111.3 (1) & 125 (9)  & 0.35 (3) & 39 (17) & 58,701 (1) &  0.014 (5) & 0.19 (2) & 1.72 (43)   & 0.40 (12) & 32.0 \\
402656423 & 1.5192999 (1)& 135.1 (3) & 212 (6)  & 0.46 (6) & 210 (10)& 58,854 (4) &  0.19 (5)  & 0.31 (3) & 4.88 (42)   & 2.14 (38) & 1.03 \\
417701432 & 9.366174 (6) & 446.8 (9) & 457 (23) & 0.46 (2) & 340 (3) & 58,910 (3) &  0.16 (7)  & 0.30 (5) & 4.51 (80)   & 1.93 (56) & 6.78 \\
420177014 &3.25858077 (1)& 845 (1)   & 471 (26) & 0.426 (7)&155.9 (6)& 59,130 (2) & 0.09 (2)   & 0.36 (1) & 1.26 (28)   & 0.71 (16) & 0.30 \\
423927178 $^d$&4.79982 (1)& 96.6 (2)  & 164 (12) & 0.04 (2) & 154 (7) & 58,716 (3) & 1.42 (59)  & 0.61 (7) & 2.46 (43)   &3.84 (1.32)& 1.41 \\
441768471 &27.295894 (3) & 915.4 (5) & 594 (5)  & 0.266 (3)&267.5 (7)& 58,806 (3) & 0.147 (8)  & 0.357 (6)& 2.16 (7)    & 1.20 (5)  & 8.40 \\
459969365 & 9.006620 (1) & 473.7 (6) & 411 (9)  & 0.53 (1) & 324 (5) & 58,676 (9) &  0.19 (4)  & 0.36 (3) & 2.67 (16)   & 1.49 (21) & 7.36 \\ 
\bottomrule
\end{tabularx}
\end{adjustwidth}
\noindent{\footnotesize{{\textbf{Notes.}} {$^a$: The fourth body parameters obtained with pure LTTE solution, hence, the corresponding parameters are $a_\mathrm{ABC}\sin i_3$ in the column $a_2$ and $\left(m_\mathrm{D}\right)_\mathrm{min}$ in the column of $m_\mathrm{B}$ calculated for that total mass of the inner triple subsystem, that is, given in the column $m_\mathrm{C}$; $^b$: Masses of the innermost pair taken from \citep{rappaportetal23}; $^c$: GZ Dra---WASP data used; $^d$: quadratic ephemeris: $\Delta P=10(1)\times10^{-7}$\,d/c$^2$. }}}

\end{table}

Similar to our previous studies of {\textit Kepler}- and TESS-observed ETVs, in contrast to the LTTE term, which was applied to the analysis of all the available ETVs, we switched on the DE terms (of which a more detailed mathematical description can be found in \citet{borkovitsetal15}\endnote{Further extracts can be found, however, in \citet{mitnyanetal24}.}) only when there were clear indicators that the $\Delta_\mathrm{DE}$ contribution to a given ETV curve is large enough to necessitate these terms for a correct analysis. The selection criteria were as follows: (i) For the tightest systems, the significance of the DE terms can often be easily seen by a casual inspection of their ETVs, which have their own characteristic shapes.  In contrast to the LTTE terms, DE-dominated ETVs are not necessarily pure sinusoids and/or, in the case of eccentric systems, the amplitude (and shape) of the ETV curve calculated for the primary eclipses may differ substantially from the ETV curve for the secondary eclipses. (ii) Moreover, we calculated and checked continuously the theoretical ratio of $\mathcal{A}_\mathrm{DE}/\mathcal{A}_\mathrm{LTTE}$ or, for triples having an outer period of $P_2\lesssim1000$\,days\endnote{This is the (somewhat arbitrary) limit for the so-called `compact' triples, which, according to \citet{tokovinin21}, were formed by disk fragmentation and, therefore, are expected to be (nearly) flat.}, its coplanar counterpart of $\mathcal{A}^{\mathrm{coplanar}}_\mathrm{DE}/\mathcal{A}_\mathrm{LTTE}$.  When we found that this ratio exceeded a value of 0.2 (that is, the expected DE contribution was at least 20\% of the LTTE contribution), we repeated our analysis and switched on the DE terms.  We did not, however, apply this latter criterion to some EBs with very uncertain LTTE solutions (categorized into our last, most uncertain subgroup) because, in these cases, a large ratio was generally caused by an unrealistically large outer eccentricity (see below, in Section~\ref{sec:discussion}, for a more detailed discussion).

\begin{table}[H]
\small
\caption{Apsidal motion and/or orientation parameters from AME and dynamical fits.} 
\label{Tab:AMEparam}
\begin{adjustwidth}{-\extralength}{0cm}
\centering
	\begin{tabularx}{\fulllength}{lccccccccccc}
\toprule
\textbf{TIC No.} & \boldmath{$P_\mathrm{anom}$} & \boldmath{$a_1$}      & \boldmath{$e_1$} & \boldmath{$\omega_1$} & \boldmath{$\tau_1$} & \boldmath{$P_\mathrm{apse}$} & \boldmath{$i_\mathrm{m}$} & \boldmath{$i_1$} & \boldmath{$i_2$} & \boldmath{$\Delta\Omega$} & \boldmath{$P_\mathrm{node}$}\\
        & \textbf{(days)}            &\textbf{(R}$_\odot$\textbf{)} &       & \textbf{(deg)}      & \textbf{(MJD)}    &   \textbf{(years)}         & \textbf{(deg)} & \textbf{(deg)} & \textbf{(deg)} &   \textbf{(deg)  } &  \textbf{(years)} \\
\midrule
198241524 $^a$&0.77515141 (3)&4.74 (7)& 0        & $-$     & $-$        & $-$         & 0      & 80 & 80 &   0     & $-64$\\
199616648    &14.078273 (2) & 29 (2) & 0.089 (4) &106.5 (8) & 58,707.87 (4)& 125         & 11 (4)  & 89 & 94 &   9 (3)  & $-85$\\
219771659    &14.494829 (1) &31.9 (8)&0.2719 (4) &   3 (3)  & 58,675.81 (1)& 344         & 21 (3)  & 89 & 87 &  21 (3)  & $-450$\\
219885468    & 7.5508 (3)   &20.5 (9)& 0.0413 (8)& 263 (1)  & 58,686.70 (3)& 13.7 (3)     & 0      & 88 & 88 &   0     & $-15$\\
219892913    & 3.5445610 (2)&15.3 (8)& 0.013 (2) & 179 (12) & 58,679.1 (1) & 331         & 8 (4)   & 88 &87.5&  $-$8 (4)& $-299$\\
229594479    &1.93753081 (9)& 7.9 (8)& 0        & $-$     & $-$        & $-$         & 0      & 87 & 87 &   0     & $-1915$ \\
229785001 $^b$&0.92949085 (1)& 5.34  & 0        & $-$     & $-$        & $-$         & 0      &87.7&87.7&   0     & $-149$\\
230002837    &4.14571430 (8)& 12 (2) & 0        & $-$     & $-$        & $-$         & 0      & 87 & 87 &   0     & $-3521$\\
230012179    & 8.9235304 (2)& 20 (3) & 0.014 (3) & 92.5 (5) & 58,670.53 (1)& 7151        & 0      & 89 & 89 &   0     & $-5804$ \\
233073872    &14.142398 (4) & 37 (1) &0.0386 (7) & 140 (2)  & 58,665.54 (6)& 219         & 0      & 89 & 89 &   0     & $-209$ \\ 
233496533    &2.60612170 (7)&13.1 (6)& 0        & $-$     & $-$        & $-$         & 0      & 89 & 89 &   0     & $-56$ \\
233530543    &1.6215399 (2) & 7.8 (3)& 0        & $-$     & $-$        & $-$         & 0      & 86 & 86 &   0     & $-291$ \\ 
233684019 $^c$&7.2839197 (5) & 18 (1) & 0.0064 (4)& 83.5 (6) & 58,680.56 (1)& 60          & 12 (1)  & 97 & 89 &   9 (1)  & $-38$ \\
233729038 $^d$&11.404906 (1) & 28 (3) &0.0127 (6) &  71 (1)  &58,706.60 (3) & 141         & 17 (1)  & 89 & 82 & $-15$ (1)& $-143$ \\
233738966    & 9.2717245 (2)&25.8 (7)& 0.123 (1) & 301.9 (3)&58,677.784 (7)& 200         & 14 (2)  & 89 &103 &$-3.3$ (4)& $-175$ \\
235687051    & 3.432503 (2) & 16 (1) & 0        & $-$     & $-$        & $-$         & 0      & 89 & 89 &   0     & $-95$ \\
235934882    & 6.8382380 (5)&20.7 (6)&0.00248 (4)& 343 (2)  & 58,680.78 (4)& 24          & 0      & 89 & 89 &   0     & $-22$ \\
236769201    & 6.3755050 (4)&21.0 (2)& 0.0089 (5)& 9 (3)    & 58,679.68 (6)& 38          & 0      & 87 & 87 &   0     & $-34$ \\
237234024    &1.89575417 (6)&9.0 (6) & 0        & $-$     & $-$        & $-$         & 0      & 86 & 86 &   0     & $-171$ \\
243337122    &10.9372960 (7)& 26 (1) & 0.14 (1)  & 234 (4)  & 58,676.7 (1) & 1446        & 21 (5)  & 89 &107 &  10 (5)  & $-1286$ \\
256514937    & 5.3298892 (1)&17.6 (4)& 0.0187 (3)& 39.2 (9) & 58,682.13 (1)& 77          & 19.6 (8)& 89 & 104& $-13$ (1)& $-54$ \\
259004910 $^d$&12.5852604 (2)&26.1 (2)&0.14772 (7)&357.1 (1) &58,678.345 (3)& $-20$       &49.83 (9)&87.5&87.0&$-49.88$ (9)&$-20$ \\
259006185    &1.93963905 (9)&10.7 (4)& 0        & $-$     & $-$        & $-$         & 0      & 87 & 87 &   0     & $-623$ \\
259271740    &11.7518598 (7)& 37 (1) & 0.28 (2)  & 222 (3)  & 58,673.87 (7)& 2035        & 32 (3)  & 89 &102 &  29 (3)  & $-2265$ \\
272679385    & 17.839962 (3)&37.5 (7)& 0.038 (2) & 62 (1)   & 58,683.00 (6)& 4296        & 0      & 87 & 87 &   0     & $-2885$\\  
272705788    & 7.2564859 (1)&19.8 (3)& 0.170 (1) & 160 (1)  & 58,678.29 (3)& $-1083$     & 45 (1)  & 89 & 76 & $-43$ (1)& $-1083$ \\
288611133 $^d$& 16.130457 (3)& 33 (2) & 0.12 (2)  & 34 (2)   & 58,678.3 (1) & $-790$      & 43.4 (3)& 89 & 61 & 34.7 (7) & $-758$ \\
288611883    & 1.0104055 (2)& 4.3 (2)& 0        & $-$     &  $-$       & $-$         & 0      & 89 & 89 &   0     & $-479$ \\
356014478 $^e$& 9.524622 (4) & 25 (2) & 0        & $-$     &  $-$       &  $-$        & 0      & 89 & 89 &   0     & $-48$ \\ 
356896561    &2.253360796 (5)&10.1 (6)&0        & $-$     &  $-$       &  $-$        & 56 (5)  &85.3&134.2&$-31$ (6)& $-3585$ \\
357686232 $^d$& 7.3089470 (1)&20.4 (3)& 0.0680 (9)& 329 (2)  & 58,680.46 (3)&  76         & 3 (2)   &89.6&92.9&$-0.8$ (6)& $-146$ \\
376976908    & 34.77026 (2) & 57 (2) & 0.074 (2) & 111.1 (7)& 58,646.46 (8)& 502         & 17.6 (5)& 89 & 106&  3.1 (8) & $-337$ \\
389966039 $^f$& 6.830297 (5) & 18 (2) & 0.08 (2)  & 14 (14)  & 58,676.4 (3) & 11          & 0      & 88 & 88 &   0     & $-19$ \\
392569978    & 39.44787 (1) & 70 (2) & 0.29 (1)  & 230 (3)  & 58,662.6 (2) & 4462        & 17 (2)  & 89 & 92 & $-17$ (2)& $-2604$ \\
392572173 $^g$& 12.151030 (1)& 33 (1) & 0.51 (6)  &  63 (14) & 58,713 (1)   & -6465       & 44 (10) & 89 &119 & $-34$ (8)& $-2432$ \\
402656423    & 1.5192070 (2)& 7.7 (7)& 0        & $-$     & $-$        & $-$         & 0      & 80 & 80 &   0     & $-80$ \\
417701432 $^d$& 9.366113 (6) & 31 (2) & 0.19 (2)  & 236 (5)  & 58,675.9 (1) &  434        & 23 (7)  & 89 & 104&$-18$ (13)& $-143$ \\
420177014    &3.25855726 (2)&10.0 (7)& 0        & 0       &  $-$       &  $-$        & 0      & 88 & 88 &   0     & $-1521$\\ 
423927178    & 4.79259 (1)  &16.2 (9)& 0        & $-$     &  $-$       &  $-$        & 0      & 88 & 88 &   0     & $-12$ \\
441768471 $^d$&27.300297 (5) &49.3 (6)& 0.0694 (6)& 47.3 (5) & 58,675.04 (4)& 485         & 20.4 (4)& 89 &98.3&$-$18.2 (4)&$-255$ \\
459969365    & 9.008074 (1) &25.3 (5)& 0.0112 (4)&  12 (5)  & 58,680.7 (1) & 154         & 1 (4)   &89.0&89.8&  1 (3)   & $-132$ \\     
\bottomrule
\end{tabularx}
\end{adjustwidth}
	\noindent{\footnotesize{{\bf Notes.} {$^a$: 2+1+1 configuration; $^b$: Third body eclipses; 2+1+1 configuration; $^c$: Grazing third body eclipses + EDV $^d$: Eclipse depth variations; $^e$: unrealistically large third body mass; $^f$: No secondary ETV curve; $^g$: EDV, secondary eclipses disappeared for Year 7. }}}

\end{table}

This DE contribution to the ETVs, for the most part, consists of `physical' or `real' period variations, as the third body perturbations may cause true variations in the instantaneous anomalistic period (even when the long-term average of these variations disappear\endnote{It is a well-known fact that there are no apse-node timescale variations in the semi-major axes and, hence, in the anomalistic period, see, e.g., \citet{brown936a,brown936b,brown936c}.}).  However, there are also some `apparent' parts of these DE variations as well.  These are caused by the changing orientation of the orbit and its plane with respect to the observer which, in turn, result from either the $P_2$-timescale, perturbed apsidal motion (AM), or the nodal precession of the EB, or both.

The discussion in this latter paragraph is also valid for the last part of the r.h.s. of expression \ref{Eq:Delta}, that is, for the term $\Delta_\mathrm{AM}$, which stands for the longer timescale apsidal motion (AM) of eccentric EBs. AM may occur in every eccentric binary, even in the absence of any additional third (or more) bodies. As is well known, for example, the non-spherical mass distributions of the binary stars (due to the tidal and rotational interactions) as well as general relativistic effects may cause a rotation (mostly prograde) of the ellipse within the orbital plane, leading to a variation in the argument of periastron ($\omega$). In turn, this results in a changing orientation of the orbit relative to the observer during consecutive eclipsing events, causing apparent period variations. Again, apart from a weak inclination-dependence of non-exactly edge-on orbits (see, e.g., in \cite{gimenezgarcia983}), the middle of the eclipses occur when $v_1+\omega_1=\pm90^\circ$. Therefore, with the use of the Kepler's equation, it can be shown that, irrespective of the origin, or even the presence, of any kinds of AM,
\begin{equation}
\Delta_\mathrm{AM}=\frac{P_1}{2\pi}\left[2\arctan\left(\frac{\pm e_1\cos\omega_1}{1+\sqrt{1-e_1^2}\mp e_1\sin\omega_1}\right)\pm\sqrt{1-e_1^2}\frac{e_1\cos\omega_1}{1\mp e_1\sin\omega_1}\right],
\label{Eq:apse-def}
\end{equation}
where the upper signs refer to primary eclipses and the lower ones to secondary eclipses.

Naturally, when $\omega_1$ remains constant (and also $e_1$ and $P_1$ are constant), $\Delta_\mathrm{AM}$ yields the same contribution to each  type of eclipse, primary or secondary.  Therefore, it does not represent any measurable period variations in an EB. In the case of classical (that is, tidally forced) as well as relativistic apsidal motions, $e_1$ and $P_1$ remain constant in time, while $\omega_1$ varies only linearly in time.  Hence, one can write that $\omega_1=(\omega_1)_0+\Delta\omega_1\times E$, where $\Delta\omega_1$ is the change in $\omega_1$ (sometimes called the `AM rate') between two consecutive primary eclipses.  For very long AM periods, $\Delta\omega_1$ can safely be considered to be constant, despite the fact that what is really constant is $\dot\omega_1$)\endnote{Strictly speaking, in the case of the classic or tidal AM, this fact remains true only insofar as the spin axes of the binary stars are considered to be parallel to the orbital spin. For different situations, see \citet{shakura985,hegedusnuspl986}.}. In the case of AM forced by a third body, however, $\omega_1$ will no longer vary linearly in time and, moreover, neither the eccentricity nor the instantaneous anomalistic period ($P_1$) will remain constant. This apse-node timescale dynamical AM (which in the currently investigated systems, has much shorter periods and larger amplitudes than the unconsidered, non-third body forced AM components) are calculated under the term $\Delta_\mathrm{AM}$ in the way that is described in \citet{borkovitsetal15}.

{

\subsection{Numerical Representation of the Fitting Process}

Depending on whether an ETV curve can be modeled by a pure LTTE solution, possibly including a polynomial term (up to third order), or rather if DE contribution(s) must also be included, the applied numerical methods to obtain physically reliable solutions are different. In the case of a pure LTTE (or LTTE + polynomial) solution, the mathematical form of the ETV curve is quite simple, as the shape of an LTTE curve can be described by a pure sinusoidal curve. Though, we note that the dependence of the correct shape on the orbital elements of the outer orbit, as well as on its period ($P_2$), is non-linear.  But, it has been found that, as long as the initial parameters are not very far from the final obtained parameter set, a pure non-linear differential correction method, such as, in the current case, a Levenberg--Marquardt non-linear least squares method, may result in realistic and correct findings, thereby identifying the absolute minimum in the phase space. The robustness of such a method and the technical details are discussed in detail in \citet{borkovitsetal15}.

In the case of DE contributions, however, the phase space becomes much more complicated and multiply degenerate and, therefore, in the case of significant DE term(s), we use a more direct and robust Markov-chain Monte Carlo (MCMC) method to obtain reliable parameters. This also enables us to explore a much larger part of the phase space, excluding of course the physically unrealistic parts of the multi-dimensional parameter space. Our method is quite similar to that which was used formerly in the work of \citet{mitnyanetal24} and, therefore, a more detailed description can be found in that paper.

Finally we address the critical parameter $P_2$, i.e., the period of the outer orbit. As we will show below, similar to our previous studies, one of the main criteria in the classification of the robustness of our solutions is how the obtained ETV period, $P_2$, is related to the duration of the entire dataset. In the case of the most robust (group or category 1) solutions, we typically require that the whole dataset must span more than two full outer orbital periods, $P_2$. In this case, the period $P_2$ should be quite certain and robust and, therefore, one can set a very realistic and accurate initial value for this parameter in both the Levenberg--Marquardt and the MCMC methods.

In the case of group or category 2 systems, i.e., when the dataset spans between 1 and 2 outer orbital periods (that is, the likely period of the third body dominated ETV curve), the input value of this parameter might also be relatively robust. However, in the case of the less robust solutions, where the outer period appears to be longer than the duration of the observed time series, the input values of $P_2$ to the fits are quite uncertain. In these cases, we usually applied different trial initial parameters for $P_2$ and, finally, selected the best fitting one from among the physically realistic solutions. That is, we dropped out any solutions, even if they were mathematically acceptable, where the amplitude of the curve became so large that the derived minimum mass was too high for a plausible companion (e.g., hundreds of solar masses).}

\section{Observational Data and Its Preparation for the Analysis}
\label{sect:dataprep}

The Transiting Exoplanet Survey Satellite (TESS, \citet{rickeretal15}) is designed primarily to discover extrasolar planets via the transit method over the entire sky, especially around brighter stars.  Such stars then become accessible to spectroscopic, especially RV, follow-up measurements, even with modest instruments located in either the northern or the southern hemispheres, or both. From the beginning of its operation in summer 2018, orbiting on a highly eccentric lunar-synchronous orbit, TESS monitors a given 24 $\times$ 90 degree-sized stripe of the sky almost continuously. The observations of one stripe (called a sector) last for $\sim27.4$\,days (apart from the current sectors 97, 98, which had durations of double those of the former sectors). At the end of a given sector, the field of view of the satellite is rotated, largely in ecliptic longitude, to the next adjacent area of the sky.  

During these first 98 sectors, with the exceptions of ten, so called `ecliptic' sectors, the surrounding area of one of the two ecliptic poles was continuously observed. These areas, called the `Northern/Southern Continuous Viewing Zone(s)' (N/SCVZs), made it possible (amongst others things) to gather several year-long, quasi continuous ETV curves (though interrupted by year-long gaps) for the EBs located in these zones. Regarding the NCVZ, this area was observed during the time intervals 18 July 2019--4 July 2024 (Year 2, Sectors 14--26), 24 June  2021--20 August  2021, and then 30 December 2021--18 January  2023 (Year 4, Sectors 40--41 and 47--55, and Year 5, Sectors 56--60), and, finally, 7 December  2023--18 December 2024 (Year 6, Sectors 73--83 and Year 7, Sectors 84--86); but no further observations are scheduled before Sector 117 of Year 9, in May 2027. Most objects were observed in full frame image (FFI) mode with cadence times 1800 s (Year 2), 600 s (Year 4), and 200 s (from Year 5). For a smaller subgroup of our targets 120 s cadence (and for an even smaller subgroup 20 s cadence), observations and light curves are also available, at least for a part of the observing sectors.

Within the framework of a former project, the light curves, and the ETV curves derived from them, of more than 3500 EBs, located in or near to the NCVZ, were collected after the Sector 60 data became available. From this huge sample, we selected 351 targets with suspected non-linear ETVs. Finally, third body ETV solutions were published for 135 hierarchical triple candidates from this latter sample in \citet{mitnyanetal24}. The technical details of the methods of the target selection, FFI photometry, and the calculations of the ETVs from the light curves are described in detail in that paper.

For this current study, we returned to the previously selected narrower sample of the above-mentioned 351 targets. We downloaded the Sector 73--86 FFIs for these EBs. For the new photometry, however, instead of the previously used pipeline {\sc FITSH} (\citet{pal12}), for purely technical reasons, now we used the publicly available package {\sc Lightkurve} \citep{lightkurve18}. Moreover, for some of our targets, two-minute cadence light curves were also available. However, we have found that for accurate calculations of the mid-eclipse times, the sampling rate of the current 200 s cadence FFIs are essentially as good as the less available 120 s cadence observations.   We therefore preferred to use more plentiful FFI-derived light curves with 200 s cadence for calculating the eclipse times and, hence, to derive the new ETV sections. The determinations of the mid-eclipse times were made, however, in the very same manner as was described formerly for the S14--S60 data in \citet{mitnyanetal24}.

It should be noted that TESS is primarily designed for observations of extrasolar planets around nearby, generally brighter stars, and it is able to use small-aperture, wide-field telescopes.  Therefore, it is not entirely suitable for observations of very faint stars. According to our experience, one can determine accurate mid-minima times and, hence, ETV curves down to $15$\,magnitude. Naturally, however, the accuracy of the mid-minima time calculations also depends on the eclipse depths. This is the reason that, in several eccentric EBs, where the primary and secondary eclipses should be fitted separately, the mid-eclipse times of the shallower secondary eclipses can be determined only with much lesser accuracies or with larger error bars. This effect can be seen on several panels of the forthcoming individual ETV plots.

\section{Discussion}
\label{sec:discussion}

As was mentioned above, the last TESS observations, which were utilized in \citet{mitnyanetal24}, were carried out more than three years ago. Since then, a newer, $\sim376$\,day-long dataset (which follows a $\sim323$\,day gap) has become available. Therefore, now the nominal length of the available data train has increased to $\sim1950$\,days. Despite the fact that, from a general astronomical point of view, neither the entire duration of the TESS observations nor the length of the available new dataset looks significant, in our opinion, a repeat study has great importance for the following reasons: (i) Now the length of the high-precision TESS photometry is about twice that of the (somewhat arbitrary) period limit of the most compact triple stars ($P_2\lesssim1000$\,days, \citep{tokovinin21}) and, therefore, we may expect that all (or at least the majority) of such compact triples which are located in the NCVZ and have an EB as the inner, close pair can be identified with great confidence. (ii) Furthermore, we can expand the likely identification of triples toward even longer timescales, which now may reach even a decade. (iii) Finall,y it is very illuminating to compare the current results with those of the former findings of \citet{mitnyanetal24}, especially in the case of the most uncertain of the probable LTTE systems, where such an extension may show clearly that there might be such ETVs that mimic LTTE orbits during a shorter timescale. This suggests again that no LTTE solution should be accepted without caution unless at least two or three outer orbital periods were observed with good coverage. In light of these cautionary remarks, we not only discuss our findings but extensively compare them with the former results.

Investigating the Sector 14--86 ETV curves of the EBs in the TESS NCVZ, we were able to obtain third body solutions for 168 EBs. The reliabilities of these solutions are very different among the sample, and, in that sense, they are similar to our former study of the NCVZ ETVs \citet{mitnyanetal24}.  There, we divided these third body ETV solutions into five subgroups (three subgroups with pure LTTE and two subgroups with LTTE + DE-type solutions). For better comparison, we followed the definitions of each subgroup, as they were described in \citet{mitnyanetal24} (though, naturally, small modifications, were applied). Here, however, we briefly repeat these definitions. First, we divided our sample into two larger groups. This main division is based on whether the third body effect can be modeled simply with the well-known LTTE (L-type solution) or the system appears to be so tight that DE should also be considered (combined, D-type solution). 

Here, we note that we found several ETV curves for which exclusive modeling with either the LTTE or the LTTE + DE terms was not sufficient. This was so mainly because the ETVs frequently, already upon a first inspection, show evidence of other longer timescale non-linearities or period variations. These effects, as described above, were taken into account by introducing additional quadratic or cubic polynomials. In a few cases, we used a four-body model, that is, we fitted simultaneously a second, longer-period LTTE solution, which was considered to be independent of either the LTTE or the LTTE + DE solution of the third body. We will return to these solutions near the end of this discussion, but first, we now describe the main groups of the categorized systems. Here, under `pure' LTTE solution (L-type) systems, we take all those candidates for which the modeling did not include any  dynamical terms. This is independent of whether quadratic or cubic polynomials or a fourth-body solution were or were not added to the LTTE solution. And, naturally, a similar idea is intended in the case of LTTE + DE (that is, D-type) systems. In other words, when using these two broad categories, we do not consider the presence or absence of any other kinds of period variations.

\subsection{Triple-Star Candidates with Pure LTTE Solution}

In the second step we divided the L-type ETV solutions into three additional subgroups, according to our confidence in the solution. Into the first, most secure, group (subgroup $L_1$), where the derived outer orbital period was found to be shorter than half of the entire datatrain (i.e., outer orbital periods $P_2\lesssim950$\,days), we selected 35 systems. The tabulated system  parameters for these systems are found in Table~\ref{Tab:Orbelem_LTTE1}, while the ETVs themselves, together with the LTTE solutions, are plotted in Figures~\ref{Fig:ETVs_L1a}--\ref{Fig:ETVs_L1c}. We note, however, that this period limit was applied only as a first-cut criterion. Due to the fact that there are large data gaps between the TESS NCVZ observations (due to the observations of the southern and the ecliptical sectors), we identified such short-period ETVs where sensitive sections of the ETV curves were missing and/or other kinds of non-linear variations at least partially masked the available LTTE patterns. Therefore, in such cases, despite the shorter outer period (which would fulfill the main selection criterion), we ranked some LTTE solutions only into the lower, less certain subgroups. In this regard, we note that we followed practically a similar and partly subjective ranking system in our previous work. A good justification for this type of decision making (at least in our opinion) is that there were no such most certain $L_1$ or $D_1$ ranked third-body ETV solutions in \citet{mitnyanetal24}, which we had to downrank now, using the new two-year longer dataset. Even in the moderately certain subgroups of $L_2$ and $D_2$, we had to downrank only six third-body solutions.

Comparing this category with the same, most certain LTTE subgroup of \citet{mitnyanetal24}, all of their 19 triple-star candidates are present in our refreshed list, and even the refined orbital parameters are also similar. Eleven additional systems were upranked from their lower categories---nine from a less certain category, and even two from the most uncertain subgroup\endnote{The exact TIC ids of all the upranked and downranked systems are listed later, in Sections~\ref{subsect:upranked} and \ref{subsect:downranked}.}. For these systems, the new sections of the ETVs demonstrated clearly that pure LTTE solutions are plausible. On the other hand, we should note that there were five additional systems which were not listed in any categories of \citet{mitnyanetal24}, while they nominally should have been. Therefore, leaving them out from that previous work (though they were formerly preselected amongst the 351 suspected non-linear ETVs) was a consequence of some bookkeeping errors rather than any objective arguments.

We classified 28 additional systems as having less certain but likely LTTE solutions (see Table~\ref{Tab:Orbelem_LTTE2} and Figures~\ref{Fig:ETVs_L2a} and \ref{Fig:ETVs_L2b}). These were put into subgroup $L_2$. The main (necessary but not sufficient) criterion for being classified in this subgroup was that the outer period obtained from the LTTE solution appears to be shorter than the length of the entire dataset (that is, for the majority of the sample, $P_\mathrm{out}\lesssim 1900--1950$\,days). Comparing these candidates, however, with the 19 systems which were ranked into the similar second category of \citet{mitnyanetal24}, we find that there are only three shared triple-star candidates. Nine of those EBs, of which the ETV solutions were put into that second, less certain, but likely pure LTTE solution systems, have now been upranked to the most certain ($L_1$) category. Moreover, at the other end, 15 such EBs for which the LTTE solutions were categorized only as uncertain in that earlier work can now be put into this second ($L_2$) subgroup. These results again confirm the strictness of our conditions for categorizing the third-body ETV solutions and, moreover, offer evident proof of the fact that even this plus two-year-long dataset carries important advantages. We note, however, that, in the opposite direction, there are five EBs for which the LTTE solutions have now been downranked from the former second to the current third (uncertain, $L_3$) subgroup. These are typically low-ETV amplitude systems, where other kinds of shorter-period ETVs may hide, or even mimic, an LTTE solution. Therefore, these ETVs again show very clearly that no LTTE solutions should be considered as certain before good coverage of at least two or three outer orbital cycles is obtained. Finally, there are ten such triple-star candidates in this subgroup for which the ETVs were not characterized by \citet{mitnyanetal24} for one reason or another. Some of these systems had been left out of the \citet{mitnyanetal24} work erroneously, as was mentioned for the formerly uncategorized $L_1$ systems. While for some others of these systems, the weak earlier coverage of the most sensitive portions of the ETV curves (mostly the lack of extrema in the curves) made it nearly impossible to find any reliable LTTE solutions before the new Sector 73--86 data.

Some further caution in regard to the $L_2$ triple-star candidates is warranted until future studies of these systems are carried out. For example, taking some examples from Figure~\ref{Fig:ETVs_L2a}, if one considers the ETV patterns of TICs 230393824 and 233719825 (right panels in the fourth and fifth rows, respectively), it looks almost certain that these ETVs reflect LTTE, though the TESS observations do not cover at least two full third-body orbital cycles, and, hence, they cannot be put into subgroup $L_1$. On the other hand, considering the roughly similar period ETVs of TICs 199632809, 229500406, 229651225, or 233056681 (see the corresponding panels in Figure~\ref{Fig:ETVs_L2a}), one cannot be certain that the ETV points will actually `turn back' a few months after the last TESS observations, as would be predicted by the current LTTE solutions.

This last cautionary note is even more valid in the case of the less certain $L_3$ systems (see Table~\ref{Tab:Orbelem_LTTE3} and Figures~\ref{Fig:ETVs_L3a}--\ref{Fig:ETVs_L3e}). During our analysis, while adding the new ETV points sector by sector, we experienced several times that our analytic LTTE fitter found third-body solutions which initially appeared to be quite acceptable, but the behavior of the ETVs in the newer sectors departed from the predicted ones. Typically, there were no extrema at the previously predicted times, that is, the ETV points did not `want' to turn back. As a direct consequence, the most uncertain ($L_3$) solutions should also be considered with even less confidence. This is true even in those cases where an initial look at the ETV does really suggest some longer-period LTTE, as in the cases of TIC 219787718 (V377 Dra, left panel of the second row in Figure~\ref{Fig:ETVs_L3a}), or of TICs 229751503, 23008057, 233680160 (see the appropriate panels of Figure~\ref{Fig:ETVs_L3b}), etc. To reduce such large uncertainties, we introduced a further rule at the selection of the appropriate candidates to put into subgroup $L_3$, which was not followed formerly.  Now, we decided to introduce an upper limit for the obtained outer period, and we dropped out all those hierarchical triple-star candidates where the obtained LTTE period (independent of how good-looking that LTTE solution was) was $P_2\gtrsim5000$\,days. We departed from this strict limit only in those cases where former times of minima obtained from ground-based observations were also considered and, hence, the length of the investigated ETV curves did exceed (or, at least, were close to) this limit.

In conclusion, putting into this category all systems for which we obtained pure LTTE solutions with periods $1950^d\lesssim P_2\lesssim5000^d$, and adding those for which we found a shorter period, but judging our solution to be very uncertain\endnote{Mostly low-amplitude ETVs, affected by other non-linearities and approximated by quadratic or cubic polynomials.}, we finally ranked 64 EBs into this ($L_3$) category. A number of these triple-star candidates are identical to those which were listed in the most uncertain category of \citet{mitnyanetal24}. (We note, however, that a quick comparison amongst the parameters obtained in that work with those in the current analysis reveals that, in most cases, the old and the new LTTE solutions are completely different.) Comparing our group $L_3$ systems directly to those of \citet{mitnyanetal24}, as we list below, besides those 17 EBs for which the LTTE solutions have now been upranked to subgroups either $L_1$ or $L_2$, we found no new LTTE solutions for 9 EBs. This is manly due to the fact that we have introduced the new, $P_2\lesssim5000$\,d rule. As mentioned above, we downranked three more systems which formerly were categorized into the second subgroup of \citet{mitnyanetal24}. Finally, we found 26 systems for which \citet{mitnyanetal24} were unable to find any kind of LTTE (or LTTE + DE) solutions, but now, after utilizing the new data, we are already able to find, at least uncertain, pure LTTE solutions.

\subsection{Triple-Star Candidates with Combined LTTE + DE Solutions}

As was mentioned above, we found several such systems where the different shapes and amplitudes of the primary and secondary ETVs already revealed at a first glance that dynamical effects (that is, DE terms) should also be taken into account during the ETV analysis.  And, even if this were not the case, sometimes the precalculated ratio of the DE and LTTE amplitudes suggested that the triple system was likely tight enough for the application of a combined LTTE + DE analysis. Similar to the previous analysis of \citet{mitnyanetal24}, for these D-type systems, we introduced only two subgroups instead of three. Categorizing hierarchical triple-star candidates into the most certain ($D_1$) category subgroup (tabulated in Table~\ref{Tab:Orbelemdyn1} and shown in Figures~\ref{Fig:ETVs_D1a} and \ref{Fig:ETVs_D1b}), we essentially follow the same strict criteria that are described above in regard to pure LTTE systems in subgroup $L_1$. Now, we have put 31 hierarchical triple-star candidates into this category. All but three of the 27 certain LTTE + DE solution systems of \citet{mitnyanetal24} are again categorized in our certain LTTE + DE group $D_1$. One missing system is the triply eclipsing triple star TIC 441738417, which is now recategorized as a certain but pure LTTE (that is, $L_1$) triple. The two other missing triples are TICs 236774836 and 420263614, where the eclipses disappeared during the recent observations. Therefore, in these two cases, we were unable to derive any new solutions, but the disappearance of the eclipses directly reveals that these systems are actually tight but inclined triples. Moreover, we upranked five additional triple-system candidates from the less certain LTTE + DE category of \citet{mitnyanetal24}. These systems in that former work were categorized only into the less certain LTTE + DE subgroup because of their longer inferred outer periods. But now, due to the extended length of the dataset, their outer periods fulfill the restrictions of the most certain, $D_1$ category, systems. We note also that there was one system, TIC 259006185 (see second row, left panel of Figure~\ref{Fig:ETVs_D1b}), which was left out of the previous work erroneously. The short-timescale period variations of this EB, likely due to a third body, were first reported by \citet{marcadonprsa24}.

\begin{figure}[H]
\begin{adjustwidth}{-\extralength}{0cm}
\centering
\includegraphics[width=0.33\textwidth]{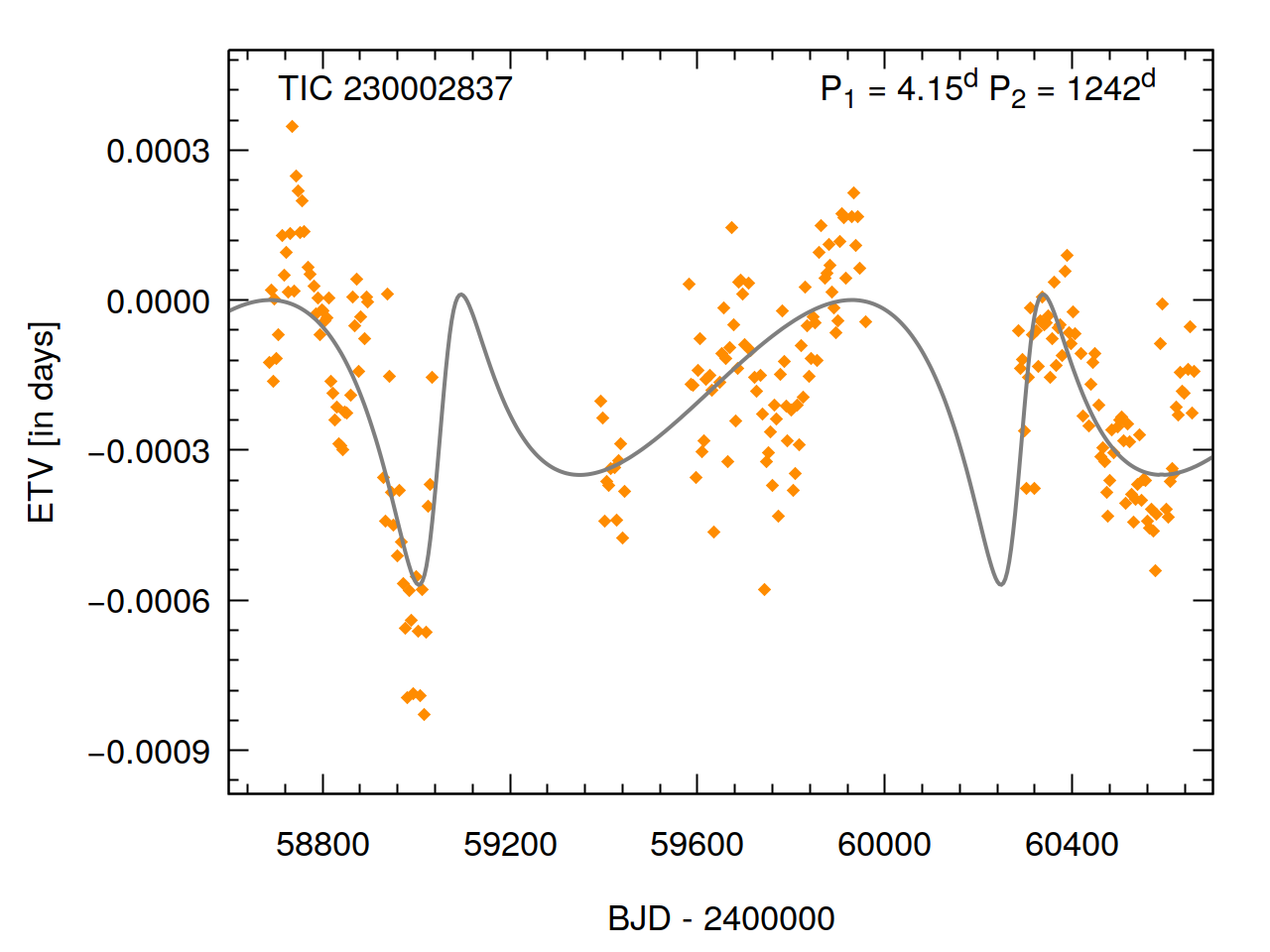}\includegraphics[width=0.33\textwidth]{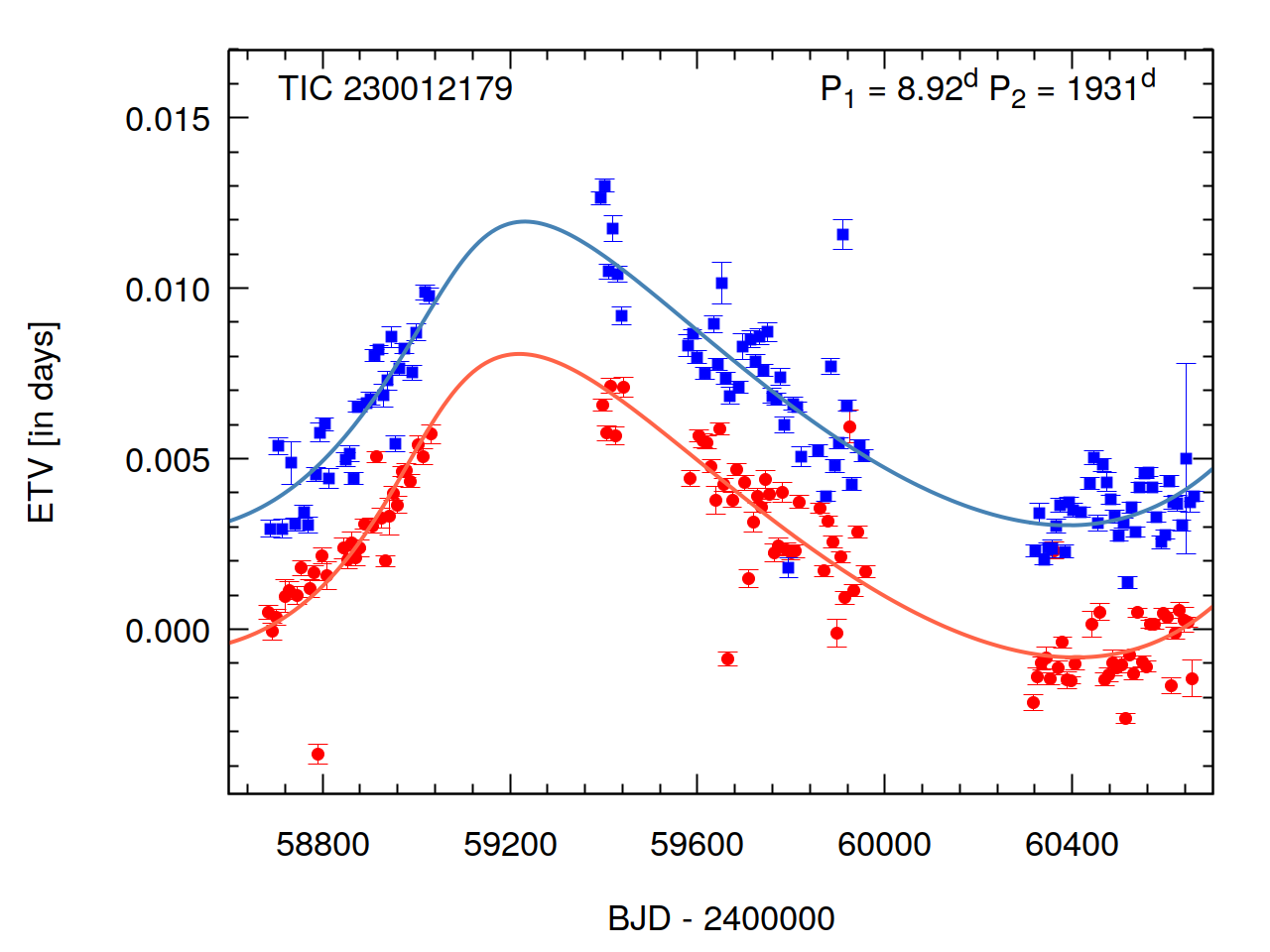}\includegraphics[width=0.33\textwidth]{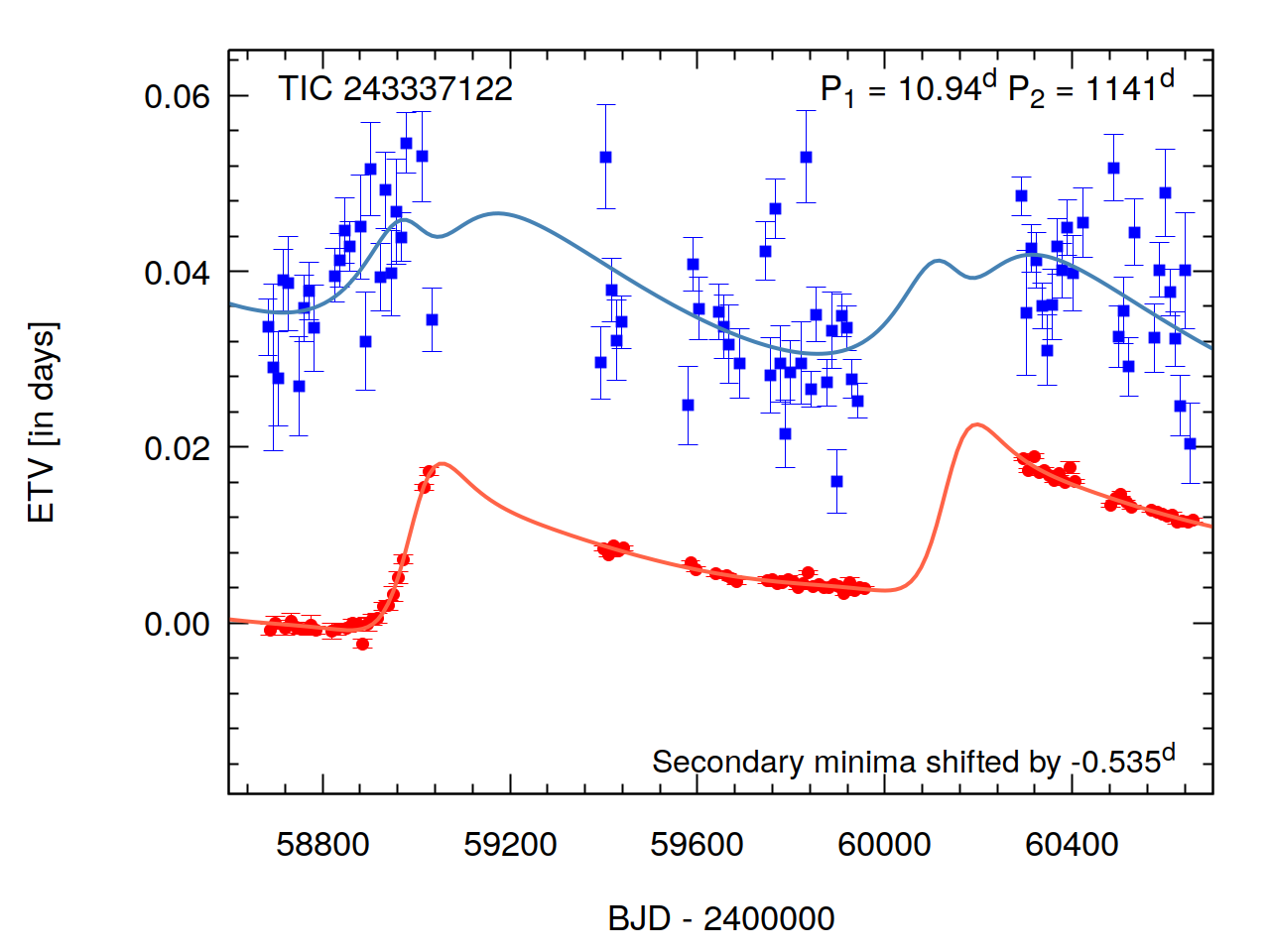}
\includegraphics[width=0.33\textwidth]{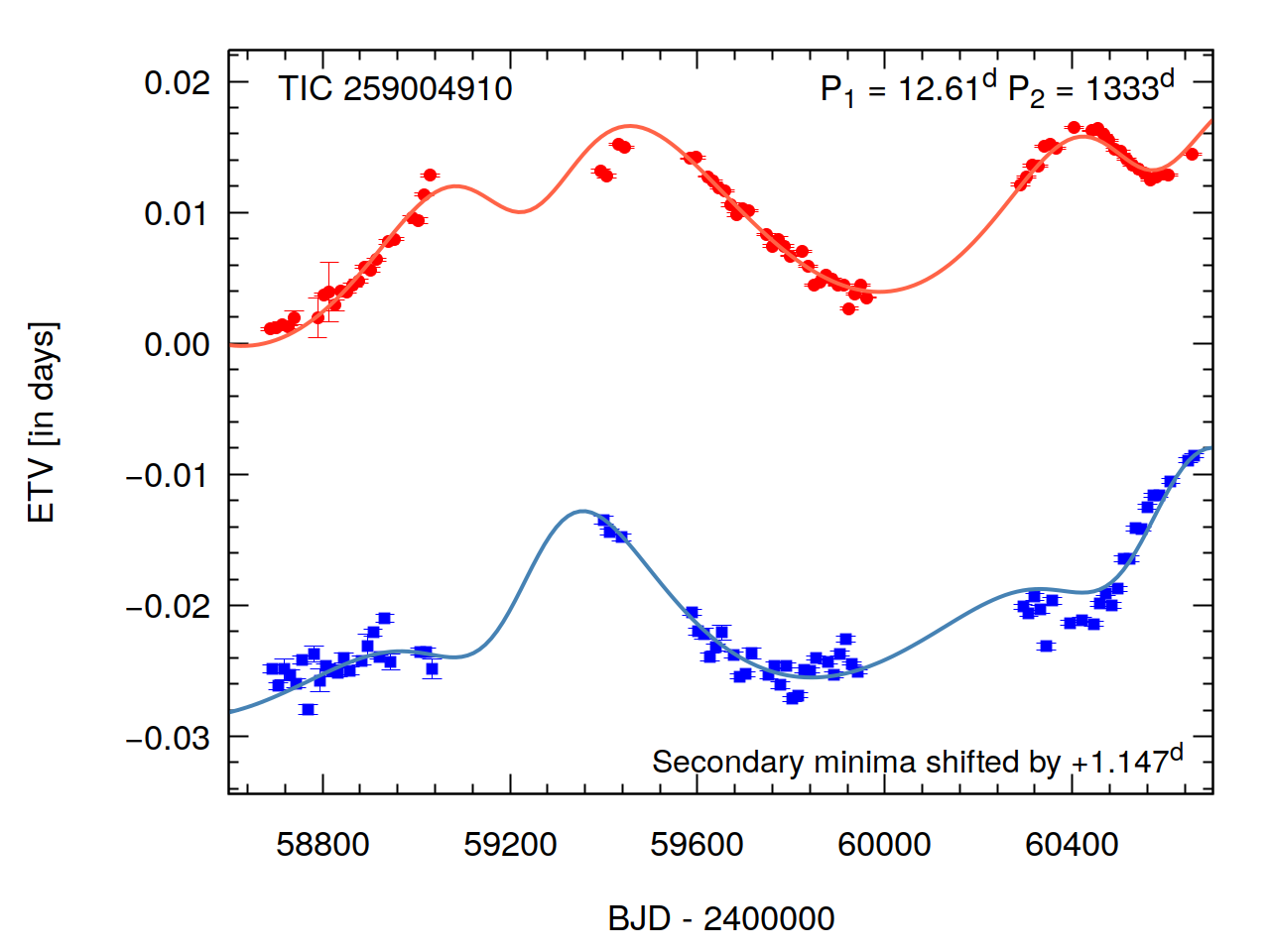}\includegraphics[width=0.33\textwidth]{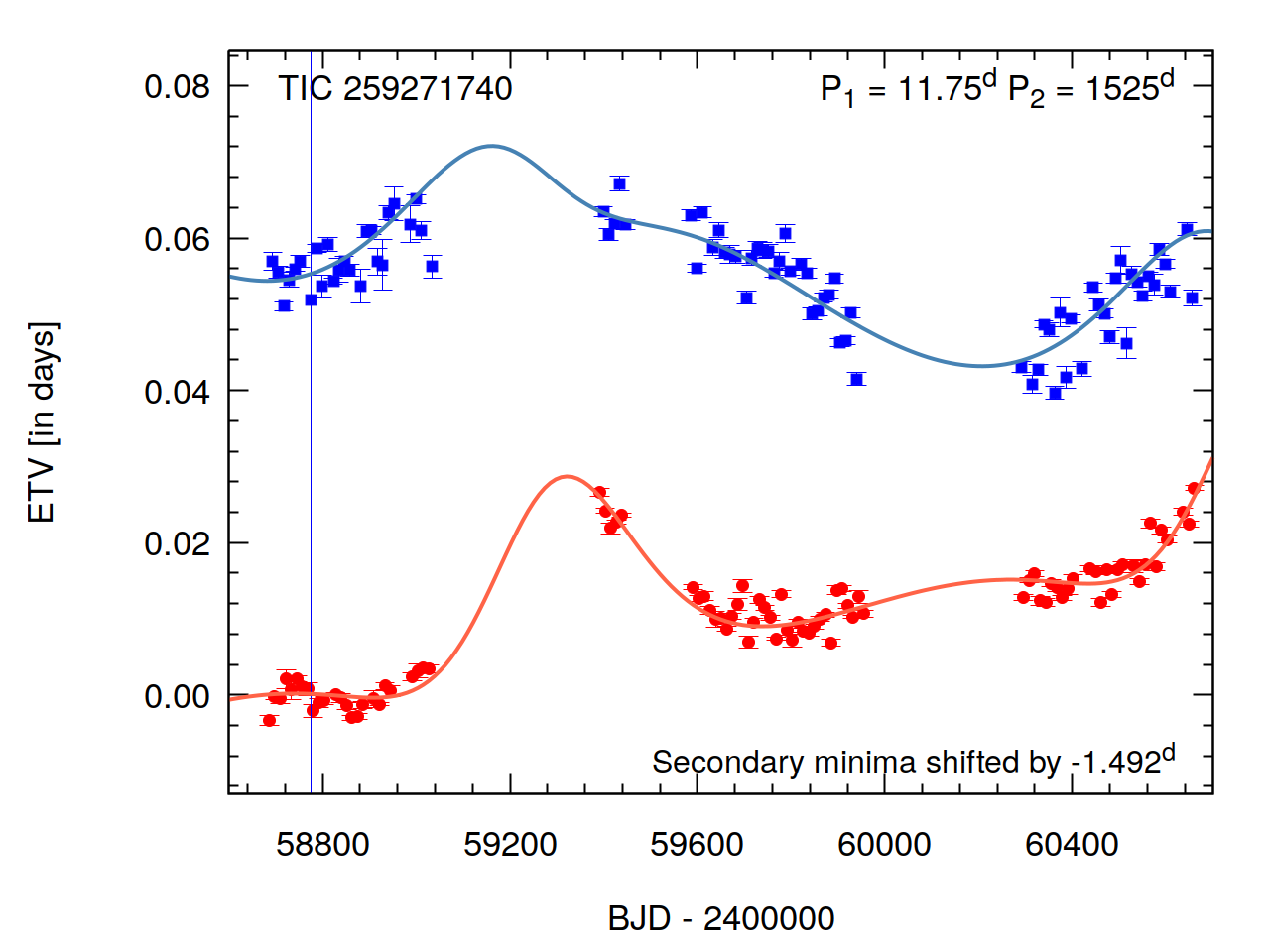}\includegraphics[width=0.33\textwidth]{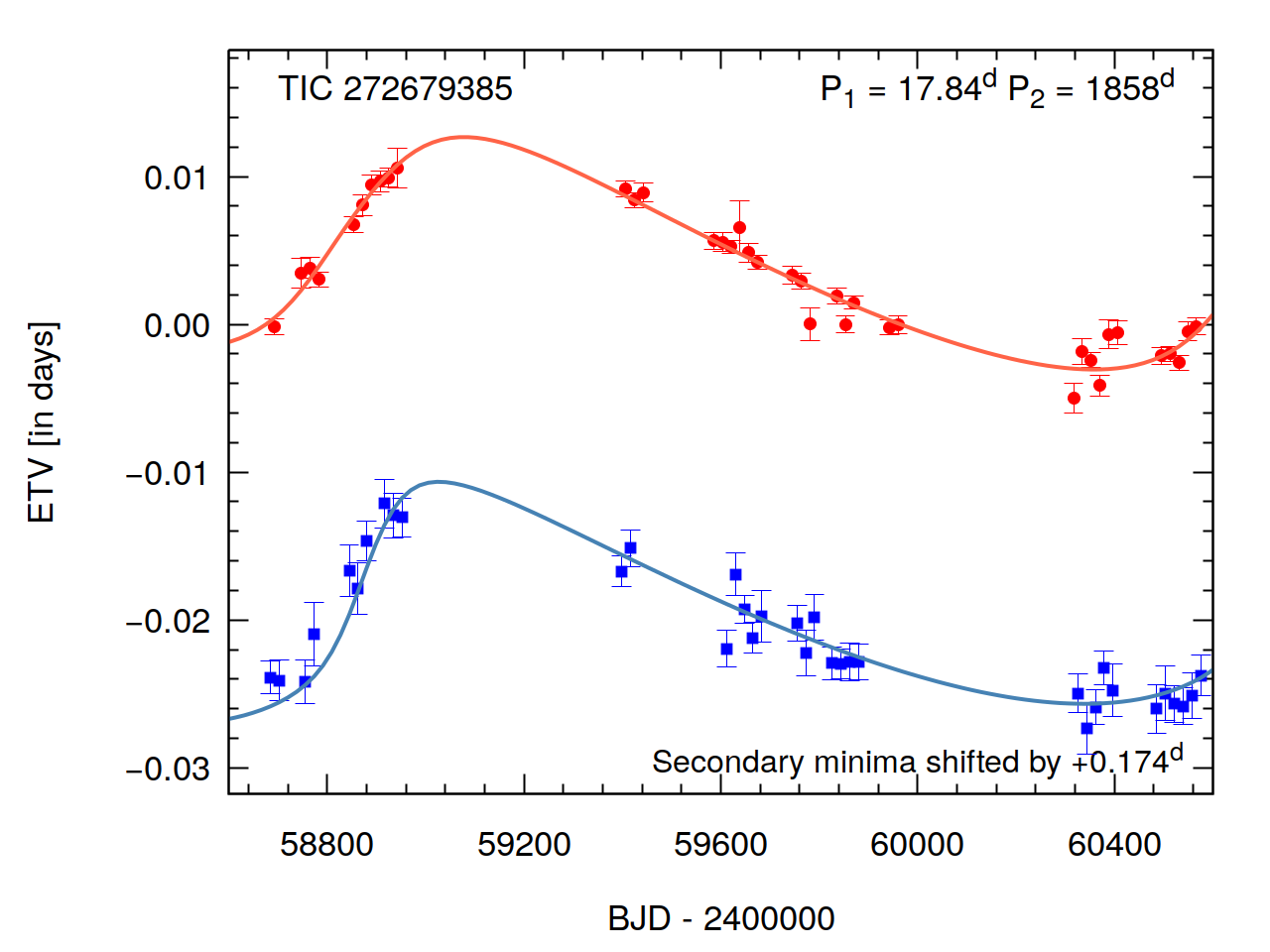}
\includegraphics[width=0.33\textwidth]{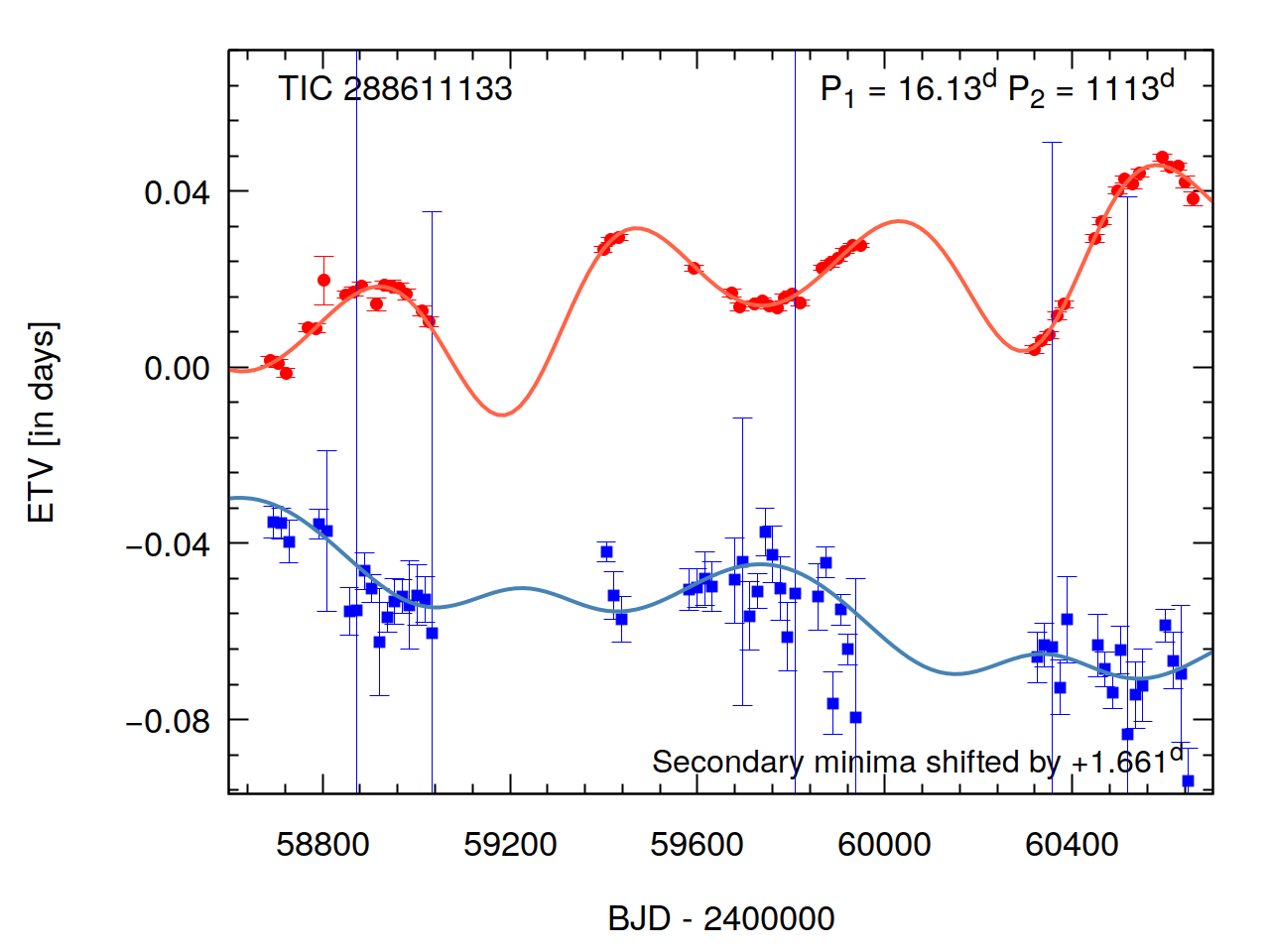}\includegraphics[width=0.33\textwidth]{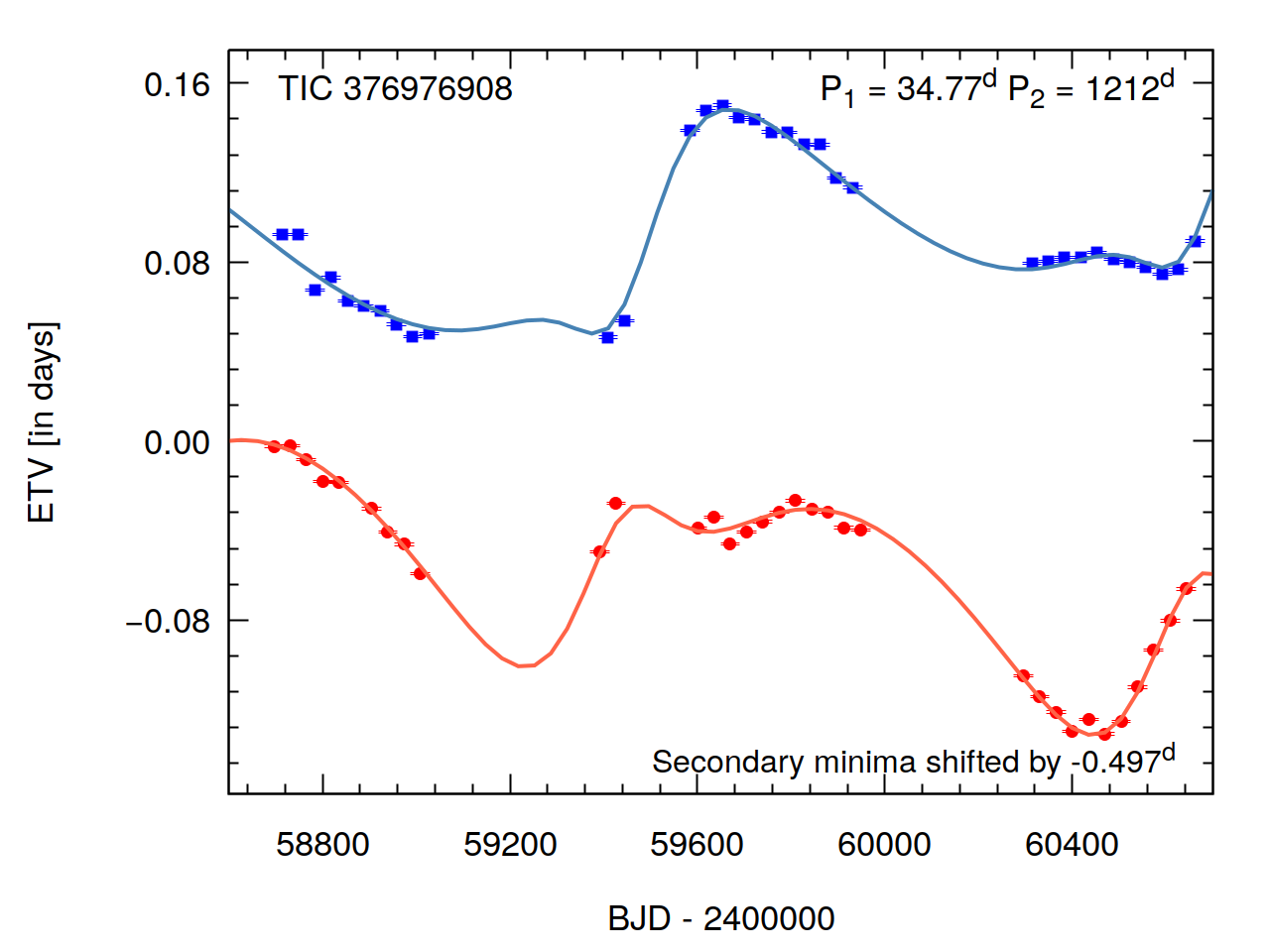}\includegraphics[width=0.33\textwidth]{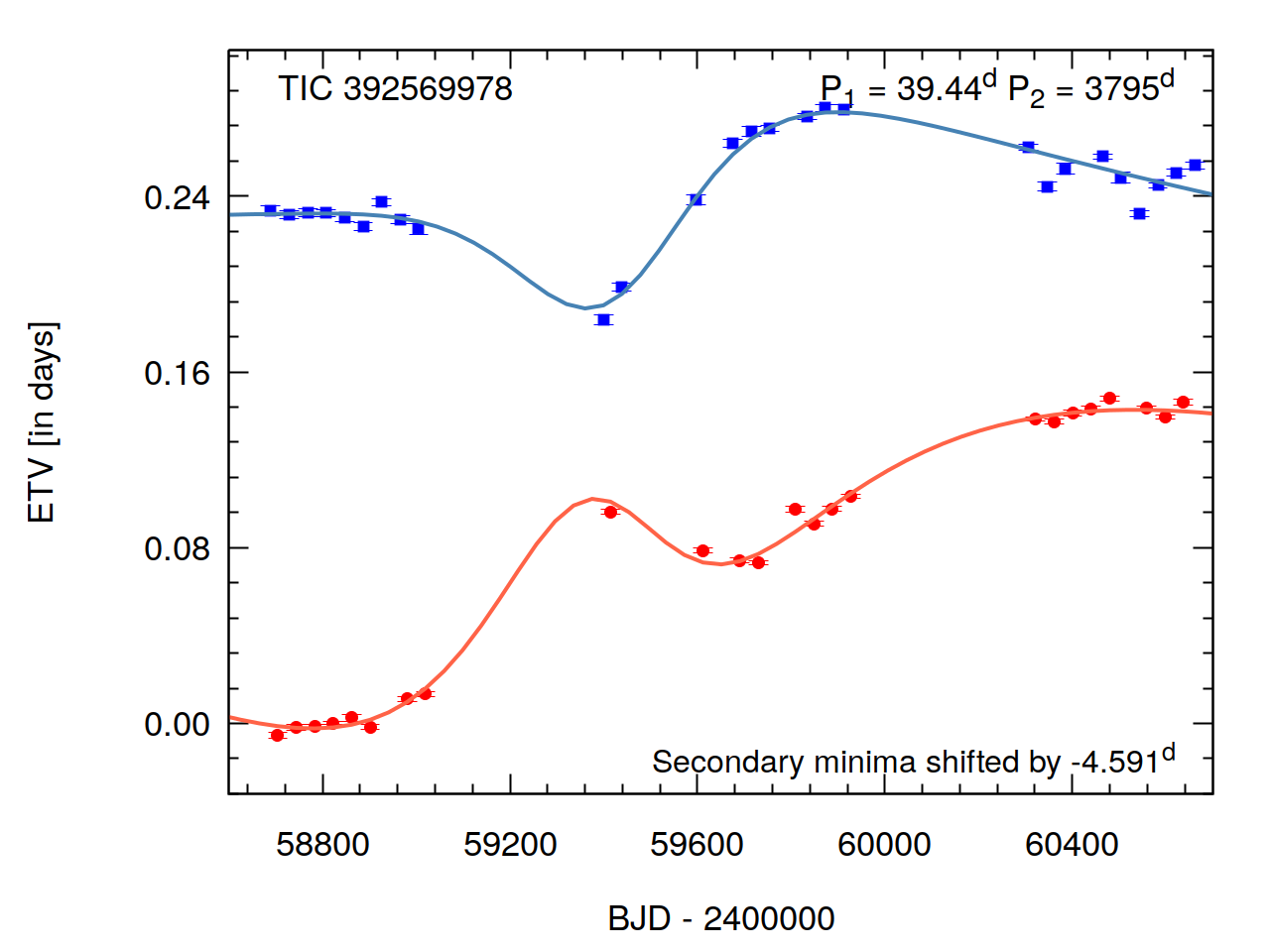}
\includegraphics[width=0.33\textwidth]{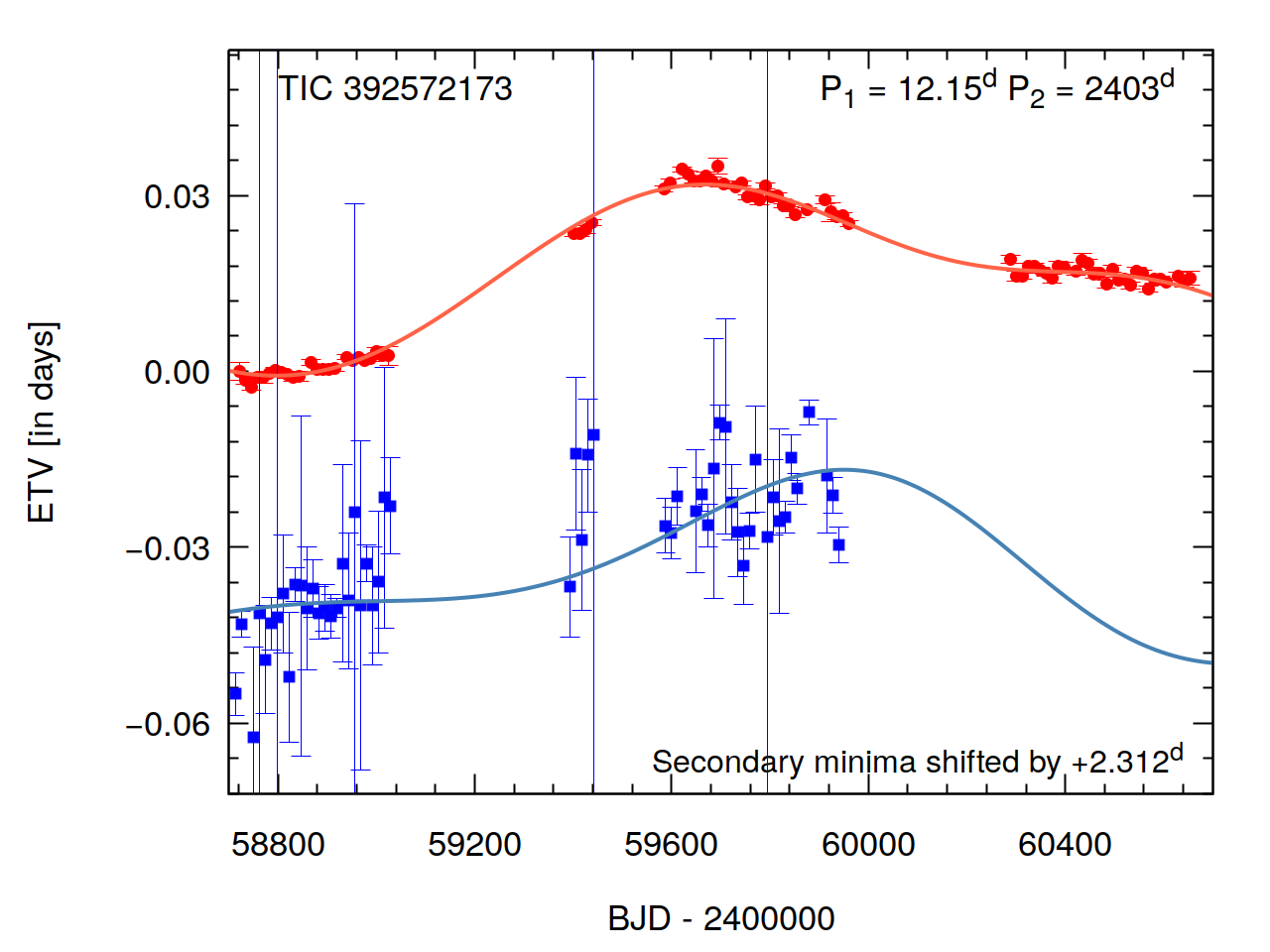}\end{adjustwidth}\caption{Systems with less certain LTTE + DE-type combined third-body ETV solutions. The meaning of each symbol is described in the former figure captions. See { Tables~\ref{Tab:AMEparam} and \ref{Tab:Orbelemdyn2}} for further details.}
\label{Fig:ETVs_D2}
\end{figure}

Our smallest subgroup, the category of the less certain LTTE + DE-type systems ($D_2$), contains only ten members (see Table~\ref{Tab:Orbelemdyn2} and Figure~\ref{Fig:ETVs_D2}). All but one system in this category are eccentric binaries where the secondary ETVs are clearly offset relative to the primaries. In seven systems of this subgroup, the secondary ETV curves have very different shapes and amplitudes than the primary ETVs, and this characteristic is a very clear indicator of third-body perturbations leading to forced dynamical ETVs. (Note also that, in the remaining two cases of TICs 230012179 and 272679382, the primary and secondary ETVs just show a very similar, but almost purely sinusoidal, shape also indicating the presence of a third companion).  Therefore, we are convinced that these ETVs are really caused by third bodies, mostly stellar components, even if the true orbital elements, due to the longer outer periods, may be more or less inaccurate. Comparing this subgroup with the analogous category of \citet{mitnyanetal24}, as was mentioned above, we upranked five of their formerly $D_2$-categorized systems into the most certain category $D_1$. Moreover, two systems which were not categorized formerly now have been added to this subgroup. There was, however, one system which was put into this $D_2$-analogous subgroup in \citet{mitnyanetal24} which we have now downranked into the most uncertain pure LTTE category ($L_3$). 

Finally, we note that, in the case of the LTTE + DE systems, several other parameters can be deduced from a combined analysis, as was described in \citet{borkovitsetal15}. These extra parameters for the LTTE + DE systems are tabulated in Table~\ref{Tab:AMEparam}.

\begin{table}[H]
\caption{Orbital elements from less certain combined dynamical and LTTE solutions. (The outer period longer than half of the length of the datasets or, critical section(s) are missing.)} 
\label{Tab:Orbelemdyn2}  
\begin{adjustwidth}{-\extralength}{0cm}
\centering

\begin{tabularx}{\fulllength}{lccccccccccc}
\toprule
\textbf{TIC No.} & \boldmath{$P_1$} & \boldmath{$P_2$} & \boldmath{$a_2$} & \boldmath{$e_2$} & \boldmath{$\omega_2$} & \boldmath{$\tau_2$} & \boldmath{$f(m_\mathrm{C})$} & \boldmath{$\frac{m_\mathrm{C}}{m_\mathrm{ABC}}$} & \boldmath{$m_\mathrm{AB}$} & \boldmath{$m_\mathrm{C}$} & \boldmath{$\frac{{\cal{A}}^\mathrm{meas}_\mathrm{dyn}}{{\cal{A}}_\mathrm{LTTE}}$}\\
        & \textbf{(day)} & \textbf{(day)} &\textbf{(R}\boldmath{$_\odot$}\textbf{)}  &       &   \textbf{(deg)}    &   \textbf{(MBJD)} & \textbf{(M}\boldmath{$_\odot$}\textbf{)}       & \textbf{(M}\boldmath{$_\odot$}\textbf{)}            &  \textbf{(M}\boldmath{$_\odot$}\textbf{)}  &   \\
\midrule
230002837 &4.14572685 (8)& 1242 (19) & 541 (82) & 0.67 (3) & 2.6 (9) & 59,048 (20)& 0.002 (1)  & 0.11 (3) & 1.23 (61)   & 0.15 (9)  & 1.02 \\
230012179 & 8.9234820 (2)& 1931 (61) & 762 (93) & 0.39 (3) & 231 (9) & 59,045 (68)& 0.007 (4)  & 0.16 (2) & 1.33 (56)   & 0.26 (12) & 0.51 \\
243337122 &10.9372965 (7)& 1141 (11) & 633 (28) & 0.42 (4) & 261 (10)& 59,001 (8) & 0.026 (8)  & 0.22 (2) & 2.03 (32)   & 0.59 (12) & 2.26 \\
259004910 &12.6073793 (2)& 1333 (4)  & 659 (3)  & 0.22 (3) & 237 (5) & 59,220 (19)& 0.057 (1)  & 0.299 (2)& 1.51 (3)    & 0.65 (1)  & 0.77 \\
259271740 &11.7515858 (7)& 1525 (6)  & 1066 (22)& 0.18 (1) & 265 (6) & 59,251 (23)& 0.21 (2)   & 0.319 (5)& 4.76 (38)   & 2.23 (19) & 1.24 \\
272679385 & 17.839787 (3)& 1858 (19) & 863 (15) & 0.48 (2) & 246 (17)&58,864 (113)& 0.0032 (6) & 0.109 (8)& 2.23 (12)   & 0.27 (3)  & 2.71 \\
288611133 & 16.130356 (3)& 1113 (3)  & 675 (23) & 0.078 (8)& 109 (2) & 59,224 (10)& 0.20 (3)   & 0.448 (8)& 1.84 (26)   & 1.49 (22) & 2.18 \\
376976908 & 34.77105 (2) & 1212 (2)  & 707 (22) & 0.253 (7)& 58.0 (6)& 58,224 (4) & 0.14 (2)   & 0.37 (1) & 2.03 (25)   & 1.19 (16) & 7.23 \\
392569978 & 39.44408 (1) & 3795 (45) & 1641 (39)& 0.411 (4)& 132 (5) & 59,418 (22)& 0.10 (2)   & 0.29 (3) & 2.93 (21)   & 1.19 (18) & 4.44 \\
392572173 &12.150809 (1) & 2403 (45) & 1331 (48)& 0.04 (2) & 333 (10)& 57,863 (82)&  0.27 (7)  & 0.42 (4) & 3.18 (36)   & 2.31 (43) & 0.46 \\
\bottomrule
\end{tabularx}
\end{adjustwidth}
\end{table}

\vspace{-12pt}

\subsection{Upranked Triple Candidates}
\label{subsect:upranked}

As was mentioned above, there were several EBs in the former sample of \citet{mitnyanetal24} which were upranked or even downranked in this new work. For the convenience of the readers, in this subsection, we tabulate the TIC numbers of all the systems, which have now been upranked, while, in the next subsection, we give the list of the downranked systems.

\begin{description}
\item $L_2\rightarrow L_1$: 159465833, 160518449, 229412530, 229789945, 229972643, 259038777, 288510753, 357596894, 441788515
\item $D_2\rightarrow D_1$:  229594479, 272705788, 288611833, 420177014, 441768471
\item $L_3\rightarrow L_1$: 394089883, 441806190
\item no solution$\rightarrow L_1$: 230394086, 232610209, 232632890, 233070708, 288301339
\item no solution$\rightarrow D_1$: 259006185
\item $L_3\rightarrow L_2$: 165500662, 199632809, 229500406, 229651225, 229799471, 229910746, 230007820, 233056681, 233719825, 259168350, 353990339, 377054541, 377251803, 389966320, 441803530
\item no solution$\rightarrow L_2$: 219111461, 230116482, 230387571, 230393824, 233657543, 237201620, 237287886, 237308045, 284800646, 288515928 
\item no solution$\rightarrow D_2$: 230002837, 272679385
\item no solution$\rightarrow L_3$: 198206417, 198383230, 198419050, 198457457, 207466859, 219091605, 219101330, 219738202, 219900027, 229713298, 229751503, 230070359, 230083136, 233008057, 233496995, 233532554, 233680757, 233742729, 256324189, 258776281, 288298154, 288407480, 334754418, 353894340, 357034252, 458479530
\end{description}

\subsection{Downranked Triple Candidates}
\label{subsect:downranked}

\begin{description}
\item $L_2\rightarrow L_3$: 159509734, 219809405, 258875507, 288734990, 377105433
\item $D_2\rightarrow L_3$: 219788156
\item $L_3\rightarrow$ no solution: 159510983, 160396455, 165527500, 198183185, 229451028, 229584508, 229791016, 258921745, 389962894
\end{description}

\subsection{Systems with Additional Period Variations}

Among the 168 EBs for which we were able to find third-body ETV solutions with greater or lesser confidence, there were 34 systems where we had to fit any other kinds of mathematical functions to describe additional period variations. In the vast majority of the investigated ETVs, this was an extra second-order polynomial term, that is, a quadratic component. Even if we exclude the 14 $L_3$-category, very uncertain solution systems, 16 quadratic ETV solutions were found in such triple systems where the presence of a third body appears to be quite certain. (On the other hand, we must note that even in the case of several shorter-period, $L_3$-categorized systems, we put them into that most uncertain subgroup just because of the presence of a substantial quadratic term, which was inevitably necessary to obtain any realistic solutions. Therefore, in these systems, while the LTTE solution and, therefore, the presence of a third body itself might be uncertain, the quadratic or cubic part of the ETVs looks quite certain.) As mentioned above, it is well known that a quadratic ETV mathematically reflects any eclipsing period variations that are constant in time, irrespective of their origin. Finding so many quadratic ETV variations in our present sample is in accord with the recent conclusions of \citet{borkovitsetal25}, who compared ETVs from the original \textit{Kepler}
 EBs with those reobserved by TESS and stated that `the longer the time window of precision ETV observations the larger the number of ETVs which exhibit longer term period variations.' Moreover, it is also well known that the large majority of those EBs which have been studied for decades or even more than a century frequently show large amplitude and parabolic period variations in their ETVs. In three cases (of which two belong to subgroup $L_1$), we found, however, that there are cubic variations. This fact, already by itself, indicates that these long-term period variations may be quite uneven or happen on not-as-long timescales, but the precisions of the former ETV curves, based mainly on heterogeneous and lower-accuracy ground-based observations, may be insufficient to detect such unevenness in the longer-timescale variations.

Another, perhaps more exciting, explanation for the quadratic and cubic terms might be (at least, in part) that they are sections of longer-timescale LTTE orbits, i.e., due to a fourth body. In this regard, it is very interesting that we found four such systems where a second LTTE solution (forming in such a way a 3+1-type hierarchical quadruple star) yields a better (sometimes much better) fit than an extra cubic polynomial term. We assumed 3+1 type quadruple solutions for three systems in our sample. Now, we omit the $L_3$ system TIC 288734990 where former, ground-based observations were also used, and the four-body solution looks quite uncertain. We consider only those three 3+1 quadruple candidates where at least the third-body solution looks highly certain. These are TIC 229751802 (an $L_1$ subgroup system) and TICs 198241524 and 229785001 from the $D_1$ systems. We should emphasize, however, that the origin of the longer-timescale period variation is not certain in any of these systems, as the deduced fourth-body period in each of the three cases is longer than the duration of the entire dataset. There are, however, some very tight 3+1 quadruple systems which are known with high confidence. For example, one of them is TIC~120362137, where the three orbital periods are known with certainty, and these are $P_1=3.28$\,d, $P_2=51.3$\,d, and $P_3=1046$\,d \citep{borkovitsetal26}, and TIC 114936199, which is the only known quadruple where fourth-body eclipses have ever been observed \citep{powelletal22}. Further follow up observations of these systems would be very important.

\subsection{Systems with Other Interesting Features}

Formerly, \citet{mitnyanetal24} reported eclipse depth variations (EDVs) for eight triple-star candidates. Amongst them, they reported the complete disappearance of the eclipses of TIC 236774836 by the beginning of Year 5 (that is, sector 56). Since then, as was mentioned above, there is another binary that is no longer eclipsing, namely TIC 420263614, where the eclipses disappeared during the gap of TESS observations between sectors 60 and 73. What is interesting, however, is the fact that this latter EB was not mentioned among the EDV systems in \citet{mitnyanetal24}. This might be a consequence of the large TESS pixel sizes and the varying amount of stray and contaminating light during all the TESS observations.  Therefore, sometimes it is difficult to decide whether the observed eclipse depth variations might be actual variations or only an instrumental effect. What is certain, however, is that after some not too marked fluctuations of the eclipse depths during sectors 14--60, there are no longer eclipses in the TESS light curves of TIC 420263614 during sectors 73--86. 

Considering the other EDV systems of \citet{mitnyanetal24}, similar to those previously reported, TICs 233684019, 233729038, 357686232, and 417701432 have shown continuously decreasing eclipse depths. In the case of TIC 233729038, this fact has led to the nearly complete disappearance of the eclipses by the end of sector 86. The shallowness of the most recent eclipses is the reason why the new ETV points are less accurate than the former ones, and one can infer this nicely from the larger error bars in the corresponding ETV plot (right panel in the fourth row of Figure~\ref{Fig:ETVs_D1a}). Regarding the case of TIC 417701432, this continuous decrease in eclipse depths led to the disappearance of the secondary eclipses during the newer sectors. This is the reason why only primary ETV points can be seen near the right end of the right panel in the fourth row of Figure~\ref{Fig:ETVs_D1b}. Conversely, continuously increasing eclipse depths were detected in TICs 259004910 and 288611133. We also call attention to TIC 441768471, which displayed increasing eclipse depths during the very first, Year 2, TESS observations then had practically constant depths during Years 4--5; and finally, during the most recent observations, it exhibits a slight decrease in the eclipse depths.

Finally, we note the case of TIC 230002837. In \citet{mitnyanetal24}, it was mentioned that this EB may also exhibit some EDVs, but no third-body solution was found. In this recent compilation, now we were able to find a new, though quite uncertain, LTTE + DE solution for this system. Moreover, we may also be able to confirm slight EDV during the new sector 73--86 observations.

{

\subsection{Long-Term Behavior of Some Formerly Known NCVZ EBs}

As was mentioned above, most of the currently investigated systems were previously unknown EBs. And, therefore, for most of them, there are no former observations, and, therefore, no minima times are available. Some of the rare exceptions are mentioned and discussed above. Here, finally we make a systematic comparison with the comprehensive work `Atlas of O-C Diagrams of Eclipsing Binary Stars' of \citet{kreineretal01}. This monumental collection contains plots of O-C diagrams for approximately 1140 such EBs which were discovered and observed more or less continuously over the last century (or, in the case of the oldest known systems, the first eclipse times can be traced back to the 19th century). A quick comparison of the online version of that catalog\endnote{\url{https://www.as.up.krakow.pl/o-c/}---last accessed on 25 March 2026.} (originally available in printed form as books) reveals that there are only two such EBs in our current NCVZ sample for which the several decade or even century-long O-C curves can be found in that book. These are RR Dra and BX Dra. For BX Dra, in the current work, we also give some historical times of minima, though the numbers of these times are substantially smaller than were used in \citet{kreineretal01}. The reason is that, in the case of the oldest times of minima, which in general were observed simply visually, involved subjective estimates of brightness differences. We therefore either cannot determine any realistic uncertainties or the originally given uncertainties are too large for us to use. Regarding the other system, RR Dra, we do not use the historical data in this work because, when we do so, we are unable to obtain any kind of LTTE solution.}

\begin{figure}[H]
\includegraphics[width=\textwidth]{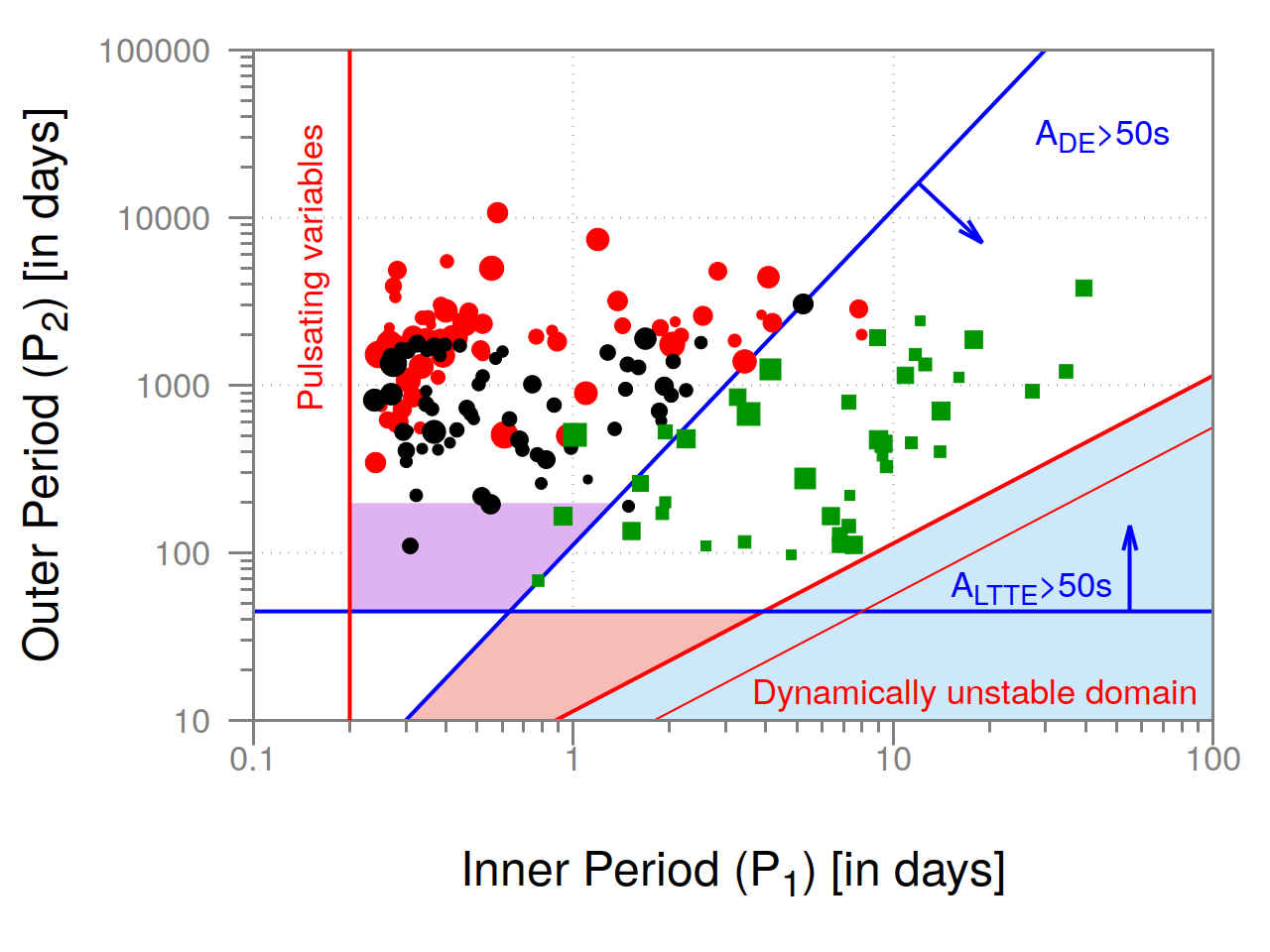}
\caption{The outer period ($P_2$) of the investigated triple-star candidates vs. the eclipsing periods ($P_1$) of the inner EBs. Black and red dots represent systems with pure LTTE solution for $L_{1,2}$ (black) and the very uncertain $L_3$ (red) group triple-star candidates, respectively. Moreover, green boxes stand for the LTTE + DE systems. The size of each symbol is related to the obtained outer eccentricity ($e_2$) of the given triple-star candidate. The horizontal and sloped blue lines represent the limits where, for a given set of $(P_1;P_2)$, the amplitude of the LTTE and the DE terms ($A_\mathrm{LTTE}$ and $A_\mathrm{DE}$) reach 50 s, which we consider to be an observational  detection limit. In calculating these amplitude limits, we assumed one solar mass for each component of a triple star and, furthermore, we used typical but arbitrary values of $e_2=0.35$, $i_2=60^\circ$, and $\omega_2=90^\circ$. The vertical red line near to the left border represents the theoretical lower period limit of H-rich EBs, while the thick and thin sloped red lines at the lower right corner stand for the dynamical stability limit, which was calculated using the usual empirical form of \citet{mardlingaarseth01} for two different outer eccentricities ($e_2=0.35$ and $e_2=0.1$ for the thick and thin lines, respectively). The meanings of the shaded areas are as follows. The light purple area represents the so-called W~UMa desert, which in turn indicates the lack of detectable short outer period triple stars around W UMa-type contact binaries. The light red area shows the region where a hypothetical, close third-body would be detectable exclusively for DE, but not for LTTE. No such close, tight system has been found. Finally, the cyan  area indicates a dynamically unstable region.} 
\label{P1vsP2}
\end{figure}

{

\section{Statistical Results}

Similar to the former work of \citet{mitnyanetal24}, we have carried out some statistical investigations of this extended sample of the triple-star candidate systems in the NCVZ. In Figure~\ref{P1vsP2}, we show the outer orbital period $P_2$ vs. the inner binary period $P_1$. The shape of the symbols and the color codes, again, represent the pure LTTE and the combined LTTE + DE solution systems. Additionally, the size of each point is related to the outer eccentricity ($e_2$).  As one can see, there is no strong correlation amongst the outer eccentricities and the outer orbital periods. There are low eccentricity third-body solutions (that is, smaller symbols) even for the largest third-body periods and, vice versa, there are large outer eccentricities even with smaller outer periods.}

\begin{figure}[H]
\includegraphics[width=0.5\textwidth]{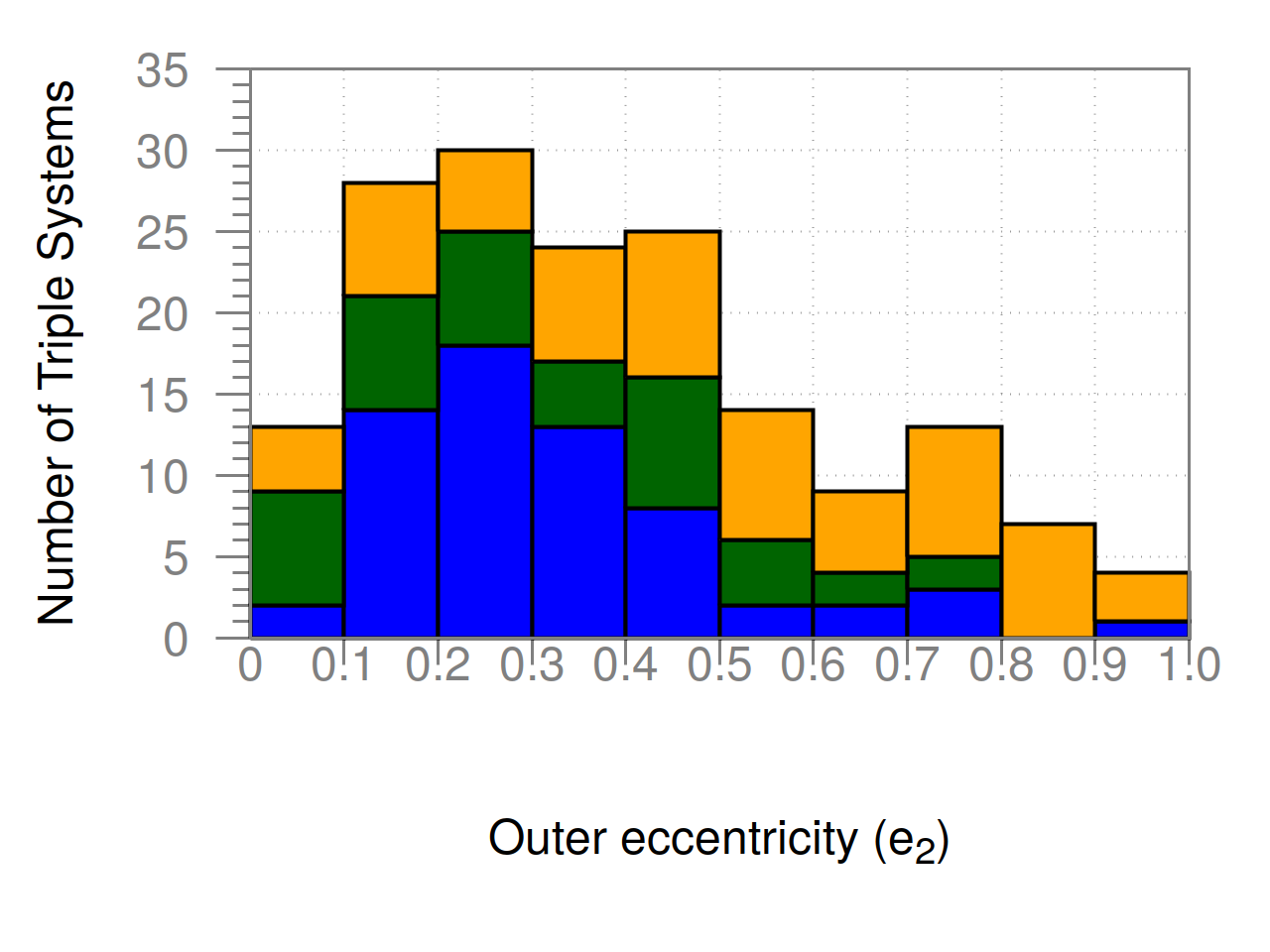}\includegraphics[width=0.5\textwidth]{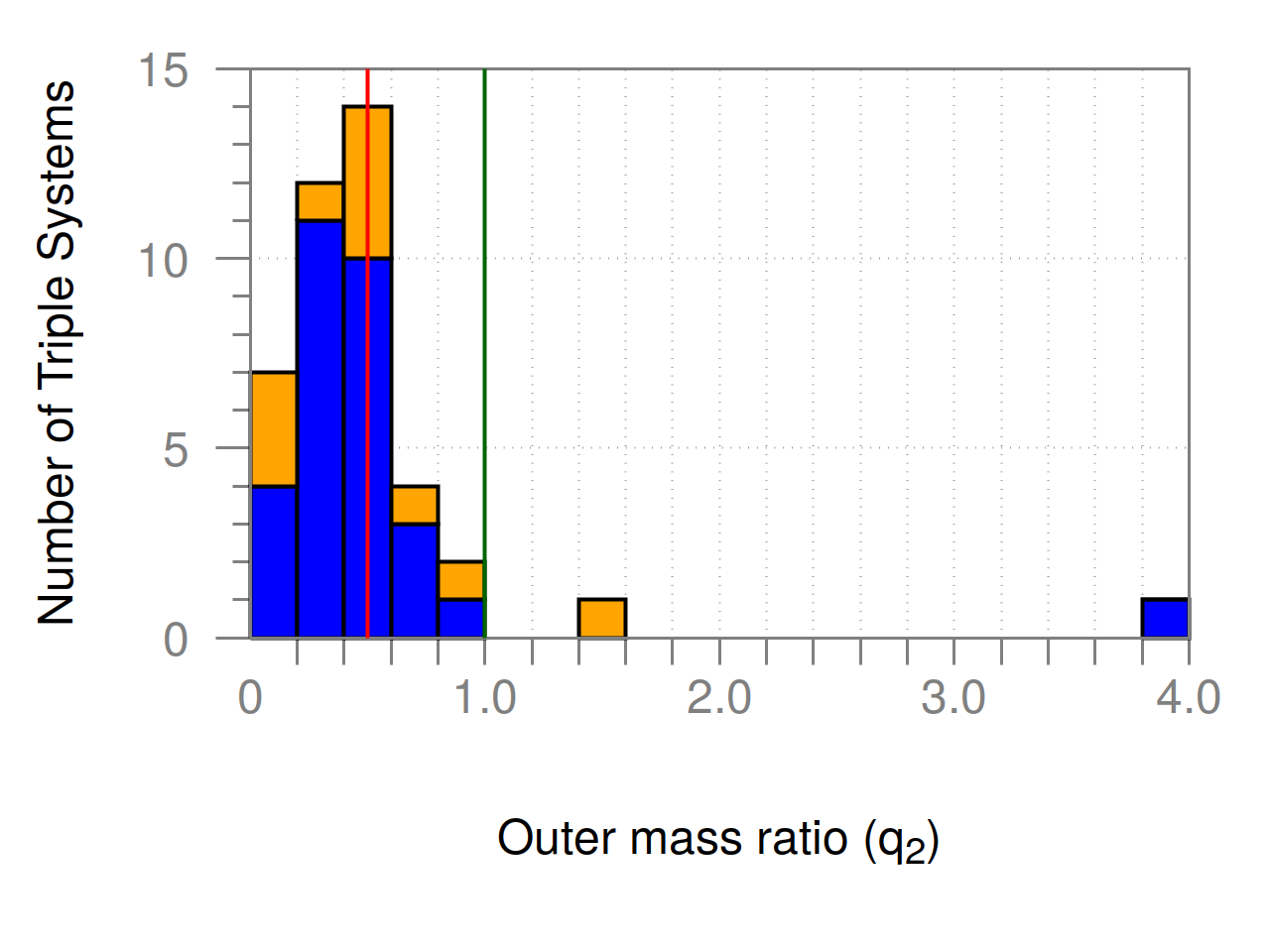}
\caption{(\textbf{\textit{Left} panel:}) 
 Distribution of the outer eccentricities ($e_2$). Blue color represents group $L_{1,2}$ pure LTTE systems. The $e_2$ distribution of the combined LTTE + DE triple-star candidates is denoted with green, while orange stands for the most uncertain (group $L_3$) pure LTTE systems.  (\textbf{\textit{Right} panel:}) The distribution of the outer mass ratio ($q_2$) amongst the LTTE + DE triple star candidates. Blue and orange stand for group D1 and D2 systems, respectively. The red vertical line is drawn at $q_2=0.5$ (this value would be the outer mass ratio in the case of three equally massive stars). The green vertical line represents the case of the so-called double twins, that is, when the mass of the outer companion is equal to the total mass of the inner EB ($q_2=1.0$). As one can see, in all but two systems, the total mass of the inner EB exceeds (in most cases, substantially) the mass of the third star. See further discussion in the text.} 
\label{fig:e2q2hist}
\end{figure}

{Finally, we also present two histogram plots in Figure~\ref{fig:e2q2hist}. As one can see in the left panel, the majority of the outer eccentricities ($e_2$) are located in the moderate outer eccentricity region of $0.1\lesssim e_2\lesssim0.5$. In some cases, however, the eccentricities reach extremely high values too. Most of the high outer eccentricity systems are, however, categorized into the most uncertain LTTE group ($L_3$), which strongly suggests that these very high eccentricities are only numerical artefacts and do not represent third bodies actually revolving on highly eccentric orbits.}

{In the right panel of Figure~\ref{fig:e2q2hist}, the outer mass ratio ($q_2$) distribution of the LTTE + DE triple-star candidates are plotted. (We have chosen to investigate only these systems because parameter $q_2$ can be realistically obtained only from a combined LTTE + DE solution.) In all but two cases, the outer mass ratios are smaller than unity. This finding, however, at least in part, may be the consequence of observational selection effect(s). That is so because for $q_2>1.0$, the distant third star is certainly more massive than both components of the inner EB (but it may happen already for $q_2>0.5$ that the third star turns out to be the most massive in the system) and, therefore, for $q_2$ values larger than unity, one may expect that the third component becomes substantially more luminous than the inner EB. This fact would result in very shallow eclipses due to the light dilution, which may lead to unnoticed eclipses or, at least, a very uncertain determination of any ETVs. In the current sample, regarding TIC~356014478, where our solution resulted in a very high outer mass ratio of $q_2=3.975$ (see Table~\ref{Tab:Orbelemdyn1}, and the corresponding ETV plot in the seventh panel of Figure~\ref{Fig:ETVs_D1b}), one may, for example, hypothesize that the third component is another binary formed by two massive and bright stars themselves.}

\vspace{9pt}

\section{Conclusions}
\label{sec:conclusion}

In this paper, we have reinvestigated the ETVs of the EBs located in or near to the NCVZ of the TESS spacecraft. Comparing our results to the former findings of \citet{mitnyanetal24}, we have expanded the number of hierarchical triple-star candidates in the NCVZ to 168 EBs from the previous 135. We discuss, however, that the difference amongst the triple candidates is not simply 33 systems.  Due to the longer observing window that is now available, we were able to detect 44 new candidates but, in parallel, we had to remove 9 EBs from the former candidates. Moreover, there are two additional tight triple systems where the inner binary no longer exhibits eclipses because of dramatic orbital inclination changes caused by a close, tight third companion on a non-aligned outer orbit, and, therefore, we were unable to determine any new ETV points.

From these 168 hierarchical triple-star candidates, we now consider 66 EBs to be the inner pairs of short-period hierarchical triple systems with full confidence. From this set of 66 secure triple-star candidates, we give pure LTTE third-body solutions for the ETV curves of 35 of them, while in the case of the remaining 31 EBs, the tightness of the triple made it necessary to calculate combined, LTTE + DE-type third-body solutions.

Regarding the remaining 102 systems, they can be considered to be hierarchical triple-star candidates with different levels of confidence.

Additionally, we found clear indicators of additional, longer-timescale period variations in 34 ETV curves (that is, in $\approx20\%$ of the entire sample). While most of these variations can be handled well at the present time with the inclusion of a quadratic or cubic polynomial term, in the case of three EBs, the LTTE caused by the presence of a fourth, not so distant companion might be a plausible explanation.

Finally, in addition to the above-mentioned disappearance of the eclipses in two former EBs, we report EDVs in eight additional EBs, which might be caused by an inclined tertiary.

We conclude that further follow-up studies of the investigated systems may inevitably be required to be certain that we have obtained the final story on the numbers of stars involved and their orbits. Note, TESS is scheduled to revisit the NCVZ, starting with sector 117 (that is, from May, 2027).

\vspace{6pt}
\authorcontributions{\textls[-15]{Conceptualization, T.B., T.M., and S.A.R.; methodology, T.B. and T.M.; software, T.B., T.M., and D.R.C.; formal analysis, T.B., T.M., and D.R.C.; investigation, T.B., T.M., and D.R.C.; resources, T.B.; writing original draft preparation, T.B. and S.A.R.; writing review and editing, T.B. and S.A.R.; funding acquisition, T.B. All authors have read and agreed to the published version of the manuscript.}}

\funding{This project has received funding from the HUN-REN Hungarian Research Network.
T.B., T.M. acknowledge the financial support of the Hungarian National Research, Development and Innovation Office---NKFIH Grant K-147131. Funding for the TESS mission is provided by the NASA Explorer Program. STScI is operated by the Association of Universities for Research in Astronomy, Inc., under NASA contract NAS5--26555. We used the Simbad service operated by the Centre des Donn\'ees Stellaires (Strasbourg, France).}

\dataavailability{The \textit{TESS} observations were provided by NASA and obtained from the Mikulski Archive for Space Telescopes (MAST) at \url{https://mast.stsci.edu/portal/Mashup/Clients/Mast/Portal.html}, accessed on 1st April 2026. The times of minima determined from these \textit{TESS} observations and used for the current study will be made available by the authors on request.}

\acknowledgments{This paper includes data collected by the \textit{TESS} mission and obtained from the MAST data archive at the Space Telescope Science Institute (STScI). This research has made use of NASA's Astrophysics Data System.}

\conflictsofinterest{The authors declare no conflicts of interest.} 


\begin{adjustwidth}{-5.0cm}{0cm}
\printendnotes[custom]
\reftitle{References}




\PublishersNote{}
\end{adjustwidth}
%


\end{document}